\documentclass{article}
\usepackage[hidelinks]{hyperref}
\usepackage{amsmath}
\usepackage{parskip} 
\usepackage{natbib}
\usepackage{mathrsfs}
\usepackage{pdfpages}
\usepackage{bm}
\usepackage{booktabs}
\usepackage{graphics}
\usepackage{multicol}
\usepackage[bf]{caption}
\usepackage{color}
\usepackage{amssymb}
\usepackage{multirow}
\usepackage{booktabs}
\usepackage{subcaption}

\usepackage{comment}
\usepackage{color}

\usepackage{amsthm}

\usepackage[margin=1in]{geometry}
\usepackage{floatrow}
\newfloatcommand{capbtabbox}{table}[][\FBwidth]
\usepackage{rotating}
\usepackage[titletoc]{appendix}
\usepackage[graphicx]{realboxes}
\usepackage{blindtext,titlefoot}

\title{
On anisotropy parameters and fluid detection in equivalent TI medium
}
\author{Filip P. Adamus\footnote{
Department of Earth Sciences, Memorial University of Newfoundland, Canada, {\tt adamusfp@gmail.com}}}

\date{}
\begin{document}
\maketitle
%%%%%%%%%%%%%%%%%%%%%%%%%%%%%%%%%%%%%%%%%%%%%%%%%%%%%%%%%%%%%%%%%%%%%%%%%%%%%%%%%%%%
\begin{abstract}
We consider a long-wave transversely isotropic (TI) medium equivalent to a series of finely parallel-layered isotropic layers, obtained using the \citet{Backus} average.
In such a TI equivalent medium, we verify the \citet{Berrymanetal} method of indicating fluids and the author's method \citep{Adamus}, using anisotropy parameter $\varphi$.
Both methods are based on detecting variations of the Lam\'e parameter, $\lambda$, in a series of thin isotropic layers, and we treat these variations as potential change of the fluid content.
To verify these methods, we use Monte Carlo (MC) simulations; for certain range of Lam\'e parameters $\lambda$ and $\mu$---relevant to particular type of rocks---we generate numerous combinations of these parameters in thin layers and, after the averaging process, we obtain their TI media counterparts. 
Subsequently, for each of the aforementioned media, we compute $\varphi$ and \citet{Thomsen} parameters $\epsilon$ and $\delta$.
We exhibit $\varphi$, $\epsilon$ and $\delta$ in a form of cross-plots and distributions that are relevant to chosen range of $\lambda$ and $\mu$.
We repeat that process for various ranges of Lam\'e parameters.
Additionally, to support the MC simulations, we consider several numerical examples of growing $\lambda$, by using scale factors.
As a result of the thorough analysis of the relations among $\varphi$, $\epsilon$ and $\delta$, we find eleven fluid detectors that compose a new fluid detection method.
Based on these detectors, we show the quantified pattern of indicating change of the fluid content.
Moreover, we show a comprehensive table consisting of aforementioned eleven fluid detectors along with the exact dependence of occurrence of these indicators on particular variations of $\mu$ and $\lambda$.
Finally, we exhibit a table containing the expected ranges and dominants of sets of $\varphi$, $\epsilon$ and $\delta$ that correspond to various ranges of Lam\'e parameters.
\end{abstract}
%%%%%%%%%%%%%%%%%%%%%%%%%%%%%%%%%%%%%%%%%%
%%%%%%%%%%%%%%%%%%%%%%%%%%%%%%%%%%%%%%%%%%
\section{Introduction}
%%%%%%%%%%%%%%%%%%%%%%%%%%%%%%%%%%%%%%%%%%
%%%%%%%%%%%%%%%%%%%%%%%%%%%%%%%%%%%%%%%%%%
The problem of fine, parallel layering and its long-wave equivalent medium approximation has been treated by a number of authors.
Among many of them there are: \citet{Postma}, \citet{Backus}, \citet{HelbigandSchoenberg}, \citet{SchoenbergandMuir}, \citet{Berrymanetal}, and \citet{Bosetal}. 
One of the first authors who stated and derived that the layered medium may be viewed as transversely isotropic (TI), was \citet{Postma}.
Nevertheless, he considered only the case of periodical structure of parallel isotropic layers.
Seven years later, fundamental work of~\citet{Backus} provided us an elegant formula of the TI medium, long-wave equivalent to isotropic or TI layers of different thicknesses, with no assumption of periodicity.
Aforementioned formula was extended to lower symmetry classes---including generally anisotropic case---by~\citet{Bosetal}.
However, in this paper, we do not consider lower symmetry classes than the TI one, only~\citeauthor{Backus} formula for isotropic layers is used.

An equivalent medium has wide application in exploration geophysics, especially in well-logging.
The frequency obtained from the sonic logs in the borehole is much higher as compared to the seismic frequency.
The \citeauthor{Backus} average allows to adjust both frequencies, which enables to establish realistic relationship between reservoir properties and seismic properties.
Among other applications, an equivalent medium may be also used in a fluid detection in layered Earth.
\citet{Berrymanetal} showed that the \citet{Thomsen} anisotropy parameters obtained for an equivalent TI medium may indicate the change of the fluid content.
Specifically, the authors state that it may be exhibited by the small positive value of $\epsilon$ and $\delta$.
Another fluid detection method is shown by the author \citep{Adamus}.
Therein a new anisotropy parameter, $\varphi$, is introduced of which certain ranges, along with its relation with $\epsilon$, may mark the change of the fluid content.
A simple pattern of fluid detection by excluding $\delta$ is shown \citep{Adamus}.
Both methods are based on the fact that, as stated by \citet{Gassmann}, the change of fluid content influences only the elasticity parameter $\lambda$, not $\mu$.
Thus, they treat the variations of $\lambda$ in layers as potential fluid variations.
These methods are valid only in the case of thin isotropic layers since they rely on Lam\'e parameters $\lambda$ and $\mu$; they are not valid for thin layers exhibiting lower symmetry. 

In this paper, we pursue the previous work of the author, investigating both methods of indicating fluids.
To do so, we perform Monte Carlo (MC) simulations for 21 different ranges of Lam\'e parameters, and also, in Appendices~\ref{sec:muvar_scale} and~\ref{sec:muconst_scale}, we consider six numerical examples using scale factors.
The MC method relies on repeated random sampling to obtain numerical results.
Specifically, for a series of thin layers, a random set of $\lambda$ and $\mu$ is chosen from the given Lam\'e parameters range.
Such a random simulation is repeated $s$ times and, after \citeauthor{Backus} averaging, we obtain $s$ different TI media.
For each medium, anisotropy parameters $\varphi$, $\epsilon$ and $\delta$ are computed.
Subsequently, we analyze cross-plots of $\varphi$ versus $\epsilon$, and $\delta$ versus $\epsilon$.
Comparing the cross-plots for different ranges of $\lambda$ and $\mu$ helps to verify the usefulness of both fluid detection methods.
Also, it permits to discuss the properties of the anisotropy parameters in equivalent media and to indicate relations among them. 
Some of them, perhaps surprisingly, occur to be useful in fluid detection in layered Earth.
Finally, the exact analysis of distributions of $\varphi$, $\epsilon$ and $\delta$, in Appendix~\ref{sec:dis}, illustrate these relations.
%%%%%%%%%%%%%%%%%%%%%%%%%%%%%%%%%%%%%%%%%%%%%%%%%%%%%%%%%%%%%%%%%%%%%%%%%%%%%%%%%%%%
\section{Background: linear elasticity}
%%%%%%%%%%%%%%%%%%%%%%%%%%%%%%%%%%%%%%%%%%%%%%%%%%%%%%%%%%%%%%%%%%%%%%%%%%%%%%%%%%%%
\subsection{Basic theory}
%%%%%%%%%%%%%%%%%%%%%%%%%%%%%%%%%%%%%%%%%%%%%%%%%%%%%%%%%%%%%%%%%%%%%%%%%%%%%%%%%%%%
In the theory of linear elasticity the forces applied to a single point are expressed in terms of a stress tensor and their resultant deformations in terms of a strain tensor.
The definition of the strain tensor for infinitesimal displacements in three dimensions is
\begin{equation}\label{strain}
\varepsilon_{ij} :=\frac{1}{2}\left(\frac{\partial u_i}{\partial x_j}+\frac{\partial u_j}{\partial x_i}\right)\,
\qquad i, j =1,2,3\,,
\end{equation} 
where, throughout this paper, subscripts $i$ and $j$, denote Cartesian coordinates, and $u_i$ are the components of the displacement vector describing the deformations in the $i$-th direction.
The constitutive equation relating stresses and strains is Hooke's law, namely,
\begin{equation}\label{hook}
\sigma_{ij}=\sum_{k=1}^{3}\sum_{\ell=1}^{3}c_{ijk\ell}\varepsilon_{k\ell}\,
\qquad i, j =1,2,3\,,
\end{equation}
which states that the applied load at a point is linearly related to the deformation by elasticity tensor, $c_{ijk\ell}$\,.
Equation~(\ref{hook}) is the fundamental equation of linear elasticity, solids that obey this equation are called Hookean solids.
For an isotropic medium, Hooke's law may be rewritten conveniently in a matrix notation as
 \begin{equation*}
\left[
\begin{array}{c}
\sigma_{11}\\
\sigma_{22}\\
\sigma_{33}\\
\sigma_{23}\\
\sigma_{13}\\
\sigma_{12}\\
\end{array}
\right]
=
\left[
\begin{array}{cccccc}
C_{11}&C_{11}-2C_{44} & C_{11}-2C_{44} & 0 & 0 & 0 \\
C_{11}-2C_{44}&C_{11} &C_{11}-2C_{44} & 0 & 0 & 0 \\
C_{11}-2C_{44}& C_{11}-2C_{44}& C_{11} & 0 & 0 & 0 \\
0 & 0 & 0 & C_{44} & 0 & 0 \\
0 & 0 & 0 & 0 & C_{44} & 0\\
0 & 0 & 0 & 0 & 0 & C_{44}
\end{array}
\right]
\left[
\begin{array}{c}
\varepsilon_{11}\\
\varepsilon_{22}\\
\varepsilon_{33}\\
2\varepsilon_{23}\\
2\varepsilon_{13}\\
2\varepsilon_{12}\\
\end{array}
\right]
\,,
\end{equation*}
which also may be expressed in terms of Lam\'e parameters
\begin{equation*}
\begin{cases}
\lambda := C_{11}-2C_{44}\\
\mu := C_{44}
\end{cases}\,.
\end{equation*}
In this paper, for simplicity, Voigt's notation---as opposed to Kelvin's notation---is used.
%%%%%%%%%%%%%%%%%%%%%%%%%%%%%%%%%%%%%%%%%%
\subsection{Backus average}
%%%%%%%%%%%%%%%%%%%%%%%%%%%%%%%%%%%%%%%%%%
As shown by~\citet{Backus}, a medium composed of parallel isotropic layers, whose individual thicknesses are much smaller than the wavelength, respond---to the wave propagation---as a single, homogeneous, transversely isotropic medium. 
The elasticity parameters of such a medium are 
\begin{equation}   \label{bac1}
    C_{11}^{\overline{\rm TI}}=\overline{\left(\frac{\lambda}{\lambda+2\mu}\right)}^2\,\overline{\left(\frac{1}{\lambda+2\mu}\right)}^{-1}+\overline{\left(\frac{4(\lambda+\mu)\mu}{\lambda+2\mu}\right)}\,,
\end{equation}
\begin{equation}   \label{bac2}
    C_{12}^{\overline{\rm TI}}=\overline{\left(\frac{\lambda}{\lambda+2\mu}\right)}^2\,\overline{\left(\frac{1}{\lambda+2\mu}\right)}^{-1}+\overline{\left(\frac{2\lambda\mu}{\lambda+2\mu}\right)}\,,
\end{equation}
\begin{equation}  \label{bac3}
    C_{13}^{\overline{\rm TI}}=\overline{\left(\frac{\lambda}{\lambda+2\mu}\right)}\,\overline{\left(\frac{1}{\lambda+2\mu}\right)}^{-1}\,,
\end{equation}
\begin{equation}  \label{bac4}
     C_{33}^{\overline{\rm TI}}=\overline{\left(\frac{1}{\lambda+2\mu}\right)}^{-1}\,,
 \end{equation}
 \begin{equation}  \label{bac5}
     C_{44}^{\overline{\rm TI}}=\overline{\left(\frac{1}{\mu}\right)}^{-1}\,,
 \end{equation}
  \begin{equation}  \label{bac6}
   C_{66}^{\overline{\rm TI}}=\overline{\mu}\,,
 \end{equation}
where $\lambda$ and $\mu$ are the Lam\'e parameters for each layer and the overbar denotes the weighted average.
The average is weighted by the layer thickness; herein, since all layers have the same thickness, we use an arithmetic average.
A TI medium, whose rotation symmetry axis is parallel to the $x_3$-axis, is~\citep[see e.g.,][p.\,134]{SlawinskiGreen}
\begin{equation*}
C^{\overline{\rm TI}}=
\left[
\begin{array}{cccccc}
C^{\overline{\rm TI}}_{11} & C^{\overline{\rm TI}}_{12} & C^{\overline{\rm TI}}_{13} & 0 & 0 & 0\\
C^{\overline{\rm TI}}_{12} & C^{\overline{\rm TI}}_{11} & C^{\overline{\rm TI}}_{13} & 0 & 0 & 0\\
C^{\overline{\rm TI}}_{13} & C^{\overline{\rm TI}}_{13} & C^{\overline{\rm TI}}_{33} & 0 & 0 & 0\\
0 & 0 & 0 & C^{\overline{\rm TI}}_{44} & 0 & 0\\
0 & 0 & 0 & 0 & C^{\overline{\rm TI}}_{44} & 0\\
0 & 0 & 0 & 0 & 0 & C^{\overline{\rm TI}}_{66}
\end{array}
\right]
\,,
\end{equation*}
where $C_{12}^{\overline{\rm TI}}=C_{11}^{\overline{\rm TI}}-2C_{66}^{\overline{\rm TI}}$. 
Consequently, expressions~(\ref{bac1})--(\ref{bac6}) consist of five independent parameters.
%%%%%%%%%%%%%%%%%%%%%%%%%%%%%%%%%%%%%%%%
\subsection{Anisotropy parameters}
%%%%%%%%%%%%%%%%%%%%%%%%%%%%%%%%%%%%%%%%
To examine the strength of anisotropy of a transversely isotropic elasticity tensor, we use the~\citet{Thomsen} parameters,
 \begin{equation}\label{aniso1}
     \gamma := \frac{C_{66}^{\overline{\rm TI}}-C_{44}^{\overline{\rm TI}}}{2C_{44}^{\overline{\rm TI}}}\,,
 \end{equation}
\begin{equation}\label{aniso2}
     \delta := \frac{\left(C_{13}^{\overline{\rm TI}}+C_{44}^{\overline{\rm TI}}\right)^2-\left(C_{33}^{\overline{\rm TI}}-C_{44}^{\overline{\rm TI}}\right)^2}{2C_{33}^{\overline{\rm TI}}\left(C_{33}^{\overline{\rm TI}}-C_{44}^{\overline{\rm TI}}\right)}\,,
 \end{equation}
\begin{equation}\label{aniso3}
     \epsilon := \frac{C_{11}^{\overline{\rm TI}}-C_{33}^{\overline{\rm TI}}}{2C_{33}^{\overline{\rm TI}}}\,.
 \end{equation}
In addition, we use a fourth anisotropy parameter,
 \begin{equation}\label{aniso4}
 \varphi := \frac{C_{12}^{\overline{\rm TI}}-C_{13}^{\overline{\rm TI}}}{2C_{12}^{\overline{\rm TI}}}\,,
 \end{equation}
 which, similarly to expressions~(\ref{aniso1})--(\ref{aniso3}), is equal to zero in the case of isotropy of an equivalent medium, and---as opposed to the~\citeauthor{Thomsen} parameters---is equal to zero also in the case of constant $\lambda$ in layers~\citep{Adamus}.
 As shown by~\citet{Adamusetal}---for the \citeauthor{Backus} average---growing anisotropy of an equivalent medium implies the growth of inhomogeneity among layers. 
 %%%%%%%%%%%%%%%%%%%%%%%%%%%%%%%%%%%%%%%%
 \subsection{Stability conditions}
%%%%%%%%%%%%%%%%%%%%%%%%%%%%%%%%%%%%%%%%
The allowable relations among the elasticity parameters are stated by the stability conditions that express the fact that it is necessary to expend energy to deform a material~\citep[e.g.][Section 4.3]{SlawinskiRed}. 
These conditions mean that every elasticity tensor must be positive-definite, wherein a tensor is positive-definite if and only if all its eigenvalues are positive.
For any isotropic elasticity tensor, the inequalities
\begin{equation*}
\label{ineq}
    C_{11}\geq\tfrac{4}{3}\,C_{44}\geq0\,\,,
\end{equation*}
or, in a different notation, using Lam\'e parameters,
\begin{equation}\label{cond}
    \lambda\geq-\tfrac{2}{3}\,\mu  
\qquad    
\mathrm{and}
\qquad
\mu\geq0
\end{equation}
ensure that all eigenvalues are positive, thus the stability conditions are satisfied.
To satisfy the stability conditions, any transversely isotropic elasticity tensor must obey the inequalities
\begin{equation*}\label{eq:stab}
C_{66}\geq0\,,
\qquad
C_{44}\geq0\,,
\qquad
C_{33}\geq0\,,
\qquad
C_{12}+C_{66}\geq0\,,
\end{equation*}
\begin{equation*}
\left(C_{12}+C_{66}\right)C_{33}\geq\left(C_{13}\right)^2\,.
\end{equation*}
%%%%%%%%%%%%%%%%%%%%%%%%%%%%%%%%%%%%%%%%%%%%%%%%%%%%%%%%%%%%%%%%%%%%%%%%%%%%%%%%%%%%%%%%
\section{Fluid detection in equivalent TI media\,: mafic rocks} \label{sec:three}
%%%%%%%%%%%%%%%%%%%%%%%%%%%%%%%%%%%%%%%%%%%%%%%%%%%%%%%%%%%%%%%%%%%%%%%%%%%%%%%%%%%%%%%%
In this section, we examine the relations among parameters, $\varphi$, $\epsilon$ and $\delta$, in equivalent medium.
We are particularly focused on possible methods of fluid detection in equivalent media using these anisotropy parameters.
During our analysis, we verify the method shown in~\citet{Berrymanetal}, by checking if small positive values of $\epsilon$, together with small positive values of $\delta$, correspond to large variations of $\lambda$ in layers.
Also, we pay attention to another method, shown in~\citet{Adamus}, which combines the information from parameters $\varphi$, $\epsilon$ and $\gamma$, according to the pattern from Table~\ref{tab:pattern}.
\renewcommand{\arraystretch}{1.4}
\begin{table}[h]
\centering
\begin{tabular}{cccc} 
\toprule
\multicolumn{2}{c}{$\gamma\approx0$} & \multicolumn{2}{c}{\hphantom{xxxx}$\gamma>0$}\\
\midrule
$\lambda\approx\rm{const}$ \,\,\,&\,\,\, $\lambda\neq\rm{const}$\,\qquad &\,\qquad $\lambda\approx\rm{const}$ \,\,\,&\,\,\, $\lambda\neq\rm{const}$\\
$|\varphi|<|\epsilon|$ \,\,\,&\,\,\, $|\varphi|>|\epsilon|$\,\qquad &\,\qquad $\varphi\approx0$  \,\,\,&\,\,\, $\varphi\neq0$\\ 
\bottomrule
\end{tabular}
\caption{\small{Pattern for an alternative way of detecting fluids by excluding $\delta$}}
\label{tab:pattern}
\end{table}
We attempt to verify and quantify that pattern.
At the same time, we look for the other indicators that characterize the change of fluid content in layered Earth, {\emph{ipso facto}}, we improve the prior methods.

We divide this section into three parts; in Section~\ref{sec:general}, we focus on general case of variations of Lam\'e parameters in layers.
In Section~\ref{sec:var}, we analyze the case of non near-constant shear modulus, $\mu$ (in Table~\ref{tab:pattern} denoted as $\gamma>0$). 
In Section~\ref{sec:mu}, we consider the case of near-constant rigidity (in Table~\ref{tab:pattern} denoted as $\gamma\approx0$), which is equivalent to nearly constant $\mu$ in layers, or $\gamma$ close to zero.
As a quantitative tool of variations of $\lambda$ and $\mu$, we use their relative standard deviations, namely,
\begin{equation*}
RSD_{\lambda}=\overline{\lambda}\,^{-1}\left(\dfrac{1}{n}\sum_{i=1}^n\left(\overline{\lambda}-\lambda_i\right)^2\right)^{\frac{1}{2}}\times100\%
\qquad
\mathrm{and}
\qquad
RSD_{\mu}=\overline{\mu}\,^{-1}\left(\dfrac{1}{n}\sum_{i=1}^n\left(\overline{\mu}-\mu_i\right)^2\right)^{\frac{1}{2}}\times100\%\,\,,
\end{equation*}
where $n$ is the number of layers, $\overline{\lambda}$ and $\overline{\mu}$ are the arithmetic mean values of $\lambda$ and $\mu$ in layers, and $\lambda_i$, $\mu_i$, are the values for each layer.

Throughout the paper, by the notion of  ``variations of $\mu$'', we understand all possible variations of $\mu$, denoted as $RSD_{\mu}\in\forall$.
``Non near-constant rigidity" or ``non near-constant $\mu$" means $RSD_{\mu}>2\,\%$, whereas $RSD_{\mu}<2\,\%$ refers to ``near-constant rigidity" or ``near-constant $\mu$".
Similarly, ``variations of $\lambda$''  mean $RSD_{\lambda}\in\forall$, ``non near-constant $\lambda$" refers to $RSD_{\lambda}>2\,\%\,$ ($\lambda\neq0$ in Table~\ref{tab:pattern}), $RSD_{\lambda}<2\,\%$ refers to ``near-constant $\lambda$" ($\lambda\approx0$ in Table~\ref{tab:pattern}). 
Additionally, notion ``moderate variations of $\lambda$" or ``moderately varying $\lambda$" refers to $RSD_{\lambda}\in(2\,\%\,,\,20\,\%\,)$, whereas ``strong variations of $\lambda$" or ``strongly varying $\lambda$" refers to $RSD_{\lambda}>20\,\%\,$.
To examine fluid detection methods and obtain distributions of the anisotropy parameters, we use the values of elasticity parameters relevant to various types of rocks in Earth's crust and upper mantle.
These values are based on the works of~\citet{JiEtAl}, \citet{CastagnaSmith} and \citet{WanniarachichiEtAl}.
To make the composition of the article clear, in this section, we only consider the ranges of elasticity parameters relevant to mafic rocks (gabbro, diabase, mafic granulite, and mafic gneiss).
The ranges corresponding to the other types of rocks, namely, felsic rocks and sandstones, are considered in Appendix~\ref{ap:one}.
As shown in Appendix~\ref{ap:one}, the analogical statements and similar relations between the anisotropy parameters, to ones shown in this section, are valid also for felsic rocks and sandstones. 
It is important to notice that the ranges corresponding to mafic rocks, felsic rocks and sandstones, are also relevant to some other types of rocks; certain limestones, basalts, shales, and many others \citep{JiEtAl}.
%%%%%%%%%%%%%%%%%%%%%%%%%%%%%%%%%%%%%%%%%%%%%%%%%%%%%%%%%%%%%%%%%%%%%%%%%%%%%%%%%%%%%%%%
\subsection{General case of variations} \label{sec:general}
%%%%%%%%%%%%%%%%%%%%%%%%%%%%%%%%%%%%%%%%%%%%%%%%%%%%%%%%%%%%%%%%%%%%%%%%%%%%%%%%%%%%%%%%
Let us perform Monte Carlo simulations to obtain examples of equivalent TI media and to compute their respective $\varphi$, $\epsilon$ and $\delta$.
To do so, we impose certain restrictions, listed in Table~\ref{tab:restr}.
\begin{table}[!htbp]
\begin{tabular}{cccc}
\toprule
range of $\lambda$ \scriptsize{[\,GPa\,]}       &       $<a\,,\,b>$     &     range of $RSD_{\mu}$ \scriptsize{[\,\%\,]}     &      $<m_1\,,\,m_2>$        \\
range of $\mu$ \scriptsize{[\,GPa\,]}              &       $<c\,,\,d>$     &     range of $RSD_{\lambda}$ \scriptsize{[\,\%\,]}     &      $<l_1\,,\,l_2>$        \\          
number of layers                   &         $n$             &           number of examples         &             $s$  \\
\bottomrule
\end{tabular}
\caption{\small{The restrictions imposed on Monte Carlo method.}}
\label{tab:restr}
\end{table}
Based on the work of \citeauthor{JiEtAl}, and following Table~\ref{tab:restr}, we set possible range of the Lam\'e parameters for mafic rocks, namely, $a=40$, $b=70$, $c=35$, and $d=60$.
We receive randomly sampled $s=10000$ examples of TI media, equivalent to $n=5$ isotropic layers of mafic rocks having the Lam\'e parameters within aforementioned ranges.
As stated by \citeauthor{Backus}, minimum of three layers is required to perform the Backus average correctly.
Also, the further increase of, $s$, influences the results only slightly.
Herein, we consider the general case of variations of the Lam\'e parameters in layers, namely, $RSD_{\mu}\in\forall$ and $RSD_{\lambda}\in\forall$.
In other words, the simulations are not additionally limited by the restricted range of relative standard deviations, as it is the case in Sections~\ref{sec:var} and~\ref{sec:mu}.

\begin{figure}[!htbp]
\centering
\begin{subfigure}{.4\textwidth}
  \centering
   \includegraphics[scale=0.35]{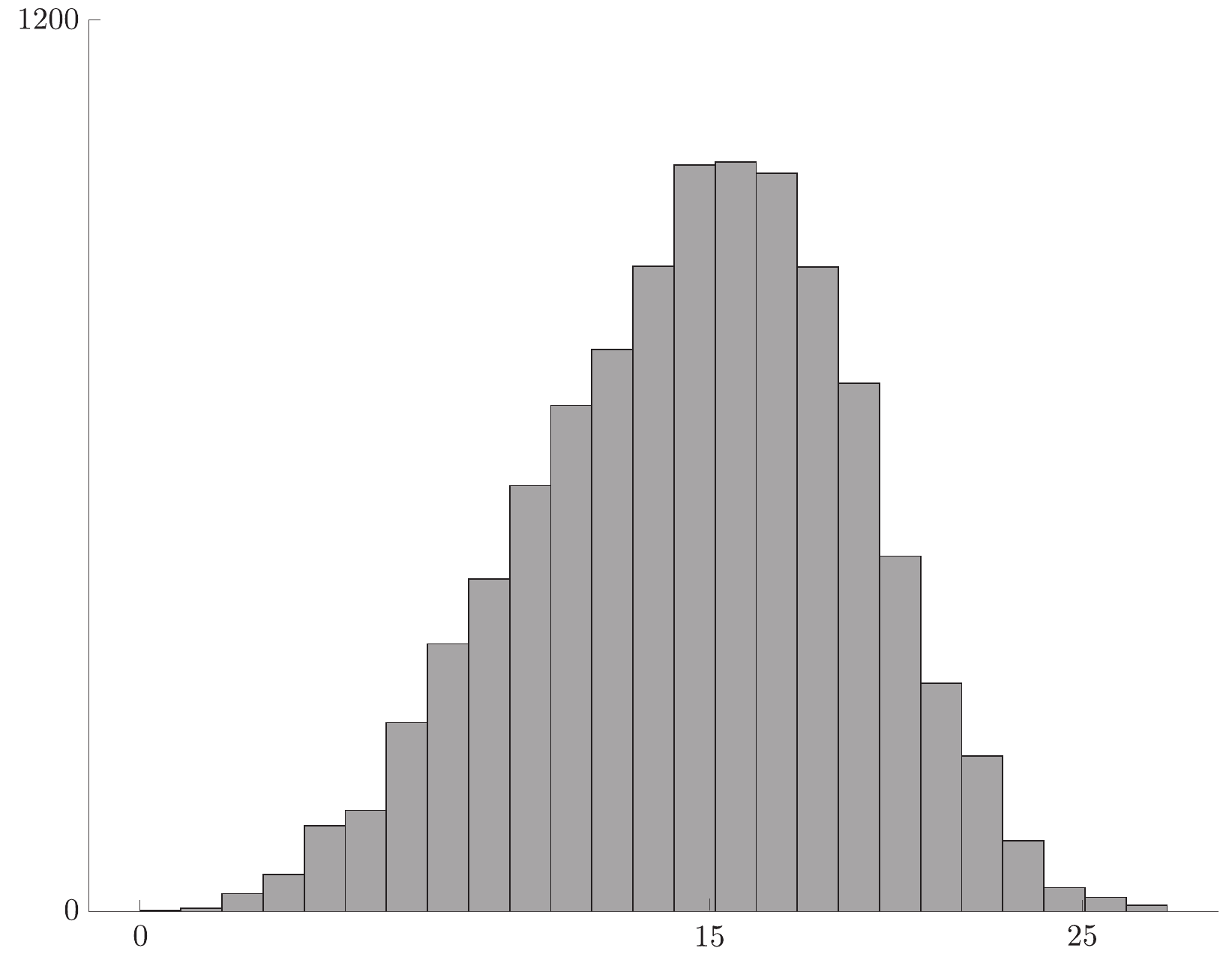}
  \caption{\footnotesize{Variations of $\mu$ $(RSD_{\mu}\,[\%])$}}
  \label{fig:rsdm}
\end{subfigure}%
\begin{subfigure}{.4\textwidth}
  \centering
   \includegraphics[scale=0.35]{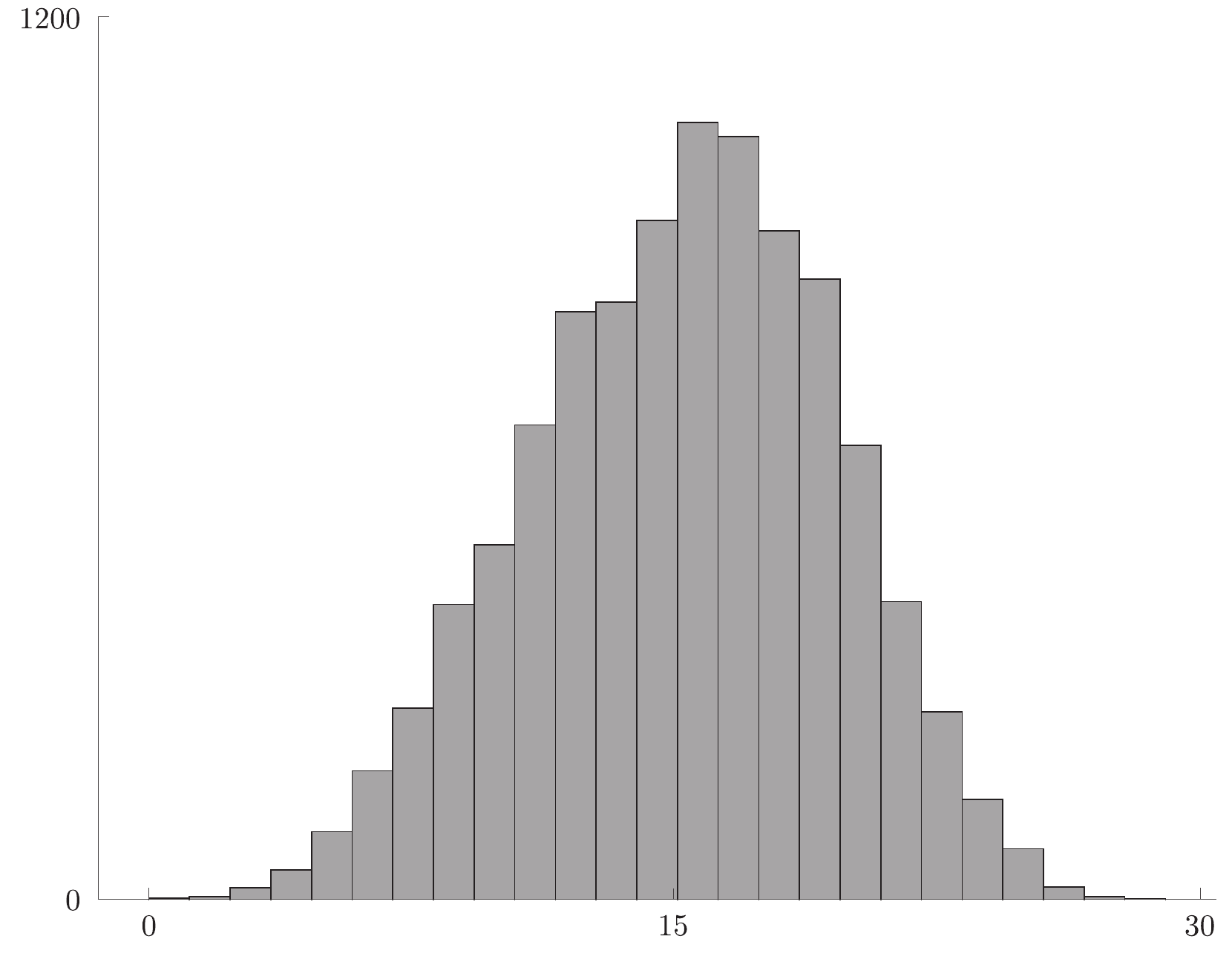}
  \caption{\footnotesize{Variations of $\lambda$ $(RSD_{\lambda}\,[\%])$}}
  \label{fig:rsdl}
\end{subfigure}%
\caption{\small{Intensity distributions of random variations of Lam\'e parameters corresponding to layered mafic rocks. Horizontal axes show percentage of $RSD_{\mu\,,\,\lambda}$\,, vertical axes show number of examples of TI media.}}
\label{fig:rsd}
\end{figure}

In Figure~\ref{fig:rsd}, we show the intensity distributions of variations of Lam\'e parameters in layers. 
%
%To estimate the most frequent value of relative standard deviations, we choose the highest bin from their distributions, subsequently, we compute the mean value of the set of $RSD_{\mu\,,\,\lambda}$ contained in that bin.
%We are aware of numerous ways of defining the dominant value in a distribution, and of inaccuracy of our method.
%Dividing the distribution in more bins might seem to give more accurate dominant, however, due to growing inaccuracy of simulations, we abandon this idea.
%Also, smaller number of bins might give too general value, thus, we believe that choice of 25 bins is a golden mean.
The most frequent variations of $\mu$ are around $15.82\,\%$, whereas $RSD_{\lambda}\approx15.66\,\%\,$; thus, their dominants have very similar values.
Analyzing both distributions in Figures~\ref{fig:rsdm} and~\ref{fig:rsdl}, we notice that their shapes and ranges are also very similar.
Using our nomenclature, we see that in most of MC examples, $\mu$ and $\lambda$ vary moderately in layers $(RSD_{\mu\,,\,\lambda}\in(2\,\%\,,\,20\,\%)\,)$.
Nevertheless, near-constant or strongly varying $\mu$, appears in $0.03\,\%$ and $9.75\,\%$ of cases, respectively.
Similarly, near-constant or strongly varying $\lambda$ occurs in $0.04\,\%$ and $13.3\,\%$ of cases, respectively.
It is important to remember that these are results obtained for random sampling and the situation of near-constant or strongly varying Lam\'e parameters in mafic rocks, might appear more often in the real data.

\begin{figure}[!htbp]
\centering
\begin{subfigure}{.4\textwidth}
  \centering
   \includegraphics[scale=0.35]{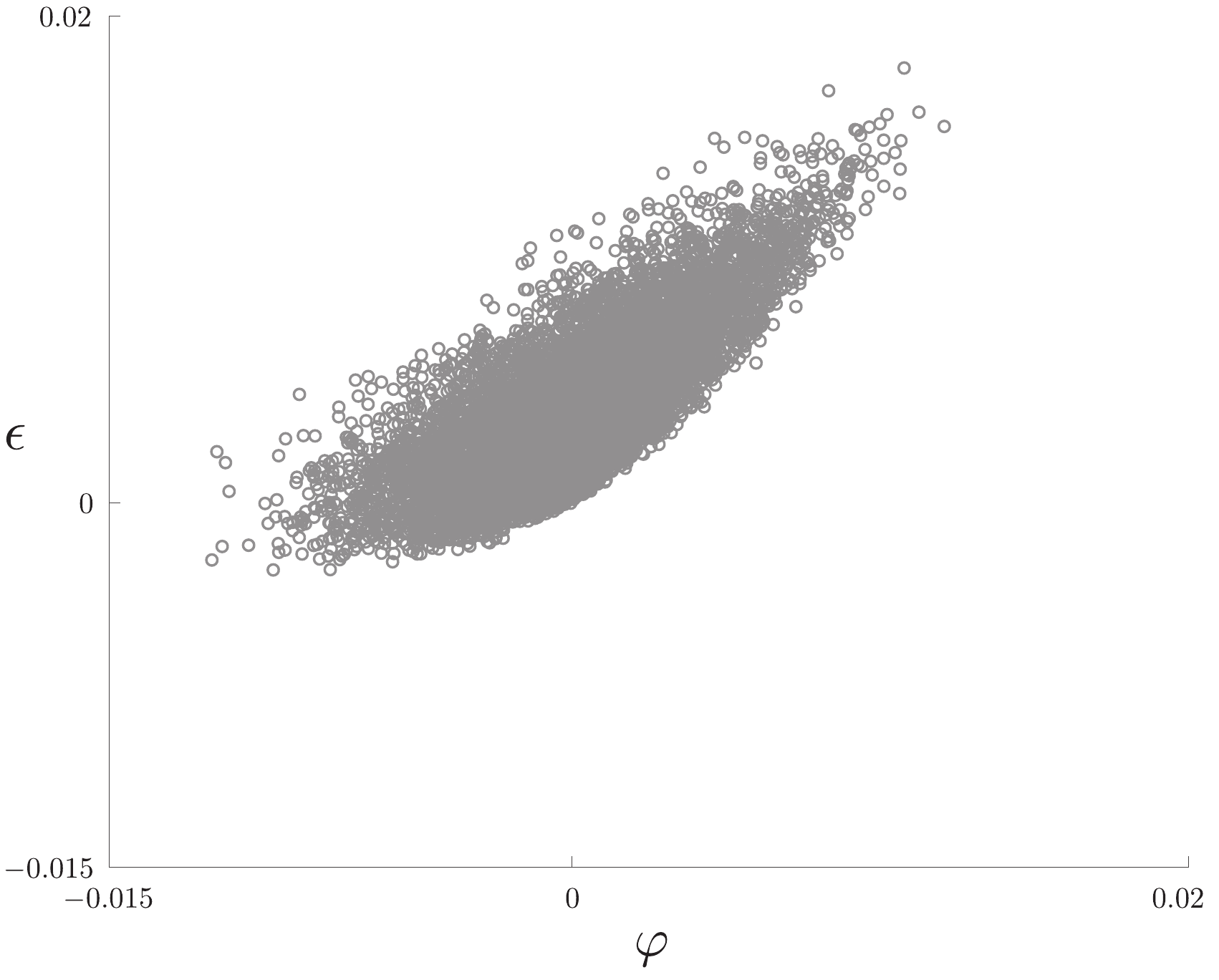}
  \caption{\footnotesize{$\varphi$ versus $\epsilon$}}
  \label{fig:rel_ml_phieps}
\end{subfigure}%
\begin{subfigure}{.4\textwidth}
  \centering
   \includegraphics[scale=0.35]{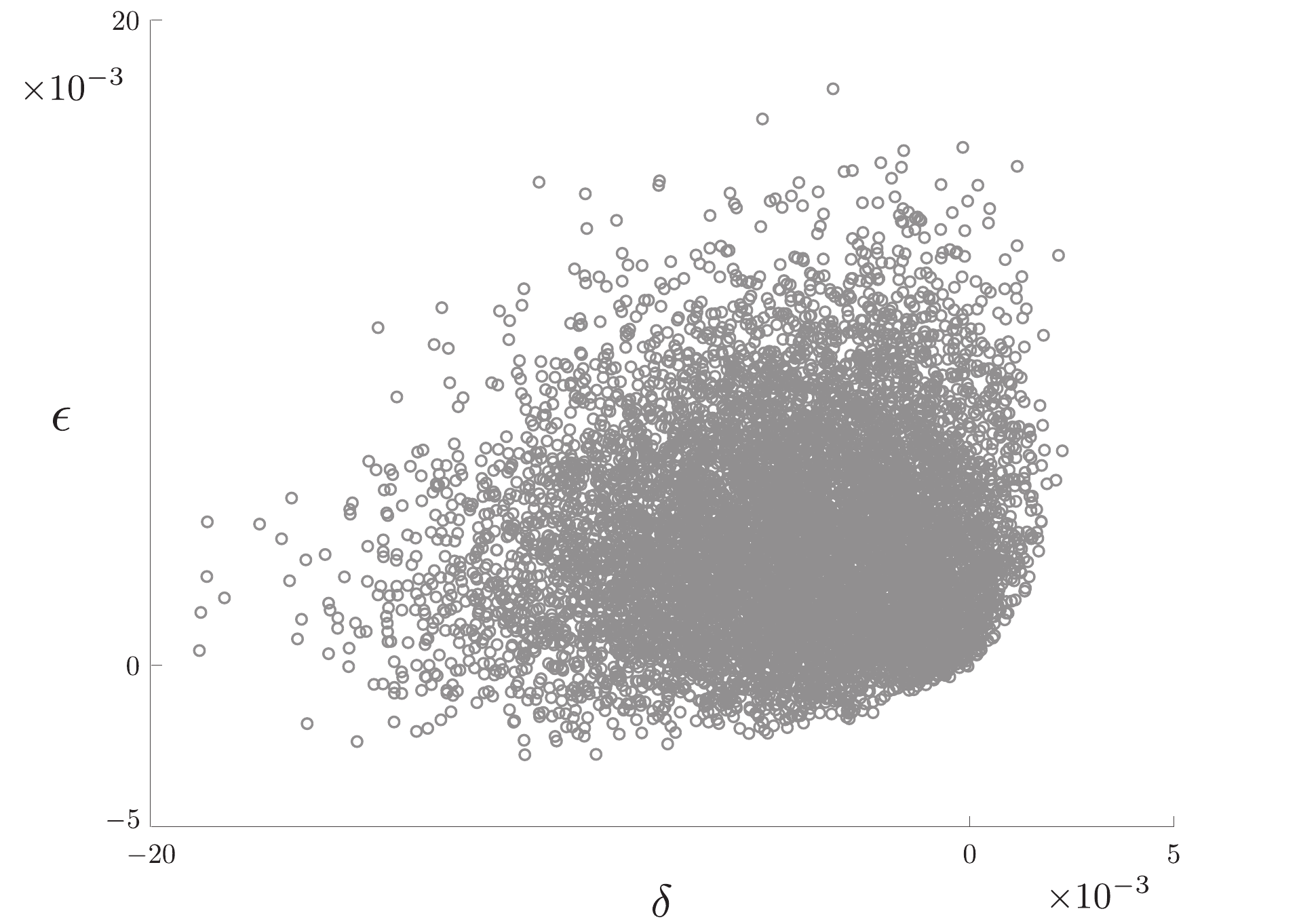}
  \caption{\footnotesize{$\delta$ versus $\epsilon$}}
  \label{fig:rel_ml_deleps}
\end{subfigure}%
\caption{\small{Cross-plots of anisotropy parameters presented for $10000$ examples of equivalent TI mafic media.}}
\label{fig:rel_ml}
\end{figure}

Let us verify the relations among the anisotropy parameters in the context of $\varphi$ and \citeauthor{Berrymanetal} fluid detection methods.
Relation between $\varphi$ and $\epsilon$, as well as, $\delta$ and $\epsilon$, are shown in Figure~\ref{fig:rel_ml}.
Analyzing Figure~\ref{fig:rel_ml_phieps}, we see that there is a similar amount of examples in which $\varphi$ is positive or negative.
Also, we notice that in a great majority of MC examples $\varphi<\epsilon$.
However, if we compare their absolute values, we see that there is a part of examples in which $|\varphi|>|\epsilon|$.
A well-known relation in layered media \citep{Berryman79}, $\epsilon>\delta$, is obviously confirmed in Figure~\ref{fig:rel_ml_deleps}.
Nevertheless, it is easy to notice that there is almost equal proportion between $|\delta|<|\epsilon|$ and $|\delta|>|\epsilon|$.
This is an interesting insight, discussed more in Appendix~\ref{sec:dis}, during the thorough analysis of distributions of these anisotropy parameters.
Taking that proportion into consideration, we might expect that proportion between $|\varphi|>|\epsilon|$ and $|\varphi|>|\delta|$ is also almost equal.
Another property resulting from the distributions of the anisotropy parameters, is the similar proportionality between $\varphi<\epsilon$ and $\varphi>\delta$. 
Aforementioned relations along with ranges and dominants of $\varphi$, $\epsilon$ and $\delta$, are shown in Table~\ref{tab:ml}.
\begin{table}[!htbp]
\centering
\begin{subtable}[b]{0.4\textwidth}
\centering
\begin{tabular}{ccc} 
\toprule 
&&\\[-0.26in]
& {\small{dominant}}  & \hphantom{x} {\small{range}}   \\ [-0.04in]
 & {\small{$\left[\cdot\,10^{-3}\right]$}} &\hphantom{x} {\small{$\left[\cdot\,10^{-3}\right]$}} \\
\midrule
$\varphi$ \hphantom{X}  & $0.296$   & \hphantom{X} $\left(-11.68\,,\,12.07\right)$    \\
$\epsilon$ \hphantom{X}  & $1.904$   &\hphantom{X}  $\left(-2.767\,,\,17.87\right)$    \\
$\delta$ \hphantom{X} & $-1.580$ \hphantom{x}  & \hphantom{X} $\left(-18.83\,,\,2.275\right)$    \\
\bottomrule
\end{tabular}
\caption{}
\label{subtab:a}
\end{subtable} 
\begin{subtable}[b]{0.35\textwidth}
\centering
\begin{tabular}{cc} 
\toprule
$\varphi>\epsilon$     \hphantom{X} & $0.32\,\%$      \\
$\varphi<\delta$         \hphantom{X} & $0.38\,\%$     \\
$|\varphi|>|\epsilon|$  \hphantom{X} & $25.66\,\%$     \\
$|\varphi|>|\delta|$      \hphantom{X}& $24.23\,\%$      \\
$|\delta|>|\epsilon|$    \hphantom{X}& $53.87\,\%$      \\
\bottomrule
\end{tabular}
\caption{}
\label{subtab:b}
\end{subtable}
\caption{\small{
\subref{subtab:a} dominants and ranges along with
\subref{subtab:b} percentage of occurrence of relations among $\varphi$, $\epsilon$ and $\delta$.
}}
\label{tab:ml}
\end{table}

Based on the results from this section, we cannot verify if the pattern from Table~\ref{tab:pattern} is true or not.
Due to certain amount of MC examples which exhibit small $\epsilon$ along with small $\delta$, the \citet{Berrymanetal} method is possible to be successful.
To verify if both methods are able to correctly detect variations of $\lambda$, and to look for the other fluid indicators, in the next sections, we require to focus on specific ranges of variations of the Lam\'e parameters.
In other words, we need to restrict the general case of variations by the limited ranges of $RSD_{\mu}$ and $RSD_{\lambda}\,$.
We conclude this section by noticing that:
\begin{itemize}
\item{$\epsilon$ mostly has positive values,}
\item{$\delta$ mostly has negative values,}
\item{the probability of $\varphi$ being negative or positive is more or less equal,}
\item{the probability of $|\delta|>|\epsilon|$ or $|\delta|<|\epsilon|$ is more or less equal,}
\item{$\varphi$ mostly has smaller absolute values than $\epsilon$ or $\delta$,}
\item{to verify $\varphi$ and ~\citeauthor{Berrymanetal} methods we should limit the variations of $\mu$ and $\lambda$.} 
\end{itemize}
%%%%%%%%%%%%%%%%%%%%%%%%%%%%%%%%%%%%%%%%%%%%%%%%%%%%%%%%%%%%%%%%%%%%%%%%%%%%%%%%%%%%%%%
\subsection{Non near-constant rigidity} \label{sec:var}
\begin{figure}[!htbp]
\centering
   \includegraphics[scale=0.35]{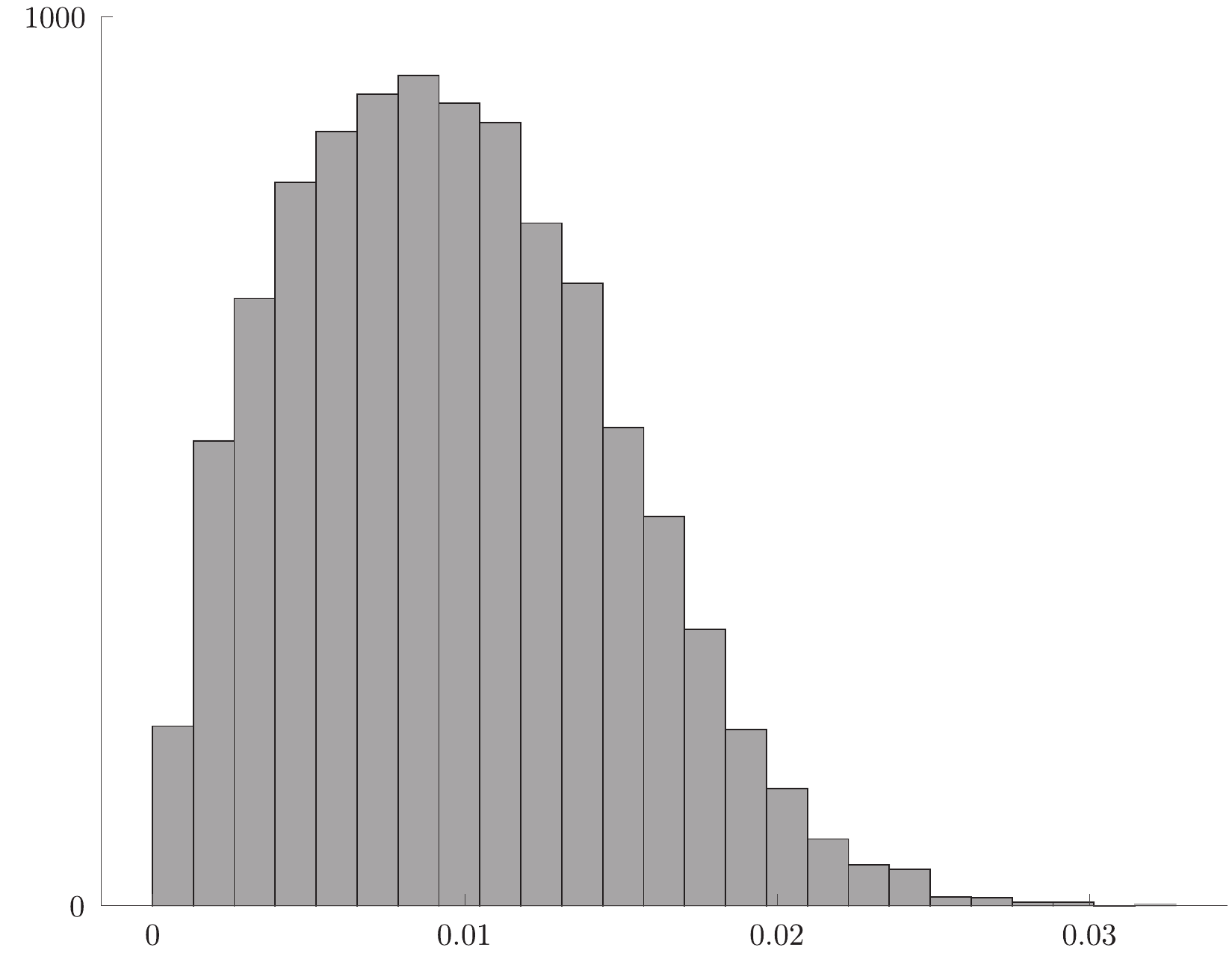}
 \caption{\small{Distribution of $\gamma$ presented for $10000$ examples of equivalent TI mafic media with $RSD_{\mu}>2\,\%\,$.}}
\label{fig:gamma_mb_maf}
\end{figure}

%%%%%%%%%%%%%%%%%%%%%%%%%%%%%%%%%%%%%%%%%%%%%%%%%%%%%%%%%%%%%%%%%%%%%%%%%%%%%%%%%%%%%%%
Let us again perform Monte Carlo simulations to obtain the values of the anisotropy parameters.
To do so, we again choose the same values of $a=40$, $b=70$, $c=35$ and $d=60$, from Table~\ref{tab:restr}.
In this section, apart from the range of $\mu$ and $\lambda$, we additionally limit the random sampling of Lam\'e parameters by $RSD_{\mu}>2\,\%\,$.
Such a restriction allow us to exclude the case of near-constant rigidity.
To receive the range of $\gamma$ for non near-constant rigidity, we perform MC simulation receiving $s=10000$ examples of TI media, equivalent to $n=5$ isotropic layers.
The resulted range of $\gamma$ is $(1.617\cdot10^{-4}\,,\,3.259\cdot10^{-2})$ and its dominant is $8.499\cdot10^{-3}$.
The distribution of $\gamma$ for $RSD_{\mu}>2\,\%\,$ is shown in Figure~\ref{fig:gamma_mb_maf}.
Subsequently, we execute three different MC simulations with three different additional restrictions (apart from $RSD_{\mu}>2\,\%$) imposed on variations of $\lambda$, namely, $RSD_{\lambda}<2\,\%$, $RSD_{\lambda}\in\left(2\,\%\,,\,20\,\%\right)$ and $RSD_{\lambda}>2\,\%$.
In other words, we verify separately three different variations of $\lambda$; near-constant, moderately varying and strongly varying case, respectively.
In each case, we receive $s=1000$ examples of TI media, equivalent to $n=5$ isotropic layers.
The decrease of number of examples is due to longer time of simulation process that is caused by the new restrictions imposed on variations of $\lambda$.
Nevertheless, that number is large enough to obtain sufficiently accurate results.

\begin{figure}[!htbp]
\centering
\begin{subfigure}{.33\textwidth}
  \centering
   \includegraphics[scale=0.3]{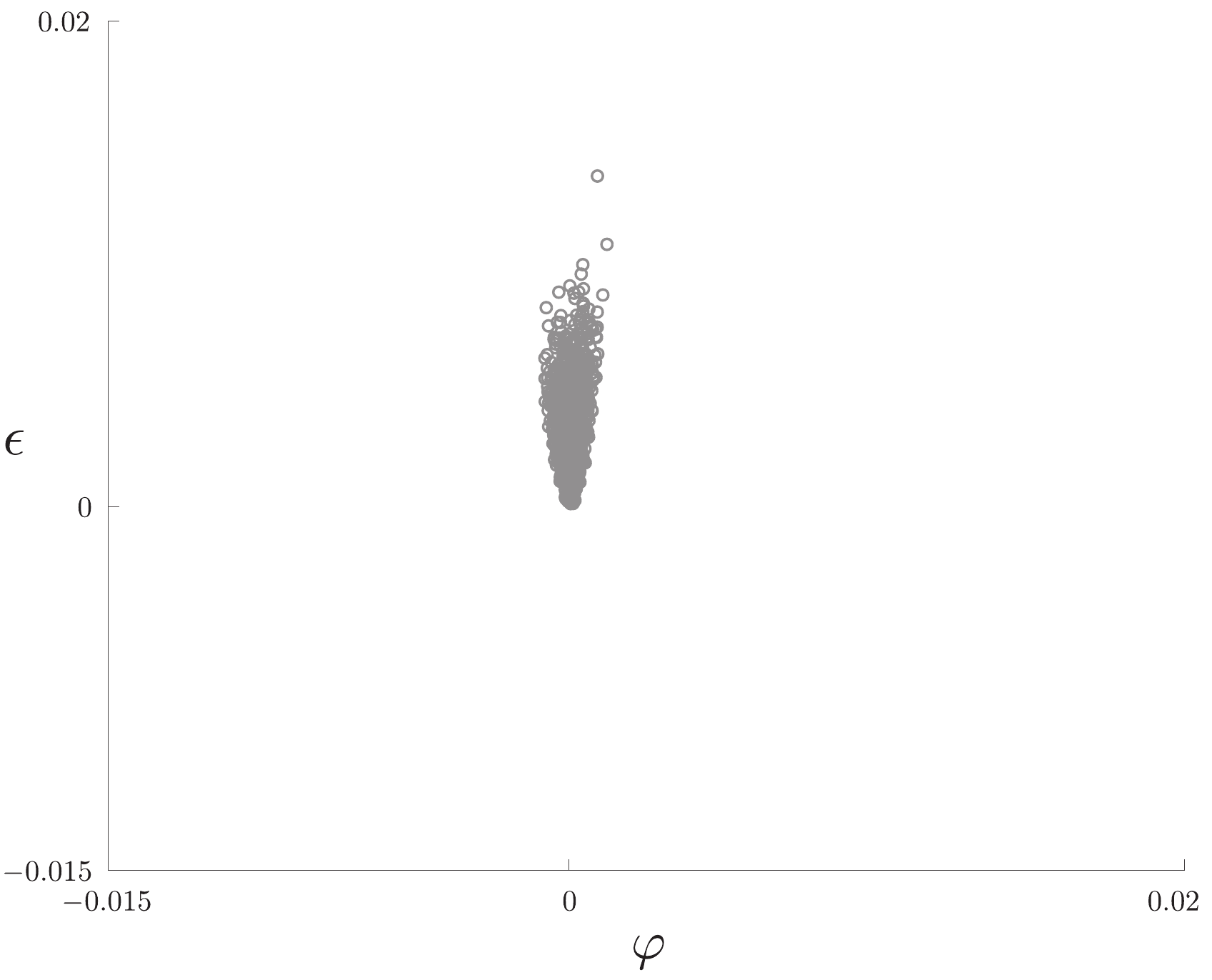}
  \caption{\footnotesize{$RSD_{\lambda}<2\,\%$}}
  \label{fig:rel_phieps_mb_ls}
\end{subfigure}%
\begin{subfigure}{.33\textwidth}
  \centering
   \includegraphics[scale=0.3]{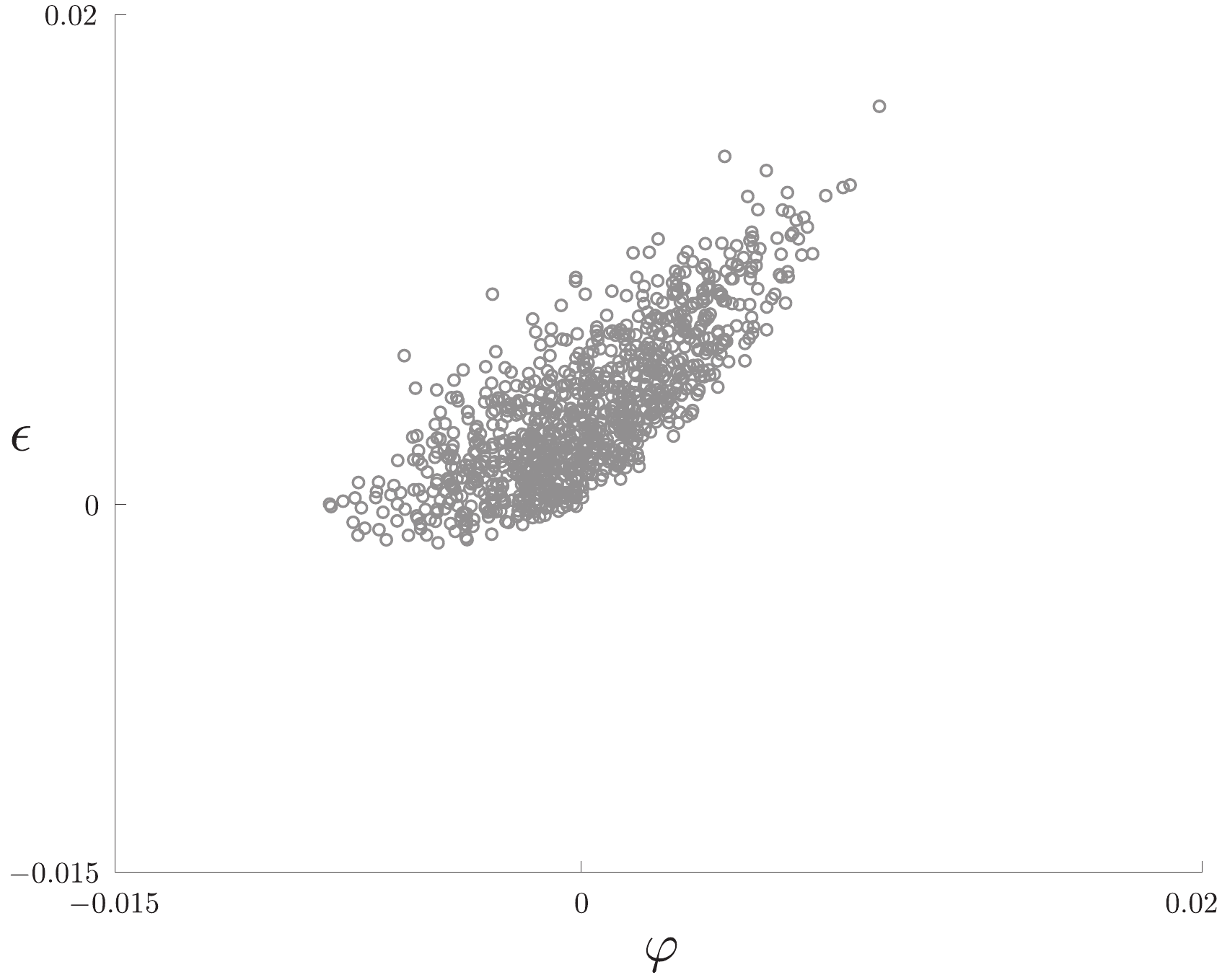}
  \caption{\footnotesize{$RSD_{\lambda}\in\left(2\,\%\,,\,20\,\%\right)$}}
  \label{fig:rel_phieps_mb_2l20}
\end{subfigure}%
\begin{subfigure}{.33\textwidth}
  \centering
   \includegraphics[scale=0.3]{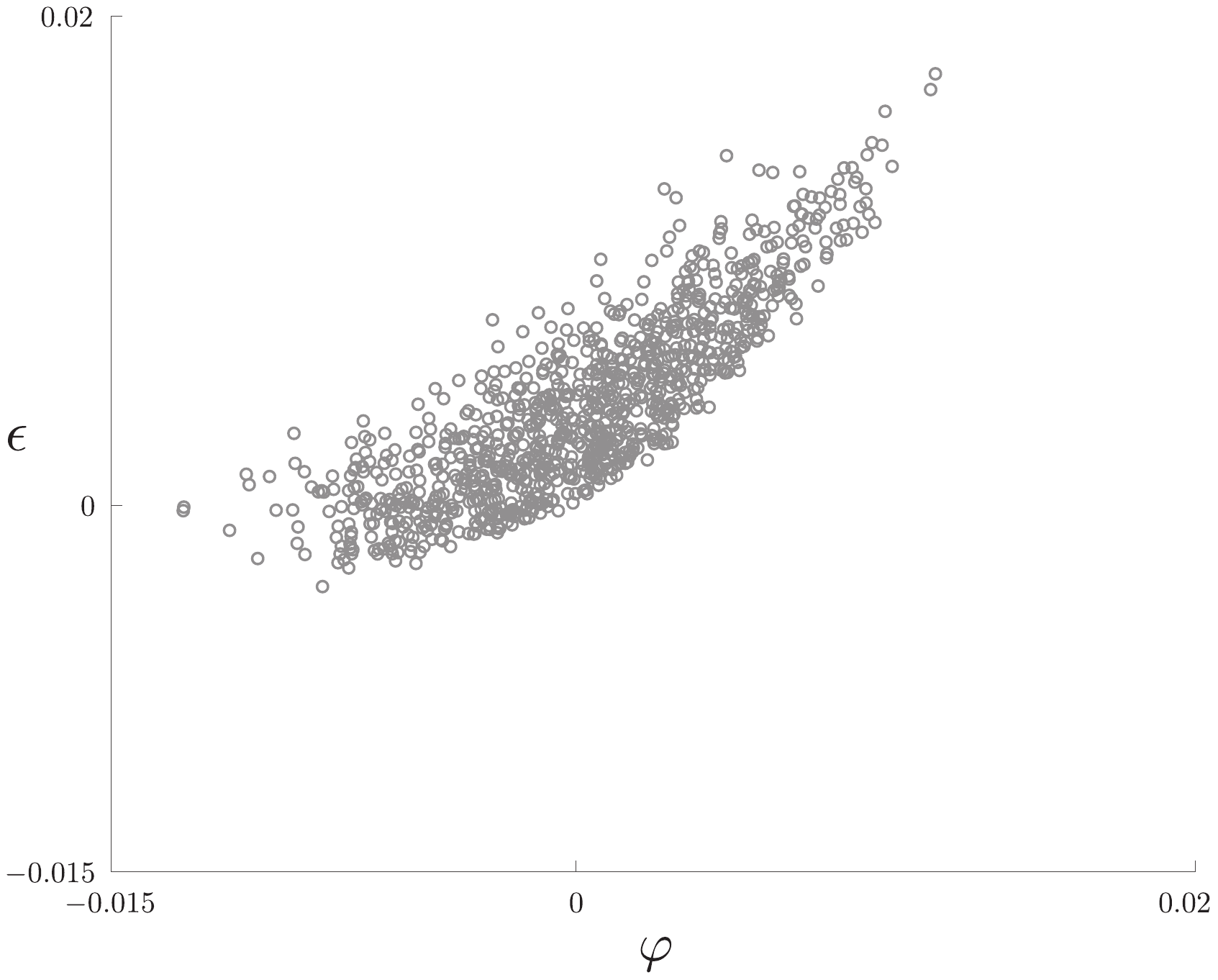}
  \caption{\footnotesize{$RSD_{\lambda}>20\,\%$}}
  \label{fig:rel_phieps_mb_lvb}
\end{subfigure}
\caption{\small{Cross-plots of $\varphi$ versus $\epsilon$ presented for $1000$ examples of equivalent TI mafic media with $RSD_{\mu}>2\,\%\,$.}}
\label{fig:rel_phieps_mb}
\end{figure}

\begin{figure}[!htbp]
\centering
\begin{subfigure}{.33\textwidth}
  \centering
   \includegraphics[scale=0.3]{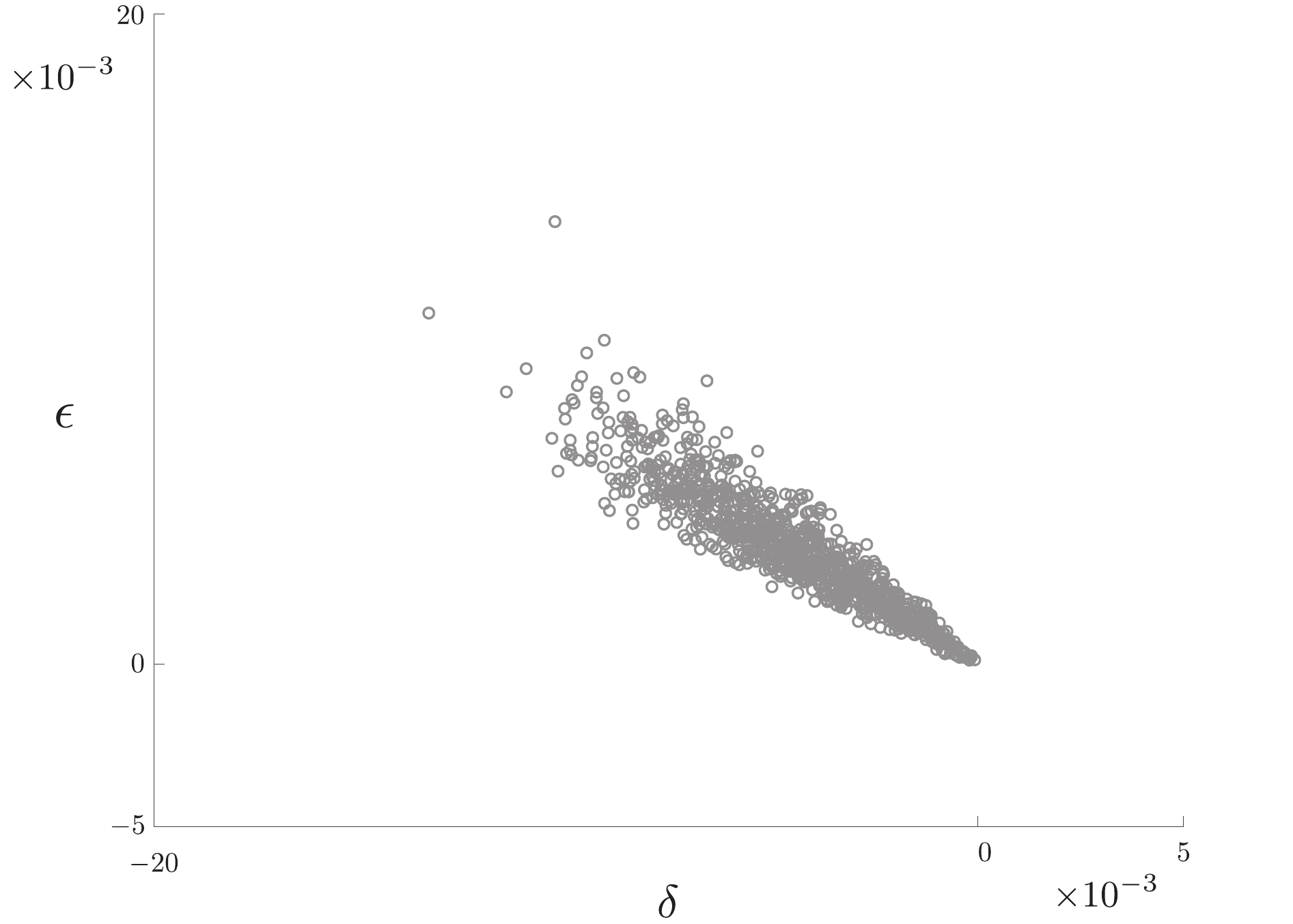}
  \caption{\footnotesize{$RSD_{\lambda}<2\,\%$}}
  \label{fig:rel_deleps_mb_ls}
\end{subfigure}%
\begin{subfigure}{.33\textwidth}
  \centering
   \includegraphics[scale=0.3]{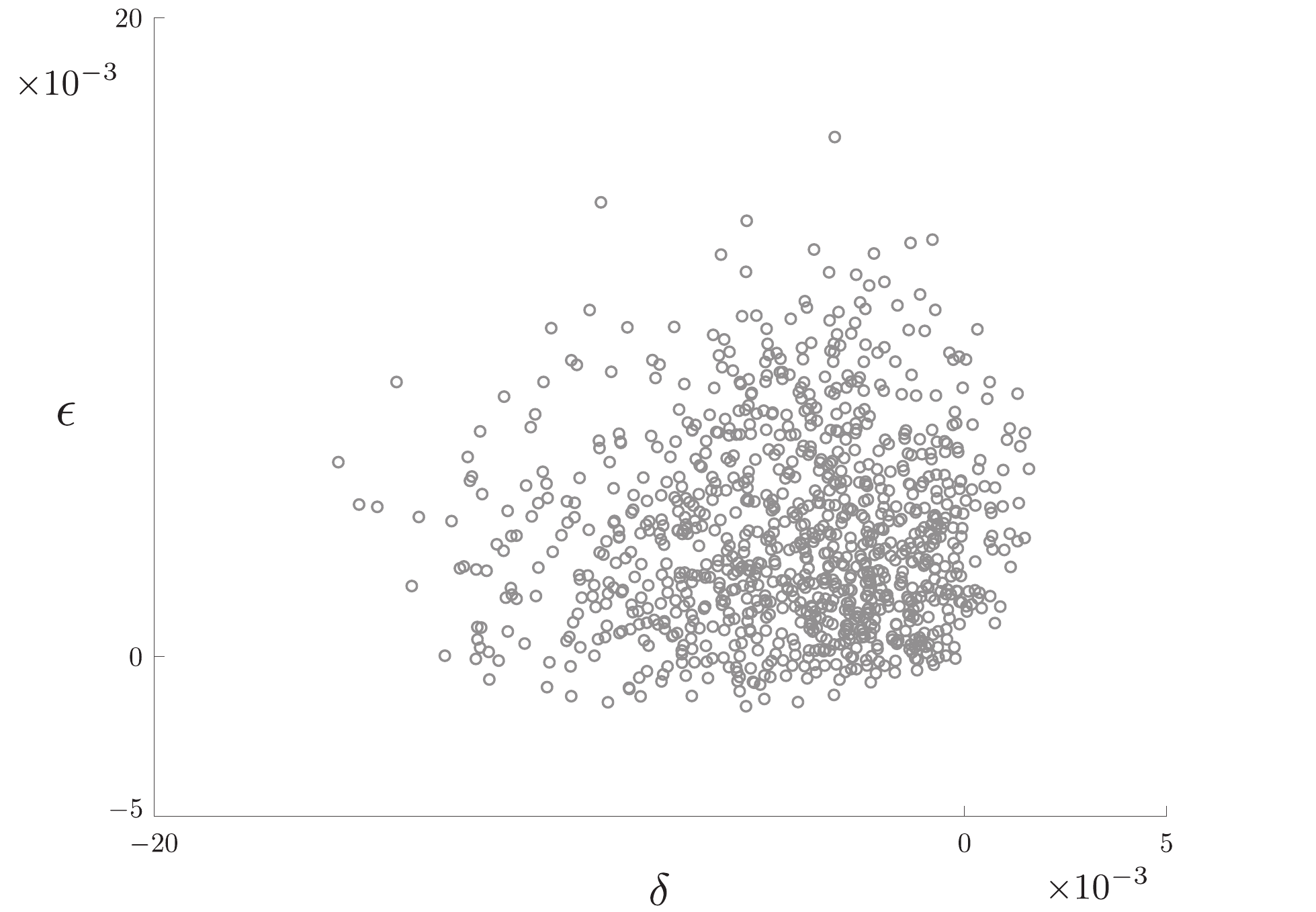}
  \caption{\footnotesize{$RSD_{\lambda}\in\left(2\,\%\,,\,20\,\%\right)$}}
  \label{fig:rel_deleps_mb_2l20}
\end{subfigure}%
\begin{subfigure}{.33\textwidth}
  \centering
   \includegraphics[scale=0.3]{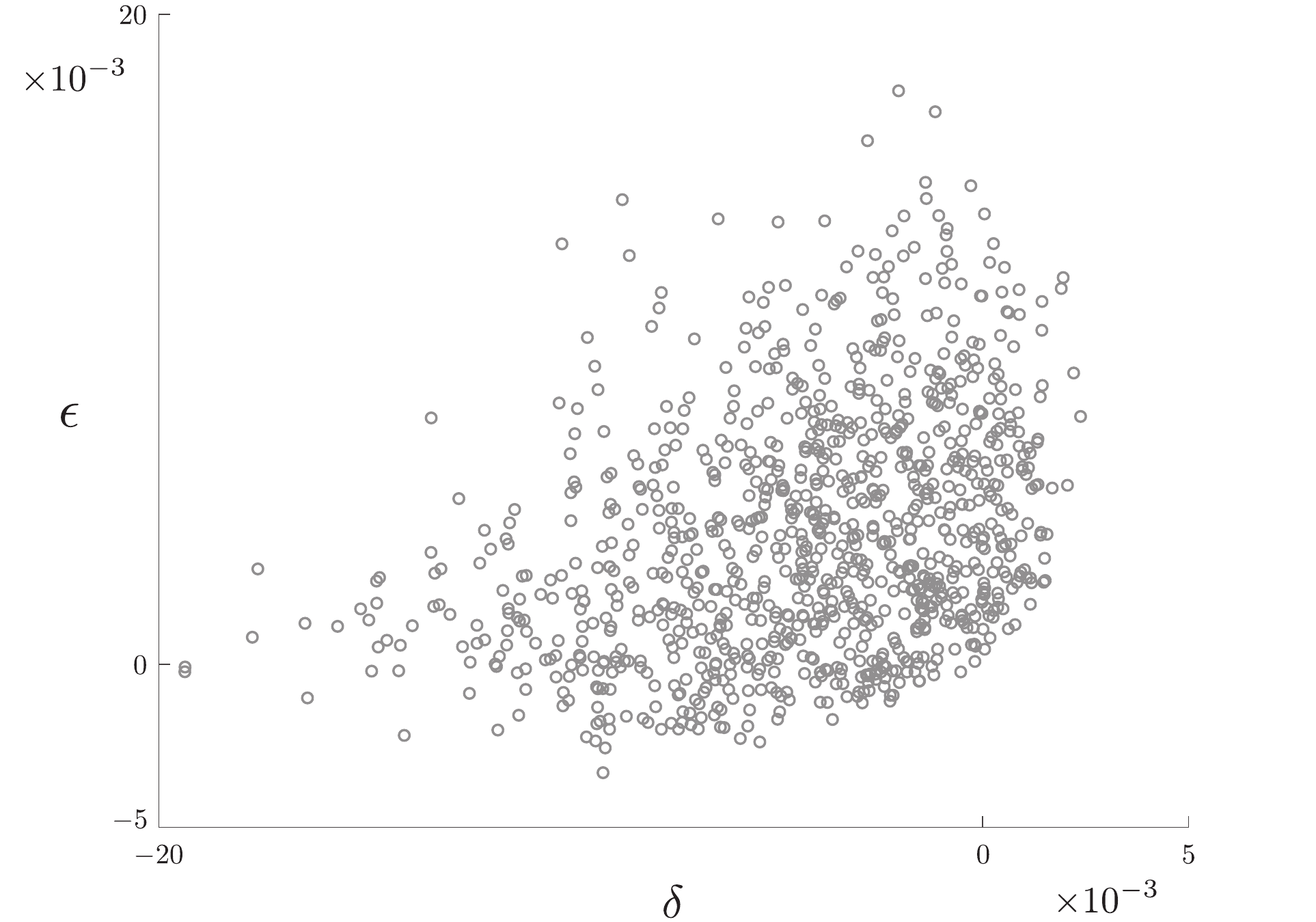}
  \caption{\footnotesize{$RSD_{\lambda}>20\,\%$}}
  \label{fig:rel_deleps_mb_lvb}
\end{subfigure}
\caption{\small{Cross-plots of $\delta$ versus $\epsilon$ presented for $1000$ examples of equivalent TI mafic media with $RSD_{\mu}>2\,\%\,$.}}
\label{fig:rel_deleps_mb}
\end{figure}

Similarly to Section~\ref{sec:general}, we again show the relations among the anisotropy parameters in a form of cross-plots.
For the case of near-constant, moderately varying, and strongly varying $\lambda$, $\varphi$ versus $\epsilon$ is exhibited in Figures~\ref{fig:rel_phieps_mb_ls},~\ref{fig:rel_phieps_mb_2l20} and~\ref{fig:rel_phieps_mb_lvb}, respectively.
$\delta$ versus $\epsilon$ is shown in Figures~\ref{fig:rel_deleps_mb_ls},~\ref{fig:rel_deleps_mb_2l20} and~\ref{fig:rel_deleps_mb_lvb}.
Dominants and ranges of $\varphi$, $\epsilon$ and $\delta$, are represented in Table~\ref{tab:ran_mb}. 
The percentages of MC examples in which certain relations among the anisotropy parameters occur, are exposed in Table~\ref{tab:rel_mb}.
\begin{table}[!htbp]
\begin{tabular}{ccccccc}
\toprule
& \multicolumn{2}{c}{$RSD_{\lambda}<\,2\%$} & \multicolumn{2}{c}{$RSD_{\lambda}\in\left(2\,\%\,,\,20\,\%\right)$} & \multicolumn{2}{c}{$RSD_{\lambda}>\,20\%$} \\
\midrule 
&&&&&&\\[-0.3in]
& \hphantom{X} {\small{dominant}}  & {\small{range}} & \hphantom{X} {\small{dominant}}  & {\small{range}} & \hphantom{X} {\small{dominant}}  & {\small{range}} \\ [-0.06in]
& \hphantom{X}{\scriptsize{$\left[\cdot\,10^{-3}\right]$}} & {\scriptsize{$\left[\cdot\,10^{-3}\right]$}} & \hphantom{X}{\scriptsize{$\left[\cdot\,10^{-3}\right]$}} & {\scriptsize{$\left[\cdot\,10^{-3}\right]$}} & \hphantom{X}{\scriptsize{$\left[\cdot\,10^{-3}\right]$}} & {\scriptsize{$\left[\cdot\,10^{-3}\right]$}}\\ [-0.02in]
\midrule
$\varphi$ \hphantom{X}   & $-0.105$ & $\left(-0.867\,,\,1.149\right)$     &\hphantom{X} $-0.002$   & $\left(-8.096\,,\,9.605\right)$  &\hphantom{X} $1.070$ &   $\left(-12.67\,,\,11.61\right)$ \\
$\epsilon$ \hphantom{X} & \hphantom{X}$4.104$  & $\left(0.118\,,\,13.62\right)$     &\hphantom{X} $1.978$  & $\left(-1.559\,,\,16.26\right)$  &\hphantom{X} $1.639$ &   $\left(-3.328\,,\, 17.64\right)$  \\
$\delta$ \hphantom{X}  & $-0.892$ & $\left(-13.33\,,\,-0.070\right)$  &\hphantom{X} $-3.514$&$\left(-15.45\,,\,1.603\right)$  &\hphantom{X} $-1.520$\hphantom{X} &   $\left(-19.37\,,\,2.377\right)$     \\
\bottomrule
\end{tabular}
\caption{\small{Dominants and ranges of anisotropy parameters relevant to layered mafic rocks with $RSD_{\mu}>2\,\%\,$.}}
\label{tab:ran_mb}
\end{table}
\begin{table}[!htbp]
\begin{tabular}{ccccccc}
\toprule
&$\varphi>\epsilon$&$\varphi<\delta$&$|\varphi|>|\epsilon|$&$|\varphi|>|\delta|$&$|\delta|>|\epsilon|$\\
\midrule
$RSD_{\lambda}<\,2\%$\hphantom{X}&$0\,\%$&$0\,\%$&$0\,\%$&$0\,\%$&$86.9\,\%$\\
$RSD_{\lambda}\in\left(2\,\%\,,\,20\,\%\right)$\hphantom{X}&$0.4\,\%$&$0.5\,\%$&$27.8\,\%$&$25.8\,\%$&$53.2\,\%$\\
$RSD_{\lambda}>\,20\%$\hphantom{X}&$0.9\,\%$&$0.7\,\%$&$32.8\,\%$&$32.4\,\%$&$48.5\,\%$\\
\bottomrule
\end{tabular}
\caption{\small{Percentage of occurrence of relations among $\varphi$, $\epsilon$ and $\delta$, relevant to layered mafic rocks with $RSD_{\mu}>2\,\%\,$.}}
\label{tab:rel_mb}
\end{table}

Let us analyze the anisotropy parameters in the context of $\varphi$ method.
Based on Figure~\ref{fig:rel_phieps_mb} and ranges from Table~\ref{tab:ran_mb}, we notice that $\varphi$ is sensitive to variations of $\lambda$.
Its range for strong variations of $\lambda$ is around ten times larger as compared to the case of near-constant $\lambda$.
If we also take a look at Figure~\ref{fig:rel_deleps_mb}, we see that $\epsilon$ and $\delta$ are also influenced by the variations of $\lambda$, however, not as much as $\varphi$.
Thus, the right-hand side of the pattern from Table~\ref{tab:pattern} is correct, and we may quantify it by stating that for the Lam\'e parameters relevant to layered mafic rocks with non near-constant $\mu$ $(RSD_{\mu}>2\,\%\,)$, moderate values of $|\varphi|\in\left(10^{-3}\,,\,10^{-2}\right)$ always indicate non near-constant $\lambda$ $(RSD_{\lambda}>2\,\%)$.
Moreover, large values of $|\varphi|>10^{-2}$ always indicate large variations of $\lambda$ $(RSD_{\lambda}>20\,\%\,)$.
In general, the larger absolute values of $\varphi$ are, the larger variations of $\lambda$ occur.
However, it is important to remember that small variations of $\lambda$ do not necessarily mean that $\varphi$ is small.
Dominants of $\varphi$ give us information that $\varphi$ most likely has small values oscillating around zero, which is in agreement with Table~\ref{subtab:a}.
The percentage of occurrence of $\varphi>\epsilon$ and $|\varphi|>|\epsilon|$ (also $\varphi<\delta$ and $|\varphi|>|\delta|$) grows along with the strength of variations of $\lambda$. 

Let us analyze $\epsilon$ and $\delta$ in the context of \citeauthor{Berrymanetal} method.
Small positive $\epsilon$ along with small positive $\delta$ do not occur in the case of near-constant $\lambda$.
Their occurrence is characteristic only for moderate and strong variations of $\lambda$.
It is important to notice that small positive $\epsilon$ is also characteristic for near-constant $\lambda$, which is not the case for $\delta$.
For small positive $\delta$, $\epsilon$ is always small positive.
Thus, small positive $\delta$ is the sufficient condition to ensure that $RSD_{\lambda}>2\,\%\,$.
However, small positive $\delta$ is not the only one possible indicator of moderate and strong variations of $\lambda$.
For instance, negative $\epsilon$, is also characteristic only for these variations. 
In general, as shown in Figure~\ref{fig:rel_deleps_mb}, and confirmed by dominants from Table~\ref{tab:ran_mb}, $\epsilon$ has positive values, whereas $\delta$ has mostly negative values, which is in agreement with general case from Table~\ref{subtab:a}.
Relation $|\delta|>|\epsilon|$ occurs very often for near-constant $\lambda$, thus, it is a special case, since it is not in agreement with general case shown in Section~\ref{sec:general}. 
This special case is further discussed in Appendix~\ref{sec:dis}.
On the other hand, the percentage of occurrence of $|\delta|>|\epsilon|$ for moderate and strong variations of $\lambda$ is similar to the one from the general case.
Especially, for moderate variations, since the general case is mostly represented by $RSD_{\lambda}\in(2\,\%\,,\,20\,\%)$.

In the context of fluid detection methods, if $RSD_{\mu}>2\,\%\,$ ($\gamma>1.6\cdot10^{-4}$), we may propose five main, and two additional, indicators of moderate and strong variations of $\lambda$ in layers. 
Firstly, moderate or large values of $|\varphi|$ only indicate moderate and strong variations of $\lambda$, respectively; thus, we propose $|\varphi|>10^{-3}$ (as it is shown in Appendix~\ref{ap:two} and mentioned in Section~\ref{sec:conclusions}, an universal indicator valid for each type of rocks is $|\varphi|>5\cdot10^{-3}$).
Another relations, $|\varphi|>|\epsilon|$ or $|\varphi|>|\delta|$, also occur only in the case of $RSD_{\lambda}>2\,\%\,$.
Finally, we propose negative $\epsilon$ or positive $\delta$, which mostly have small values.
Additional relations, which occur very rarely, but only in the case of $RSD_{\lambda}>2\,\%\,$, are $\varphi>\epsilon$ or $\varphi<\delta$.
If any of the above conditions is obeyed it means that there is a change of the fluid content in layered Earth.
If none of the above conditions is satisfied it does not necessarily mean that the moderate or strong variations of $\lambda$ in layers do not occur.
The usage of these fluid indicators is exemplified in Appendix~\ref{sec:muvar_scale}.
The exact percentages of MC examples in which aforementioned indicators do occur, are presented in Table~\ref{tab:bigtable2}.
Based on that table, in non near-constant rigidity case, for mafic rocks, $|\varphi|>10^{-3}$, occurs to be the most effective in detecting fluids among all of the indicators.
Probably the safest way to detect moderate or strong variations of $\lambda$ is to compare values of $\varphi$, $\epsilon$ and $\delta$, obtained for mafic rocks from the real data, to the ones from Figures~\ref{fig:rel_phieps_mb} and~\ref{fig:rel_deleps_mb}.
Nevertheless, the values of anisotropy parameters that occur for near-constant $\lambda$, also occur for its moderate or strong variations, thus, in cases of these particular values, fluid detection based on anisotropy parameters is probably impossible.
To conclude, we notice that:
\begin{itemize}
\item{the larger the variations of $\lambda$, the larger the values of $\varphi$, $\epsilon$ and $\delta$,}
\item{$\varphi$ is very sensitive on variations of $\lambda$, as opposed to $\epsilon$ and $\delta$.}
\item{for $RSD_{\lambda}>2\,\%\,$, $\delta$ mostly have negative values, thus, \citeauthor{Berrymanetal} method is inaccurate,}
\item{$|\varphi|>10^{-3}$, $|\varphi|>|\epsilon|$, $|\varphi|>|\delta|$, $\epsilon<0$ and $\delta>0$, are possible fluid indicators for mafic rocks.}
\end{itemize}
%%%%%%%%%%%%%%%%%%%%%%%%%%%%%%%%%%%%%%%%%%%%%%%%%%%%%%%%%%%%%%%%%%%%%%%%%%%%%%%
\subsection{Near-constant rigidity} \label{sec:mu}
\begin{figure}[!htbp]
\centering
   \includegraphics[scale=0.35]{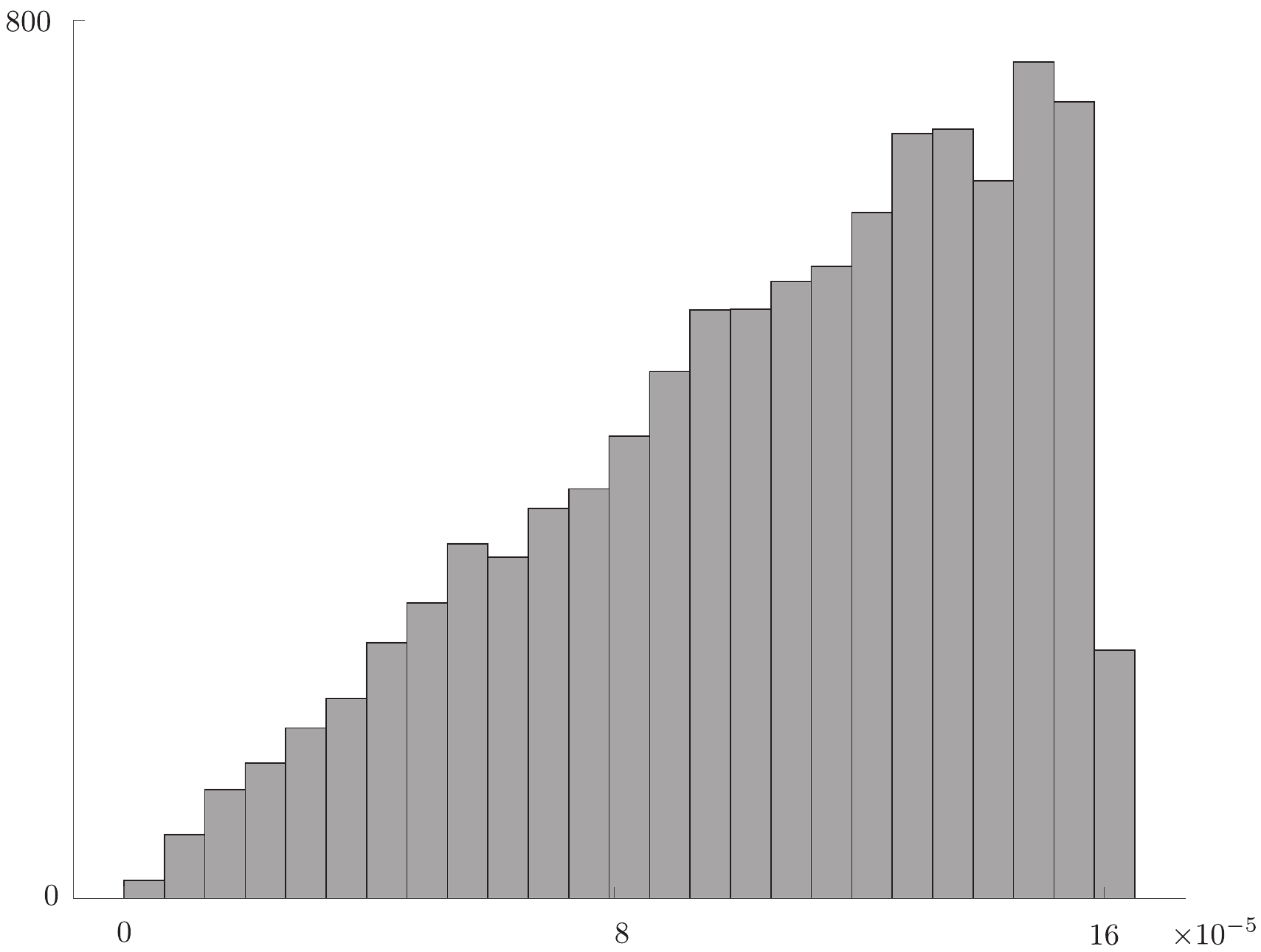}
 \caption{\small{Distribution of $\gamma$ for $10000$ examples of equivalent TI mafic media with $RSD_{\mu}<2\,\%\,$.}}
\label{fig:gamma_ms}
\end{figure}

%%%%%%%%%%%%%%%%%%%%%%%%%%%%%%%%%%%%%%%%%%%%%%%%%%%%%%%%%%%%%%%%%%%%%%%%%%%%%%%%%%%%%%%
Let us repeat the MC procedure from Section~\ref{sec:var}.
The only one significant change is the restriction imposed on variations of $\mu$; instead of $RSD_{\mu}>2\,\%\,$, we impose $RSD_{\mu}<2\,\%\,$.
In other words, we focus on particular case of Section~\ref{sec:general}, namely, case of near-constant rigidity.
For elasticity parameters relevant to layered mafic rocks, $RSD_{\mu}<2\,\%\,$, refers to $\gamma\in(2.650\cdot10^{-6}\,,\,1.635\cdot10^{-4})$.
The most frequent value of $\gamma$ is $1.485\cdot10^{-4}$, and its distribution is shown in Figure~\ref{fig:gamma_ms}.
Another, and last, change in MC restrictions, as compared to simulations from Section~\ref{sec:var}, is that for the case of $RSD_{\mu}<2\,\%\,$ and $RSD_{\lambda}<2\,\%\,$, we obtain $s=1000$ examples of TI media, equivalent to only $n=3$ isotropic layers.
It is caused by the very long time of the simulation process, since we look for a rare and particular case.

\begin{figure}[!htbp]
\centering
\begin{subfigure}{.33\textwidth}
  \centering
   \includegraphics[scale=0.3]{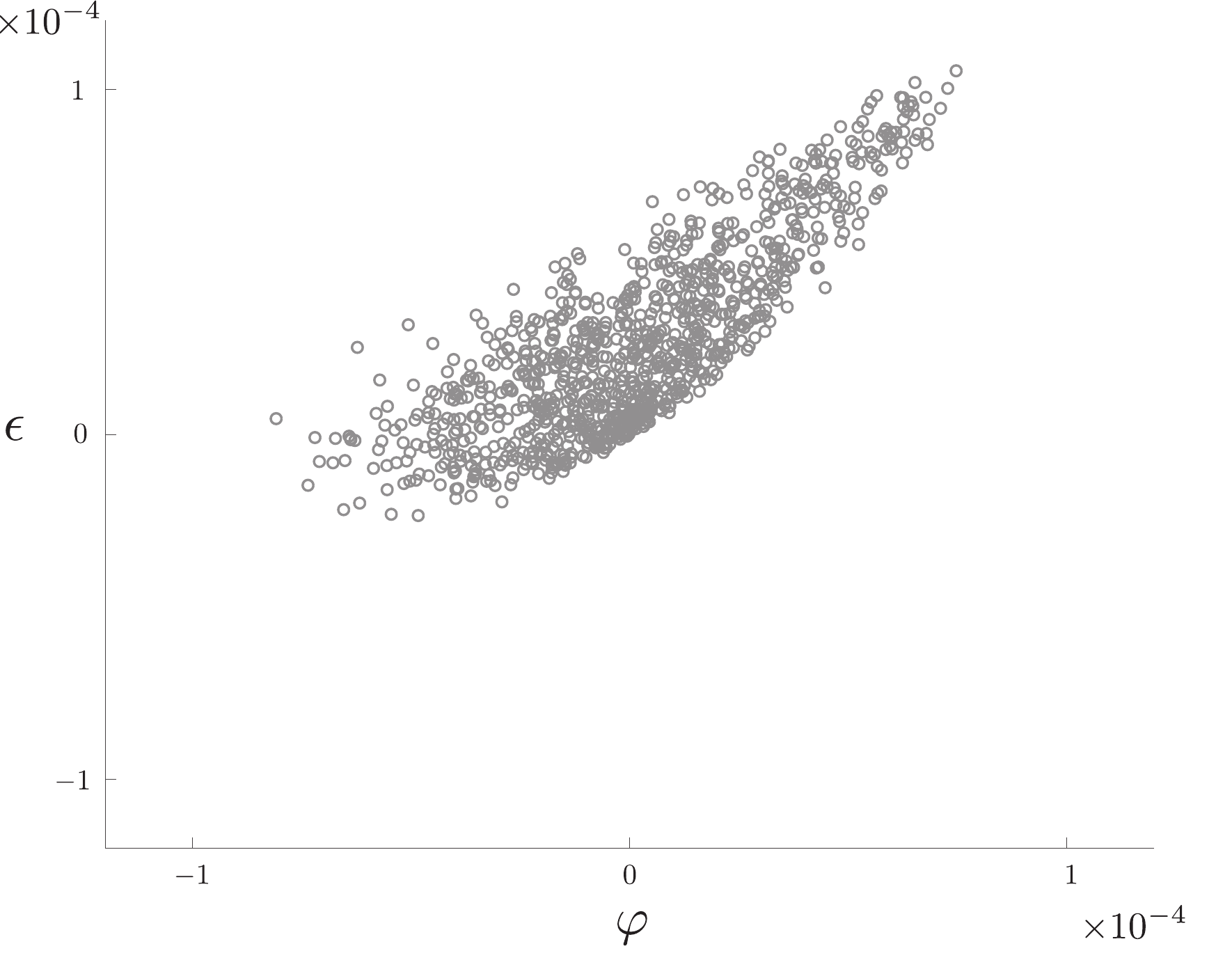}
  \caption{\footnotesize{$RSD_{\lambda}<2\,\%$}}
  \label{fig:rel_phieps_ms_ls}
\end{subfigure}%
\begin{subfigure}{.33\textwidth}
  \centering
   \includegraphics[scale=0.3]{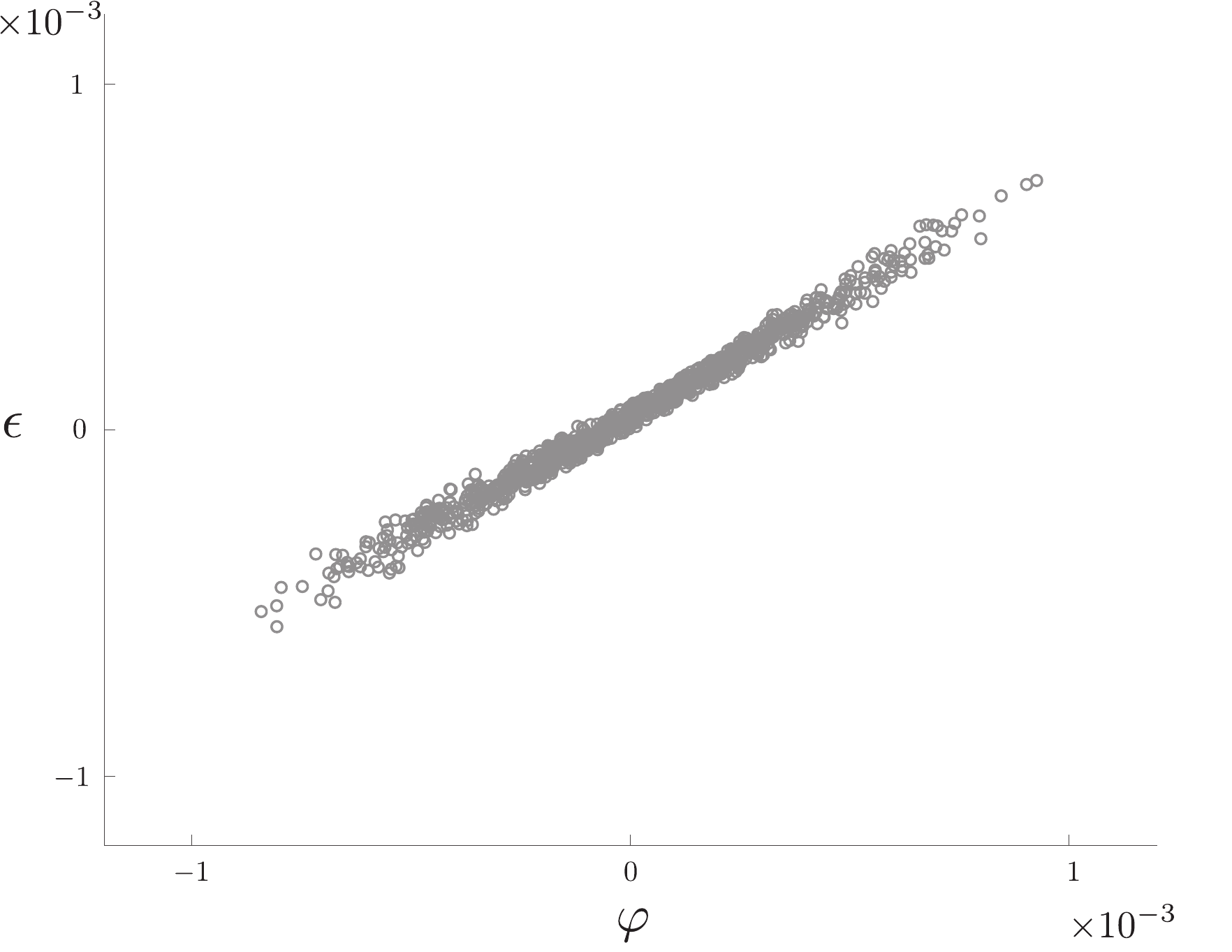}
  \caption{\footnotesize{$RSD_{\lambda}\in\left(2\,\%\,,\,20\,\%\right)$}}
  \label{fig:rel_phieps_ms_2l20}
\end{subfigure}%
\begin{subfigure}{.33\textwidth}
  \centering
   \includegraphics[scale=0.3]{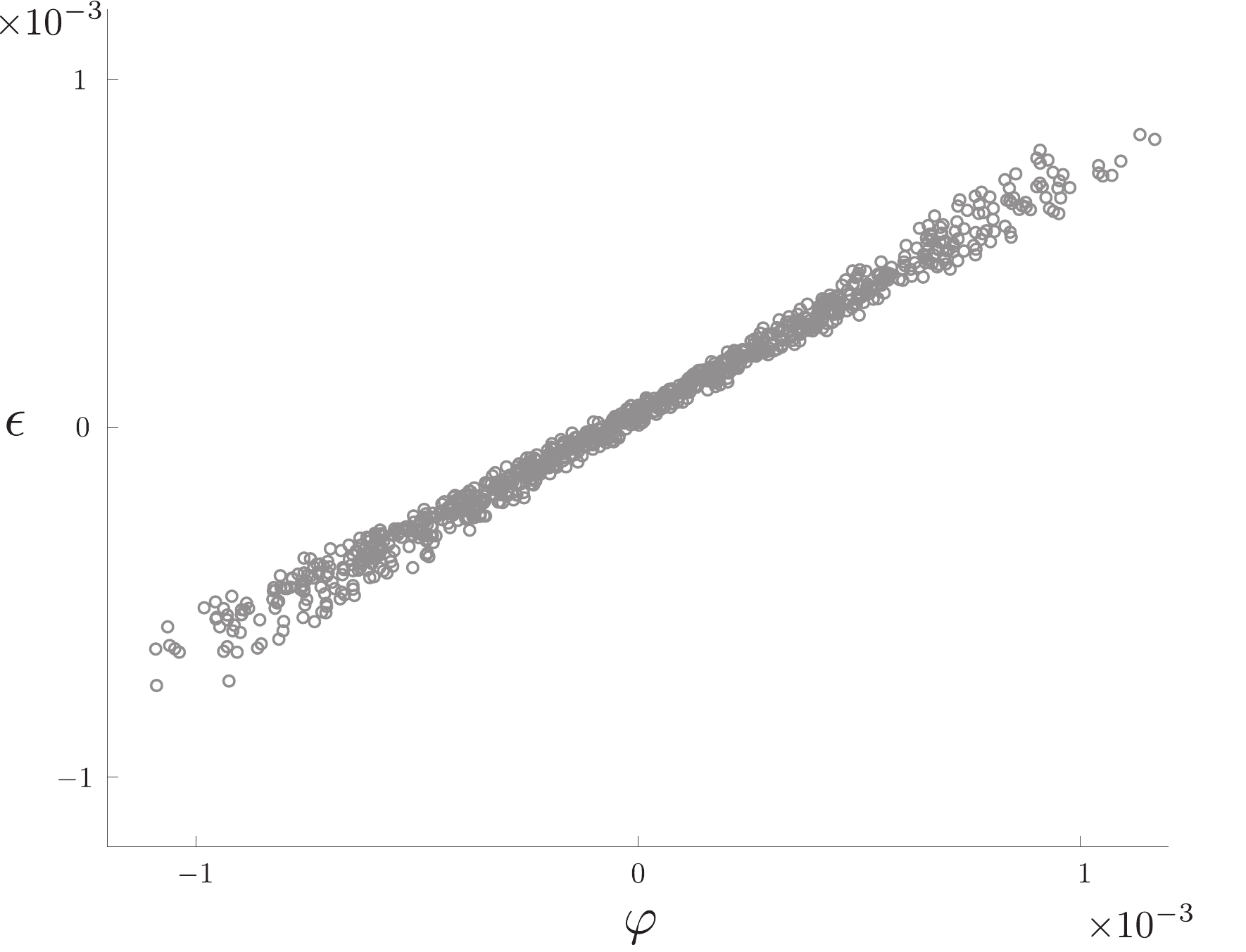}
  \caption{\footnotesize{$RSD_{\lambda}>20\,\%$}}
  \label{fig:rel_phieps_ms_lvb}
\end{subfigure}
\caption{\small{Cross-plots of $\varphi$ versus $\epsilon$ presented for $1000$ examples of equivalent TI mafic media with $RSD_{\mu}<2\,\%\,$.}}
\label{fig:rel_phieps_ms}
\end{figure}

\begin{figure}[!htbp]
\centering
\begin{subfigure}{.33\textwidth}
  \centering
   \includegraphics[scale=0.3]{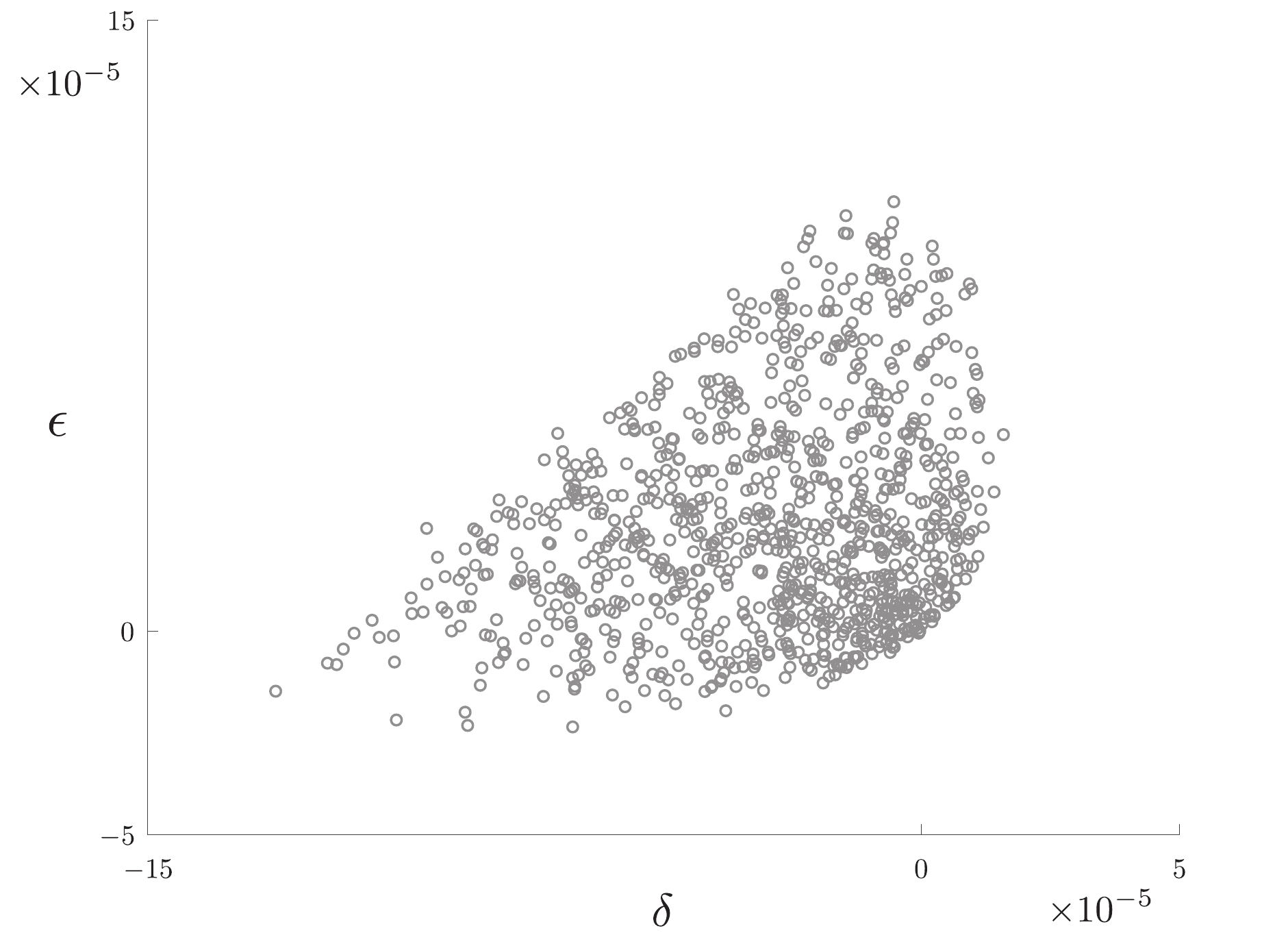}
  \caption{\footnotesize{$RSD_{\lambda}<2\,\%$}}
  \label{fig:rel_deleps_ms_ls}
\end{subfigure}%
\begin{subfigure}{.33\textwidth}
  \centering
   \includegraphics[scale=0.3]{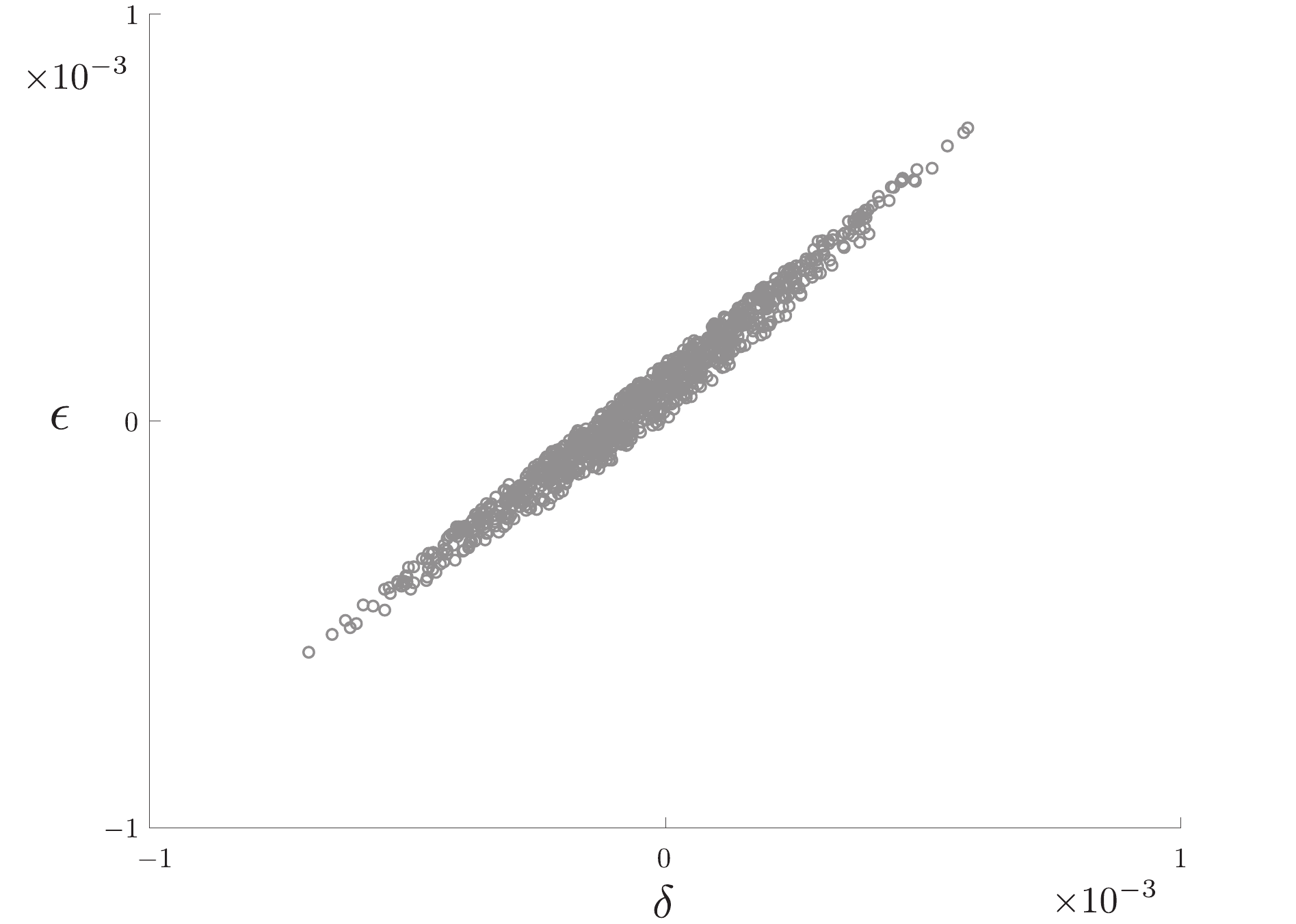}
  \caption{\footnotesize{$RSD_{\lambda}\in\left(2\,\%\,,\,20\,\%\right)$}}
  \label{fig:rel_deleps_ms_2l20}
\end{subfigure}%
\begin{subfigure}{.33\textwidth}
  \centering
   \includegraphics[scale=0.3]{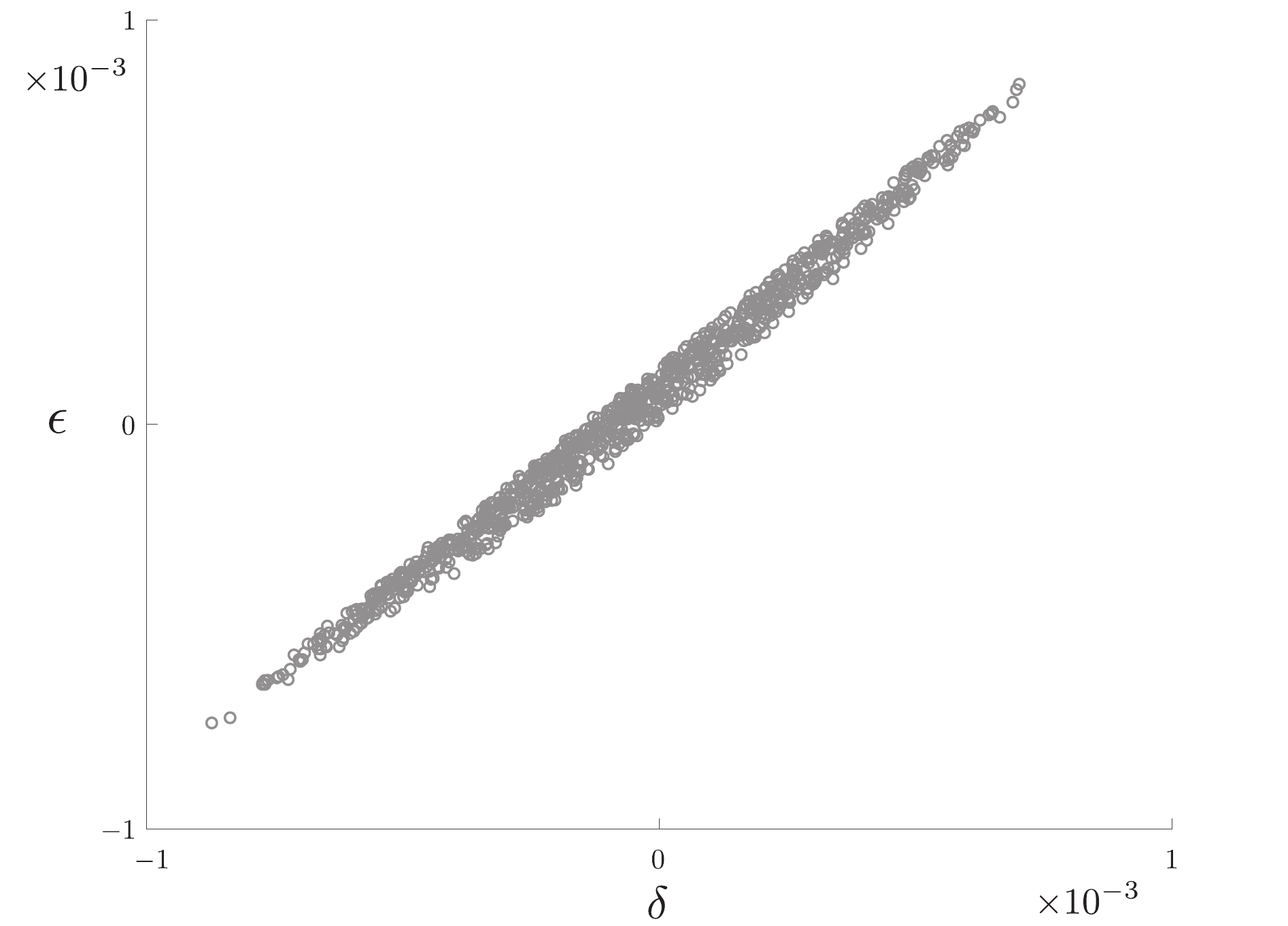}
  \caption{\footnotesize{$RSD_{\lambda}>20\,\%$}}
  \label{fig:rel_deleps_ms_lvb}
\end{subfigure}
\caption{\small{Cross-plots of $\delta$ versus $\epsilon$ presented for $1000$ examples of equivalent TI mafic media with $RSD_{\mu}<2\,\%\,$.}}
\label{fig:rel_deleps_ms}
\end{figure}

For the case of near-constant, moderately varying, and strongly varying $\lambda$, cross-plots of $\varphi$ versus $\epsilon$, are exhibited in Figures~\ref{fig:rel_phieps_ms_ls},~\ref{fig:rel_phieps_ms_2l20} and~\ref{fig:rel_phieps_ms_lvb}, respectively.
$\delta$ versus $\epsilon$ is shown in Figures~\ref{fig:rel_deleps_ms_ls},~\ref{fig:rel_deleps_ms_2l20} and~\ref{fig:rel_deleps_ms_lvb}.
Dominants and ranges of $\varphi$, $\epsilon$ and $\delta$, are represented in Table~\ref{tab:ran_ms}. 
The percentages of MC examples in which certain relations among the anisotropy parameters occur, are presented in Table~\ref{tab:rel_ms}.

Let us analyze Figure~\ref{fig:rel_phieps_ms} and Table~\ref{tab:ran_ms}.
Similarly to Section~\ref{sec:var}, the absolute values of $\varphi$, $\epsilon$ and $\delta$ grow along with the strength of variations of $\lambda$.
Again, the range of $\varphi$ for moderate and strong variations is around ten times larger as compared to the near-constant case.
Perhaps surprisingly, the ranges of $\epsilon$ and $\delta$ for $RSD_{\lambda}>2\,\%$ are also around ten times larger.
%We notice that the range for $RSD_{\mu}<2\,\%$ and $RSD_{\lambda}>20\,\%$ is very similar to its range for $RSD_{\mu}>2\,\%$ and $RSD_{\lambda}<2\,\%\,$.
However, they are still smaller than the one of $\varphi$, which is reflected in the percentage of occurrence of $|\varphi|>|\epsilon|$ and $|\varphi|>|\delta|$.
Recalling Table~\ref{tab:ran_mb}, we notice that the anisotropy parameters have larger values in the case of $RSD_{\mu}>2\,\%$\,, than in the case of $RSD_{\mu}<2\,\%$\,.

Analyzing Table~\ref{tab:rel_ms} in the context of the left-hand side of the pattern from Table~\ref{tab:pattern}, we see that it is true that for the near-constant $\lambda$, in most of MC examples $|\varphi|<|\epsilon|$, and for stronger variations of $\lambda$, we mostly have $|\varphi|>|\epsilon|$.
However, the method from the pattern seems to be quite inaccurate, since in the case of near-constant $\lambda$, the percentage of occurrence of $|\varphi|>|\epsilon|$ is $34.3\,\%$, thus, it is quite high.
If $|\varphi|>|\epsilon|$, there is large probability that we encounter case of $RSD_{\lambda}>2\,\%\,$ (but, we are not sure, as it is the case in Section~\ref{sec:var}).
Another indicator is the relation, $\varphi>\epsilon$, which occurs very rarely in the case of near-constant $\lambda$, but is quite probable to appear in the case of moderate or strong variations of $\lambda$.
In analogous way, we may treat $|\varphi|>|\delta|$ and $\varphi<\delta$, which also seem to be good indicators of the change of fluid content in layered Earth.
In general, as in Section~\ref{sec:var}, the percentage of occurrence of $\varphi>\epsilon$ and $|\varphi|>|\epsilon|$ (also $\varphi<\delta$ and $|\varphi|>|\delta|$) grows along with the strength of variations of $\lambda$. 
Based of Figures~\ref{fig:rel_phieps_ms_2l20} and \ref{fig:rel_phieps_ms_lvb}, we notice that, in the case of $RSD_{\lambda}>2\,\%\,$, $\varphi$ and $\epsilon$ have similar absolute values.
Also, only for $RSD_{\lambda}>2\,\%\,$, both $|\varphi|$ and $|\epsilon|$ have values $>10^{-4}$; that might be another fluid indicator. 

Let us also analyze Figure~\ref{fig:rel_deleps_ms}.
We see that small positive values of $\epsilon$ and $\delta$, occur regardless the strength of variations of $\lambda$, thus, it is not a good indicator, and, in the case of near-constant rigidity, \citeauthor{Berrymanetal} method fails.
As compared to the general case from Section~\ref{sec:general}, $\epsilon$, is more likely to have negative values, and $\delta$ to have positive values.
Also, we notice an interesting relation between $\epsilon$ and $\delta$---they always have very similar values (of course $\epsilon>\delta$).
This property does not appear in Sections~\ref{sec:general} and~\ref{sec:var}.
Absolute very similar values of $\epsilon$ and $\delta$ that are larger than $10^{-4}$ might be a very good detector of the fluid in layered Earth.
The aforementioned property, obvious condition of $\epsilon>\delta$, and their distributions (discussed in Appendix~\ref{sec:dis}), result in a percentage of occurrence of $|\delta|>|\epsilon|$ to be around $50\,\%\,$.

\begin{table}[!htbp]
\begin{tabular}{ccccccc}
\toprule
& \multicolumn{2}{c}{$RSD_{\lambda}<\,2\%$} & \multicolumn{2}{c}{$RSD_{\lambda}\in\left(2\,\%\,,\,20\,\%\right)$} & \multicolumn{2}{c}{$RSD_{\lambda}>\,20\%$} \\
\midrule 
&&&&&&\\[-0.3in]
& \hphantom{X} {\small{dominant}}  & {\small{range}} & \hphantom{X} {\small{dominant}}  & {\small{range}} & \hphantom{X} {\small{dominant}}  & {\small{range}} \\ [-0.06in]
& \hphantom{X}{\scriptsize{$\left[\cdot\,10^{-5}\right]$}} & {\scriptsize{$\left[\cdot\,10^{-3}\right]$}} & \hphantom{X}{\scriptsize{$\left[\cdot\,10^{-5}\right]$}} & {\scriptsize{$\left[\cdot\,10^{-3}\right]$}} & \hphantom{X}{\scriptsize{$\left[\cdot\,10^{-5}\right]$}} & {\scriptsize{$\left[\cdot\,10^{-3}\right]$}}\\ [-0.02in]
\midrule
$\varphi$ \hphantom{X}  &\hphantom{X}$0.227$& $\left(-0.081\,,\,0.075\right)$&\hphantom{X}$-13.17$&$\left(-0.842\,,\,0.926\right)$&$-8.566$&$\left(-1.091\,,\,1.169\right)$ \\
$\epsilon$ \hphantom{X} &\hphantom{X}$0.916$&$\left(-0.024\,,\,0.105\right)$&\hphantom{X}$-4.327$&$\left(-0.569\,,\,0.721\right)$&$-3.503$&$\left(-0.738\,,\,0.841\right)$ \\
$\delta$ \hphantom{X}  &$-0.892$&$\left(-0.125\,,\,0.016\right)$&\hphantom{X}$-10.17$&$\left(-0.691\,,\,0.588\right)$&$-21.79$&$\left(-0.874\,,\,0.702\right)$  \\
\bottomrule
\end{tabular}
\caption{\small{Dominants and ranges of anisotropy parameters relevant to layered mafic rocks with $RSD_{\mu}<2\,\%\,$.}}
\label{tab:ran_ms}
\end{table}
\begin{table}[!htbp]
\begin{tabular}{ccccccc}
\toprule
&$\varphi>\epsilon$&$\varphi<\delta$&$|\varphi|>|\epsilon|$&$|\varphi|>|\delta|$&$|\delta|>|\epsilon|$\\
\midrule
$RSD_{\lambda}<\,2\%$\hphantom{X}&$2.00$&$2.30$&$34.3$&$33.4$&$53.8$\\
$RSD_{\lambda}\in\left(2\,\%\,,\,20\,\%\right)$\hphantom{X}&$32.4$&$32.1$&$80.5$&$77.2$&$51.3$\\
$RSD_{\lambda}>\,20\%$\hphantom{X}&$37.3$&$40.6$&$87.2$&$86.1$&$52.2$\\
\bottomrule
\end{tabular}
\caption{\small{Percentage of occurrence of relations among $\varphi$, $\epsilon$ and $\delta$ relevant to layered mafic rocks with $RSD_{\mu}<2\,\%\,$.}}
\label{tab:rel_ms}
\end{table}

In conclusion, we have found four indicators that mostly occur in the case of $RSD_{\lambda}>2\,\%\,$ ($\gamma<1.6\cdot10^{-4}$), namely, $\varphi>\epsilon$, $|\varphi|>|\epsilon|$, $\varphi<\delta$ and $|\varphi|>|\delta|$.
Another two indicators, which occur only in the case of $RSD_{\lambda}>2\,\%\,$, are $|\epsilon|\approx|\delta|>10^{-4}$ and $|\varphi|>10^{-4}$.
We propose to treat the indicator $|\varphi|>10^{-4}$ separately, since it is more sensitive to variations of $\lambda$ than $\epsilon$ or $\delta$.
The usage of these fluid indicators is exemplified in Appendix~\ref{sec:muconst_scale}.
The exact percentages of occurrence of the aforementioned indicators, for the case of $RSD_{\mu}<2\,\%$, are presented in Table~\ref{tab:bigtable2}.
Based on that table, in the near-constant rigidity case, for mafic rocks, $|\epsilon|\approx|\delta|>10^{-4}$, and, $|\varphi|>10^{-4}$, occur to be the most effective in detecting fluids. 
It is important to notice that certain indicators that are valid for $RSD_{\mu}>2\,\%\,$, namely, $\epsilon<0$ or $\delta>0$, are less efficient for $RSD_{\mu}<2\,\%\,$.
Let us indicate the most important points stated in this section, namely:
\begin{itemize}
\item{the larger the variations of $\lambda$, the larger the values of $\varphi$, $\epsilon$ and $\delta$,}
\item{compared to $RSD_{\mu}>2\,\%$ case, $\varphi$, $\epsilon$ and $\delta$, have much smaller values,}
\item{compared to general case, $\epsilon$, is more likely to have negative values,}
\item{compared to general case, $\delta$, is more likely to have positive values,}
\item{the probability of $\epsilon$ and $\delta$ having negative or positive values is more or less the same, thus, \citeauthor{Berrymanetal} method is inaccurate,}
\item{$\epsilon$ is larger than $\delta$, but they always have very similar values,}
\item{$\varphi>\epsilon$, $\varphi<\delta$, $|\varphi|>|\epsilon|$, $|\varphi|>|\delta|$, $|\epsilon|\approx|\delta|>10^{-4}$, and $|\varphi|>10^{-4}$, are the possible fluid indicators for mafic rocks.}
\end{itemize}
%%%%%%%%%%%%%%%%%%%%%%%%%%%%%%%%%%%%%%%%%%%%%%%%%%%%%%%%%%%%%%%%%%%%%%%%%%%%%%%%%%%%%%%%
 \section{Conclusions}\label{sec:conclusions}
%%%%%%%%%%%%%%%%%%%%%%%%%%%%%%%%%%%%%%%%%%%%%%%%%%%%%%%%%%%%%%%%%%%%%%%%%%%%%%%%%%%%%%%%
We conclude this paper by indicating general rules that govern the relations among the anisotropy parameters, $\varphi$, $\epsilon$ and $\delta$, describing the induced anisotropy of layered media. 
These rules are valid for the Lam\'e parameters relevant to various types of rocks (mafic rocks, felsic rocks, sandstones, certain basalts, limestones, and many others), verified by us; thus, we take into consideration also the content from the appendices.
In general ($RSD_{\mu}\in\forall$ and $RSD_{\lambda}\in\forall$),
\begin{itemize}
\item{$\epsilon$ mostly has positive values (see Figures~\ref{fig:rel_ml_phieps} and~\ref{fig:diseps}),}
\item{$\delta$ mostly has negative values (see Figures~\ref{fig:rel_ml_deleps} and~\ref{fig:disdel}),}
\item{the probability of $\varphi$ being negative or positive is more or less equal (see Figures~\ref{fig:rel_ml_phieps} and~\ref{fig:disphi}),}
\item{the probability of $|\delta|>|\epsilon|$ or $|\delta|<|\epsilon|$ is more or less equal (see Table~\ref{subtab:b}),}
\item{$\varphi$ mostly has smaller absolute values than $\epsilon$ or $\delta$ (see Table~\ref{subtab:b}),}
\item{the larger variations of $\lambda$ are, the larger values $\varphi$, $\epsilon$ and $\delta$ have (see Figures~\ref{fig:rel_phieps_mb}--\ref{fig:rel_deleps_mb} and~\ref{fig:rel_phieps_ms}--\ref{fig:rel_deleps_ms}),}
\item{the larger variations of $\lambda$ are, the larger efficiency of fluid indicators (see Table~\ref{tab:bigtable2}),}
\item{fluid indicators are less efficient for stronger variations of $\mu$ (see Table~\ref{tab:bigtable2}).}
\end{itemize}
First five points are the conclusions of Section~\ref{sec:general}, the sixth one is the conclusion of Sections~\ref{sec:var} and~\ref{sec:mu}, while the last two points are based on the appendices.
In the case of non near-constant both $\mu$ and $\lambda$ ($RSD_{\mu}>2\,\%$ and $RSD_{\lambda}>2\,\%$), thus, of possible change of fluid content in layered Earth:
\begin{itemize}
\item{$\delta$ mostly have negative values, thus, \citeauthor{Berrymanetal} method is inaccurate (see Figure~\ref{fig:rel_deleps_mb}),}
\item{$\varphi$ is very sensitive on variations of $\lambda$, as opposed to $\epsilon$ and $\delta$ (see Table~\ref{tab:ran_mb}),}
\item{$\varphi>5\cdot10^{-3}$ always indicates the variations of $\lambda$ (see Table~\ref{tab:bigtable2}),}
\end{itemize}
which are the conclusions from Section~\ref{sec:var} (the last point is slightly modified due to taking into consideration various rocks, not only mafic ones).
In the case of near-constant rigidity and non near-constant $\lambda$ in thin layers ($RSD_{\mu}<2\,\%$ and $RSD_{\lambda}>2\,\%$), thus, of possible change of fluid content in layered Earth:
\begin{itemize}
\item{compared to $RSD_{\mu}>2\,\%$ case, $\varphi$, $\epsilon$ and $\delta$, have much smaller values (compare Tables~\ref{tab:ran_mb} and~\ref{tab:ran_ms}),}
\item{compared to general case, $\epsilon$, is more likely to have negative values (compare Figures~\ref{fig:rel_deleps_mb} and~\ref{fig:rel_deleps_ms}),}
\item{compared to general case, $\delta$, is more likely to have positive values (compare Figures~\ref{fig:rel_deleps_mb} and~\ref{fig:rel_deleps_ms}),}
\item{the probability of $\epsilon$ and $\delta$ having negative or positive values is more or less the same, thus, \citeauthor{Berrymanetal} method is inaccurate (see Figure~\ref{fig:rel_deleps_ms}),}
\item{$\epsilon$ is larger than $\delta$, but they always have very similar values (see Figure~\ref{fig:rel_deleps_ms}),}
\end{itemize}
which are the conclusions from Section~\ref{sec:mu}.
  
The indicated relations may be useful, for instance, in the inverse problems, where we only know the elasticity parameters of TI media, and we want to estimate the variations of $\lambda$.
The \citeauthor{Berrymanetal} method occurs to be at least inaccurate. 
The pattern from Table~\ref{tab:pattern} is correct, however, we propose a new, universal pattern of fluid detection in TI media that is exposed in a summary Table~\ref{tab:summary}.
In non near-constant rigidity case, we propose to use five indicators to increase chances of detecting fluids, namely, $|\varphi|>5\cdot10^{-3}$, $|\varphi|>|\epsilon|$, $|\varphi|>|\delta|$, $\epsilon<0$ and $\delta>0$.
In near-constant rigidity case, we propose to check following relations, $\varphi>\epsilon$, $\varphi<\delta$, $|\varphi|>|\epsilon|$, $|\varphi|>|\delta|$, $|\epsilon|\approx|\delta|>10^{-4}$, and $|\varphi|>10^{-4}$.
The occurrence of this relations give us very large probability of detection the change of fluid content in layered Earth.
Nevertheless, the lack of occurrence, does not necessarily mean that there are no variations of $\lambda$ in layers.
\renewcommand{\arraystretch}{1.4}
\begin{table}[!htbp]
\centering
\begin{tabular}{cccccc} 
\toprule
\multicolumn{3}{c}{$RSD_{\mu}<2\,\%$ \, $(\approx\gamma<1.5\cdot10^{-4})$}& \multicolumn{3}{c}{$RSD_{\mu}>2\,\%$ \, $(\approx\gamma>1.5\cdot10^{-4})$}\\
\midrule
$\varphi>\epsilon$  \hphantom{X}  &  $|\varphi|>|\epsilon|$   \hphantom{X}   &      $|\varphi|>10^{-4}$          \hphantom{xxxx}  &   $|\varphi|>|\epsilon|$     \hphantom{X} &     $\epsilon<0$     \hphantom{X} &       $|\varphi|>5\cdot10^{-3}$  \\
$\varphi<\delta$    \hphantom{X}   &   $|\varphi|>|\delta|$       \hphantom{X}  &     $|\epsilon|\approx|\delta|>10^{-4}$      \hphantom{xxxx}  &  \multicolumn{3}{c}{  $|\varphi|>|\delta|$  \hphantom{X}         $\delta>0$      \hphantom{X}     \hphantom{X}} \\
\bottomrule
\end{tabular}
\caption{\small{Universal pattern of relations indicating fluids, valid for various layered rocks.}}
\label{tab:summary}
\end{table}

Based on the appendices, we notice that fluid indicators are less efficient in the case of stronger variations of $\mu$.
Distributions of $\epsilon$ and $\delta$ become more skewed and have less negative and less positive values, respectively.
It causes the decrease in efficiency of indicators $\epsilon<0$ and $\delta>0$.
Also, ranges of $\epsilon$ and $\delta$ become larger in comparison to the range of $\varphi$, since they are more sensitive to variations of $\mu$.
As a result, relations, $|\varphi|>|\epsilon|$, $|\varphi|>|\delta|$ and $\varphi>\epsilon$, $\varphi<\delta$, are less effective.
The only indicator that always is effective, is the absolute value of $\varphi$.
Depending on the growing variations of $\mu$, we should verify $|\varphi|>10^{-4}$ (relevant to $RSD_{\mu}<2\,\%$ for all type of rocks) , $|\varphi|>0.5\cdot10^{-3}$ (the most effective for felsic rocks), $|\varphi|>10^{-3}$ (the most effective for mafic rocks), and for the strongest variations, $|\varphi|>5\cdot10^{-3}$ (the most effective for sandstones).
Absolute values of $\epsilon$ or $\delta$ are also less efficient indicators, since they are not as sensitive to variations of $\lambda$ as $|\varphi|$ is.
We propose simplified method of fluid detection, which is based on the most effective fluid indicator $|\varphi|$, as shown in Table~\ref{tab:summary_simplified}.
\renewcommand{\arraystretch}{1.4}
\begin{table}[!htbp]
\centering
\begin{tabular}{ccc} 
\toprule
&$RSD_{\mu}<2\,\%$ & $RSD_{\mu}>2\,\%$\\
\midrule
mafic rocks &    $|\varphi|>10^{-4}$   & $|\varphi|>10^{-3}$ \\
felsic rocks &     $|\varphi|>10^{-4}$   & $|\varphi|>0.5\cdot10^{-3}$\\
sandstones &    $|\varphi|>10^{-4}$   & $|\varphi|>5\cdot10^{-3}$ \\
\bottomrule
\end{tabular}
\caption{\small{Simplified pattern of fluid detection using $|\varphi|$.}}
\label{tab:summary_simplified}
\end{table}

In conclusion, all of the fluid indicators should be examined in the future, using real data.
Probably the most efficient in detecting fluids is the absolute value of $\varphi$ that is very sensitive to variations of $\lambda$ in layers.
A comprehensive table containing values of $\varphi$, $\epsilon$ and $\delta$ that are probable to occur in layered mafic rocks, felsic rocks or sandstones, is shown in Appendix~\ref{ap:two}.
Useful relations among anisotropy parameters, along with probabilities of occurrence of these relations in layered mafic rocks, felsic rocks or sandstones, are shown in a summary table in Appendix~\ref{ap:three}.
%%%%%%%%%%%%%%%%%%%%%%%%%%%%%%%%%%%%%%%%%%%%%%%%%%%%%%%%%%%%%%%%%%%%%%%%%%%%%%%%%%%%%%%%
%%%%%%%%%%%%%%%%%%%%%%%%%%%%%%%%%%%%%%%%%%%%%%%%%%%%%%%%%%%%%%%%%%%%%%%%%%%%%%%%%%%%%%%%
\section*{Acknowledgements}
%%%%%%%%%%%%%%%%%%%%%%%%%%%%%%%%%%%%%%%%%%%%%%%%%%%%%%%%%%%%%%%%%%%%%%%%%%%%%%%%%%%%%%%%
We wish to acknowledge discussions with supervisor Michael A. Slawinski, and consultations with Tomasz Danek and Izabela Kudela.
Also, we thank Theodore Stanoev for the editorial work along with fruitful discussions, and Elena Patarini for the graphical support.
This research was performed in the context of The Geomechanics Project supported by Husky Energy. 
%%%%%%%%%%%%%%%%%%%%%%%%%%%%%%%%%%%%%%%%%%%%%%%%%%%%%%%%%%%%%%%%%%%%%%%%%%%%%%%%%%%%%%%%
%%%%%%%%%%%%%%%%%%%%%%%%%%%%%%%%%%%%%%%%%%%%%%%%%%%%%%%%%%%%%%%%%%%%%%%%%%%%%%%%%%%%%%%%
\bibliographystyle{apa}
\bibliography{bibliography}
%%%%%%%%%%%%%%%%%%%%%%%%%%%%%%%%%%%%%%%%%%%%%%%%%%%%%%%%%%%%%%%%%%%%%%%%%%%%%%%%%%%%%%%%
%%%%%%%%%%%%%%%%%%%%%%%%%%%%%%%%%%%%%%%%%%%%%%%%%%%%%%%%%%%%%%%%%%%%%%%%%%%%%%%%%%%%%%%%
\appendix
\addcontentsline{toc}{section}{Appendices}
%%%%%%%%
\section{Mafic rocks\,: further analysis}
\subsection{Scale factor examples for $RSD_{\mu}>2\,\%$} \label{sec:muvar_scale}
%%%%%%%%%%%%%%%%%%%%%%%%%%%%%%%%%%%%%%%%%%%%%%%%%%%%%%%%%%%%%%%%%%%%%%%%%%%%%%%%%%%%%%%
Let us exemplify the usefulness of fluid indicators from Section~\ref{sec:var} by analyzing the numerical, non-random examples.
To do so, we choose five layers and their initial five Lam\'e parameters from the range relevant to mafic rocks; while $\lambda$ has the same value in each layer.
Subsequently, we use scale factor, $x$, to increase the variation of $\lambda$, whereas $\mu$ is not influenced by $x$, and does not change.
The \citeauthor{Backus} average, and then $\varphi$, $\epsilon$ and $\delta$, are computed for each value of $x$.
Along with growing variations of $\lambda$, we observe the tendency of change of anisotropy parameters, and we verify at which strength of variations of $\lambda$---thus, at which value of $RSD_{\lambda}$---the main fluid indicators from Section~\ref{sec:var}, namely, $|\varphi|>10^{-3}$, $|\varphi|>|\epsilon|$, $|\varphi|>|\delta|$, $\epsilon<0$ and $\delta>0$, are satisfied.

\begin{figure}[!htbp]
\begin{floatrow}
\capbtabbox{
    \begin{tabular}{ccc} 
\toprule 
$\lambda$\,\scriptsize{[GPa]} & \hphantom{x} & $\mu$\,\scriptsize{[GPa]}\\ 
\toprule 
$50\,x$ && 50.2 \\ 
50 && 44.5 \\ 
$50\,x$ && 46.2 \\ 
50 && 39.9 \\ 
$50\,x$ && 42.9 \\ 
\bottomrule
$RSD_{\mu}=17.14\,\%$ & \hphantom{x}&$\gamma=2.922\cdot10^{-3}\,$ \\
 \end{tabular}
}{
\captionsetup{width=0.45\textwidth}
  \caption{\small{Five layers relevant to mafic rocks with variations of $\mu$ and growing variations of $\lambda$. Case I; increasing anisotropy parameters, water saturation.}}
  \label{tab:mb_inc}
  }
\qquad
  \ffigbox{%
  \includegraphics[scale=0.35]{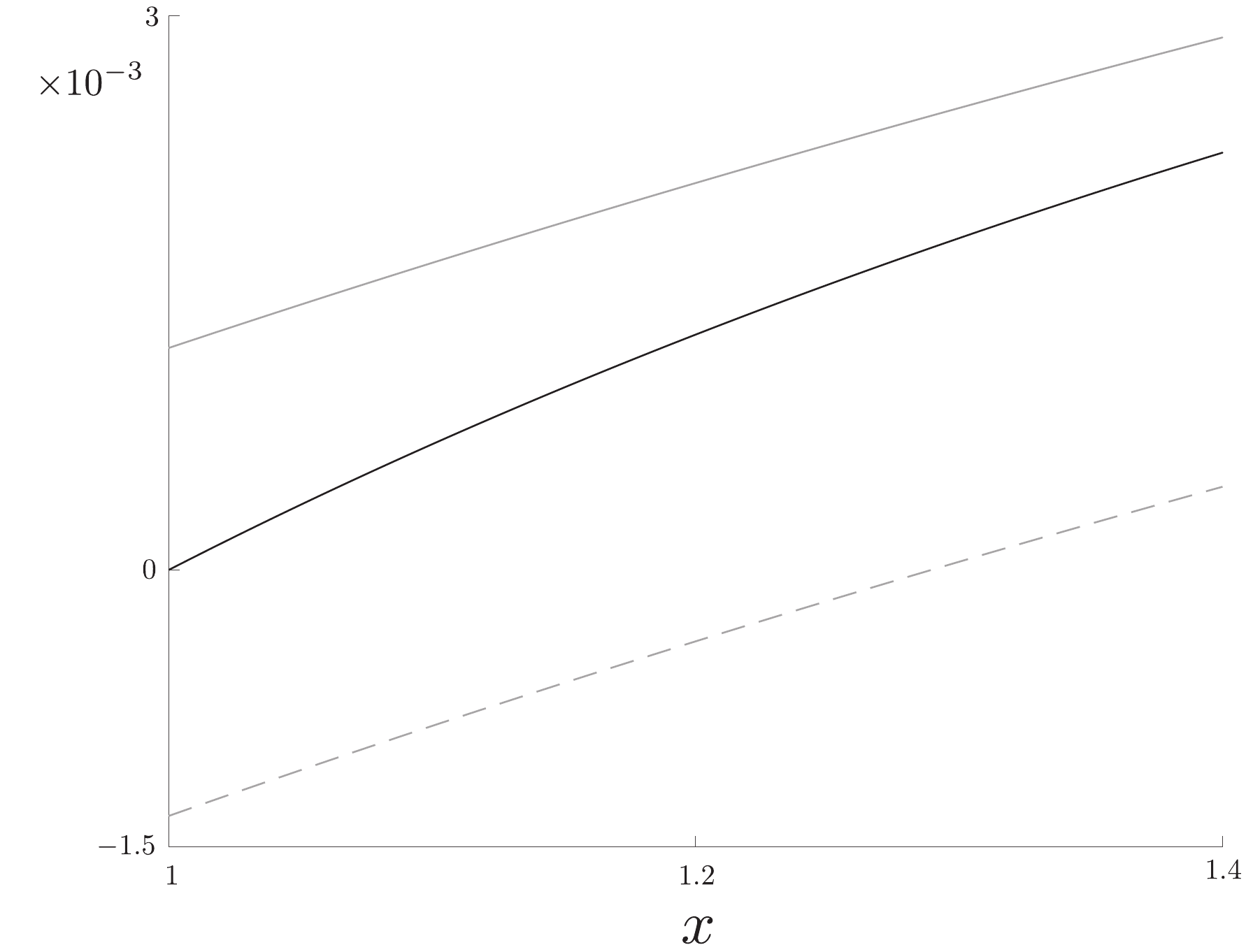}%
}{%
\captionsetup{width=0.45\textwidth}
  \caption{\small{Horizontal axis exhibits values of $x$. Values of $\varphi$ are shown by a black line, $\epsilon$ by a grey line, $\delta$ by a dashed grey line.}}%
\label{fig:mb_inc}}
\end{floatrow}
\end{figure}

\begin{figure}[!htbp]
\begin{floatrow}
\capbtabbox{
    \begin{tabular}{ccc} 
\toprule 
$\lambda$\,\scriptsize{[GPa]}& \hphantom{x}&$\mu$\,\scriptsize{[GPa]} \\ 
\toprule 
$50\,(1\,/\,x)$ && 50.2 \\ 
50 && 44.5 \\ 
$50\,(1\,/\,x)$ && 46.2 \\ 
50 && 45.0 \\ 
$50\,(1\,/\,x)$ && 49.0 \\ 
\bottomrule
$RSD_{\mu}=10.67\,\%$ & \hphantom{x}& $\gamma=1.125\cdot10^{-3}\,$\\
 \end{tabular}
}{
\captionsetup{width=0.45\textwidth}
  \caption{\small{Five layers relevant to mafic rocks with variations of $\mu$ and growing variations of $\lambda$. Case II; decreasing anisotropy parameters, gas saturation.}}
  \label{tab:mb_dec}
  }
\qquad
  \ffigbox{%
  \includegraphics[scale=0.35]{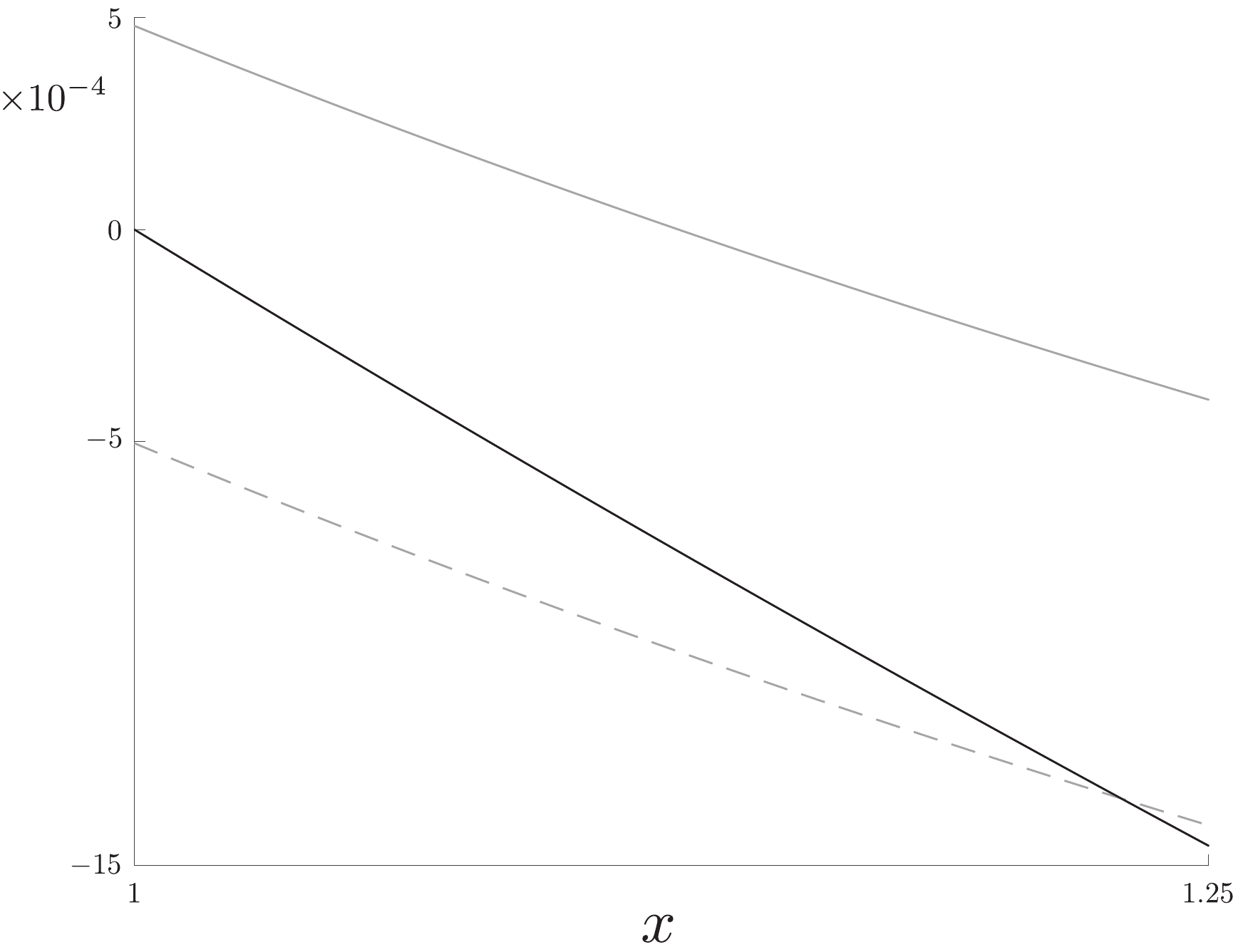}%
}{%
\captionsetup{width=0.45\textwidth}
  \caption{\small{Horizontal axis exhibits values of $x$. Values of $\varphi$ are shown by a black line, $\epsilon$ by a grey line, $\delta$ by a dashed grey line.}}%
\label{fig:mb_dec}}
\end{floatrow}
\end{figure}

\begin{figure}[!htbp]
\begin{floatrow}
\capbtabbox{
    \begin{tabular}{ccc} 
\toprule 
$\lambda$\,\scriptsize{[GPa]}& \hphantom{x}&$\mu$\,\scriptsize{[GPa]} \\ 
\toprule 
$50\,(1\,/\,x)$ && 50.2 \\ 
50 && 44.5 \\ 
$50\,x$ && 46.2 \\ 
50 && 40.9 \\ 
$50\,x$ && 50.9 \\ 
\bottomrule
$RSD_{\mu}=17.78\,\%$& \hphantom{x}&$\gamma=3.253\cdot10^{-3}\,$\\
 \end{tabular}
}{
\captionsetup{width=0.45\textwidth}
  \caption{\small{Five layers relevant to mafic rocks with variations of $\mu$ and growing variations of $\lambda$. Case III; near-constant anisotropy parameters, gas and water saturation.}}
  \label{tab:mb_eq}
  }
\qquad
  \ffigbox{%
  \includegraphics[scale=0.35]{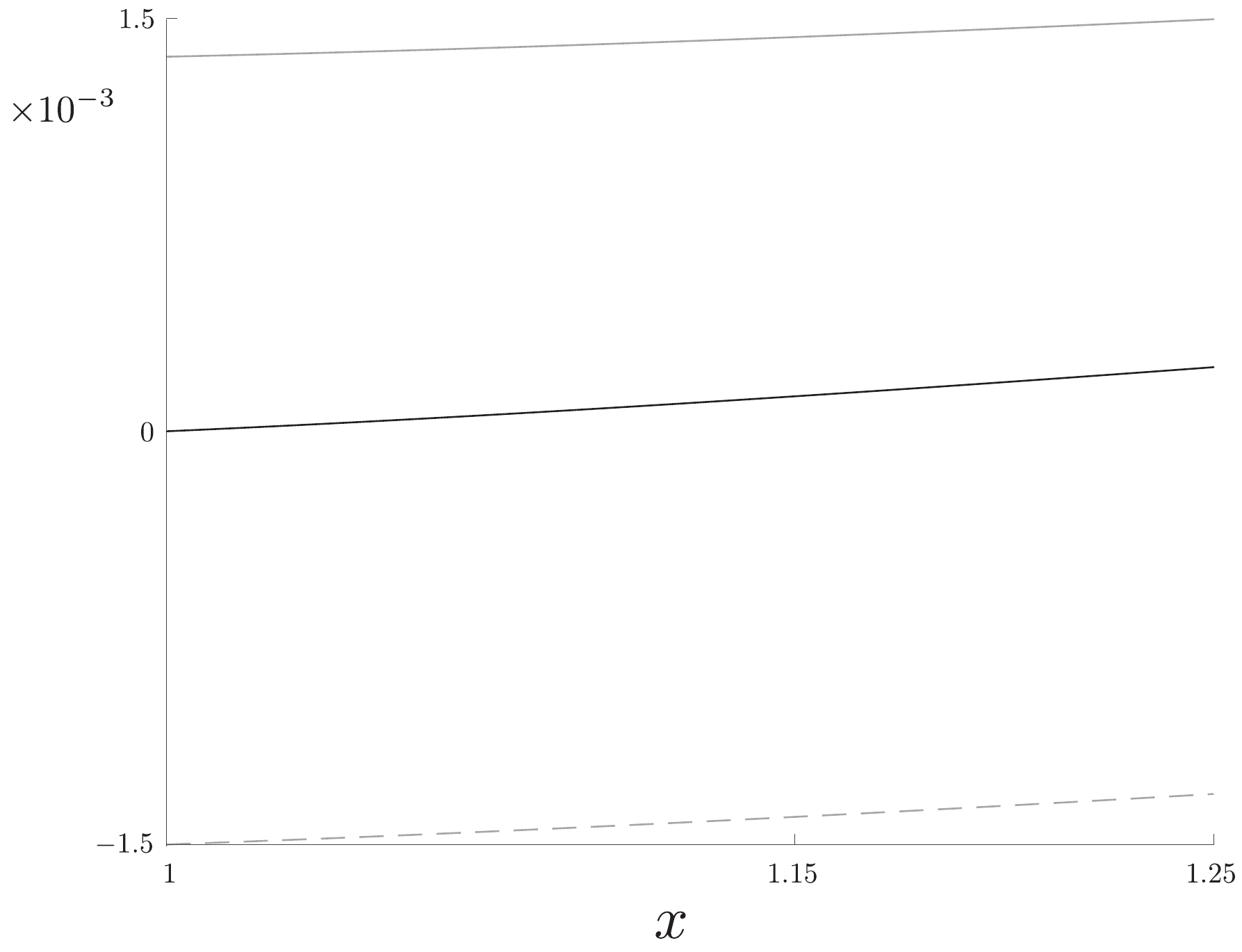}%
}{%
\captionsetup{width=0.45\textwidth}
  \caption{\small{Horizontal axis exhibits values of $x$. Values of $\varphi$ are shown by a black line, $\epsilon$ by a grey line, $\delta$ by a dashed grey line.}}%
\label{fig:mb_eq}}
\end{floatrow}
\end{figure}

We examine three tendencies of changing anisotropy parameters; increasing, decreasing and near-constant, respectively. 
Also, we verify three different behaviours of $\lambda$; we check its growth in layers, decrease, or growth along with decrease.
In each example the second and forth layer has constant $\lambda$, while in the rest of layers, $\lambda$, is influenced by the scale factor.
In general, rocks filled with water have larger $\lambda$ then the same rocks filled with gas, while $\mu$ remain constant \citep{Goodway}.
Thus, the growth of $\lambda$ may exhibit the situation of saturating the rocks by water, while its decrease could refer to the gas saturation.
We try not to limit the analysis to only of one type, for instance to growth of $\lambda$ and increase of anisotropy parameters.
Finally, we want to show that in certain examples the moderate or strong variations of $\lambda$ are detected by the aforementioned fluid indicators, while in the other examples they are not detected.
Tables~\ref{tab:mb_inc}, \ref{tab:mb_dec} and \ref{tab:mb_eq}, show three cases of different elasticity parameters in layers.
The changing values of $\varphi$, $\epsilon$ and $\delta$ that are influenced by the scale factor, $x$, are exhibited in Figures~\ref{fig:mb_inc}, \ref{fig:mb_dec} and \ref{fig:mb_eq}, respectively.

To consider the elasticity parameters relevant to mafic rocks $(\lambda\in(40\,,\,70)\,,\,\mu\in(35\,,\,60)\,)$, in Figure~\ref{fig:mb_inc}, we verify only $x\in(1\,,\,1.4)$.
In the second and third case from Figure~\ref{fig:mb_dec} and Figure~\ref{fig:mb_eq}, due to decrease of $\lambda$, thus, $1\,/\,x$, we consider only $x\in(1\,,\,1.25)$.
In each example, we analyze the situation of $\gamma>10^{-3}$, hence, the situation of non near-constant rigidity.
In Table~\ref{tab:mb_inc}, the maximum variations of $\lambda$ (for $x=1.4$) are $RSD_{\lambda}=15.80\,\%$, in Table~\ref{tab:mb_dec} (for $x=1.25$), $RSD_{\lambda}=11.13\,\%$, and in Table~\ref{tab:mb_eq} (for $x=1.25$), $RSD_{\lambda}=16.18\,\%\,$.
We notice that due to smaller variations of $\mu$ and $\lambda$, as in Case II from Table~\ref{tab:mb_dec}, parameters $\varphi$, $\epsilon$ and $\delta$, have also smaller values; as compared to the ones from Cases I and III.

Certain fluid indicators do detect moderate or strong variations of $\lambda$ in Case I, but do not detect them in Case II, and vice versa.
In Case III, none of fluid detectors indicate these variations. 
The dependence of fluid indicators on $RSD_{\lambda}$, for each case, is shown in Table~\ref{tab:mb_scale}.
\begin{table}[!htbp]
\begin{tabular}{ccccccc}
\toprule
&                                 $|\varphi|>10^{-3}$     &$|\varphi|>|\epsilon|$    &$|\varphi|>|\delta|$       &$\epsilon<0$                    &$\delta>0$\\
\midrule
Case I\hphantom{X}       &$>6.836\,\%$                &$\in\emptyset$        &$>5.270\,\%$             &$\in\emptyset$             &$>12.09\,\%$     \\
Case II\hphantom{X}      &$>7.736\,\%$                &$>2.288\,\%$           &$>10.31\,\%$             &$>5.910\,\%$               &$\in\emptyset$       \\
Case III\hphantom{X}     &$\in\emptyset$              &$\in\emptyset$         &$\in\emptyset$           &$\in\emptyset$             &$\in\emptyset$           \\
\bottomrule
\end{tabular}
\caption{\small{Occurrence of fluid indicators and its dependence on $RSD_{\lambda}$. Scale factor examples with $RSD_{\mu}>2\,\%\,$.}}
\label{tab:mb_scale}
\end{table}
In Case I, we notice that variations $RSD_{\lambda}<5.2\,\%$ are not detected, as opposed to $RSD_{\lambda}>12.1\,\%$ that are indicated by three fluid detectors.
Remembering that for layered media $\epsilon>\delta$, maximally four out of five indicators may detect the change of the fluid content simultaneously.
The larger the variations, the larger the possibility of detecting fluids.
In Case II, $|\varphi|>|\epsilon|$ is very sensitive, since it recognizes even $RSD_{\lambda}>2.3\,\%$.
Variations $RSD_{\lambda}>10.4\,\%$ are detected by the maximum possible number of fluid indicators.
The last case, confirms the fact that our five indicators do not always detect the variations of $\lambda$.
We should be aware of that deficiency, while using them in the inverse problems, in which we do not know the Lam\'e parameters.

Finally, it is important to notice that not only the saturation (water or gas) is responsible for the tendency of anisotropy parameters's change (growth, decrease or near-constant), but also the choice of $\mu$ in layers.
In each of three cases the tendency was changed due to different scale factors ($x$ or $1\,/\,x$) and different $\mu$ in last two layers.
Nevertheless, there exist cases in which the tendency is changed only because of different choice of scale factor or only because of different choice of $\mu$ (see Appendix~\ref{sec:muconst_scale}).
For instance, decreasing $\lambda$ do not impose decreasing anisotropy parameters as in Figure~\ref{fig:mb_dec}; there exist other examples in which different choice of $\mu$ causes the decrease of $\lambda$ and increase of $\varphi$, $\epsilon$ and $\delta$, but we do not show them herein.
%%%%%%%%%%%%%%%%%%%%%%%%%%%%%%%%%%%%%%%%%%%%%%%%%%%%%%%%%%%%%%%%%%%%%%%%%%%%%%%%%%%%%%%
\subsection{Scale factor examples for $RSD_{\mu}<2\,\%$} \label{sec:muconst_scale}
%%%%%%%%%%%%%%%%%%%%%%%%%%%%%%%%%%%%%%%%%%%%%%%%%%%%%%%%%%%%%%%%%%%%%%%%%%%%%%%%%%%%%%%
Let us again, as in Section~\ref{sec:mu}, analyze near-constant rigidity case of elasticity parameters relevant to mafic rocks.  
Herein, instead of MC method, we examine numerical examples using scale factor, $x$, responsible for the variations of $\lambda$ in layers.
Similarly to Appendix~\ref{sec:muvar_scale}, we consider three cases of variations of $\lambda$ in certain layers, while $\lambda$ in the rest of the layers remains constant.
Again, we analyze cases of growing, decreasing and near-constant values of $\varphi$, $\epsilon$ and $\delta$, which are changing along with growing scale factor.
These tendencies of $\varphi$, $\epsilon$ and $\delta$, may occur for rocks saturated with water, gas or with water and gas, as shown in Appendix~\ref{sec:muvar_scale}.
Herein, in the first two examples, we only consider cases relevant to water saturation, which have different values of $\mu$ in last two layers.
The second and third case have the same $\mu$ in all five layers, but different saturations (water and gas-water).
Each case have different tendency of anisotropy parameters's change.
This way, we can clearly show that the values of anisotropy parameters depend on the change of $\mu$, and also on the change of $\lambda$ (change of scale factor).
We verify the examples also in the context of six fluid indicators mentioned in Section~\ref{sec:mu}, namely, $\varphi>\epsilon$, $\varphi<\delta$, $|\varphi|>|\epsilon|$, $|\varphi|>|\delta|$, $|\epsilon|\approx|\delta|>10^{-4}$ and $|\varphi|>10^{-4}$.

\begin{figure}[!htbp]
\begin{floatrow}
\capbtabbox{
    \begin{tabular}{ccc} 
\toprule 
$\lambda$\,\scriptsize{[GPa]} & \hphantom{x} & $\mu$\,\scriptsize{[GPa]} \\ 
\toprule 
$50\,x$ && 46.8 \\ 
50 && 47.1 \\ 
$50\,x$ && 46.9 \\ 
50 && 45.9 \\ 
$50\,x$ && 46.7 \\ 
\bottomrule
$RSD_{\mu}=1.973\,\%$ &\hphantom{x} & $\gamma=3.931\cdot10^{-5}\,$\\
 \end{tabular}
}{
\captionsetup{width=0.45\textwidth}
  \caption{\small{Five layers relevant to mafic rocks with near-constant $\mu$ and growing variations of $\lambda$. Case I; increasing anisotropy parameters, water saturation.}}
  \label{tab:ms_inc}
  }
\quad\,\,
  \ffigbox{%
  \includegraphics[scale=0.35]{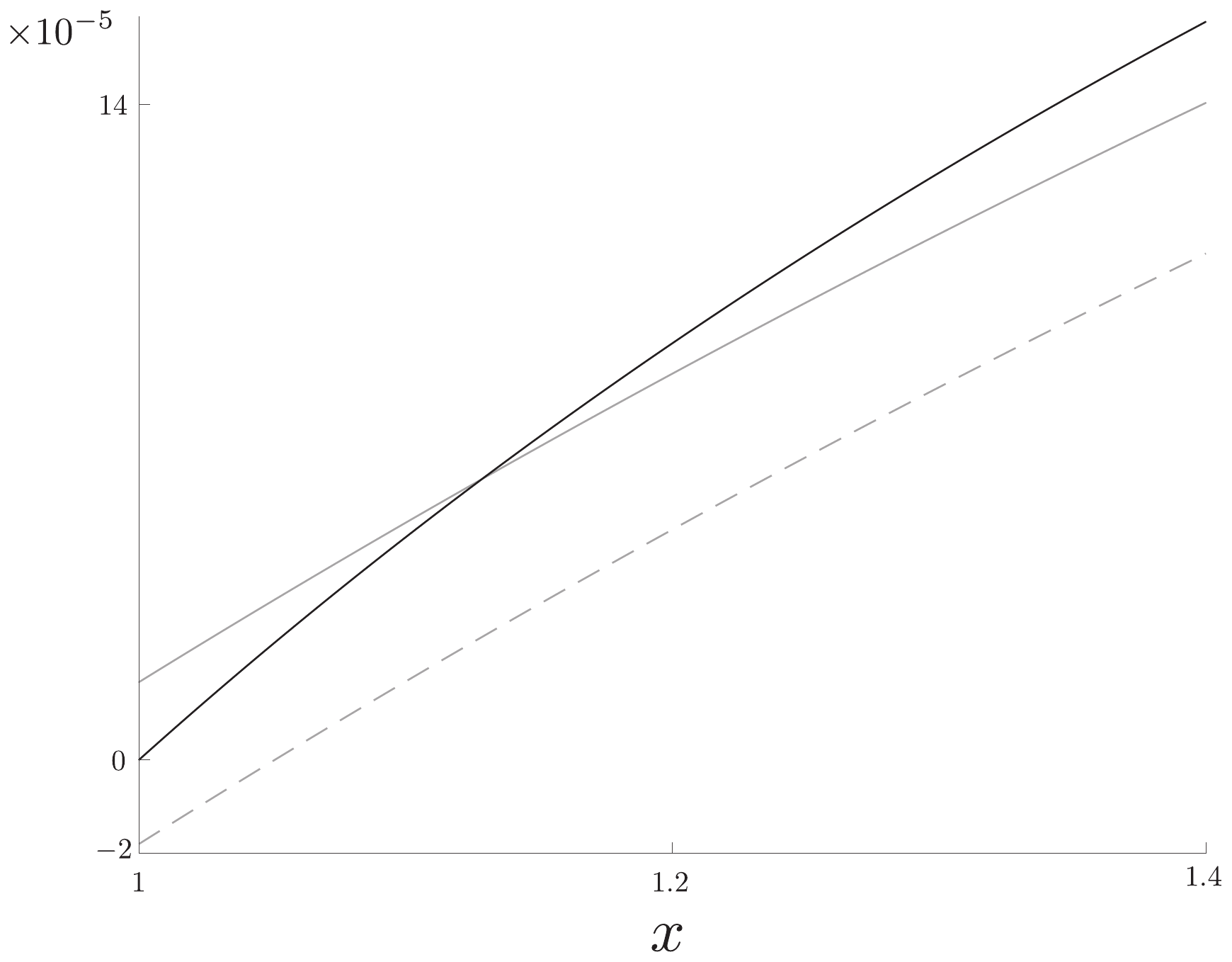}%
}{%
\captionsetup{width=0.45\textwidth}
  \caption{\small{Horizontal axis exhibits values of $x$. Values of $\varphi$ are shown by a black line, $\epsilon$ by a grey line, $\delta$ by a dashed grey line.}}%
\label{fig:ms_inc}}
\end{floatrow}
\end{figure}

\begin{figure}[!htbp]
\begin{floatrow}
\capbtabbox{
  \begin{tabular}{ccc} 
\toprule 
$\lambda$\,\scriptsize{[GPa]}& \hphantom{x} & $\mu$\,\scriptsize{[GPa]} \\ 
\toprule 
$50\,x$ && 46.8 \\ 
50 && 47.1 \\ 
$50\,x$ && 46.9 \\ 
50 && 47.4 \\ 
$50\,x$ && 47.2 \\ 
\bottomrule
$RSD_{\mu}=1.014\,\%$ &\hphantom{x}& $\gamma=1.028\cdot10^{-5}\,$ \\
  \end{tabular}
}{
\captionsetup{width=0.45\textwidth}
  \caption{\small{Five layers relevant to mafic rocks with near-constant $\mu$ and growing variations of $\lambda$. Case II; decreasing anisotropy parameters, water saturation.}}
  \label{tab:ms_dec}
}
\quad\,\,
\ffigbox{%
  \includegraphics[scale=0.35]{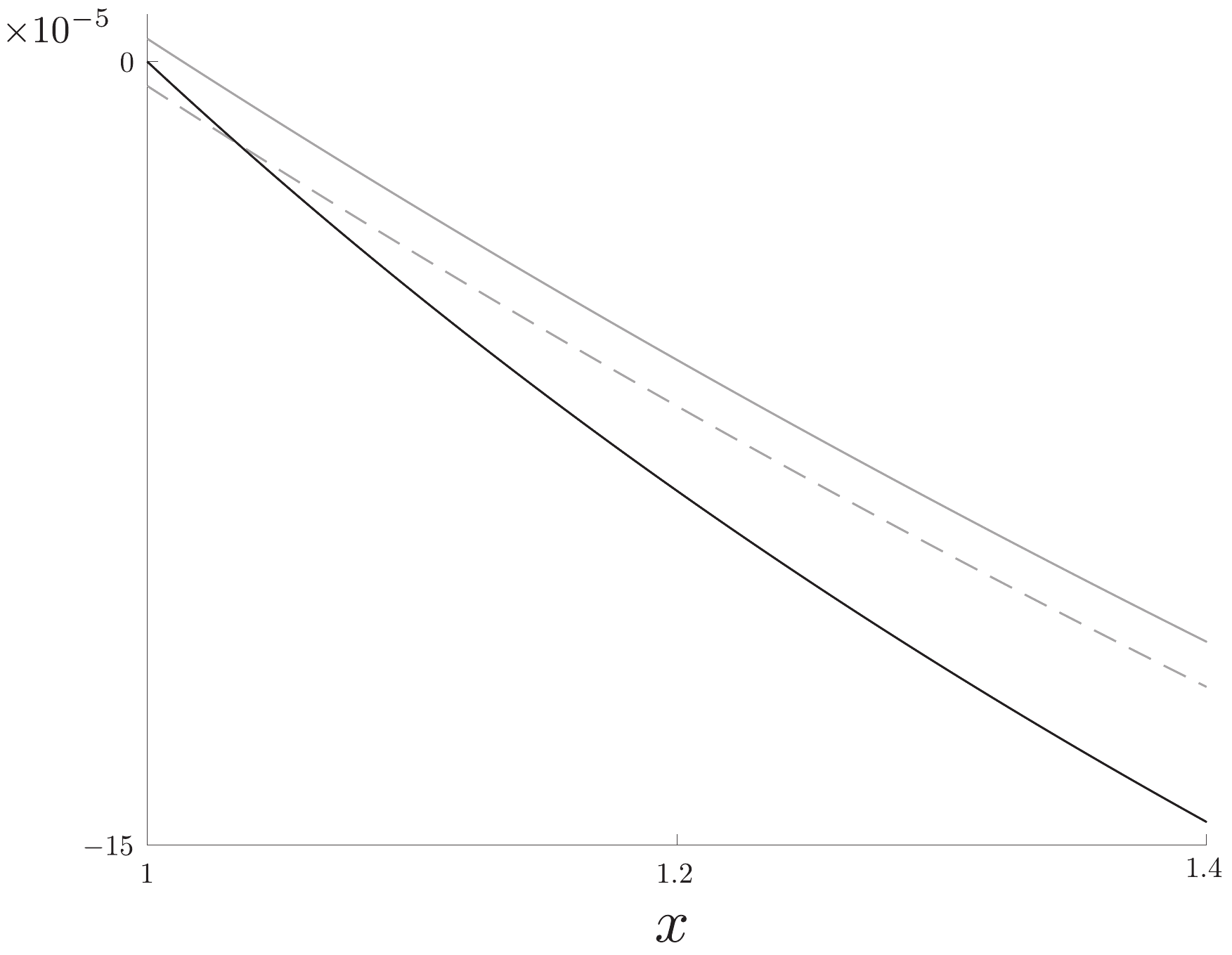}%
}{%
\captionsetup{width=0.45\textwidth}
  \caption{\small{Horizontal axis exhibits values of $x$. Values of $\varphi$ are shown by a black line, $\epsilon$ by a grey line, $\delta$ by a dashed grey line.}}%
\label{fig:ms_dec}}
\end{floatrow}
\end{figure}

\begin{figure}[!htbp]
\begin{floatrow}
\capbtabbox{
    \begin{tabular}{ccc} 
\toprule 
$\lambda$\,\scriptsize{[GPa]}& \hphantom{x}&$\mu$\,\scriptsize{[GPa]}\\ 
\toprule 
$50\,x$ && 46.8 \\ 
50 && 47.1 \\ 
$50\,(1\,/\,x)$ && 46.9 \\ 
50 && 47.4 \\ 
$50\,x$ && 47.2 \\ 
\bottomrule
$RSD_{\mu}=1.014\,\%$ &\hphantom{x}&$\gamma=1.028\cdot10^{-5}\,$ \\
 \end{tabular}
}{
\captionsetup{width=0.45\textwidth}
  \caption{\small{Five layers relevant to mafic rocks with near-constant $\mu$ and growing variations of $\lambda$. Case III; near-constant anisotropy parameters, gas and water saturation.}}
  \label{tab:ms_eq}
  }
\qquad
  \ffigbox{%
  \includegraphics[scale=0.35]{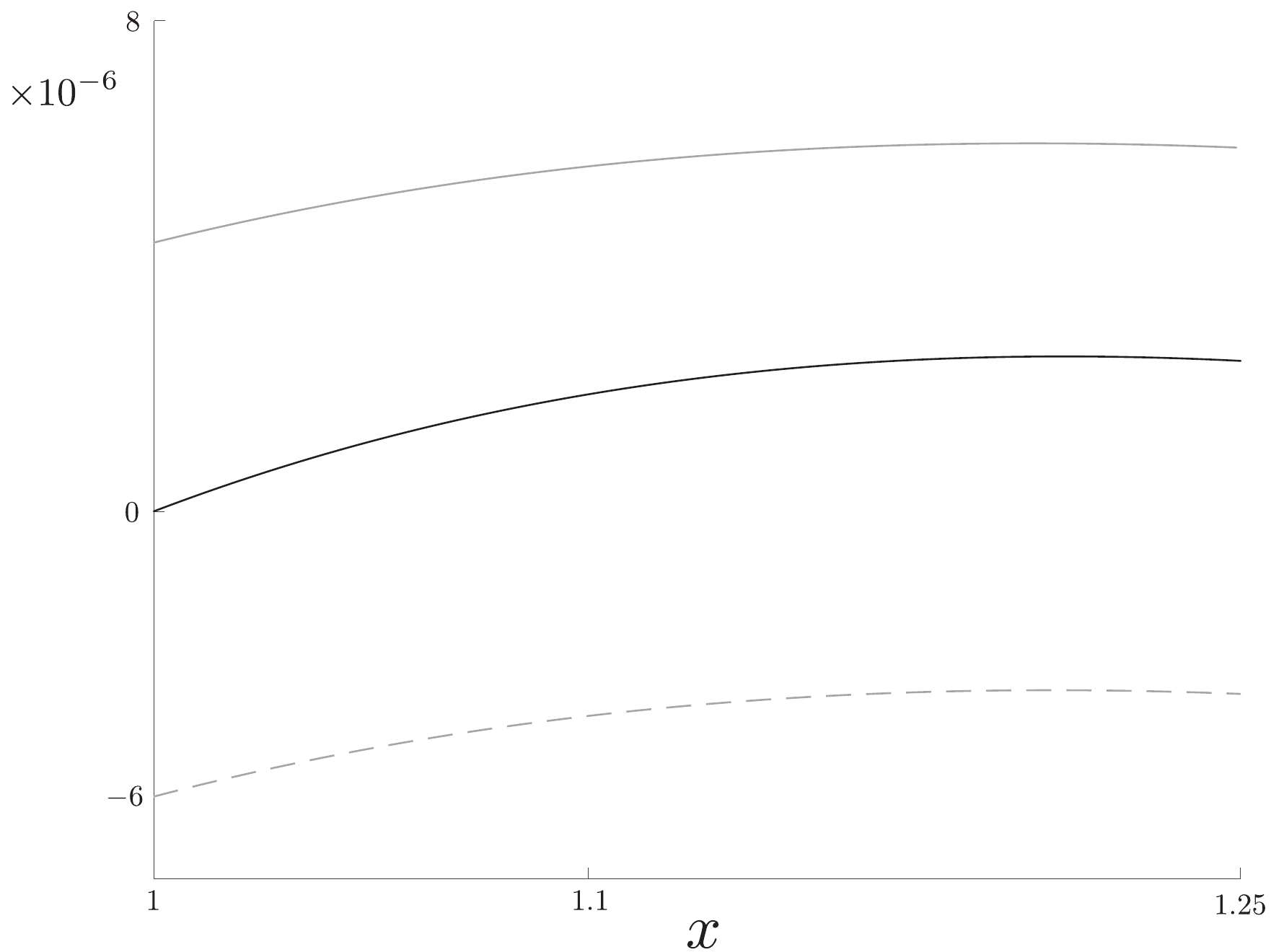}%
}{%
\captionsetup{width=0.45\textwidth}
  \caption{\small{Horizontal axis exhibits values of $x$. Values of $\varphi$ are shown by a black line, $\epsilon$ by a grey line, $\delta$ by a dashed grey line.}}%
\label{fig:ms_eq}}
\end{floatrow}
\end{figure}

Tables~\ref{tab:ms_inc}, \ref{tab:ms_dec} and \ref{tab:ms_eq}, show three cases of different elasticity parameters in layers.
The changing values of $\varphi$, $\epsilon$ and $\delta$ that are influenced by the scale factor, $x$, are exhibited in Figures~\ref{fig:ms_inc}, \ref{fig:ms_dec} and \ref{fig:ms_eq}, respectively. In all cases $\gamma<4\cdot10^{-5}$.
\begin{table}[!htbp]
\begin{tabular}{cccccccc}
\toprule
&$\varphi>\epsilon$&$\varphi<\delta$&$|\varphi|>|\epsilon|$&$|\varphi|>|\delta|$&$|\epsilon|\approx|\delta|>10^{-4}$&$|\varphi|>10^{-4}$\\
\midrule
Case I\hphantom{X}&$>5.853\,\%$&$\in\emptyset$&$>5.853\,\%$&$>1.014\,\%$&$>14.86\,\%$&$>9.860\,\%$\\
Case II\hphantom{X}&$\in\emptyset$&$>2.602\,\%$&$>0.424\,\%$&$>2.602\,\%$&$>14.40\,\%$&$>10.71\,\%$\\
Case III\hphantom{X}&$\in\emptyset$&$\in\emptyset$&$\in\emptyset$&$\in\emptyset$&$\in\emptyset$&$\in\emptyset$\\
\bottomrule
\end{tabular}
\caption{\small{Occurrence of fluid indicators and its dependence on $RSD_{\lambda}$. Scale factor examples with $RSD_{\mu}<2\,\%\,$.}}
\label{tab:rel_ms_scale}
\end{table}

In each of three cases, as opposed to Appendix~\ref{sec:muvar_scale}, $\varphi$, $\epsilon$ and $\delta$, have very close values to each other.
In general, $\varphi$, $\epsilon$ and $\delta$ are much smaller in comparison to the examples of $RSD_{\mu}>2\,\%$, from Appendix~\ref{sec:muvar_scale}.
In Tables~\ref{tab:ms_inc} and \ref{tab:ms_dec}, the maximum variations of $\lambda$ (for $x=1.4$) are $RSD_{\lambda}=15.80\,\%$, in Table~\ref{tab:ms_eq} (for $x=1.25$), are $RSD_{\lambda}=16.18\,\%\,$.
Certain fluid indicators do detect moderate or strong variations of $\lambda$ in Case I, but do not detect them in Case II, and vice versa.
In Case III, none of fluid detectors indicate these variations. 
The dependence of fluid indicators on $RSD_{\lambda}$, for each case, is shown in Table~\ref{tab:rel_ms_scale}.

In Cases I and II, $RSD_{\lambda}>14.86\,\%$ is detected by the maximum number of fluid indicators. 
In layered media, $\epsilon>\delta$, thus, maximally four out of five indicators may detect the change of the fluid content simultaneously, since if $\varphi>\epsilon$ is satisfied, $\varphi<\delta$ is not, and vice versa.
In Case I, the relation, $|\varphi|>|\delta|$, is present in both near-constant $\lambda$, and non near-constant $\lambda$ case, which might be expected, based on Table~\ref{tab:rel_ms}.
In Case II, $|\varphi|>|\epsilon|$, is also misleading, since it is occurs for $RSD_{\lambda}>0.424\,\%$.
Nevertheless, in both Cases, $|\varphi|>|\epsilon|$ and $|\varphi|>|\delta|$ occur together only for the case of moderate or strong variations of $\lambda$.
Thus, it is probable that in order to avoid confusion in fluid detection, it is better not to consider these two parameters separately.
This assumption has to be verified on more examples and real data.
We see that indicator $|\epsilon|\approx|\delta|>10^{-4}$ is the least sensible on variations, since it detects them for $RSD_{\lambda}>14.86\,\%$ and $RSD_{\lambda}>14.40\,\%$.
Also, these two examples show that small change of $\mu$ in layers may greatly influence the anisotropy parameters (they have same scale factor, but different $\mu$).
Similarly to Appendix~\ref{sec:muvar_scale}, we notice that the the larger variations are, the larger possibility of detecting fluids is.
The last case, again confirms the fact that our five indicators are not the only ones possible indicators of the change of fluid content, and that there are cases in which they do not detect the variations of $\lambda$.
Finally, that example show that change of $\lambda$ in layers may significantly influence $\varphi$, $\epsilon$ and $\delta$ (different scale factor, but the same $\mu$, as compared to Case II).
%%%%%%%%%%%%%%%%%%%%%%%%%%%%%%%%%%%%%%%%%%%%%%%%%%%%%%%%%%%%%%%%%%%%%%%%%%%%%%%%%%%%%%%
\subsection{Distributions of $\varphi$\,, $\epsilon$ and $\delta$} \label{sec:dis}
\begin{figure}[!htbp]
\centering
\begin{subfigure}{.33\textwidth}
  \centering
   \includegraphics[scale=0.3]{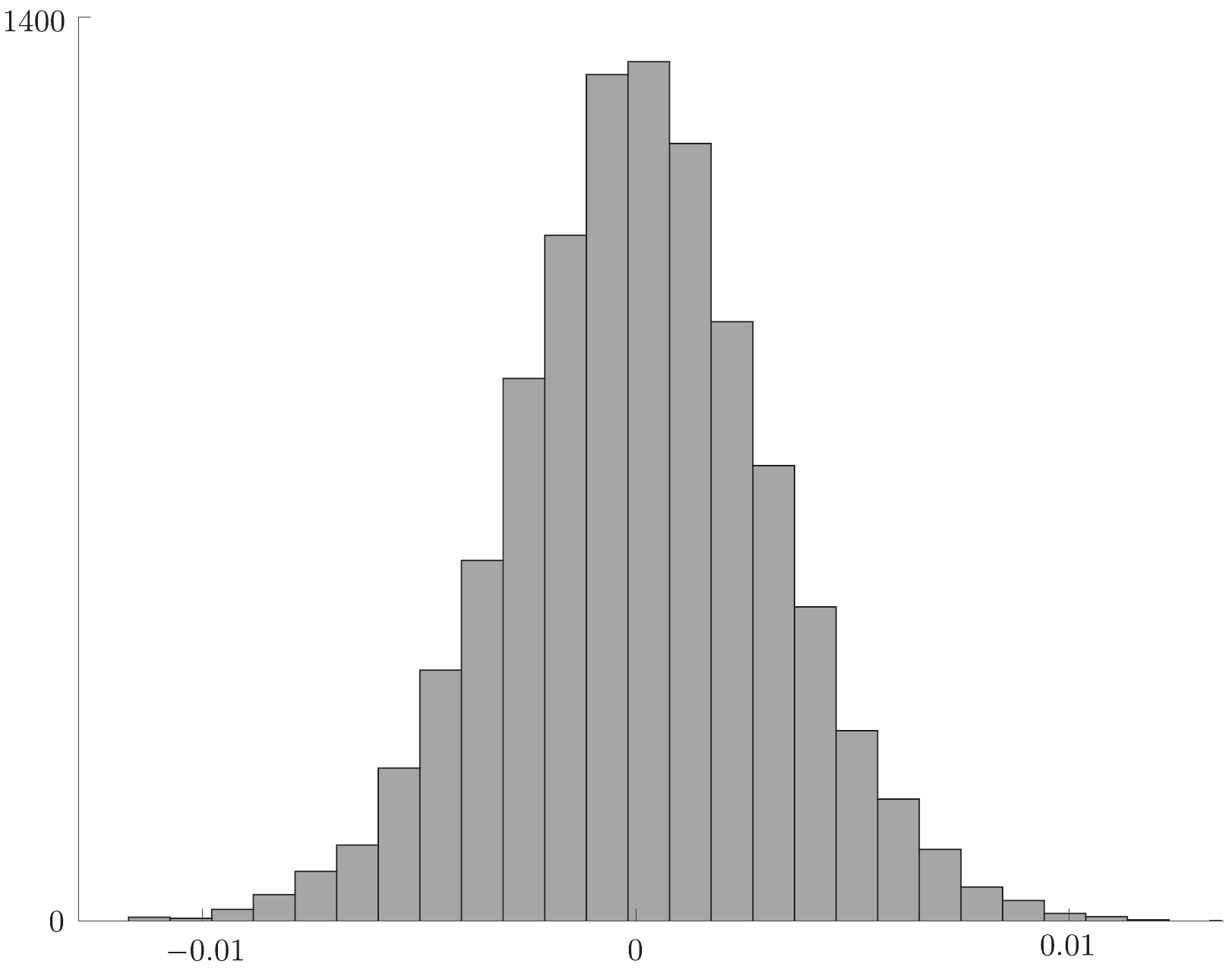}
  \caption{\footnotesize{Distribution of $\varphi$ }}
  \label{fig:disphi}
\end{subfigure}%
\begin{subfigure}{.33\textwidth}
  \centering
   \includegraphics[scale=0.3]{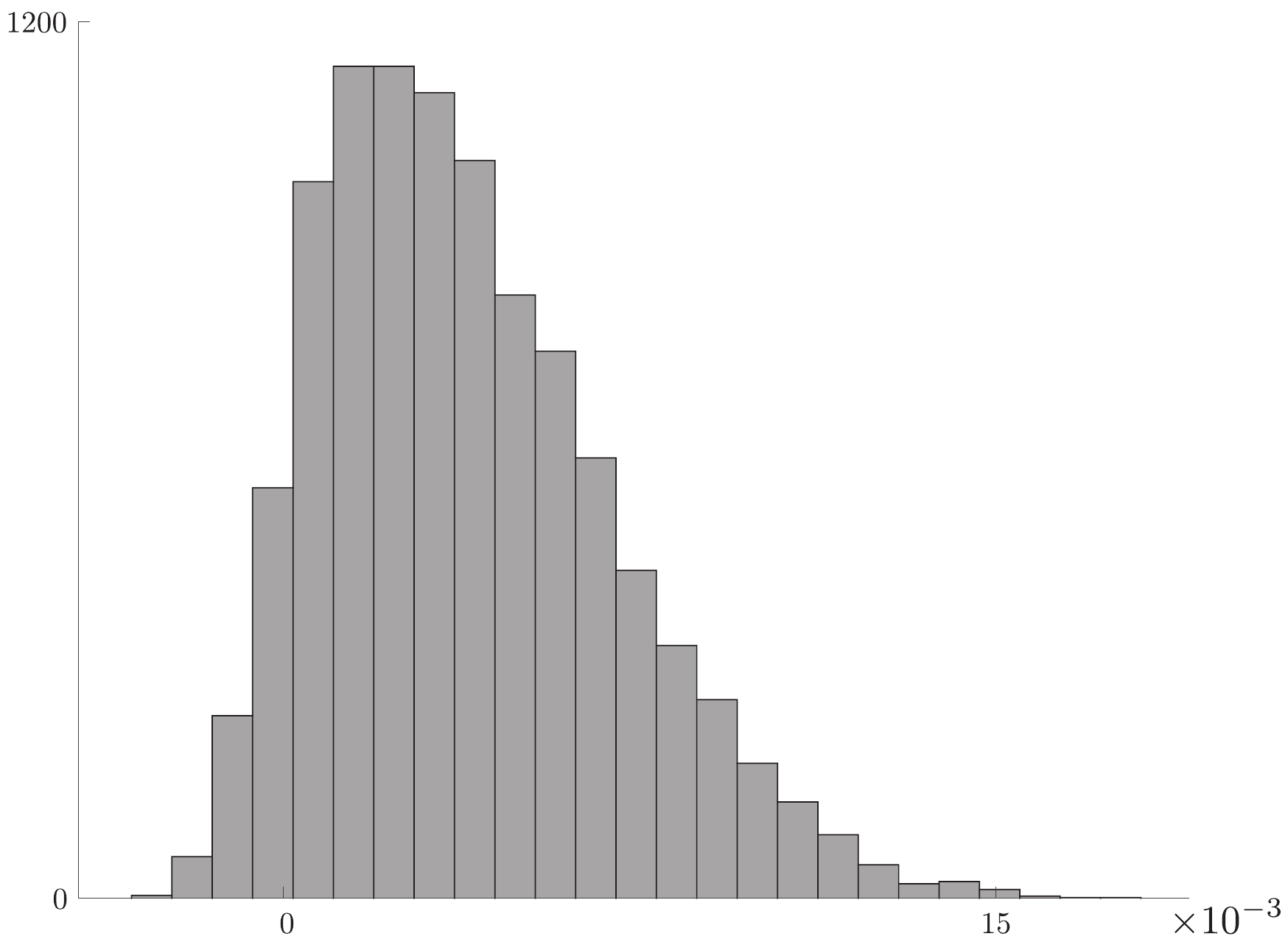}
  \caption{\footnotesize{Distribution of $\epsilon$ }}
  \label{fig:diseps}
\end{subfigure}%
\begin{subfigure}{.33\textwidth}
  \centering
   \includegraphics[scale=0.3]{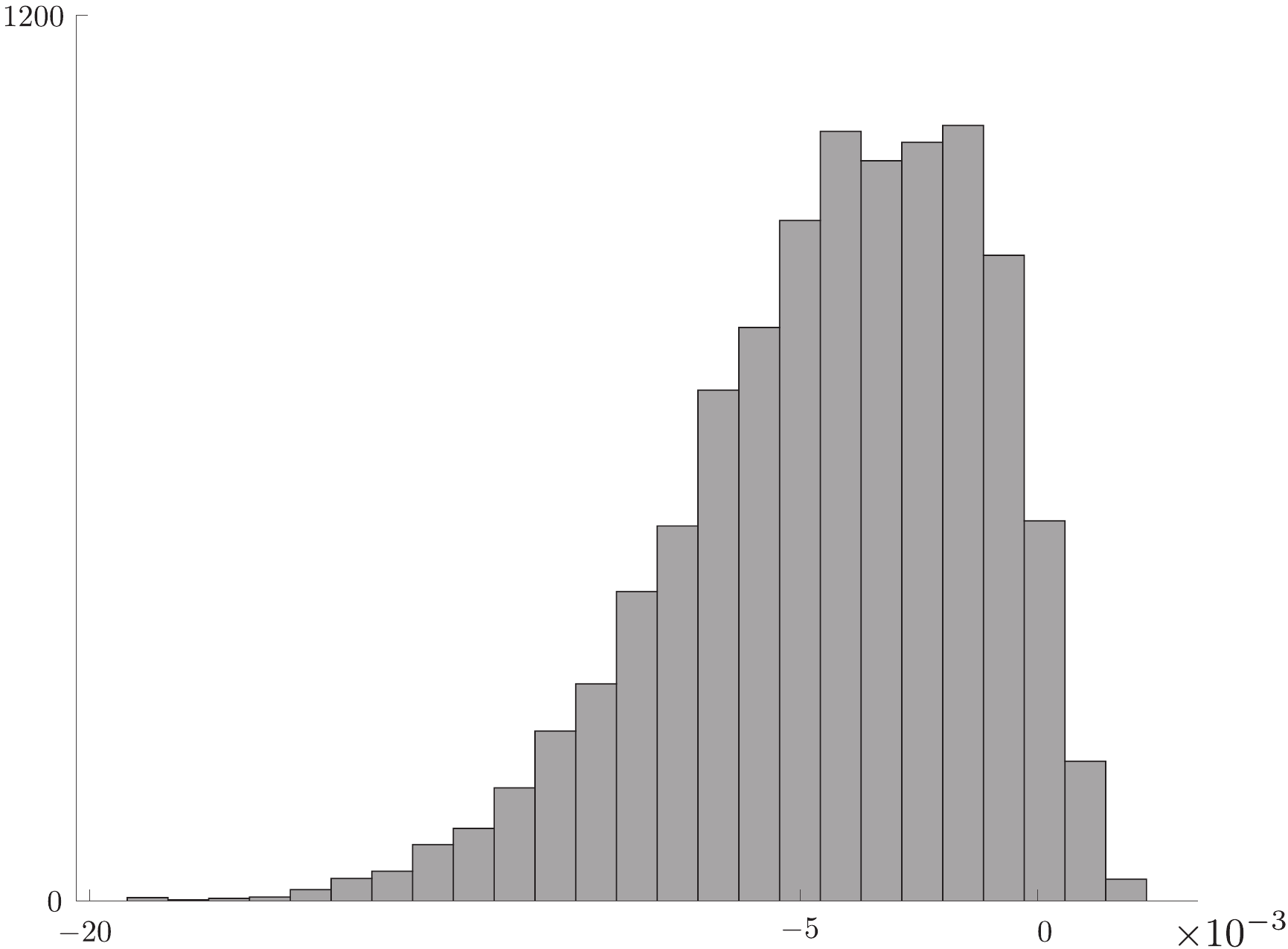}
  \caption{\footnotesize{Distribution of $\delta$ }}
  \label{fig:disdel}
\end{subfigure}
\caption{\small{Distribution of anisotropy parameters for $10000$ TI media, relevant to layered mafic rocks.}}
\label{fig:dis}
\end{figure}

%%%%%%%%%%%%%%%%%%%%%%%%%%%%%%%%%%%%%%%%%%%%%%%%%%%%%%%%%%%%%%%%%%%%%%%%%%%%%%%%%%%%%%%%
Let us discuss the distributions of $\varphi$\,, $\epsilon$ and $\delta$, resulting from MC simulations of TI media, equivalent to thin layers with elasticity parameters corresponding to mafic rocks.
First, we analyze the simulations not restricted by the limits in variations of $\lambda$ or $\mu$, discussed in Section~\ref{sec:general}. 
In other words, we consider distributions of $\varphi$\,, $\epsilon$ and $\delta$, obtained for $RSD_{\lambda}\in\forall$ and $RSD_{\mu}\in\forall$, shown in Figure~\ref{fig:dis}.
%No limits in variations of Lam\'e parameters significantly reduces the time of the simulation process.
%That enables us to set large number of TI examples, namely, $s=10000$, which smoothes the distributions and gives even more accuracy, as compared to $s=1000$. 
%We again set $n=5$, which is the same number of layers as in the great majority of prior MC simulations, performed in this paper.
%
We notice that the distribution of $\varphi$, from Figure~\ref{fig:disphi}, is almost symmetric and has shape of a normal distribution.
Its middle, and dominant, are very close to zero.
On the other hand, the distributions of $\epsilon$ and $\delta$, from Figures~\ref{fig:diseps} and~\ref{fig:disdel}, are not symmetric.
The distribution of $\epsilon$ has a positive skew, whereas the one of $\delta$, has a negative skew.
All three distributions have different ranges of similar magnitude.
However, as shown in Table~\ref{tab:ml}, we notice that the range of $\varphi$ is slightly larger from the others. 
The dominant of $\epsilon$ is positive and much larger than the one of $\varphi$, since the great majority of values of $\epsilon$ are positive.
The dominant of $\delta$ is negative and has similarly large absolute value to the one of $\epsilon$.
In most of the cases, $\delta$, has negative value.
Recording well-known relation in layered media, $\epsilon>\delta$, unsurprisingly, in great majority of cases, $\delta$ is negative and $\epsilon$ positive.
Also, we notice that both distributions of $\epsilon$ and $\delta$ are almost symmetric to each other with a symmetry axis set at zero.
It explains the fact discussed in Section~\ref{sec:general} that occurrence of $|\delta|>|\epsilon|$ and $|\delta|<|\epsilon|$ is almost equally frequent.
If we put all three distributions of $\varphi$\,, $\epsilon$ and $\delta$ on one common axis, we notice that $|\varphi|>|\epsilon|$ and $|\varphi|>|\delta|$, or, $\varphi>\epsilon$ and $\varphi<\delta$, occur with similar intensity.
It comes from the symmetry of these distributions.

\begin{figure}[!htbp]
\centering
\begin{subfigure}{.33\textwidth}
  \centering
   \includegraphics[scale=0.3]{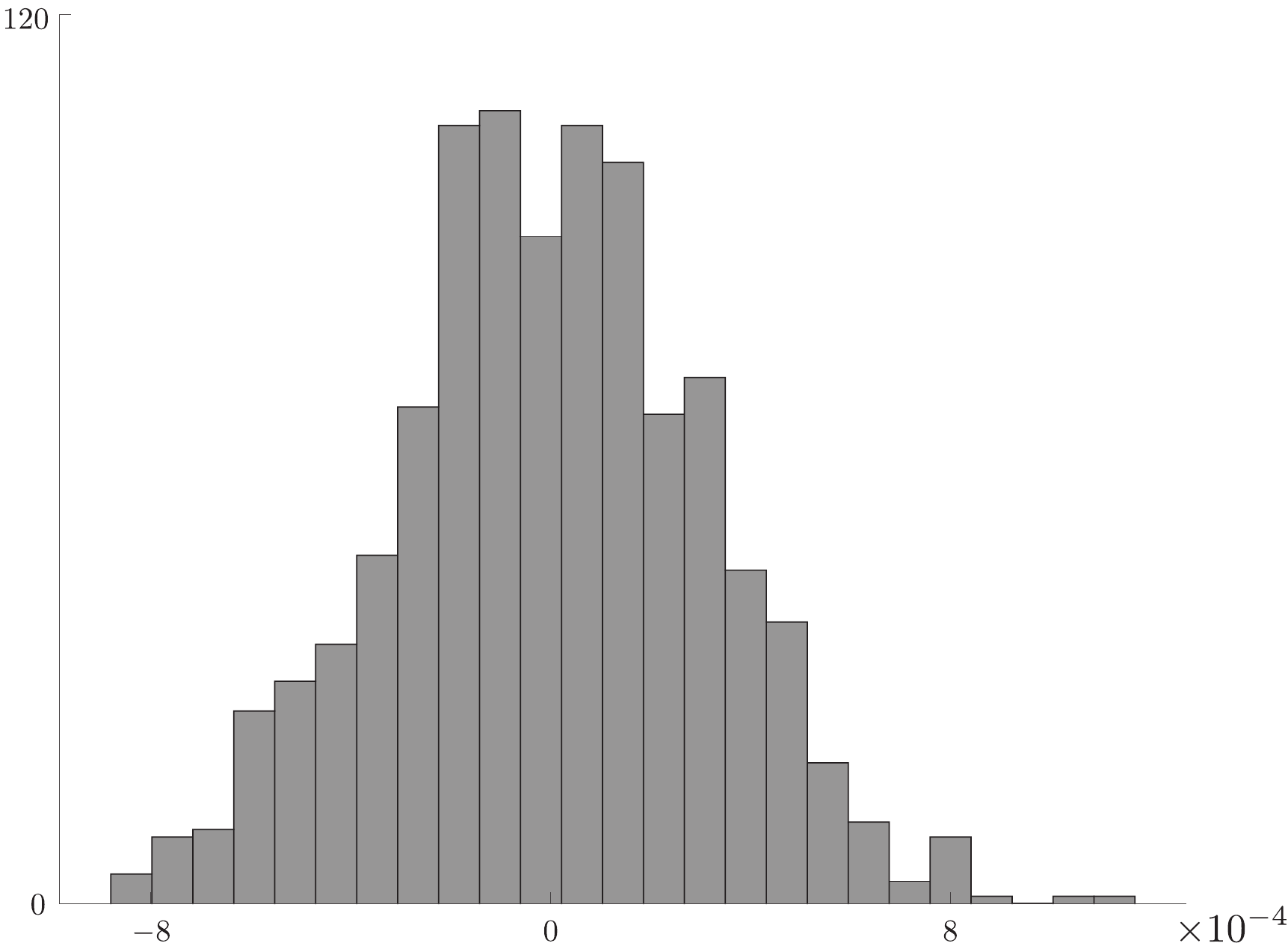}
  \caption{\footnotesize{$\varphi$\, for $RSD_{\lambda}<2\,\%$}}
  \label{fig01a}
\end{subfigure}%
\begin{subfigure}{.33\textwidth}
  \centering
   \includegraphics[scale=0.3]{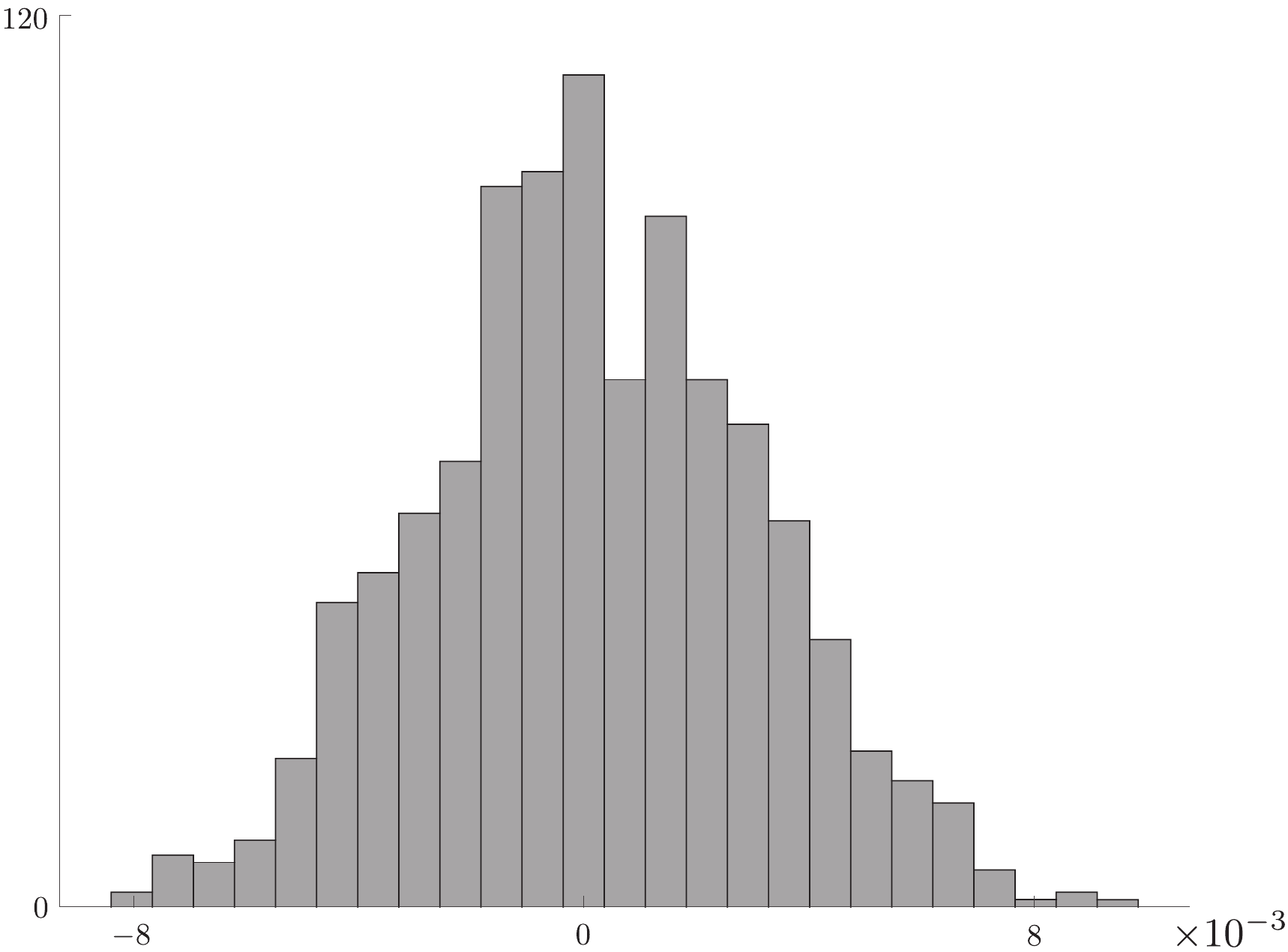}
  \caption{\footnotesize{$\varphi$\, for $RSD_{\lambda}\in(2\,\%\,,\,20\,\%)$}}
  \label{fig01b}
\end{subfigure}%
\begin{subfigure}{.33\textwidth}
  \centering
   \includegraphics[scale=0.3]{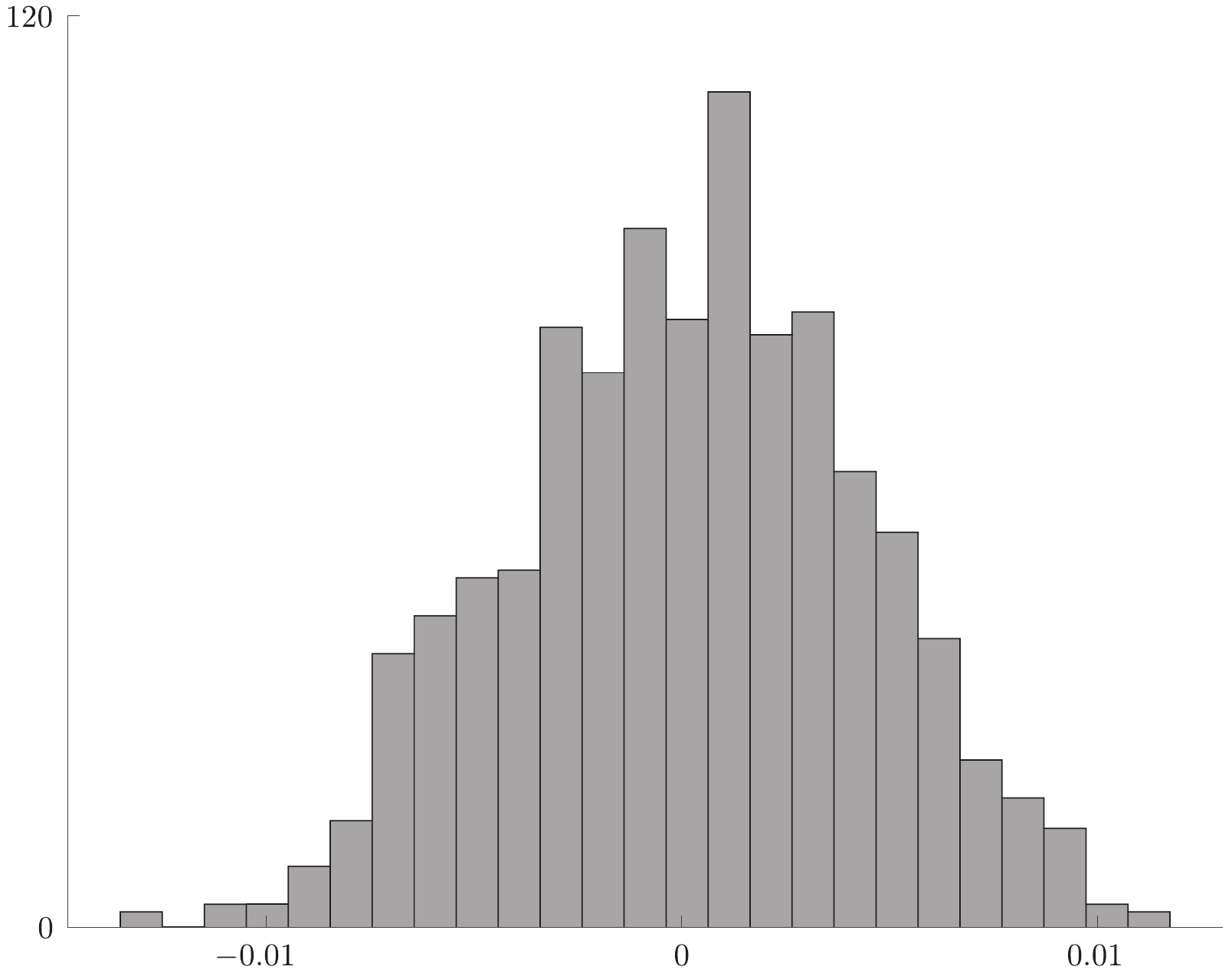}
  \caption{\footnotesize{$\varphi$\, for $RSD_{\lambda}>20\,\%$}}
  \label{fig01c}
\end{subfigure}

\qquad

\centering
\begin{subfigure}{.33\textwidth}
  \centering
   \includegraphics[scale=0.3]{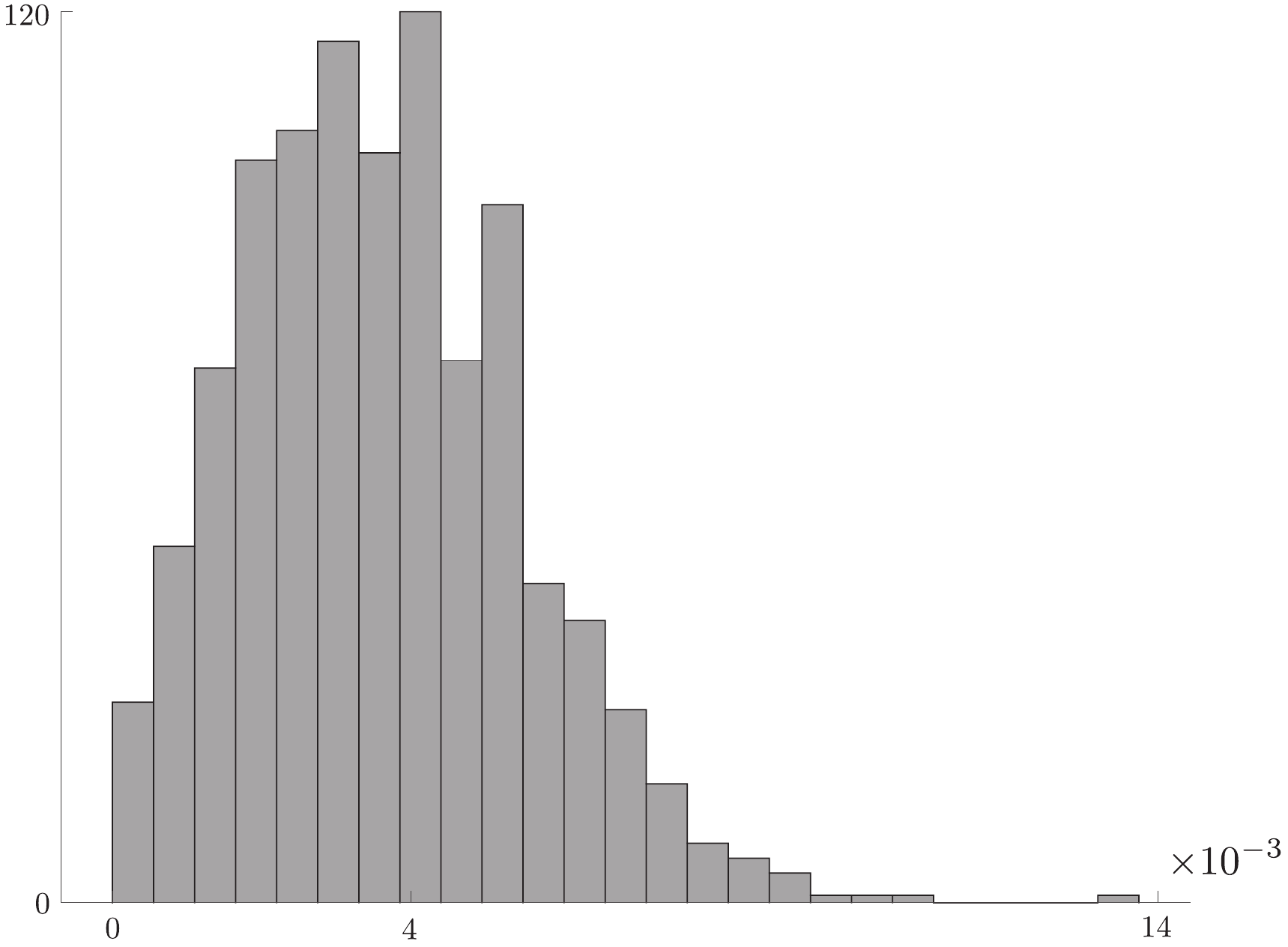}
  \caption{\footnotesize{$\epsilon$\, for $RSD_{\lambda}<2\,\%$}}
  \label{fig01d}
\end{subfigure}%
\begin{subfigure}{.33\textwidth}
  \centering
   \includegraphics[scale=0.3]{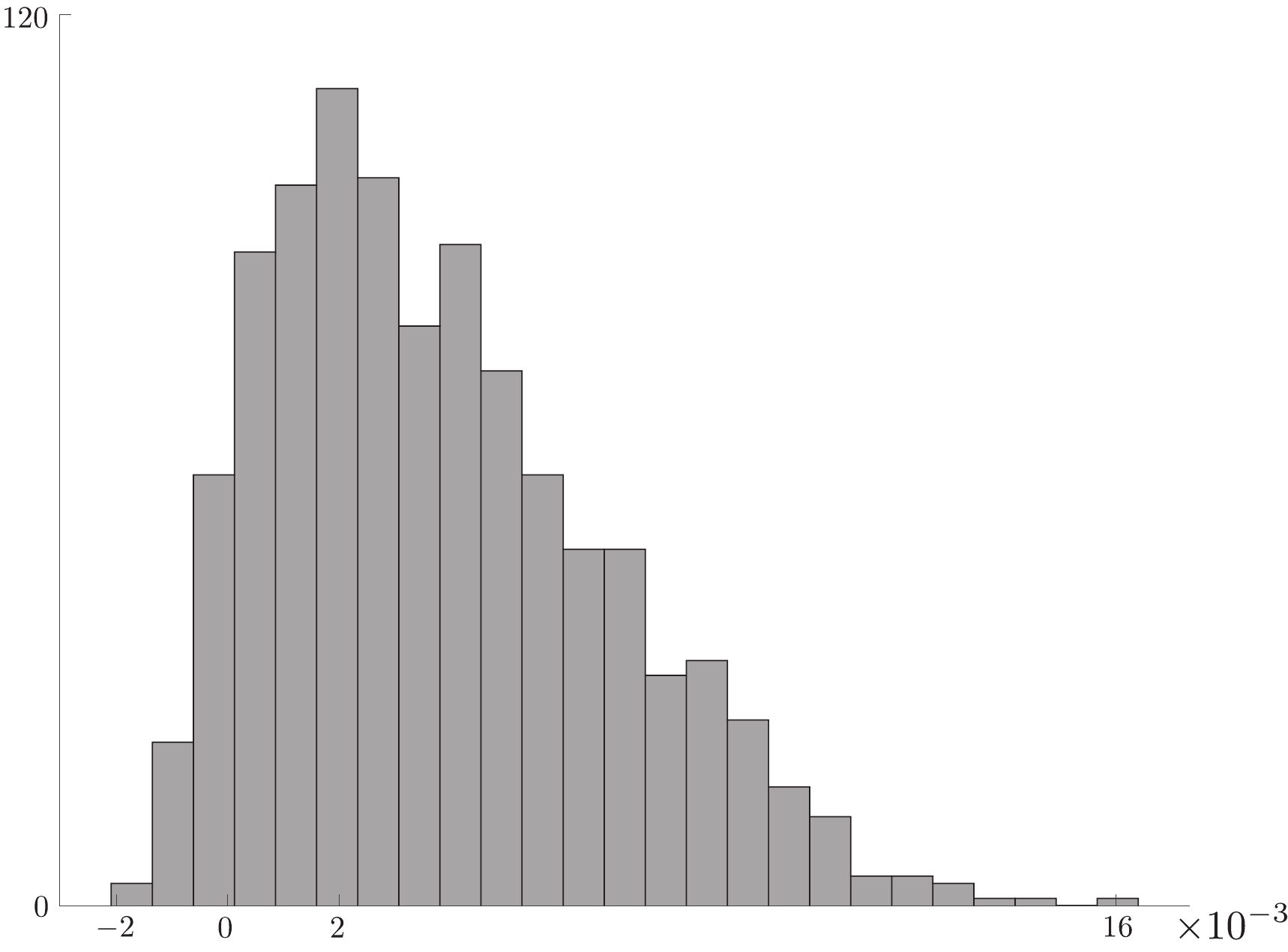}
  \caption{\footnotesize{$\epsilon$\, for $RSD_{\lambda}\in(2\,\%\,,\,20\,\%)$}}
  \label{fig01e}
\end{subfigure}%
\begin{subfigure}{.33\textwidth}
  \centering
   \includegraphics[scale=0.3]{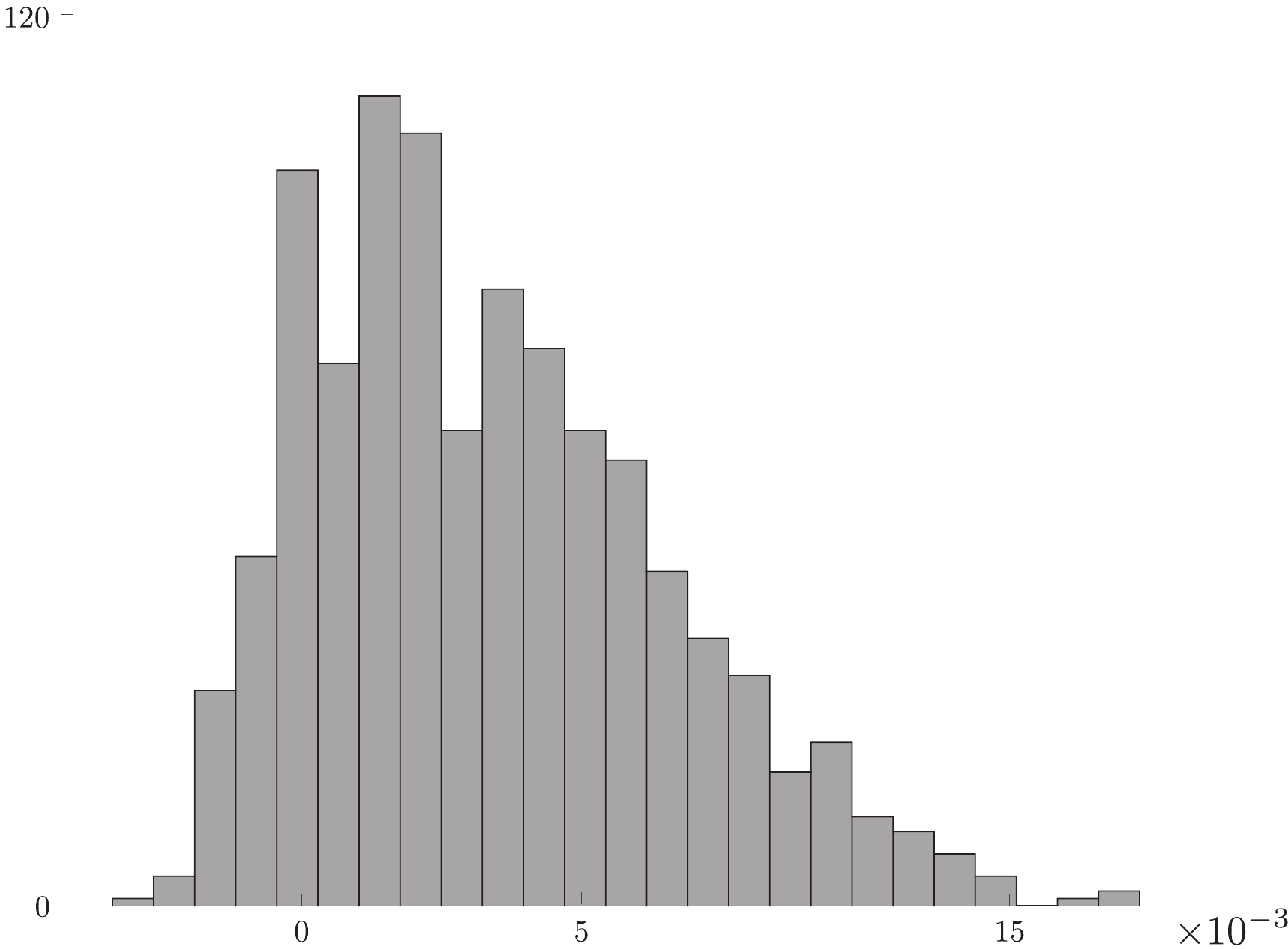}
  \caption{\footnotesize{$\epsilon$\, for $RSD_{\lambda}>20\,\%$}}
  \label{fig01f}
\end{subfigure}

\qquad

\centering
\begin{subfigure}{.33\textwidth}
  \centering
   \includegraphics[scale=0.3]{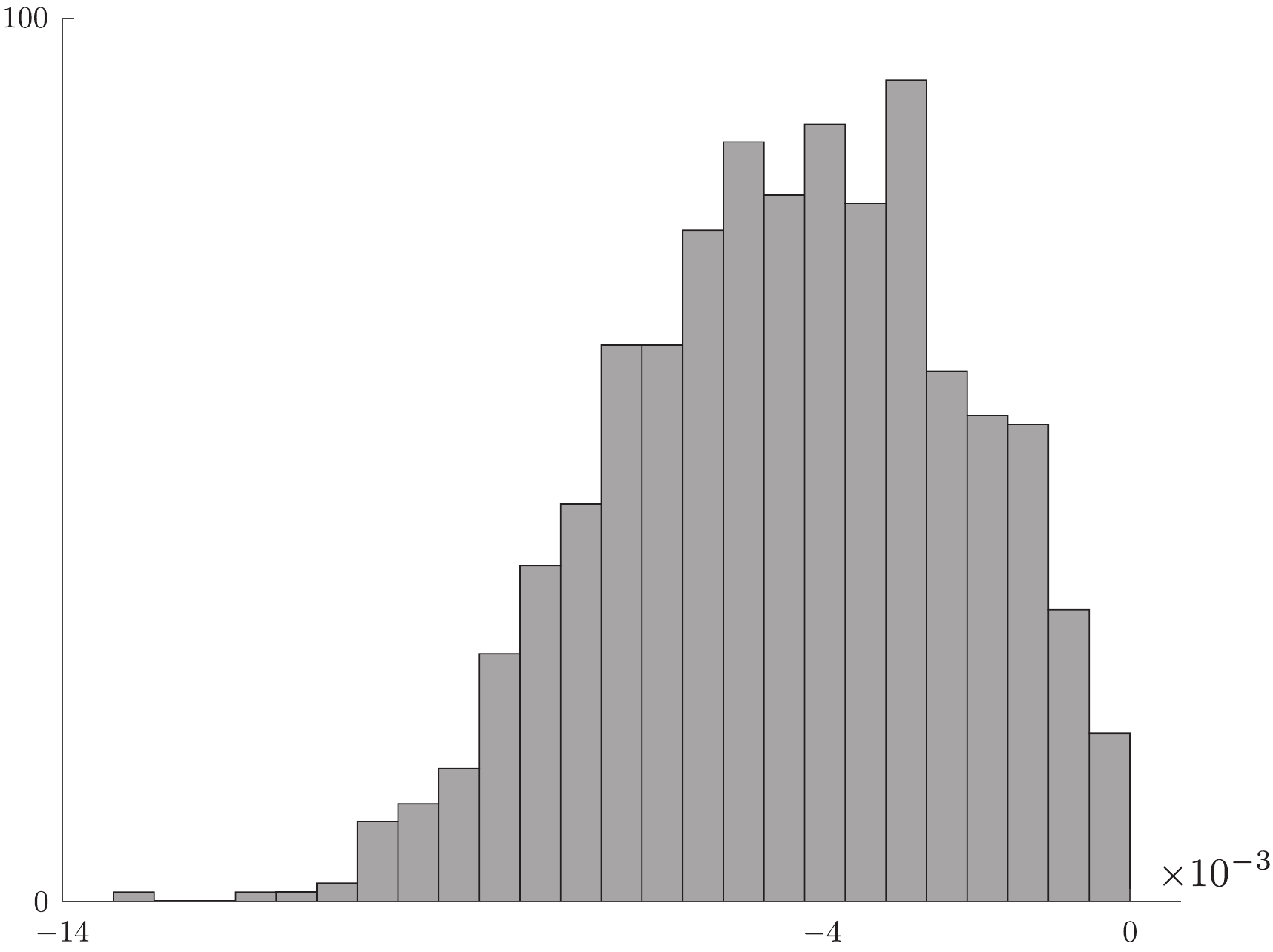}
  \caption{\footnotesize{$\delta$\, for $RSD_{\lambda}<2\,\%$}}
  \label{fig01g}
\end{subfigure}%
\begin{subfigure}{.33\textwidth}
  \centering
   \includegraphics[scale=0.3]{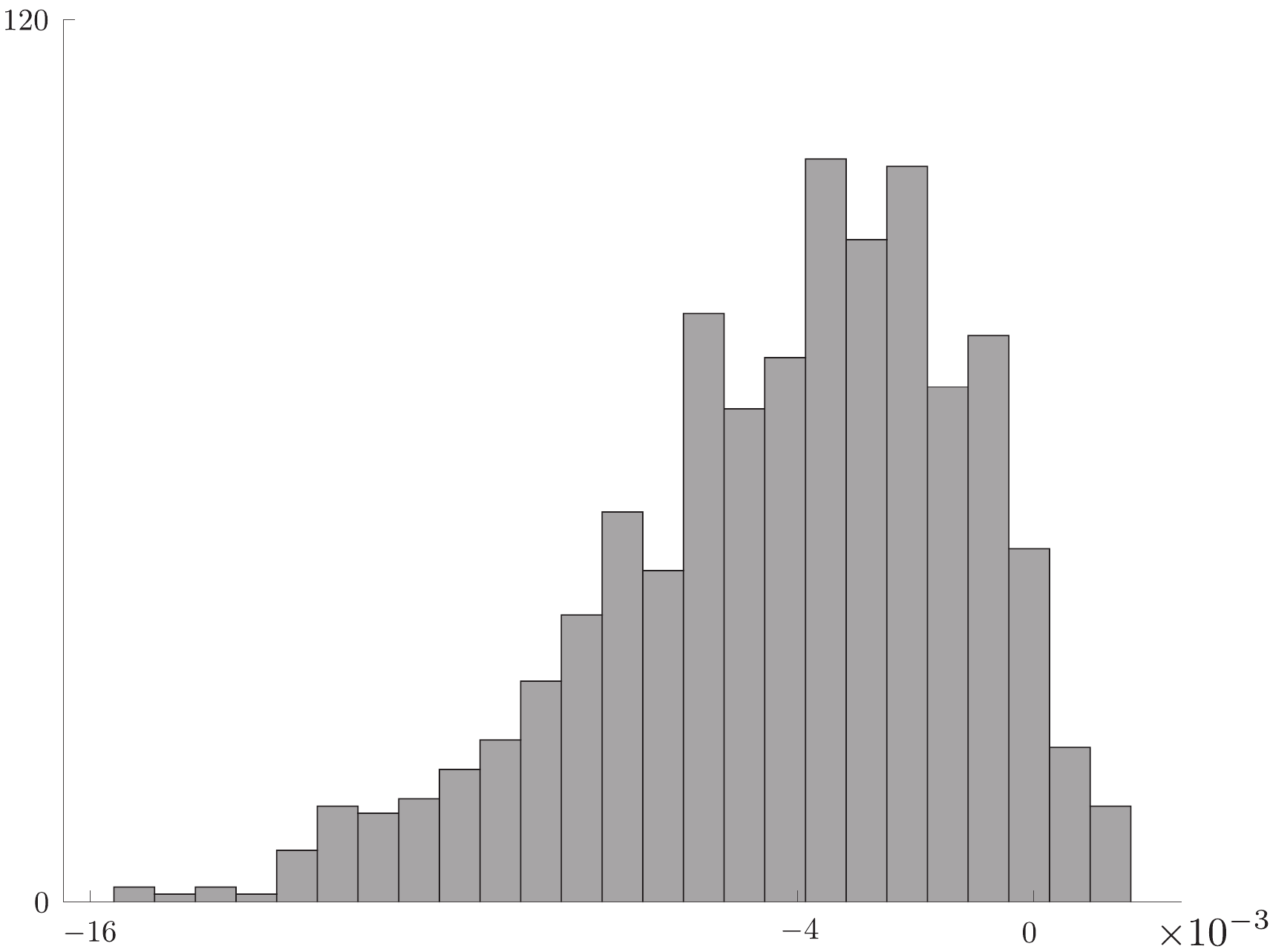}
  \caption{\footnotesize{$\delta$\, for $RSD_{\lambda}\in(2\,\%\,,\,20\,\%)$}}
  \label{fig1h}
\end{subfigure}%
\begin{subfigure}{.33\textwidth}
  \centering
   \includegraphics[scale=0.3]{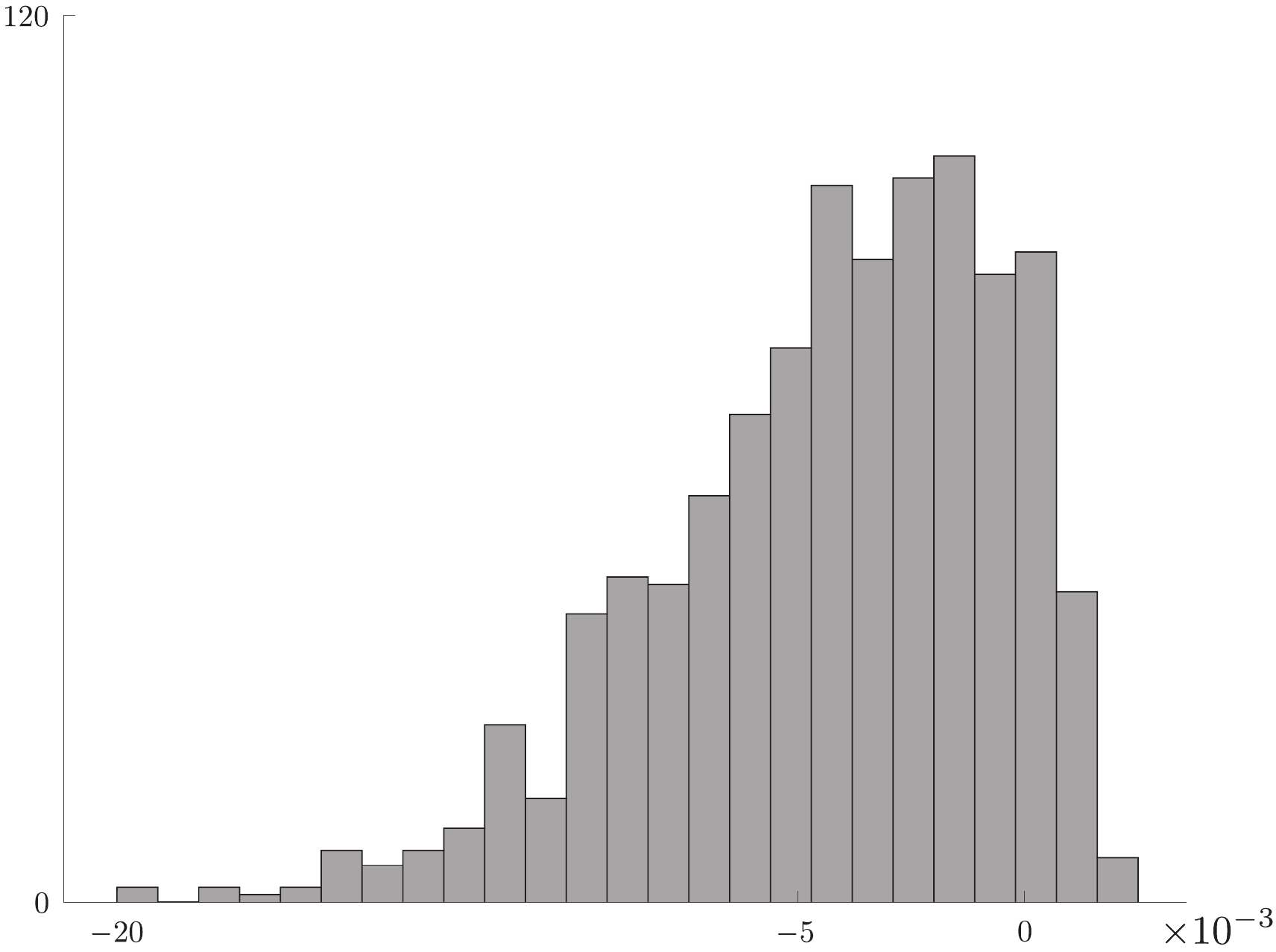}
  \caption{\footnotesize{$\delta$\, for $RSD_{\lambda}>20\,\%$}}
  \label{fig01i}
\end{subfigure}
\caption{\small{Distribution of $\varphi$, $\epsilon$ and $\delta$ for $1000$ examples of equivalent TI mafic media with $RSD_{\mu}>2\,\%\,$. }}
\label{fig:dis_mb}
\end{figure}

\begin{figure}[!htbp]
\centering
\begin{subfigure}{.33\textwidth}
  \centering
   \includegraphics[scale=0.3]{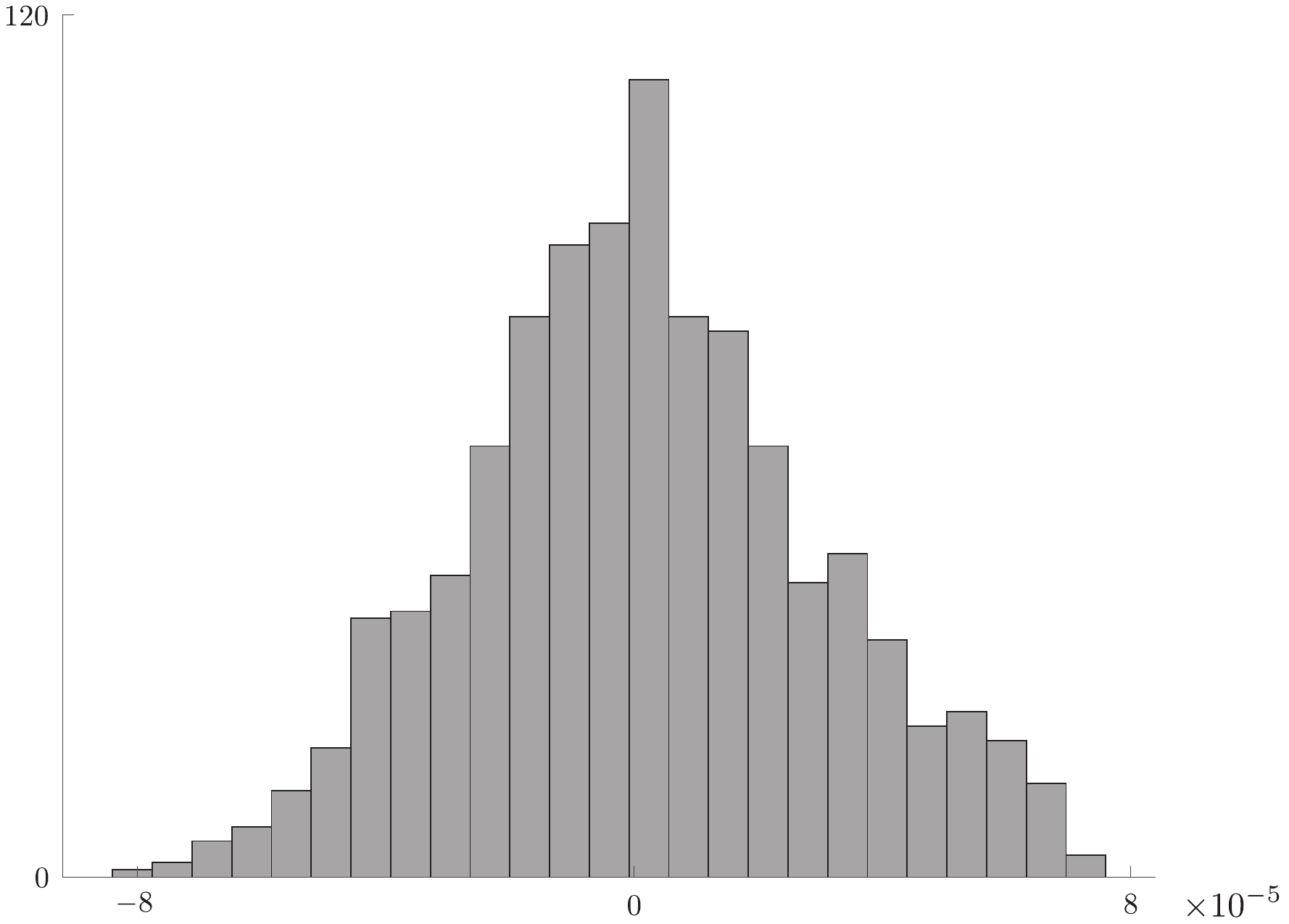}
  \caption{\footnotesize{$\varphi$\, for $RSD_{\lambda}<2\,\%$}}
  \label{fig02a}
\end{subfigure}%
\begin{subfigure}{.33\textwidth}
  \centering
   \includegraphics[scale=0.3]{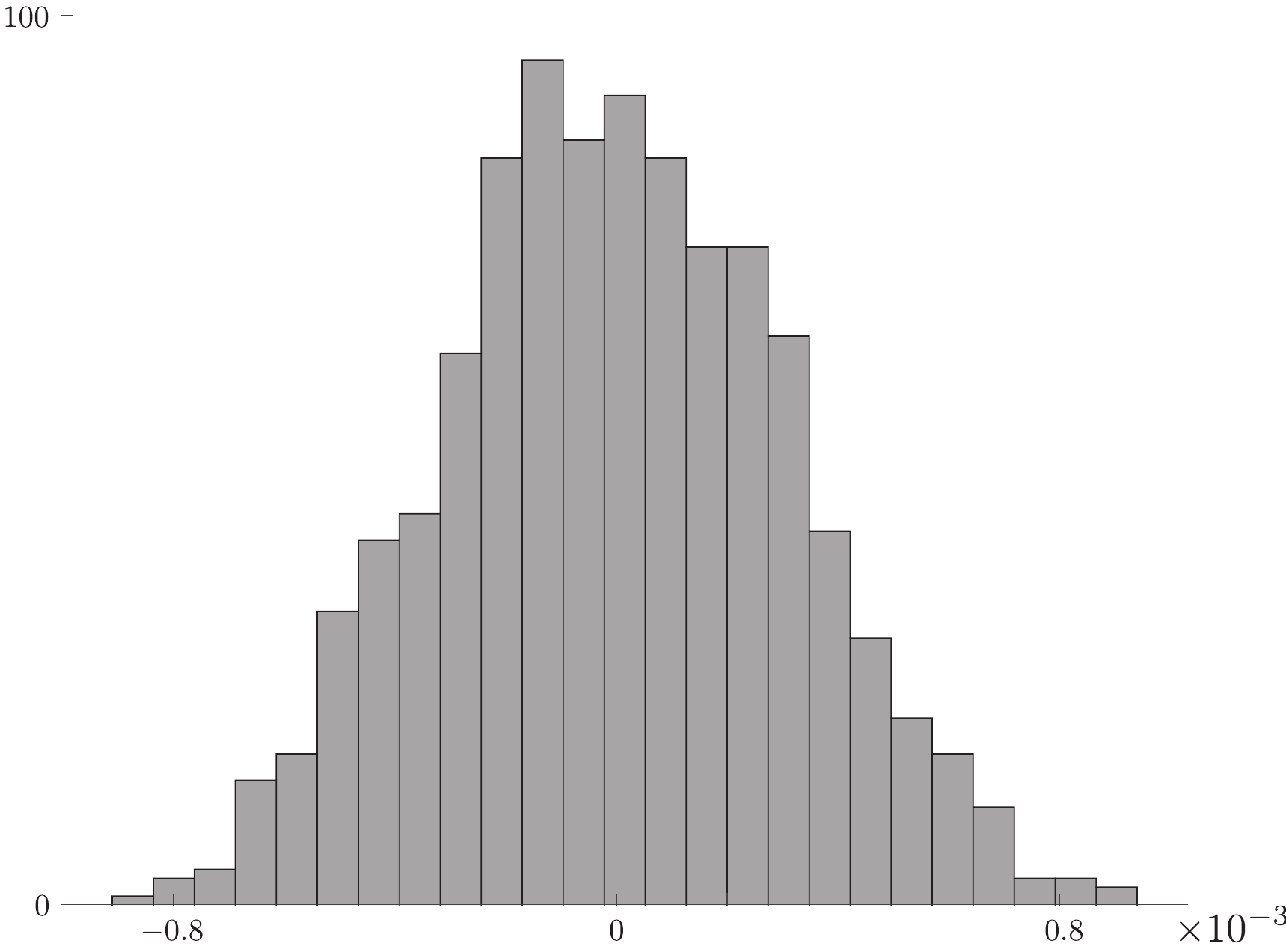}
  \caption{\footnotesize{$\varphi$\, for $RSD_{\lambda}\in(2\,\%\,,\,20\,\%)$}}
  \label{fig02b}
\end{subfigure}%
\begin{subfigure}{.33\textwidth}
  \centering
   \includegraphics[scale=0.3]{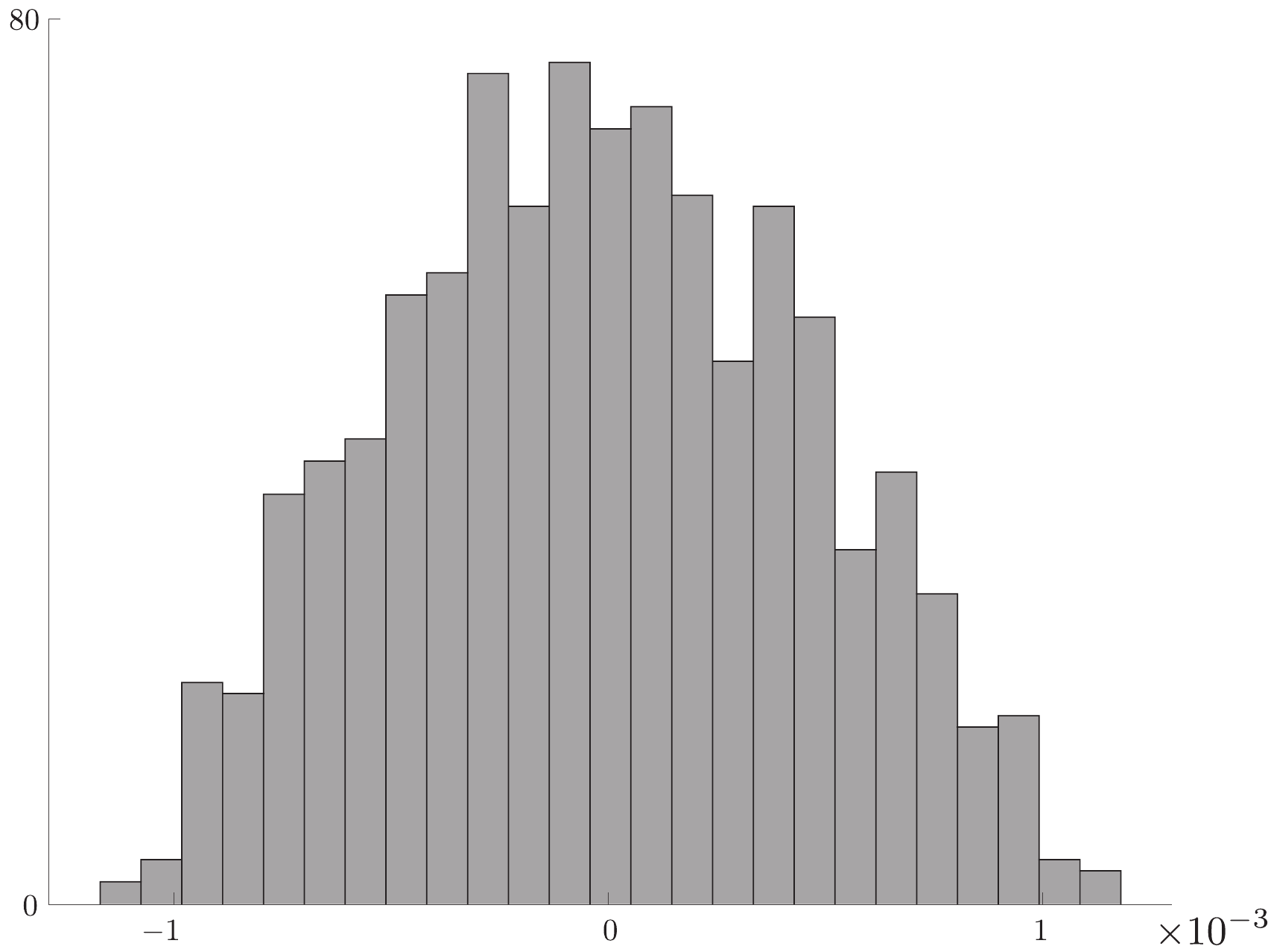}
  \caption{\footnotesize{$\varphi$\, for $RSD_{\lambda}>20\,\%$}}
  \label{fig02c}
\end{subfigure}

\qquad

\centering
\begin{subfigure}{.33\textwidth}
  \centering
   \includegraphics[scale=0.3]{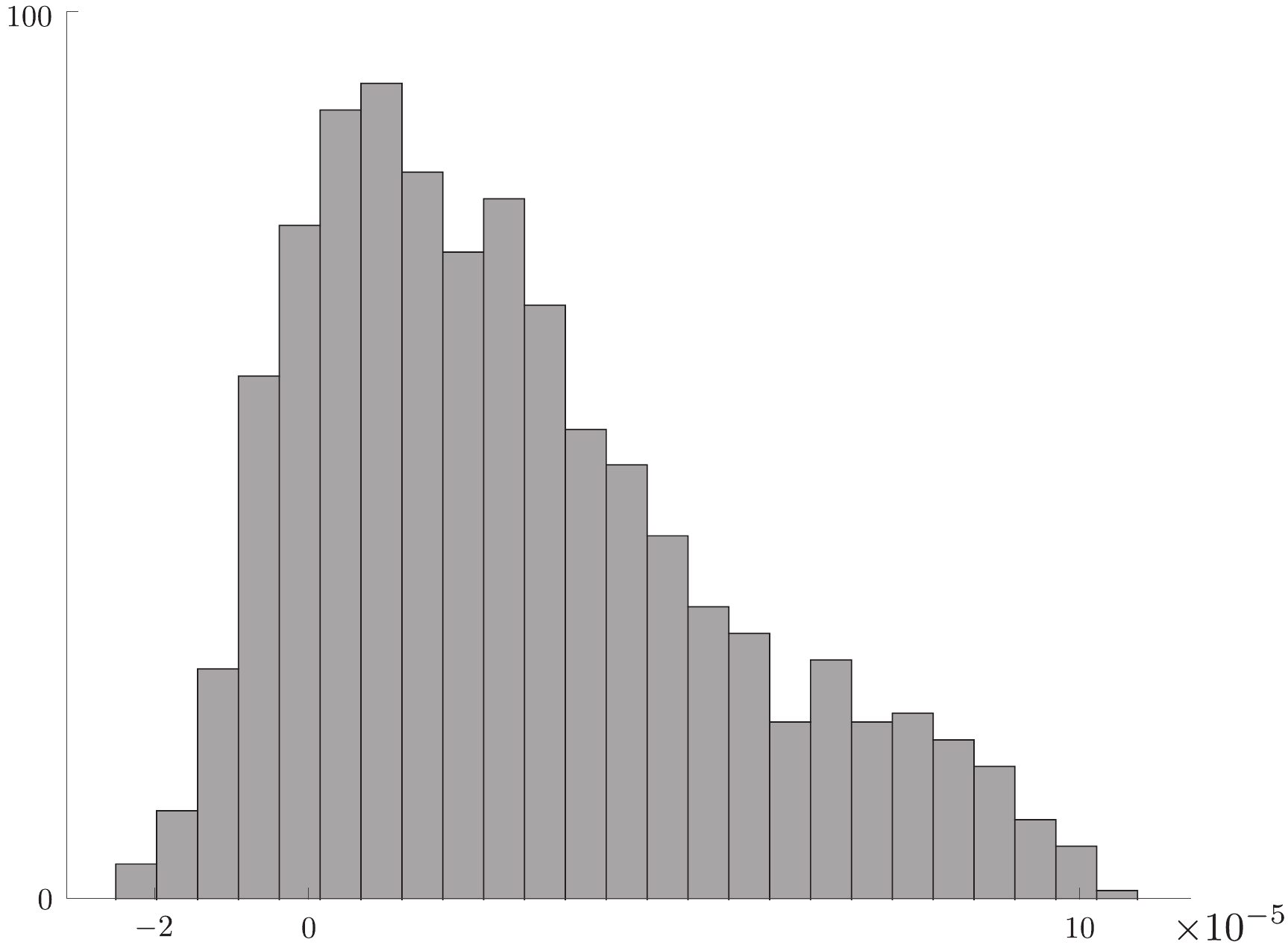}
  \caption{\footnotesize{$\epsilon$\, for $RSD_{\lambda}<2\,\%$}}
  \label{fig02d}
\end{subfigure}%
\begin{subfigure}{.33\textwidth}
  \centering
   \includegraphics[scale=0.3]{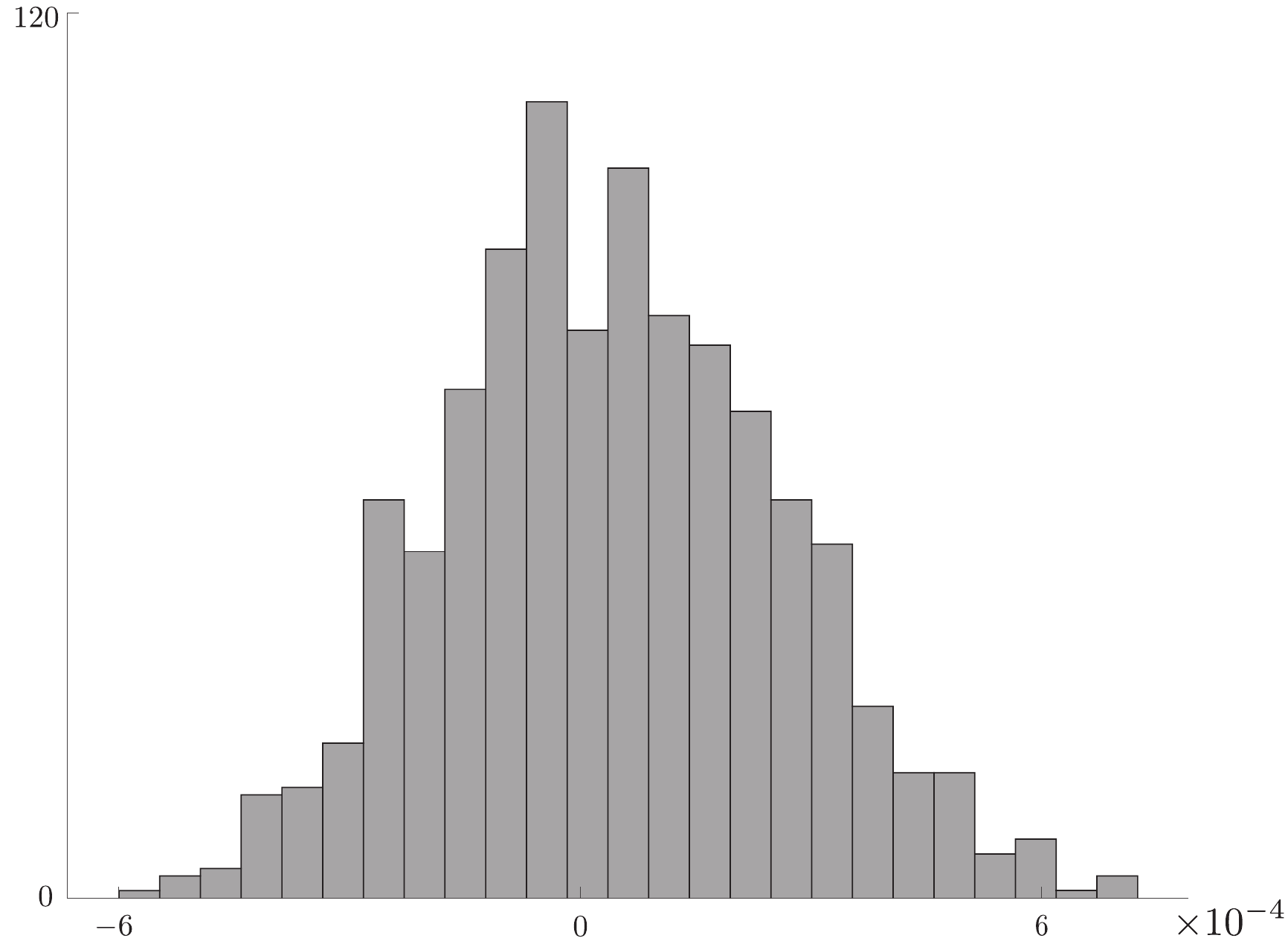}
  \caption{\footnotesize{$\epsilon$\, for $RSD_{\lambda}\in(2\,\%\,,\,20\,\%)$}}
  \label{fig02e}
\end{subfigure}%
\begin{subfigure}{.33\textwidth}
  \centering
   \includegraphics[scale=0.3]{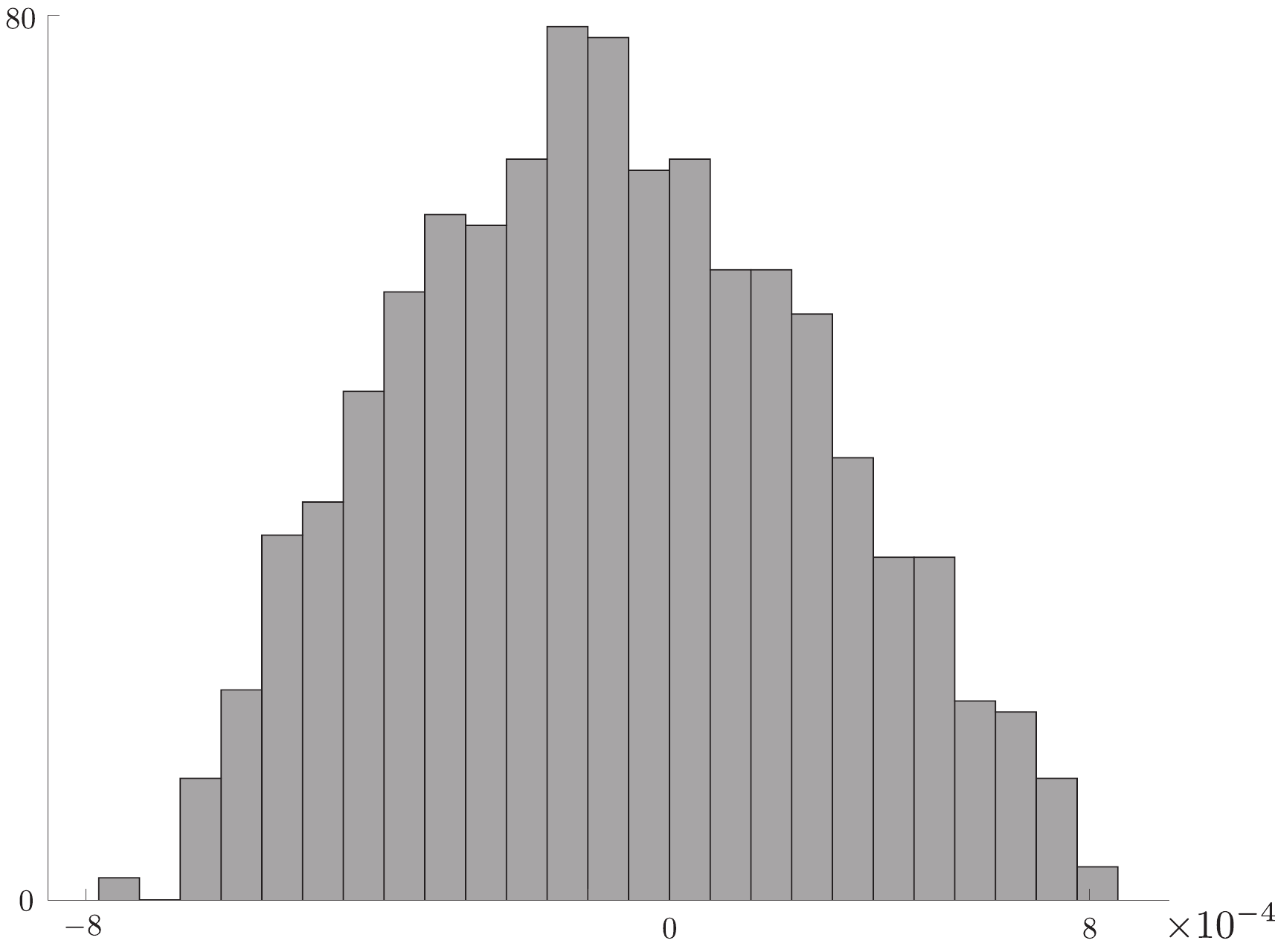}
  \caption{\footnotesize{$\epsilon$\, for $RSD_{\lambda}>20\,\%$}}
  \label{fig02f}
\end{subfigure}

\qquad

\centering
\begin{subfigure}{.33\textwidth}
  \centering
   \includegraphics[scale=0.3]{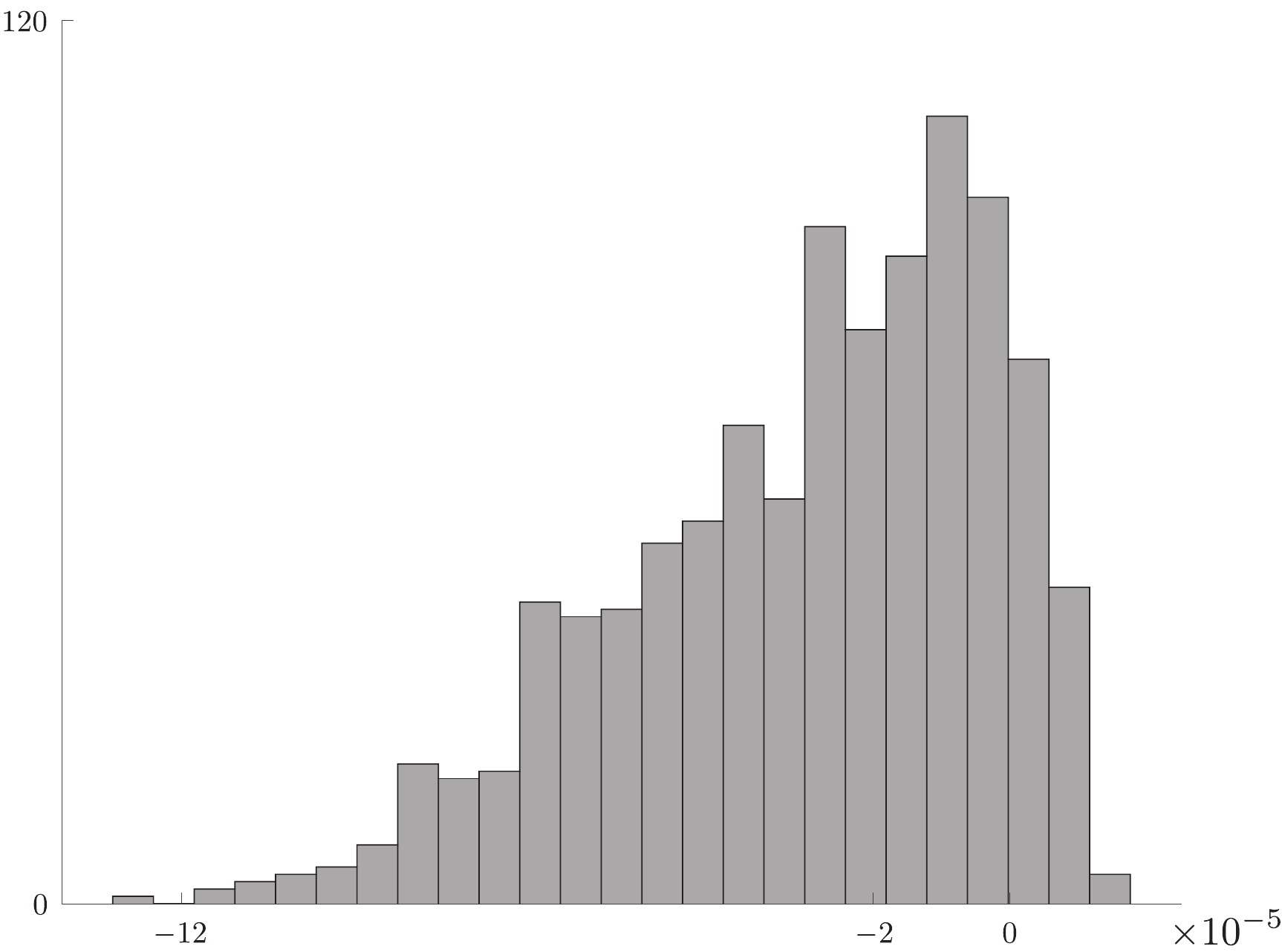}
  \caption{\footnotesize{$\delta$\, for $RSD_{\lambda}<2\,\%$}}
  \label{fig02g}
\end{subfigure}%
\begin{subfigure}{.33\textwidth}
  \centering
   \includegraphics[scale=0.3]{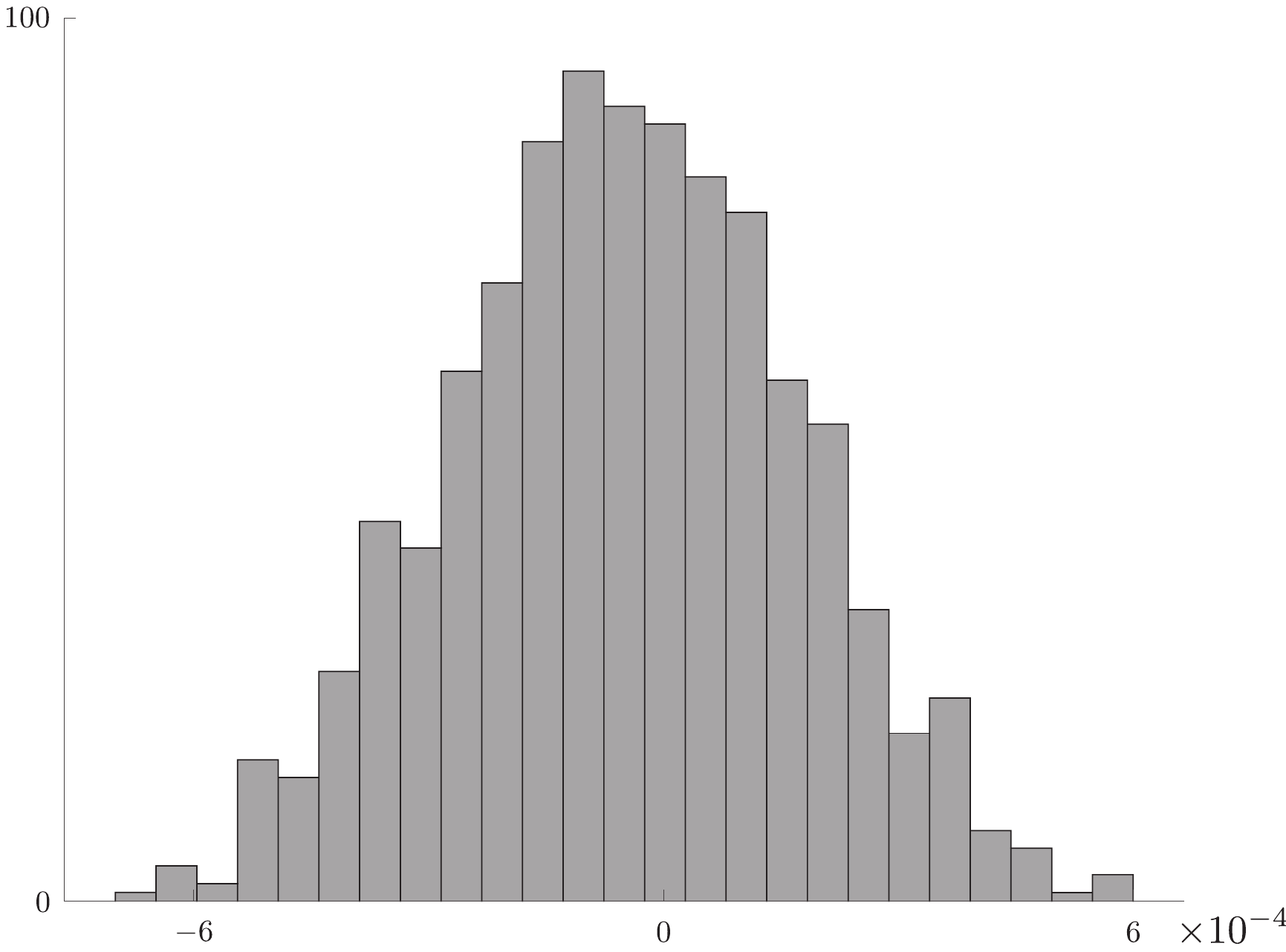}
  \caption{\footnotesize{$\delta$\, for $RSD_{\lambda}\in(2\,\%\,,\,20\,\%)$}}
  \label{fig02h}
\end{subfigure}%
\begin{subfigure}{.33\textwidth}
  \centering
   \includegraphics[scale=0.3]{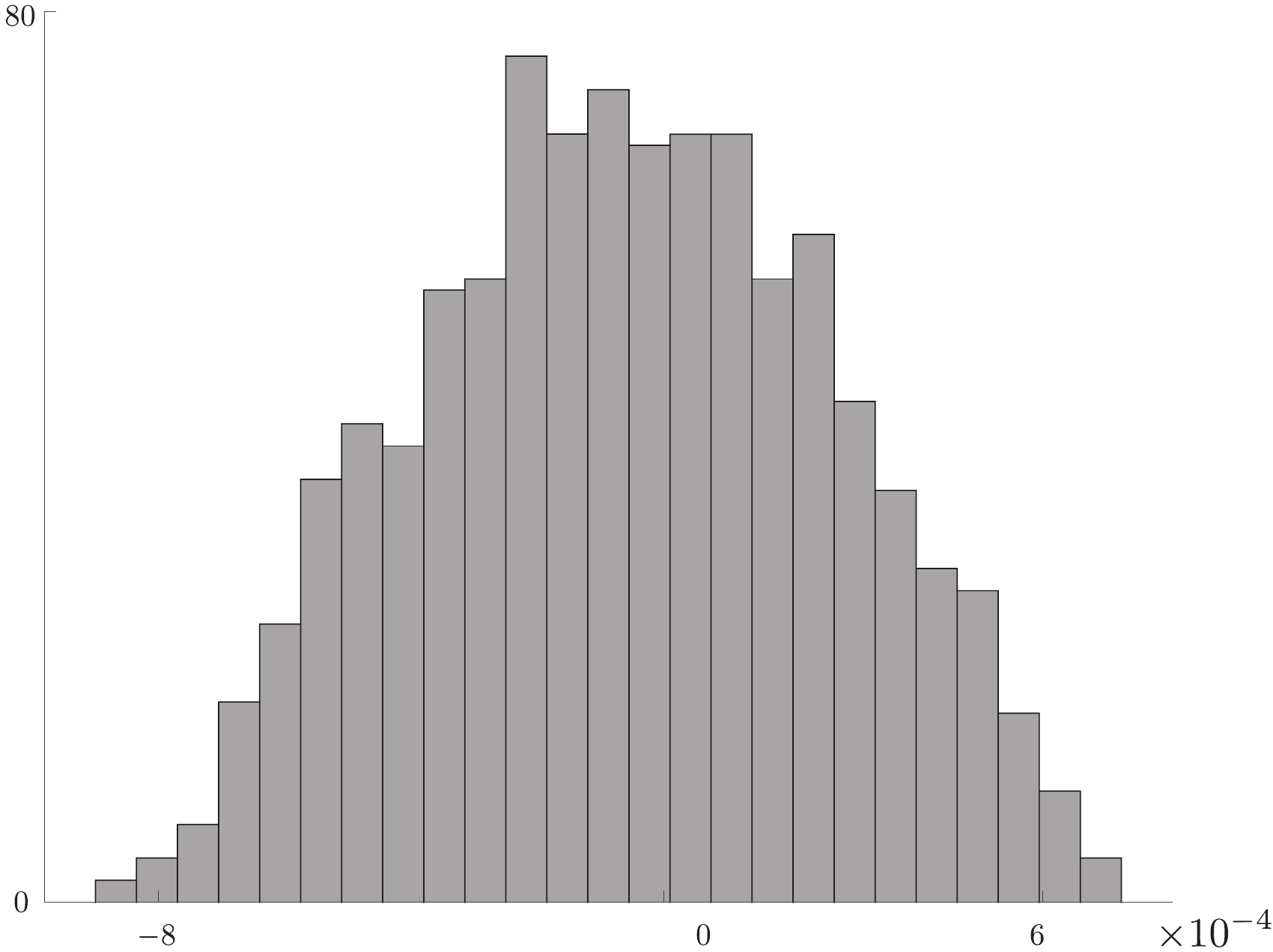}
  \caption{\footnotesize{$\delta$\, for $RSD_{\lambda}>20\,\%$}}
  \label{fig02i}
\end{subfigure}
\caption{\small{Distribution of $\varphi$, $\epsilon$ and $\delta$ for $1000$ examples of equivalent TI mafic media with $RSD_{\mu}<2\,\%\,$.}}
\label{fig:dis_ms}
\end{figure}

Now, let us analyze the distributions for the case of non near-constant rigidity, from Section~\ref{sec:var} and Appendix~\ref{sec:muvar_scale}, and case of near-constant rigidity, from Section~\ref{sec:mu} and Appendix~\ref{sec:muconst_scale}.
We focus on the change of their shapes and ranges, caused by the change of limits imposed on variations of Lam\'e parameters.
Distributions of $\varphi$\,, $\epsilon$ and $\delta$, for non near-constant rigidity, are shown in Figure~\ref{fig:dis_mb}, whereas for near-constant rigidity, are exposed in Figure~\ref{fig:dis_ms}.
The shape of the distribution of $\varphi$ is very similar for each case of variations of Lam\'e parameters.
In every example, its distribution is of Gaussian type.
However, for $RSD_{\mu}<2\,\%$ and $RSD_{\lambda}\in(2\,\%\,,\,20\,\%\,)$, or for $RSD_{\mu}<2\,\%$ and $RSD_{\lambda}>20\,\%\,$, it is of platykurtic type.
In other words, it has thicker tails, which means that extreme values are more likely to take place.
The thickest tail has distribution for $RSD_{\mu}<2\,\%$ and $RSD_{\lambda}>20\,\%\,$, thus, the largest values of $\varphi$ usually occur for the largest variations of $\lambda$.
In Figure~\ref{fig:dis_ms}, the range of $\varphi$ is much larger for $RSD_{\lambda}>20\,\%\,$, than for $RSD_{\lambda}<2\,\%\,$; the fluid indicator, $|\varphi|>10^{-4}$, is much more likely to appear for largest variations.

The distributions of $\epsilon$ and $\delta$, in most of the cases, have positive skew and negative skew, respectively.
Nevertheless, in Figure~\ref{fig:dis_ms}, they have a different shape for the case of $RSD_{\mu}<2\,\%$ and $RSD_{\lambda}>2\,\%\,$.
In that particular case, their distribution have normal shape; the tails become thicker along with stronger variations of $\lambda$, similarly to the tails in distribution of $\varphi$.
Also, in this case, the shapes of distributions of $\varphi$, $\epsilon$ and $\delta$, are very similar.
They have very similar ranges, which results in a fluid indicator, $|\epsilon|\approx|\delta|>10^{-4}$.
Another interesting issue raises if we compare Figures~\ref{fig01d} and~\ref{fig01g}.
The largest intensity of $\epsilon$ and $\delta$, is presented by the values of $\epsilon\in(2\cdot10^{-3}\,,\,5\cdot10^{-3})$ and $\delta\in(-3\cdot10^{-3}\,,\,6\cdot10^{-3})$.
Thus, in this case, mostly $|\delta|>|\epsilon|$, which is reflected in Table~\ref{tab:rel_mb}.

To conclude, the ranges of all three anisotropy parameters become much smaller for the near-constant rigidity case.
By comparing the distributions, we confirm the statement from previous sections that the anisotropy parameters have larger values in the case of stronger variations of $\lambda$ along with stronger variations of $\mu$.
In general, normal shape of $\varphi$ distribution seems to be very useful in finding relationships between this anisotropy parameter and $\epsilon$, of which distribution has positive skew, and $\delta$, of which distribution has negative skew.
Also, in general, all three parameters are of similar magnitude, thus, the comparison of their distributions does make sense.
In the case of near-constant rigidity and moderate or strong variations of $\lambda$, distribution of $\varphi$ has similar shape, but slightly larger range than distributions of $\epsilon$ or $\delta$.
As a result, $|\varphi|>|\epsilon|$, $|\varphi|>|\delta|$ and $\varphi>\epsilon$, $\varphi<\delta$ occur often, which may be good fluid indicators.
Distributions of $\varphi$, $\epsilon$ and $\delta$ for felsic rocks and sandstones, are presented in Appendix~\ref{ap:one}.
%%%%%%%%%%%%%%%%%%%%%%%%%%%%%%%%%%%%%%%%%%%%%%%%%%%%%%%%%%%%%%%%%%%%%%%%%%%%%%%%%%%%%%%%
\section{Felsic rocks and sandstones} \label{ap:one}
\begin{figure}[!htbp]
\centering
\begin{subfigure}{.4\textwidth}
  \centering
   \includegraphics[scale=0.35]{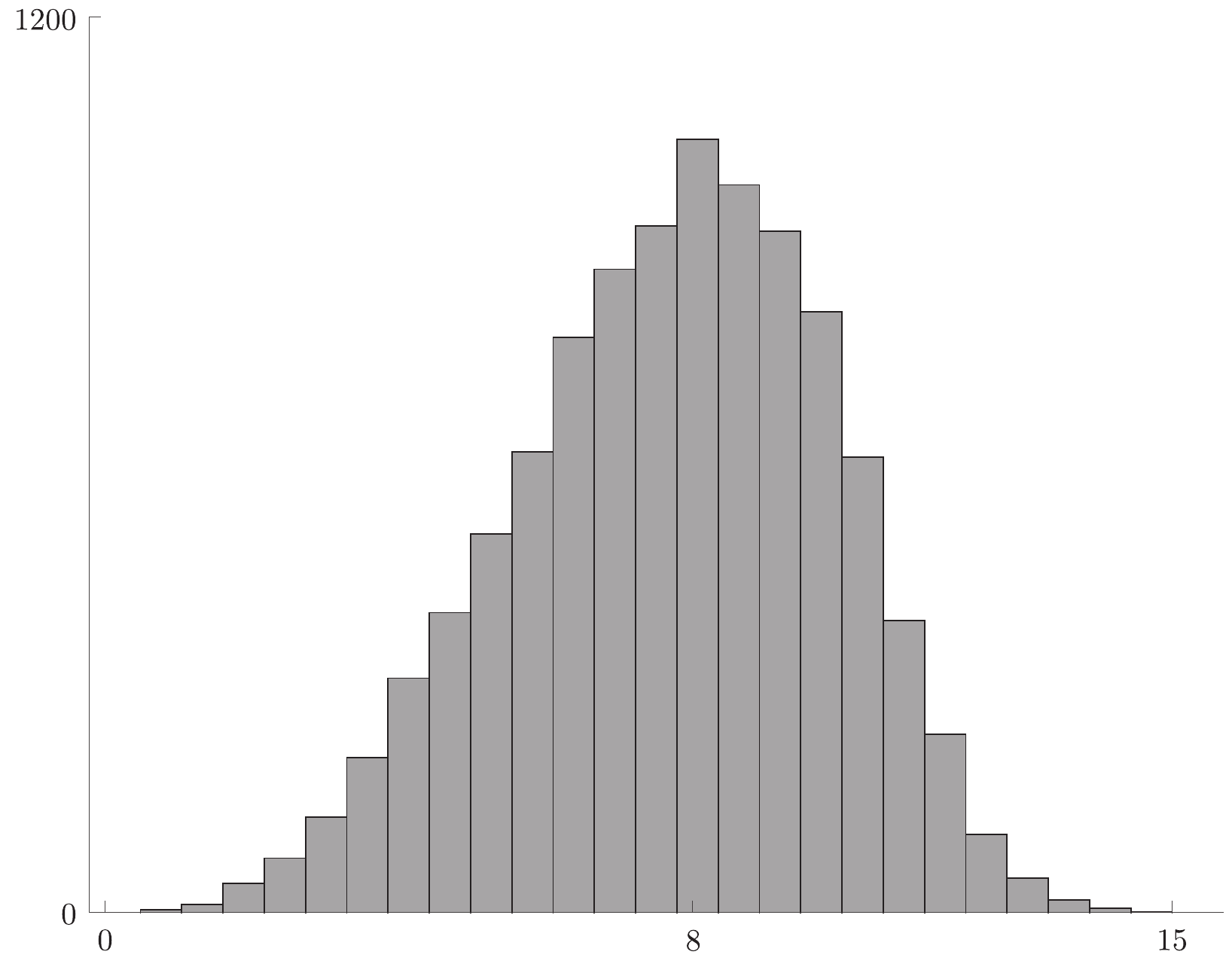}
  \caption{\footnotesize{Variations of $\mu$ $(RSD_{\mu}\,[\%])$}}
  \label{fig:rsdmfel}
\end{subfigure}%
\begin{subfigure}{.4\textwidth}
  \centering
   \includegraphics[scale=0.35]{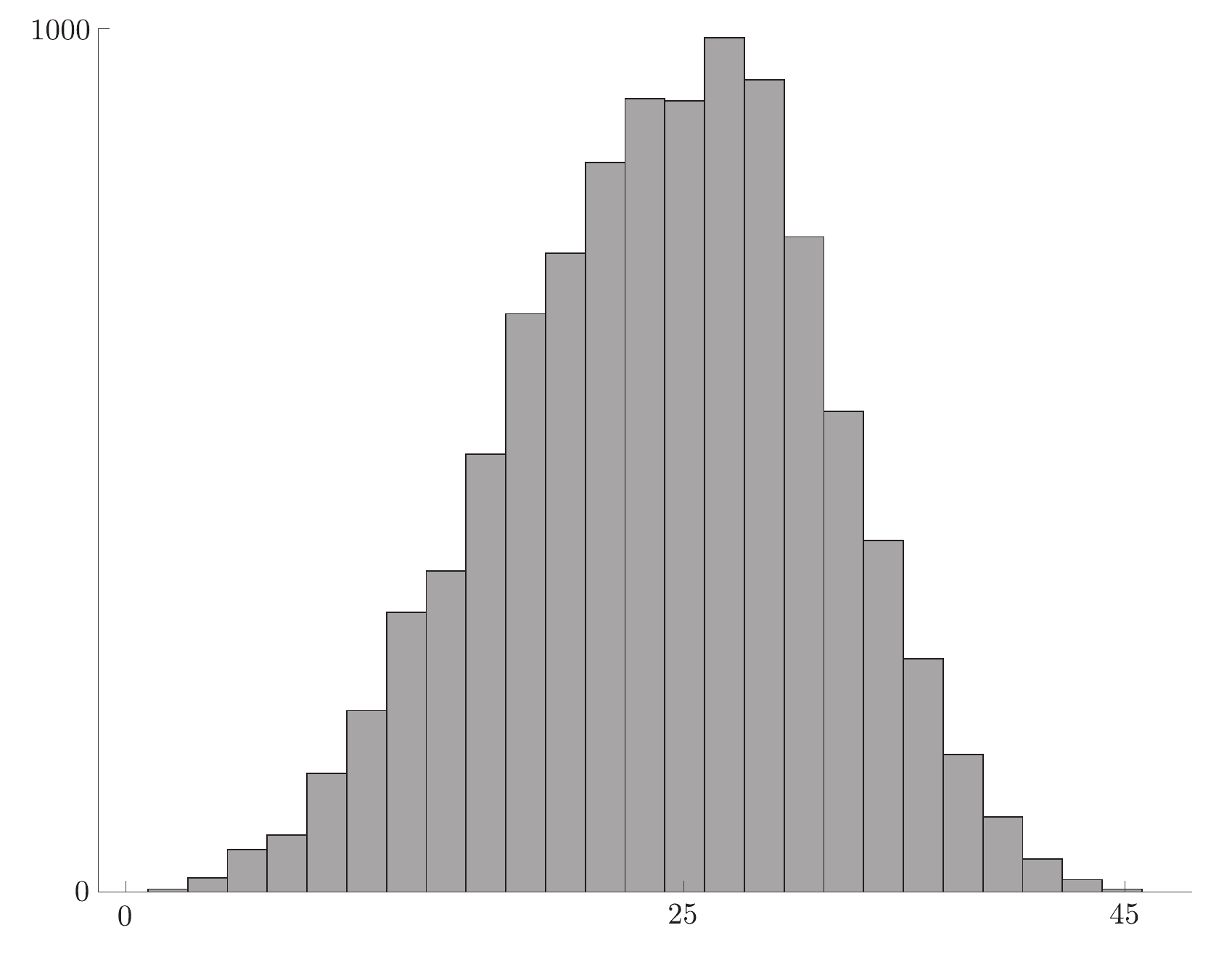}
  \caption{\footnotesize{Variations of $\lambda$ $(RSD_{\lambda}\,[\%])$}}
  \label{fig:rsdlfel}
\end{subfigure}%
\caption{\small{Distributions of random variations of $\mu$ and $\lambda$ for $10000$ TI media relevant to layered felsic rocks.}}
\label{fig:rsdfel}
\end{figure}

%%%%%%%%%%%%%%%%%%%%%%%%%%%%%%%%%%%%%%%%%%%%%%%%%%%%%%%%%%%%%%%%%%%%%%%%%%%%%%%%%%%%%%%%
Let us examine elasticity parameters relevant to felsic rocks (granite, diorite, felsic gneiss, intermediate gneiss, and metasediments). 
We set the ranges of these parameters to be $\lambda\in<20\,,\,50>$ and $\mu\in<30\,,\,40>$~\citep{JiEtAl}.
We repeat the same procedure of MC method, as in Section~\ref{sec:three}.
Herein, we compare the results obtained for felsic rocks to the ones for mafic rocks, and we indicate the differences between them (clearly exposed in Appendices~\ref{ap:two} and~\ref{ap:three}). 
In general, very limited range of $\mu$ entails small range of its variations in layers, as shown in Figure~\ref{fig:rsdmfel}.
Consequently, obtained values of anisotropy parameters, for $RSD_{\mu}<2\,\%$ or $RSD_{\mu}>2\,\%$, do not differ as much in both cases, as it is happens for mafic rocks, where the variations of $\mu$ have much larger range. 
The percentage of examples in which fluid indicators are satisfied, for $RSD_{\mu}<2\,\%$ or $RSD_{\mu}>2\,\%$, also do not differ as much in both cases.
Ranges of $\varphi$, $\epsilon$ and $\delta$, for $RSD_{\mu}>2\,\%$, are smaller.
Due to small values of anisotropy parameters, $|\varphi|>0.5\cdot10^{-3}$, is more effective in fluid detection than $|\varphi|>10^{-3}$.
According to Table~\ref{tab:bigtable2}, in case of $RSD_{\mu}>2\,\%$, fluid indicators are more efficient for felsic rocks than for mafic rocks, which might be caused by the smaller range of variations of $\mu$.
Thus, we may risk a statement that, in general, fluid indicators are more efficient for moderate variations of $\mu$, than for strong variations of $\mu$.

\begin{figure}[!htbp]
\centering
\begin{subfigure}{.4\textwidth}
  \centering
   \includegraphics[scale=0.35]{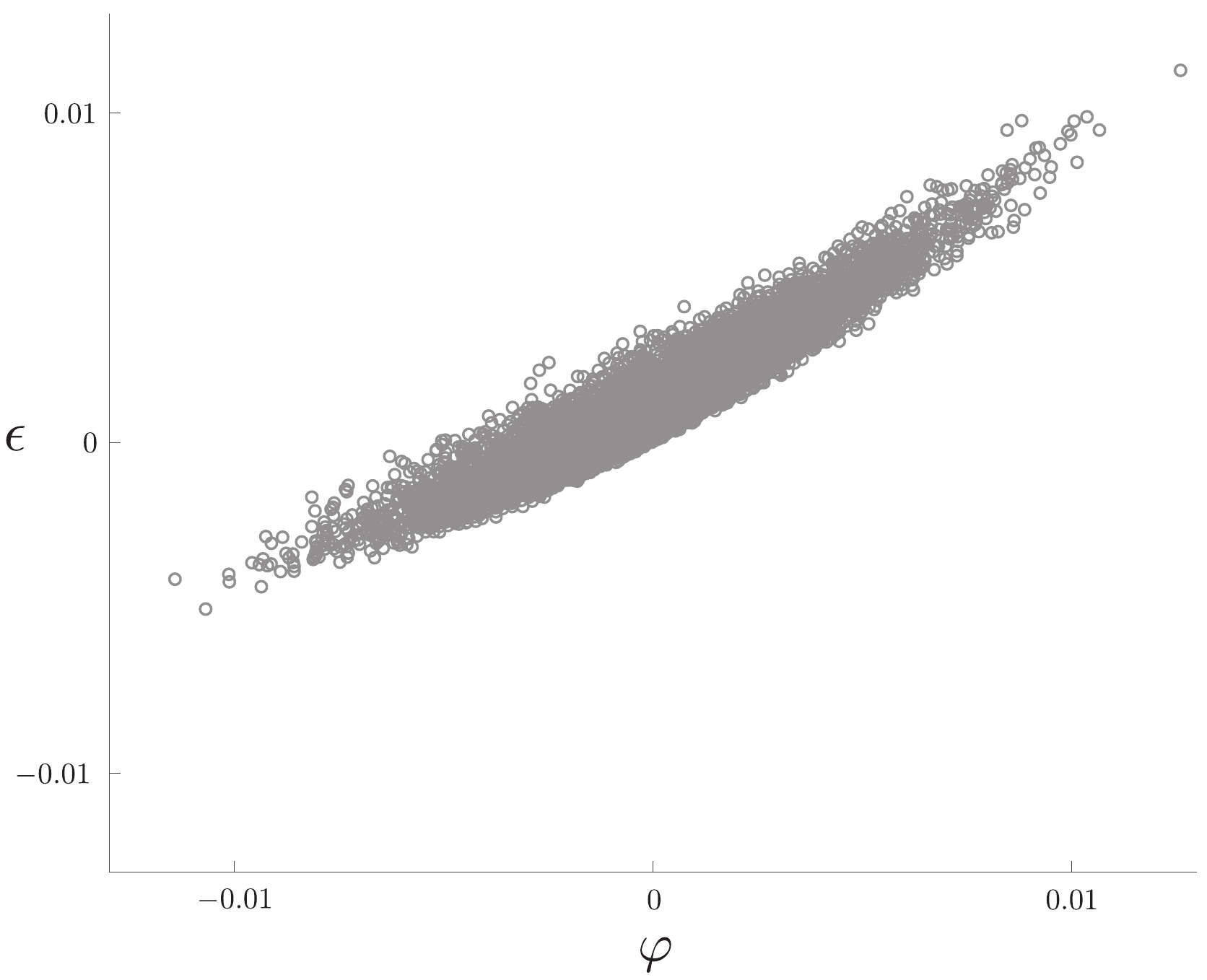}
  \caption{\footnotesize{$\varphi$ versus $\epsilon$}}
  \label{fig:rel_ml_phieps_fel}
\end{subfigure}%
\begin{subfigure}{.4\textwidth}
  \centering
   \includegraphics[scale=0.35]{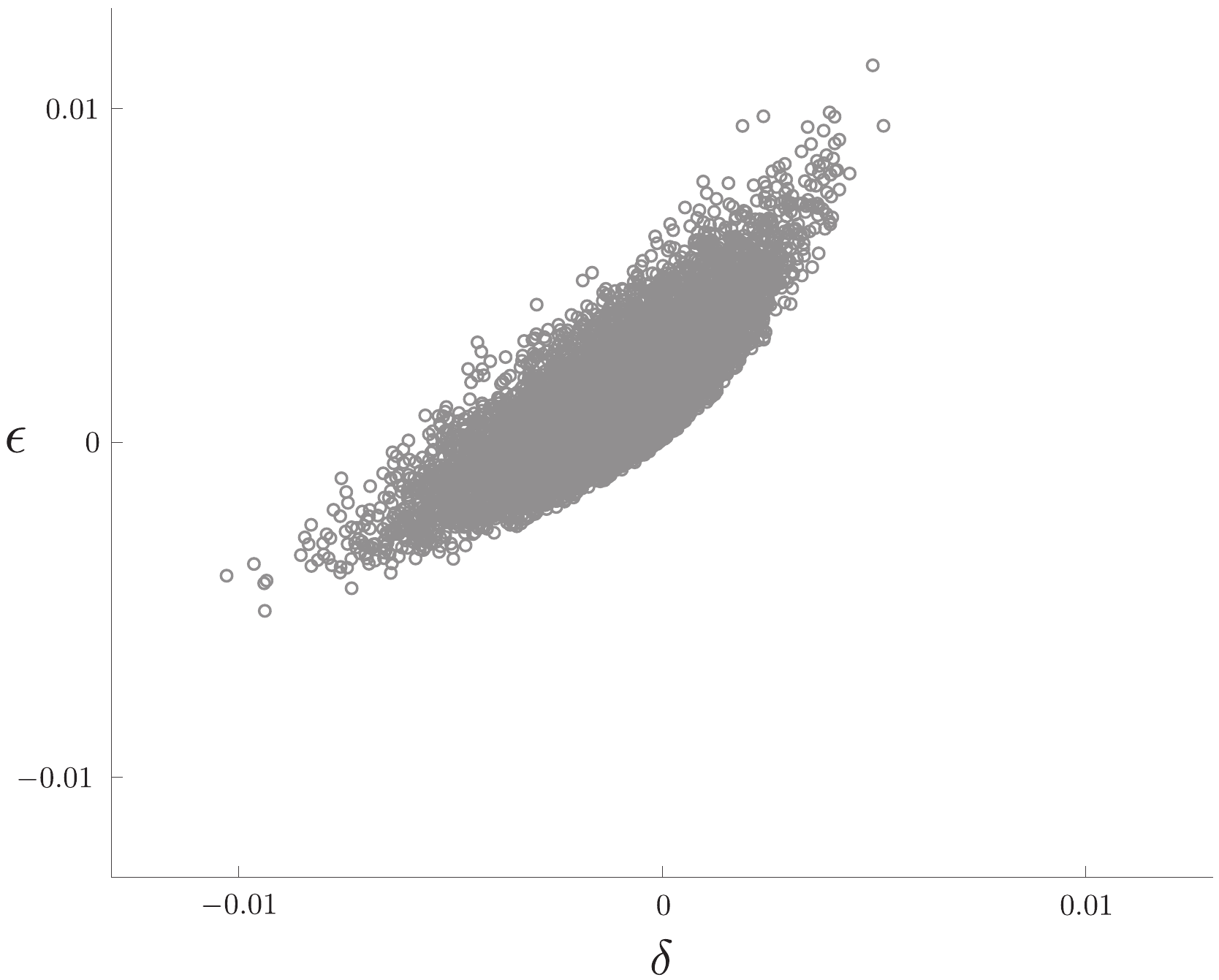}
  \caption{\footnotesize{$\delta$ versus $\epsilon$}}
  \label{fig:rel_ml_deleps_fel}
\end{subfigure}%
\caption{\small{Cross-plots of anisotropy parameters for $10000$ examples of equivalent TI felsic media.}}
\label{fig:rel_ml_fel}
\end{figure}

On the other hand, as shown in Figure~\ref{fig:rsdlfel}, for felsic rocks, strong variations of $\lambda$ occur to be much larger. 
Thus, all of the fluid indicators containing, $\varphi$, which is very sensitive to variations of $\lambda$, become even more effective.
In the case of $RSD_{\mu}>2\,\%$ and $RSD_{\lambda}>20\,\%$, anisotropy parameters are influenced by quite low variations of $\mu$ and very strong variations of $\lambda$.
As a consequence, $\varphi$ has a very similar range for both type of rocks.
Let us explain how the small range of variations of $\mu$ and large range of variations of $\lambda$ for felsic rocks, influence the anisotropy parameters $\epsilon$ and $\delta$. 
These anisotropy parameters are not as sensitive to variations of $\lambda$ as $\varphi$ is, therefore low variations of $\mu$ cause their ranges to be significantly smaller. 
Even though their ranges are smaller, we notice that $\epsilon$ has more negative values and $\delta$ more positive values, which cause their distributions to approach the normal shape (Figures~\ref{fig01e_fel}--\ref{fig01f_fel} and Figures~\ref{fig01h_fel}--\ref{fig01i_fel}).
For mafic rocks, the normal shape of distributions of $\epsilon$ and $\delta$ is characteristic only for $RSD_{\mu}<2\,\%$ and $RSD_{\lambda}>2\,\%$.
For felsic rocks, however, due to the small range of variations of $\mu$ and large range of variations of $\lambda$, the quasi-normal shape is additionally visible in the case of $RSD_{\mu}>2\,\%$ and $RSD_{\lambda}>2\,\%$. 
Since in felsic rocks the most frequently occurring variations are $RSD_{\mu}>2\,\%$ and $RSD_{\lambda}>20\,\%$, we see that also in general case of variations, distributions from Figures~\ref{fig:diseps_fel}--\ref{fig:disdel_fel} incline towards normal shape, which is reflected in long and narrow shape of a cloud of points from Figures~\ref{fig:rel_ml_phieps_fel},~\ref{fig:rel_ml_deleps_fel},~\ref{fig:rel_phieps_mb_lvb_fel} or~\ref{fig:rel_deleps_mb_lvb_fel}.

Let us analyze elasticity parameters for sandstones (brine sands, gas sands and others) and compare the results to mafic rocks.
Based on works of \cite{CastagnaSmith} and \cite{WanniarachichiEtAl}, the approximate ranges of these parameters are $\lambda\in<-3\,,\,20>$ and $\mu\in<1\,,\,30>$.
However, to perform MC simulations, we set $a=3$, $b=20$, $c=1$ and $d=30$.
The positive sign of, $a$, instead of negative one, is motivated by the fact that negative, zero or small positive values of $\lambda$ might lead to issues within Backus average~\citep{KudelaStanoev}.
Such a choice of, $a$, ensures that the resulting simulations are not influenced by the improperly used Backus average.

As shown in Figure~\ref{fig:rsdsan}, in general, sandstones represent strong variations of both Lam\'e parameters.
Consequently, $\varphi$, $\epsilon$ and $\delta$, for $RSD_{\mu}<2\,\%$ and $RSD_{\lambda}>20\,\%$, and for $RSD_{\mu}>2\,\%$ and $RSD_{\lambda}\in\forall$, have much larger ranges.
In the case of $RSD_{\mu}<2\,\%$ and $RSD_{\lambda}>20\,\%$., fluid indicators are more effective.
For $RSD_{\mu}>2\,\%$ and $RSD_{\lambda}\in\forall$, due to larger absolute values of anisotropy parameters, we should consider $|\varphi|>5\cdot10^{-3}$, instead of $|\varphi|>10^{-3}$.
As we have mentioned above, in general, our fluid indicators are more efficient in case of smaller variations of $\mu$, therefore for sandstones with $RSD_{\mu}>2\,\%$, they are slightly less effective.
For instance, indicators $\epsilon<0$ and $\delta>0$, detect only strong variations of $\lambda$, they are insensitive to moderate variations.
For near-constant rigidity, the most effective indicator is $|\varphi|>10^{-4}$, for $RSD_{\mu}>2\,\%$, the most accurate occurs, $|\varphi|>5\cdot10^{-3}$. 

\begin{figure}[!htbp]
\centering
\begin{subfigure}{.4\textwidth}
\centering
   \includegraphics[scale=0.35]{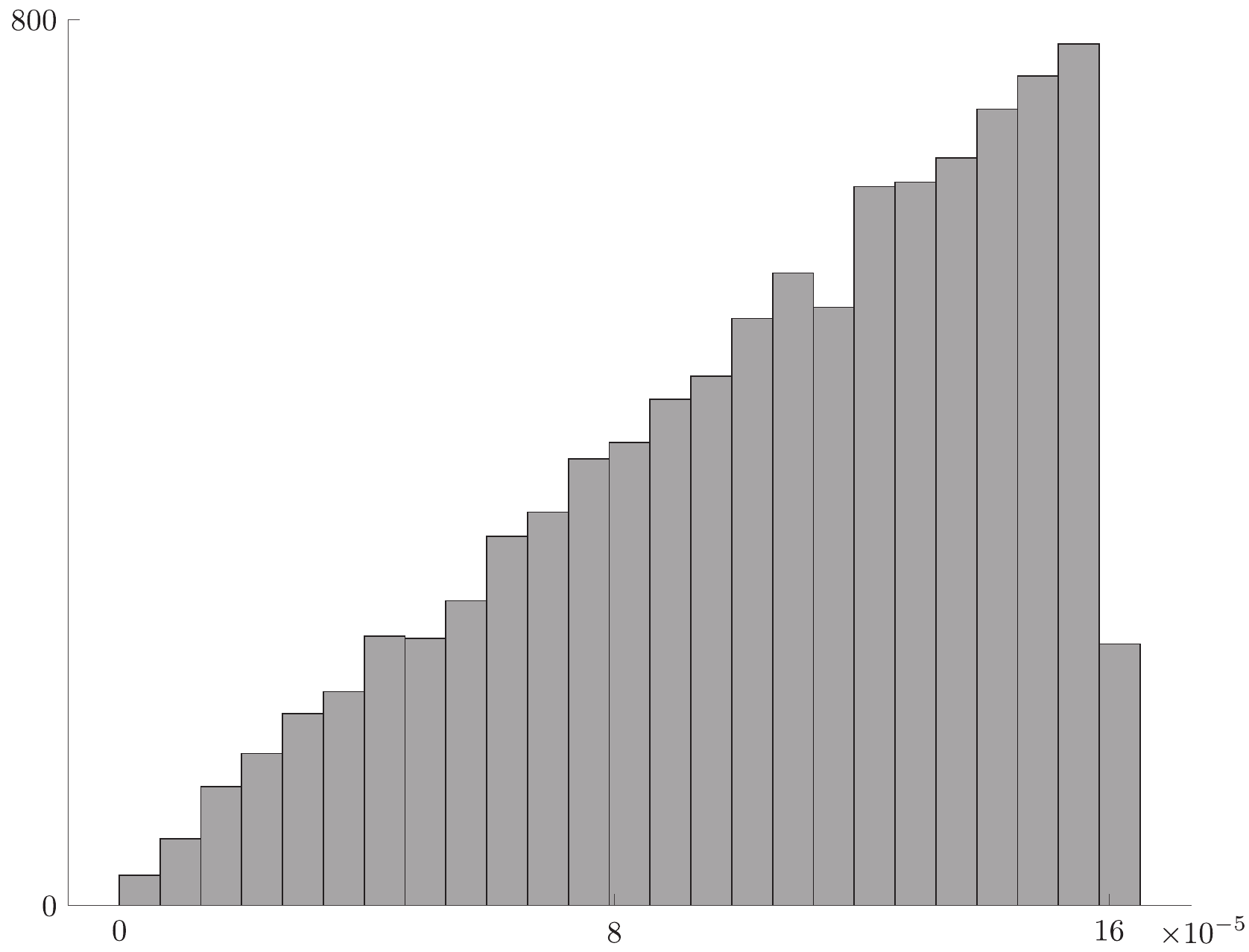}
 \caption{\small{$RSD_{\mu}<2\,\%\,$}}
\label{fig:gamma_ms_fel}
\end{subfigure}
\begin{subfigure}{.4\textwidth}
\centering
   \includegraphics[scale=0.35]{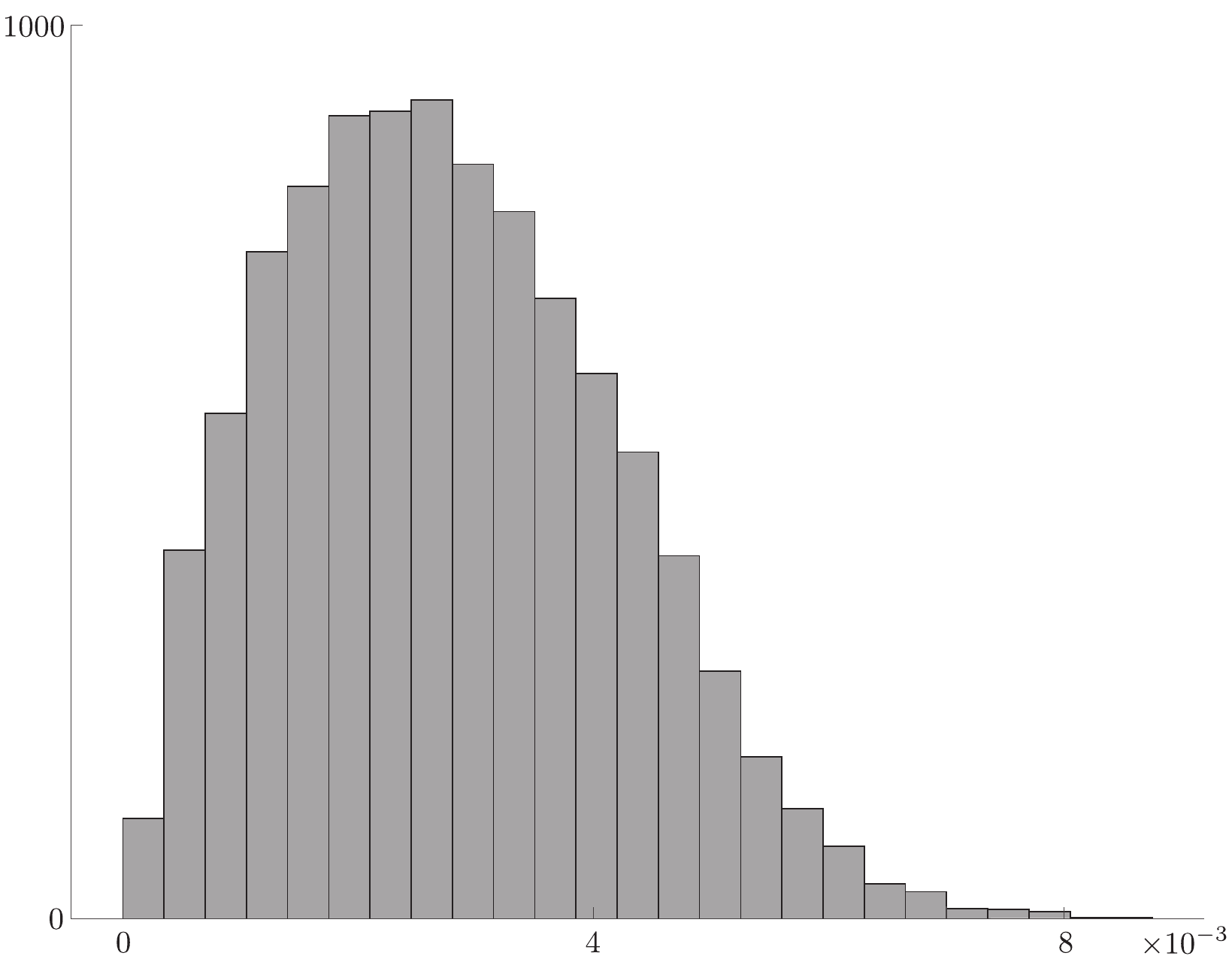}
 \caption{\small{$RSD_{\mu}>2\,\%\,$}}
\label{fig:gamma_mb_fel}
\end{subfigure}
\caption{\small{Distribution of $\gamma$ for $10000$ examples of equivalent TI felsic media.}}
\label{fig:gamma_fel}
\end{figure}

\begin{figure}[!htbp]
\centering
\begin{subfigure}{.33\textwidth}
  \centering
   \includegraphics[scale=0.3]{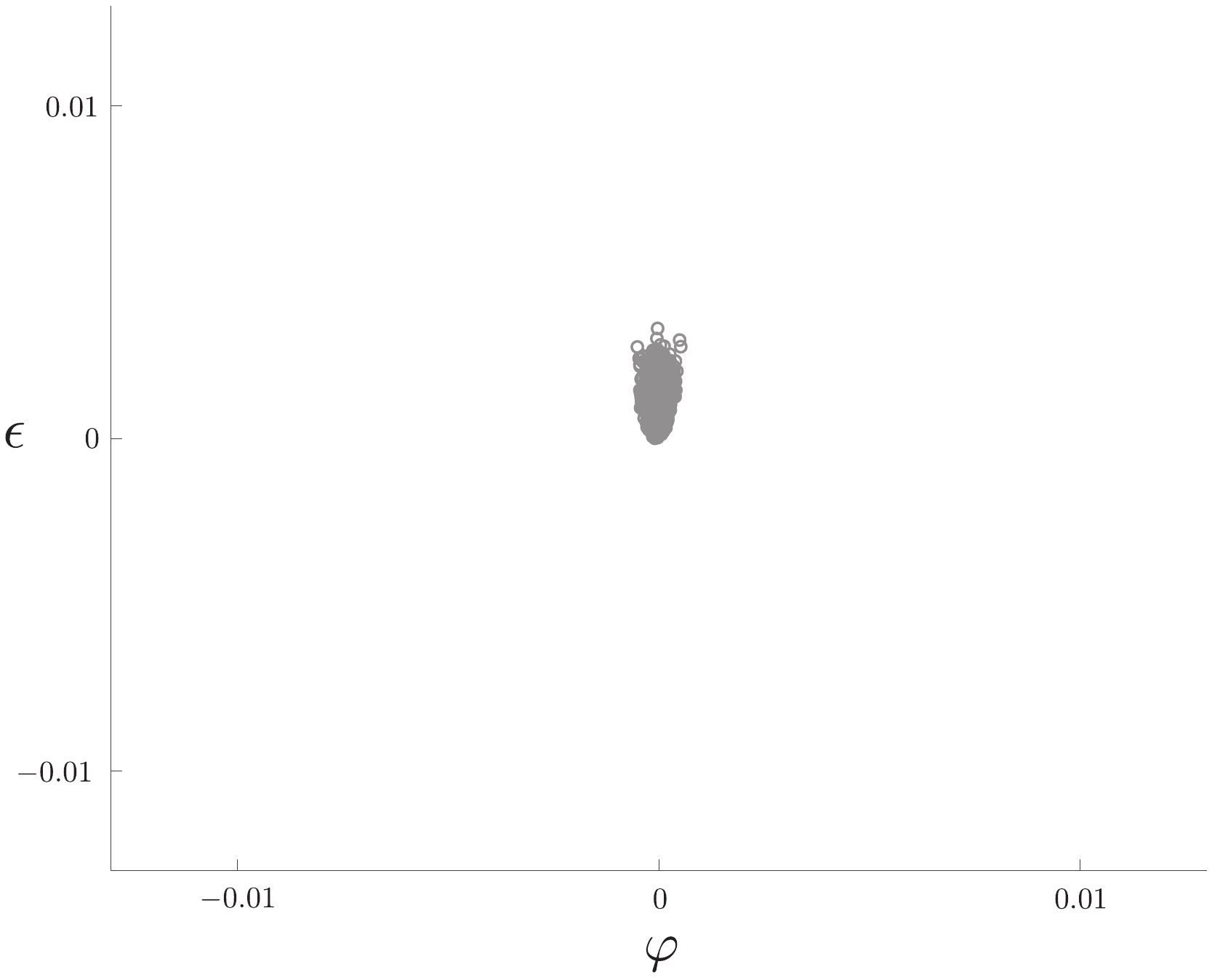}
  \caption{\footnotesize{$RSD_{\lambda}<2\,\%$}}
  \label{fig:rel_phieps_mb_ls_fel}
\end{subfigure}%
\begin{subfigure}{.33\textwidth}
  \centering
   \includegraphics[scale=0.3]{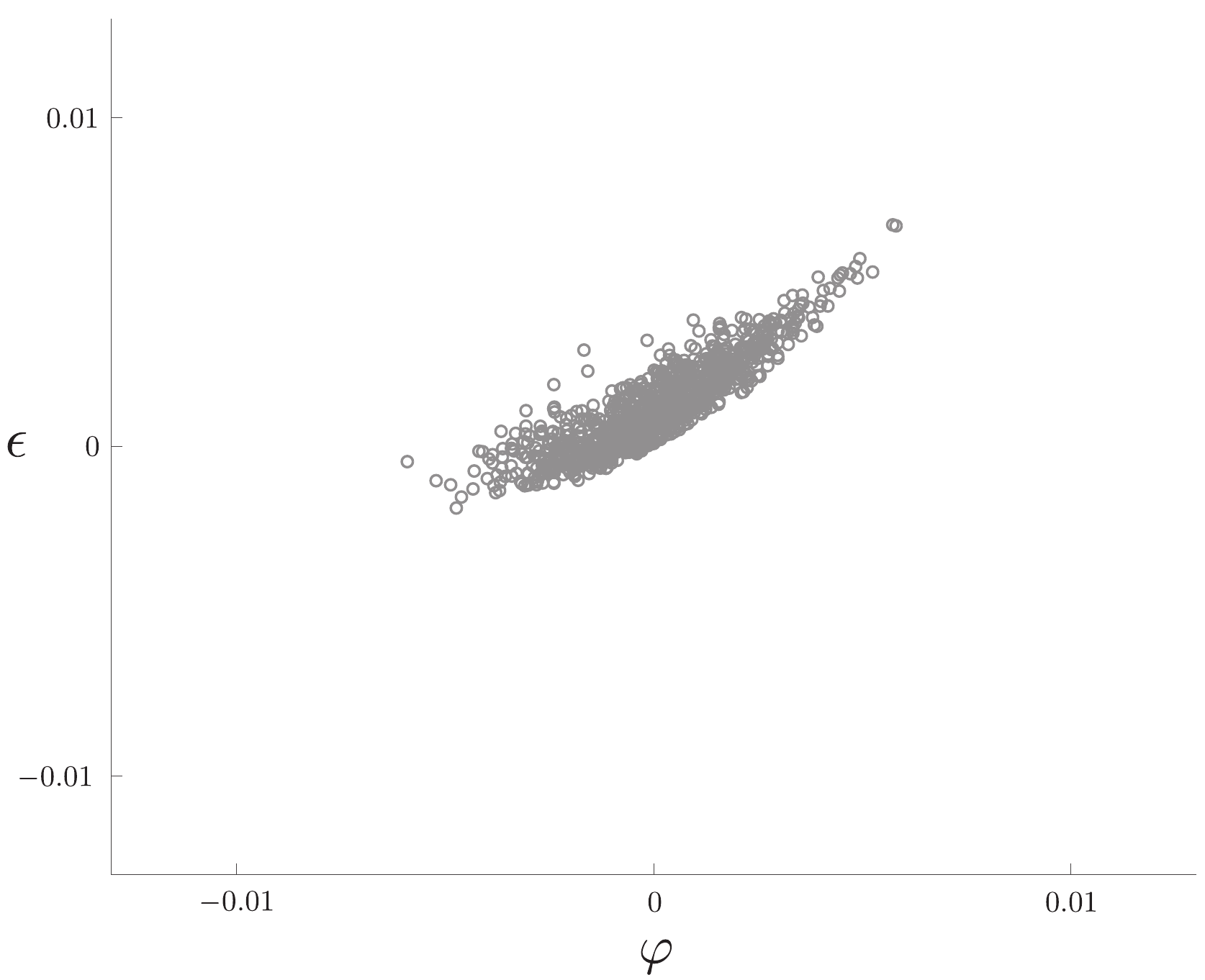}
  \caption{\footnotesize{$RSD_{\lambda}\in\left(2\,\%\,,\,20\,\%\right)$}}
  \label{fig:rel_phieps_mb_2l20_fel}
\end{subfigure}%
\begin{subfigure}{.33\textwidth}
  \centering
   \includegraphics[scale=0.3]{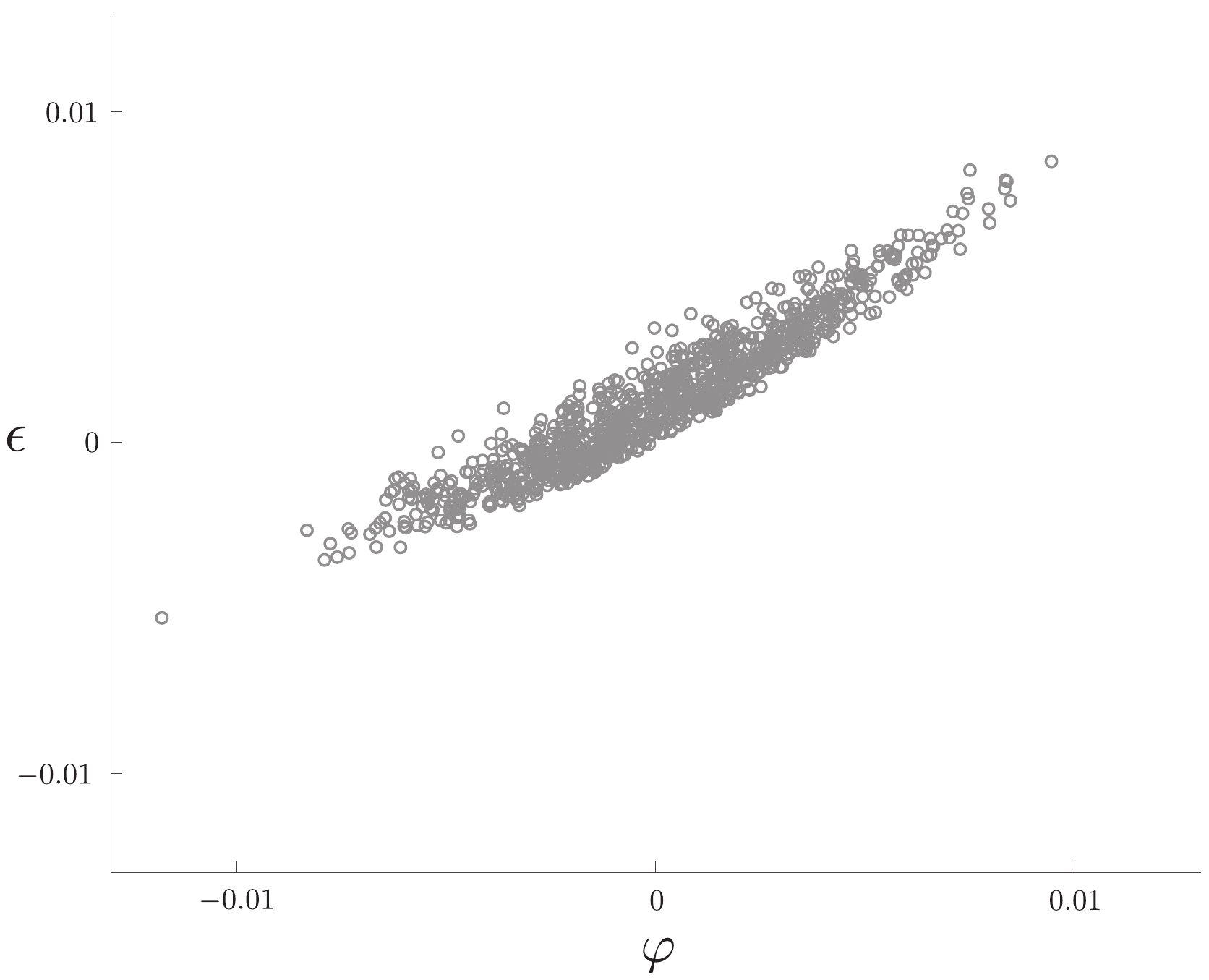}
  \caption{\footnotesize{$RSD_{\lambda}>20\,\%$}}
  \label{fig:rel_phieps_mb_lvb_fel}
\end{subfigure}
\caption{\small{Cross-plots of $\varphi$ versus $\epsilon$ for $1000$ examples of equivalent TI felsic media with $RSD_{\mu}>2\,\%\,$.}}
\label{fig:rel_phieps_mb_fel}
\end{figure}

\begin{figure}[!htbp]
\centering
\begin{subfigure}{.33\textwidth}
  \centering
   \includegraphics[scale=0.3]{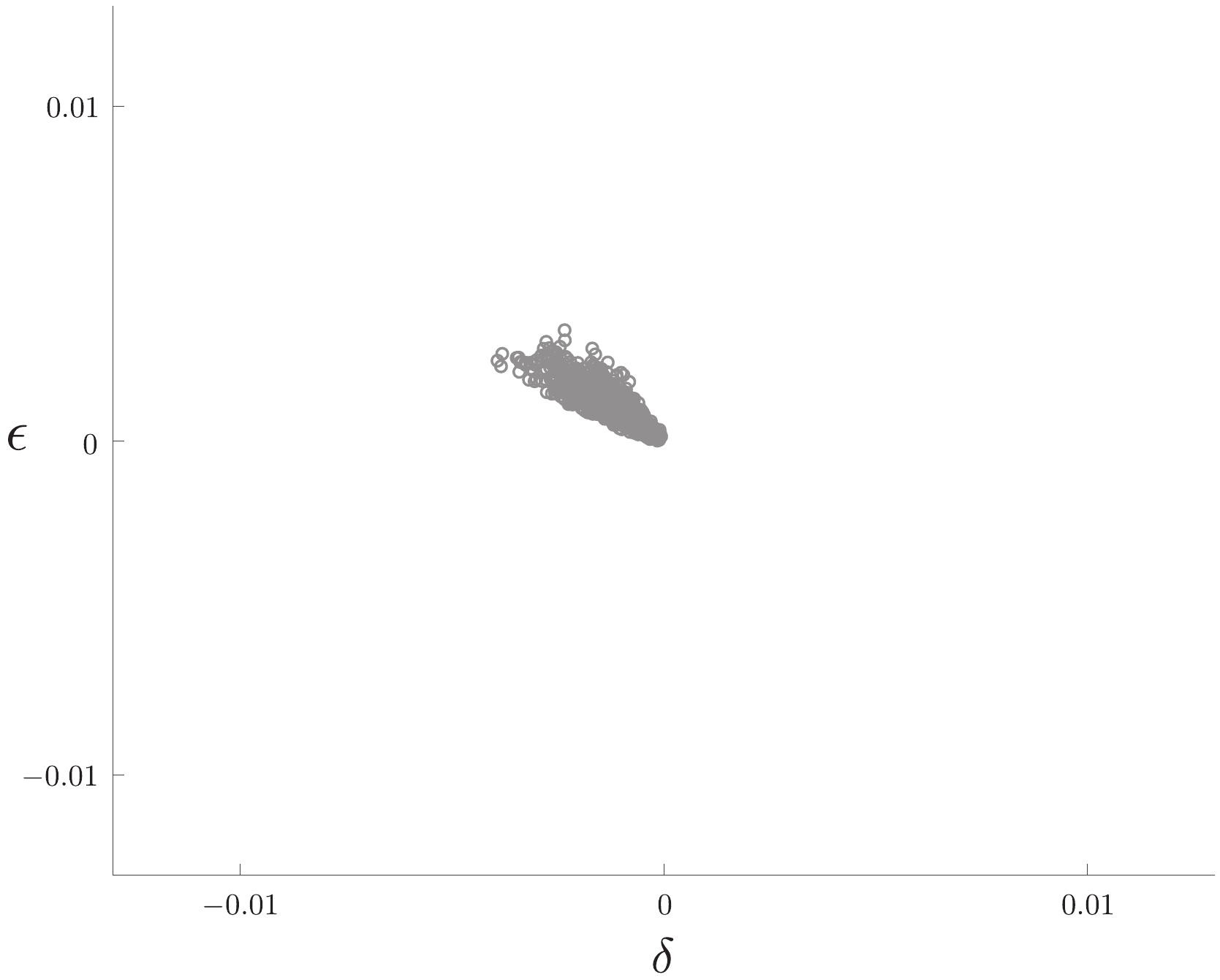}
  \caption{\footnotesize{$RSD_{\lambda}<2\,\%$}}
  \label{fig:rel_deleps_mb_ls_fel}
\end{subfigure}%
\begin{subfigure}{.33\textwidth}
  \centering
   \includegraphics[scale=0.3]{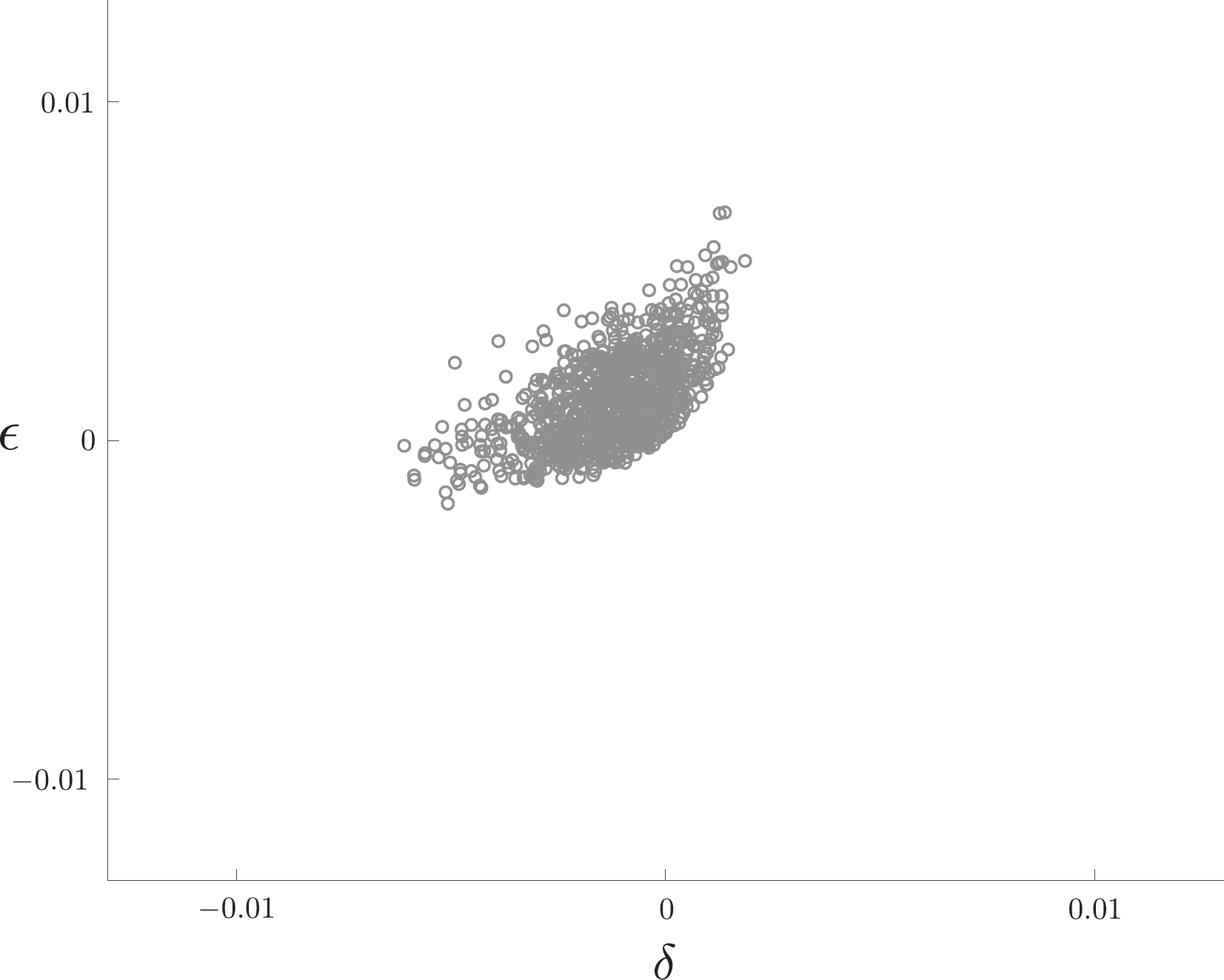}
  \caption{\footnotesize{$RSD_{\lambda}\in\left(2\,\%\,,\,20\,\%\right)$}}
  \label{fig:rel_deleps_mb_2l20_fel}
\end{subfigure}%
\begin{subfigure}{.33\textwidth}
  \centering
   \includegraphics[scale=0.3]{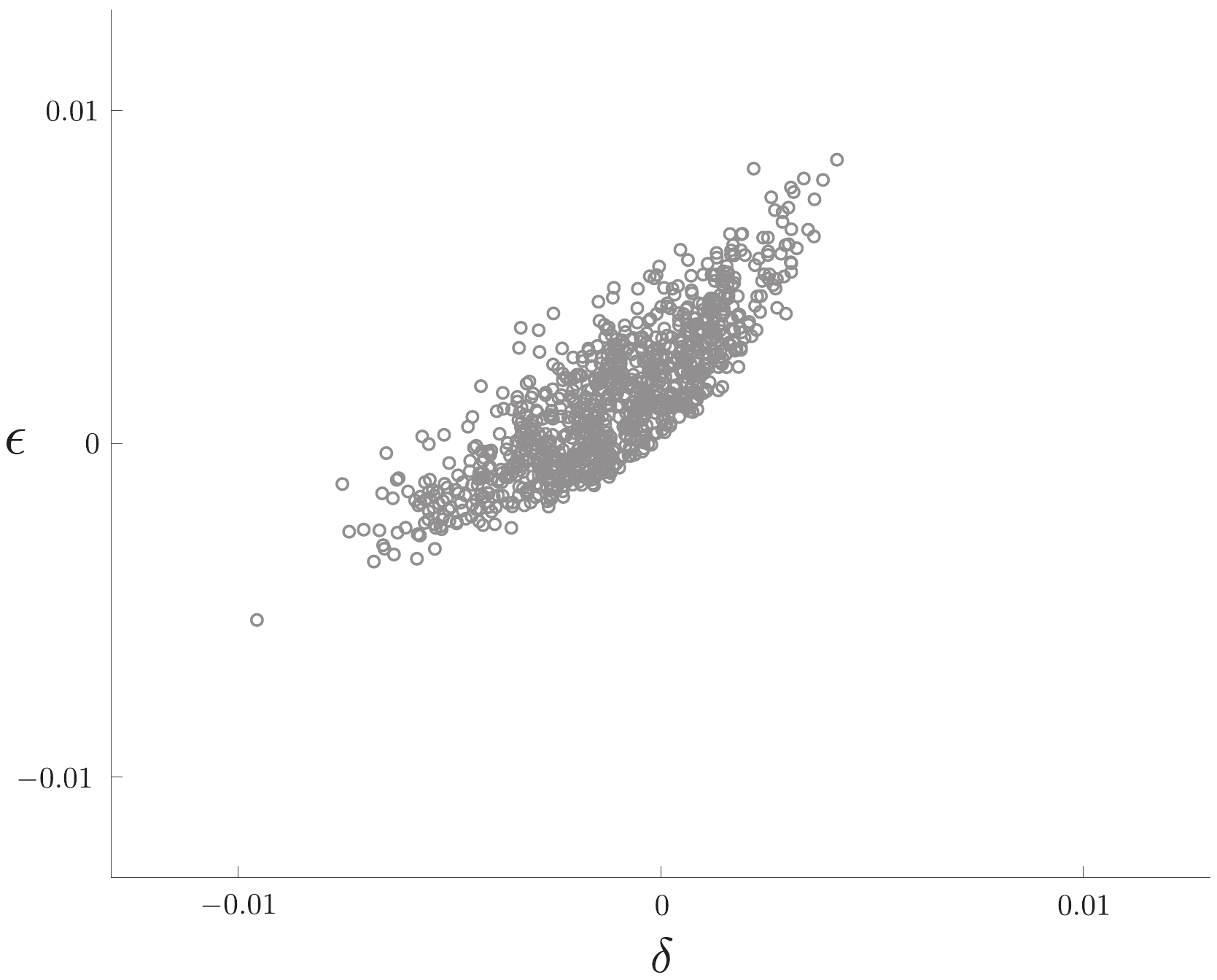}
  \caption{\footnotesize{$RSD_{\lambda}>20\,\%$}}
  \label{fig:rel_deleps_mb_lvb_fel}
\end{subfigure}
\caption{\small{Cross-plots of $\delta$ versus $\epsilon$ for $1000$ examples of equivalent TI felsic media with $RSD_{\mu}>2\,\%\,$.}}
\label{fig:rel_deleps_mb_fel}
\end{figure}

\begin{figure}
\centering
\begin{subfigure}{.33\textwidth}
  \centering
   \includegraphics[scale=0.29]{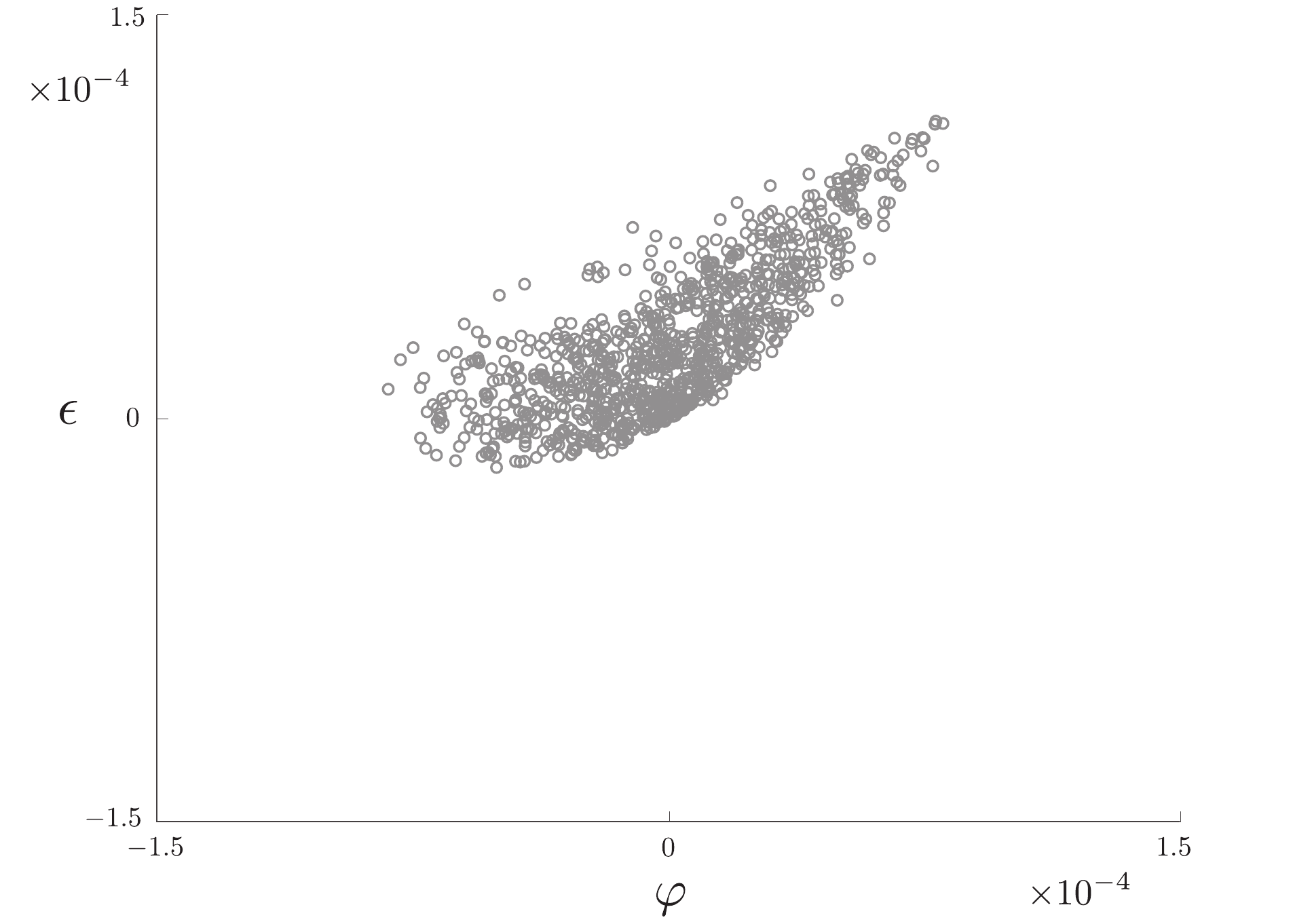}
  \caption{\footnotesize{$RSD_{\lambda}<2\,\%$}}
  \label{fig:rel_phieps_ms_ls_fel}
\end{subfigure}%
\begin{subfigure}{.33\textwidth}
  \centering
   \includegraphics[scale=0.29]{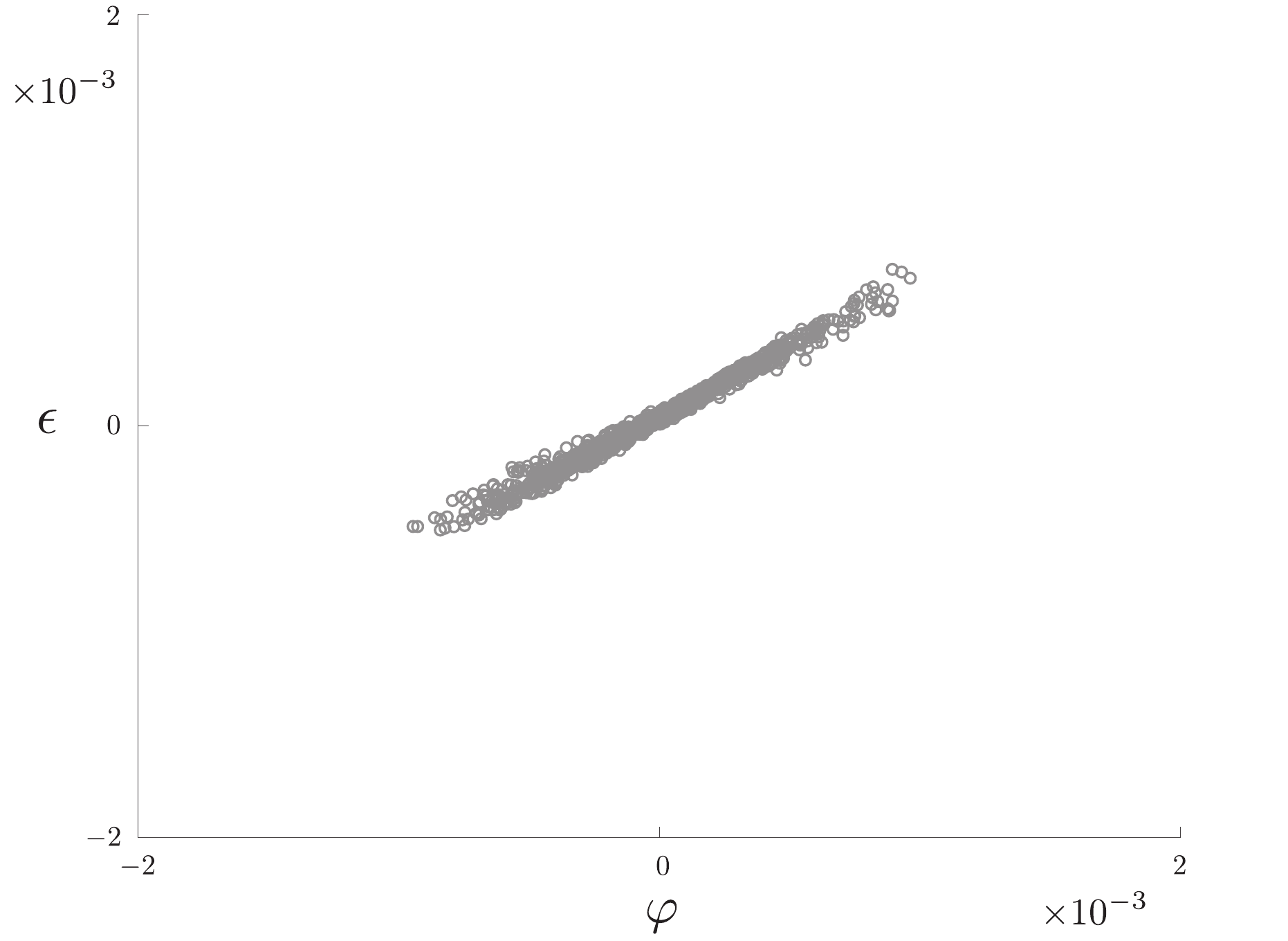}
  \caption{\footnotesize{$RSD_{\lambda}\in\left(2\,\%\,,\,20\,\%\right)$}}
  \label{fig:rel_phieps_ms_2l20_fel}
\end{subfigure}%
\begin{subfigure}{.33\textwidth}
  \centering
   \includegraphics[scale=0.29]{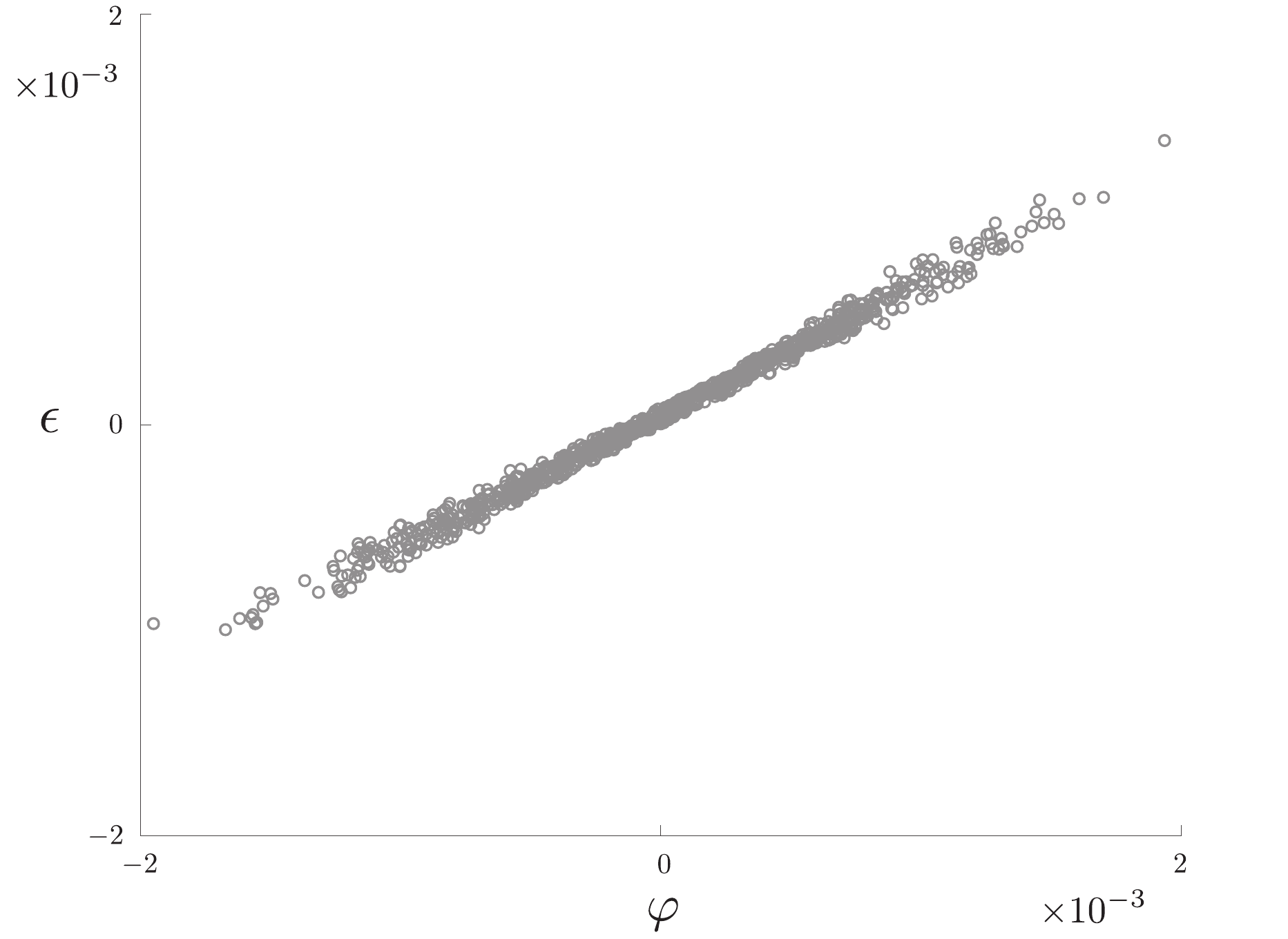}
  \caption{\footnotesize{$RSD_{\lambda}>20\,\%$}}
  \label{fig:rel_phieps_ms_lvb_fel}
\end{subfigure}
\caption{\small{Cross-plots of $\varphi$ versus $\epsilon$ for $1000$ examples of equivalent TI felsic media with $RSD_{\mu}<2\,\%\,$.}}
\label{fig:rel_phieps_ms_fel}
\end{figure}

\begin{figure}
\centering
\begin{subfigure}{.33\textwidth}
  \centering
   \includegraphics[scale=0.29]{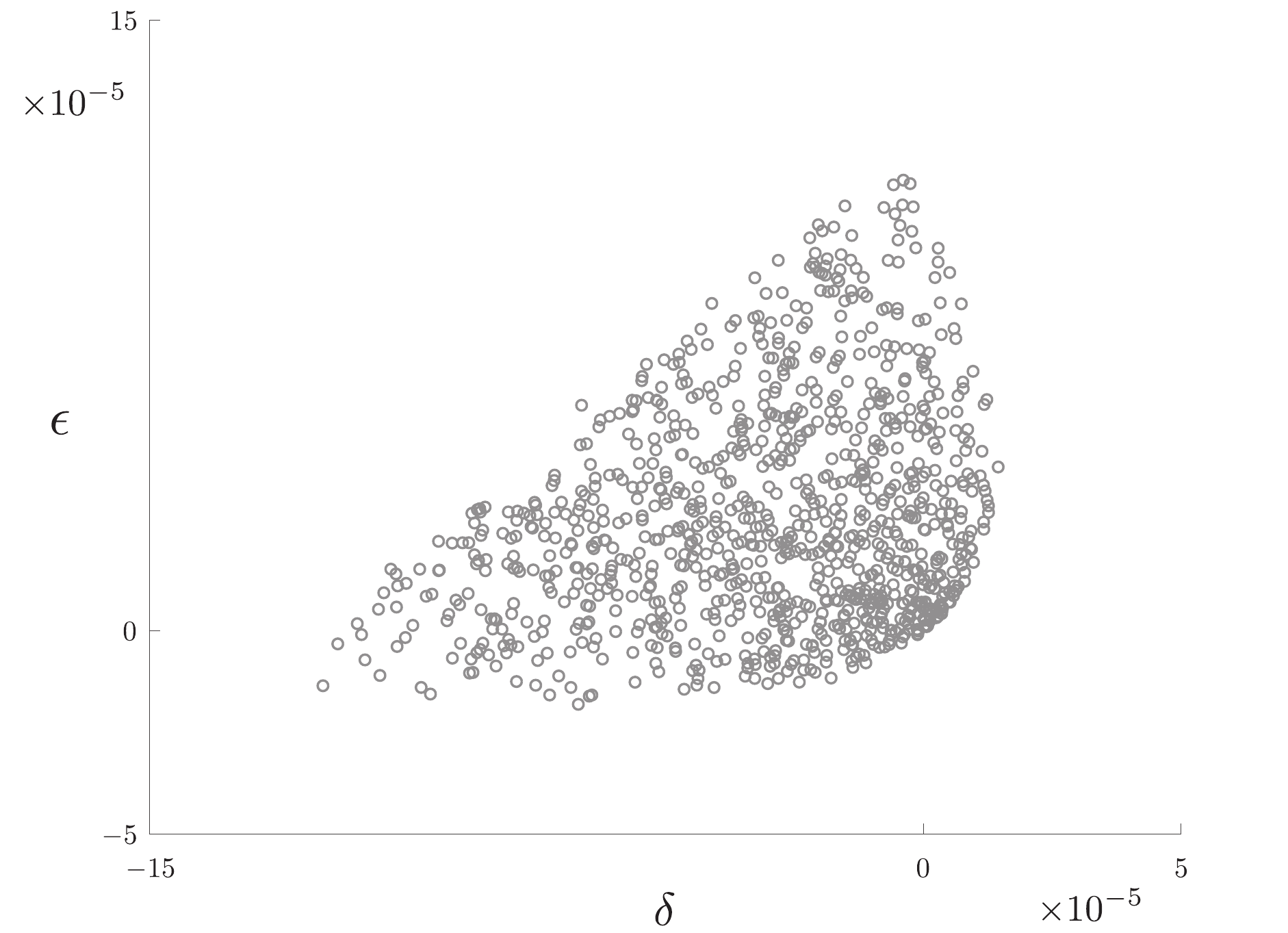}
  \caption{\footnotesize{$RSD_{\lambda}<2\,\%$}}
  \label{fig:rel_deleps_ms_ls_fel}
\end{subfigure}%
\begin{subfigure}{.33\textwidth}
  \centering
   \includegraphics[scale=0.29]{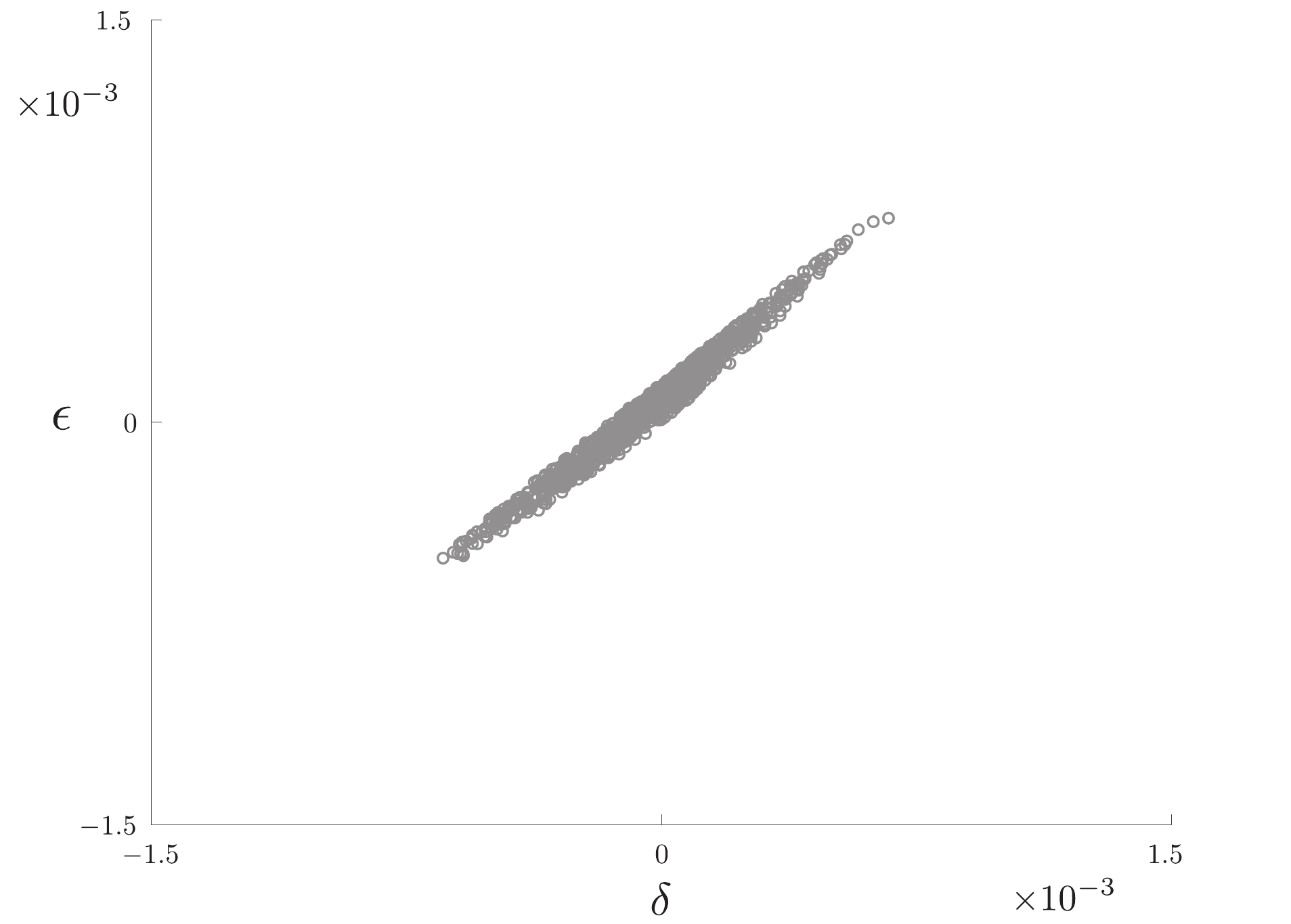}
  \caption{\footnotesize{$RSD_{\lambda}\in\left(2\,\%\,,\,20\,\%\right)$}}
  \label{fig:rel_deleps_ms_2l20_fel}
\end{subfigure}%
\begin{subfigure}{.33\textwidth}
  \centering
   \includegraphics[scale=0.29]{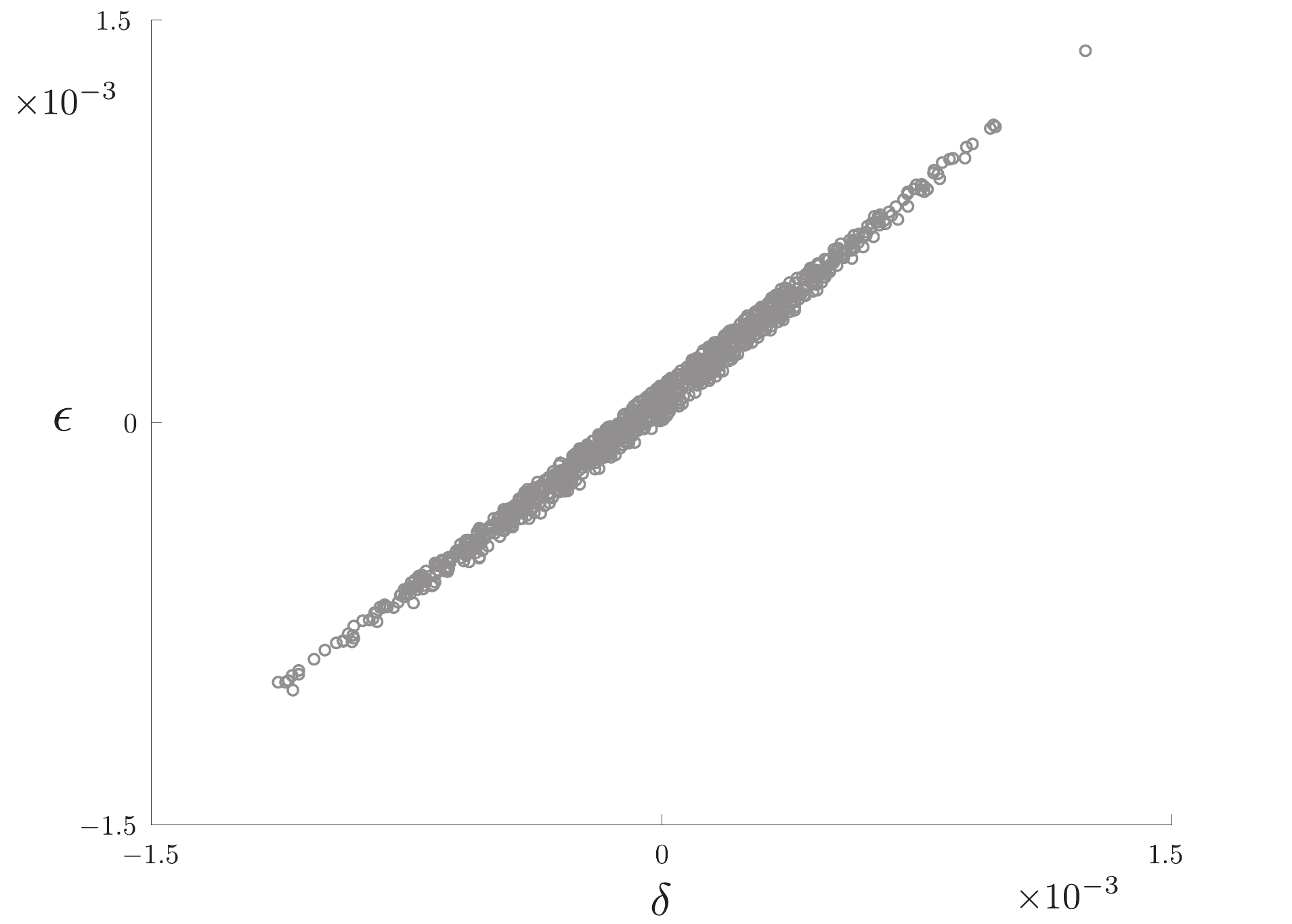}
  \caption{\footnotesize{$RSD_{\lambda}>20\,\%$}}
  \label{fig:rel_deleps_ms_lvb_fel}
\end{subfigure}
\caption{\small{Cross-plots of $\delta$ versus $\epsilon$ for $1000$ examples of equivalent TI felsic media with $RSD_{\mu}<2\,\%\,$.}}
\label{fig:rel_deleps_ms_fel}
\end{figure}

\begin{figure}[!htbp]
\centering
\begin{subfigure}{.33\textwidth}
  \centering
   \includegraphics[scale=0.3]{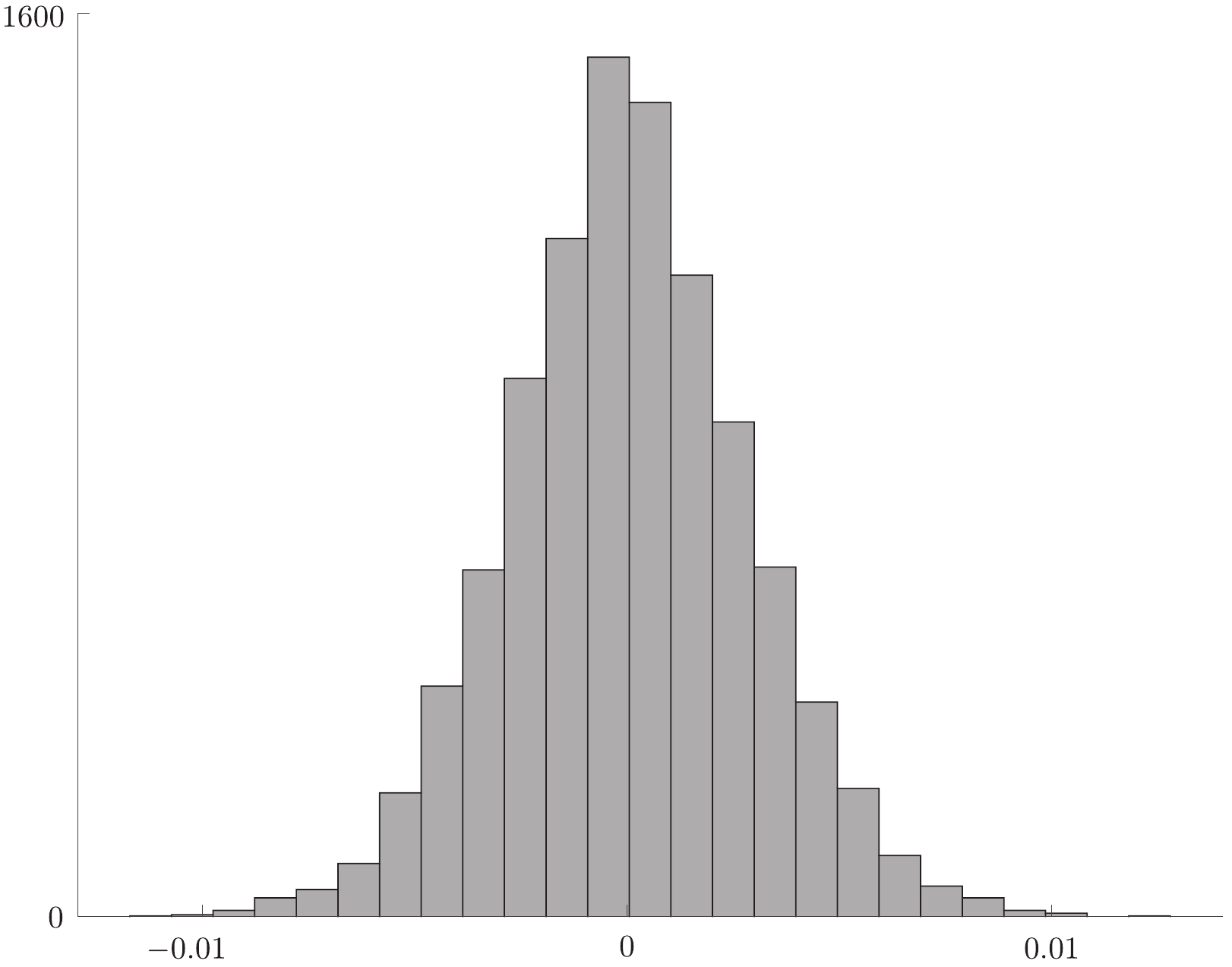}
  \caption{\footnotesize{Distribution of $\varphi$ }}
  \label{fig:disphi_fel}
\end{subfigure}%
\begin{subfigure}{.33\textwidth}
  \centering
   \includegraphics[scale=0.3]{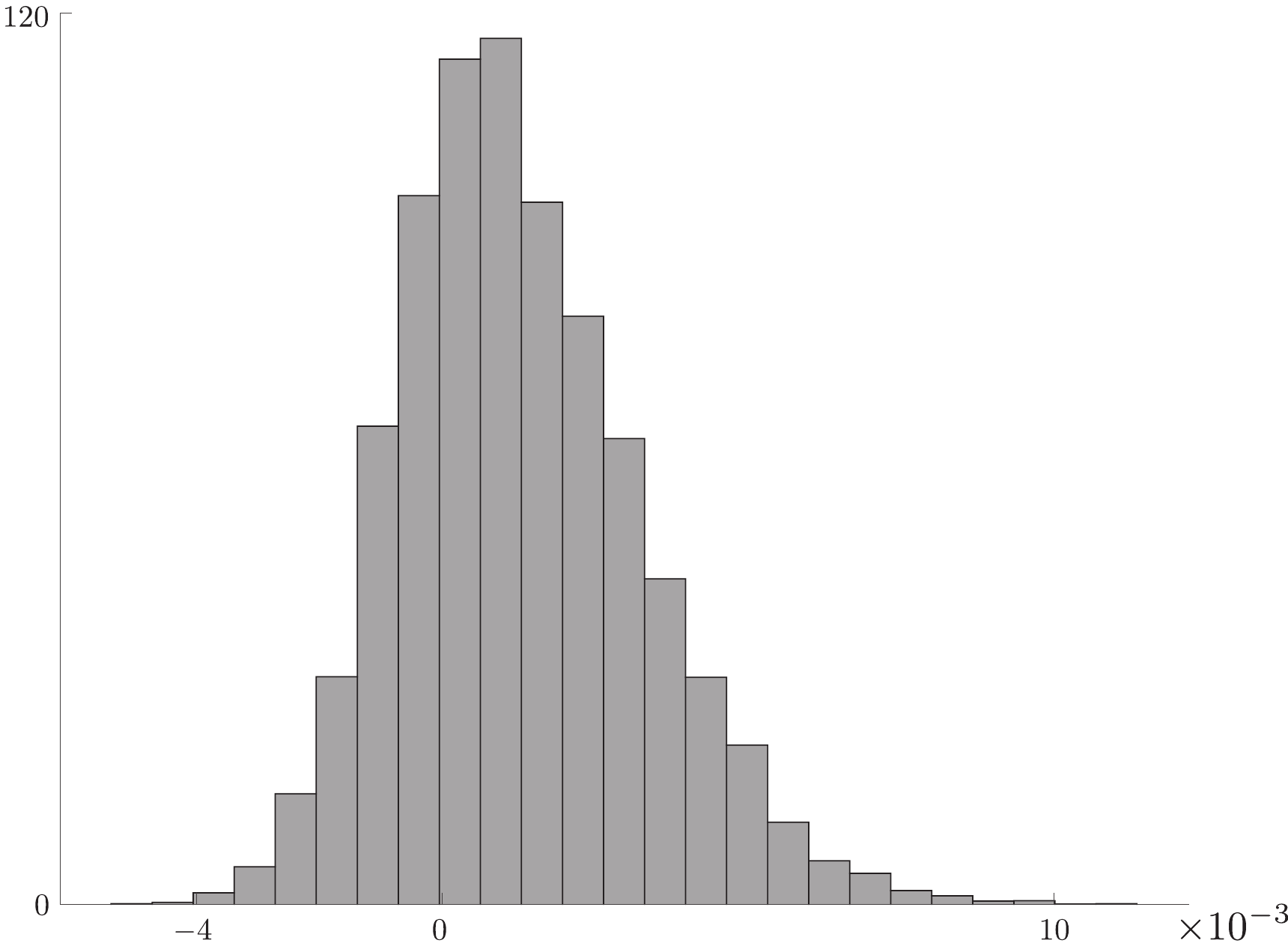}
  \caption{\footnotesize{Distribution of $\epsilon$ }}
  \label{fig:diseps_fel}
\end{subfigure}%
\begin{subfigure}{.33\textwidth}
  \centering
   \includegraphics[scale=0.3]{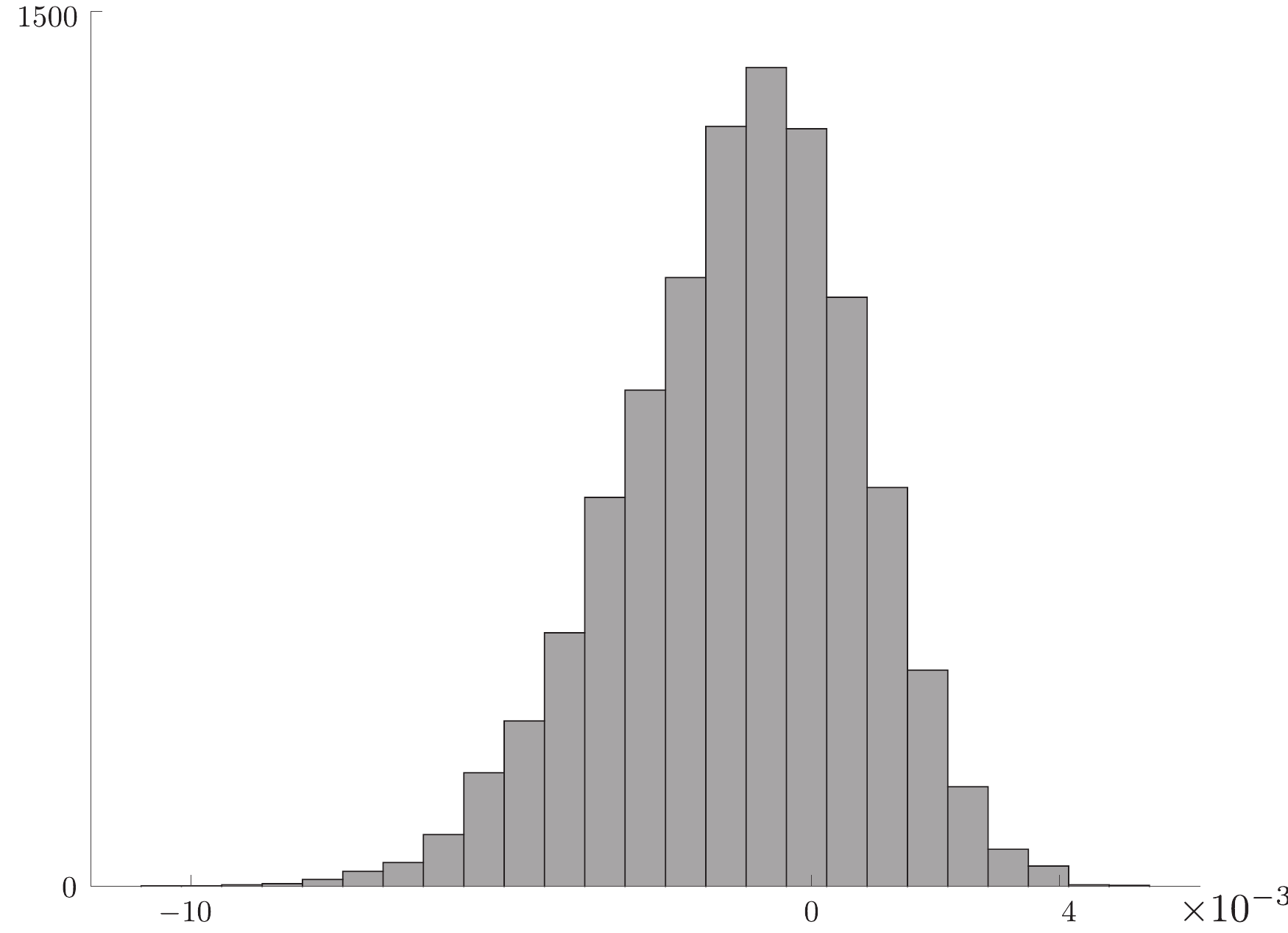}
  \caption{\footnotesize{Distribution of $\delta$ }}
  \label{fig:disdel_fel}
\end{subfigure}
\caption{\small{Distribution of anisotropy parameters for $10000$ TI media, relevant to layered felsic rocks.}}
\label{fig:dis_fel}
\end{figure}

\begin{figure}[!htbp]
\centering
\begin{subfigure}{.33\textwidth}
  \centering
   \includegraphics[scale=0.3]{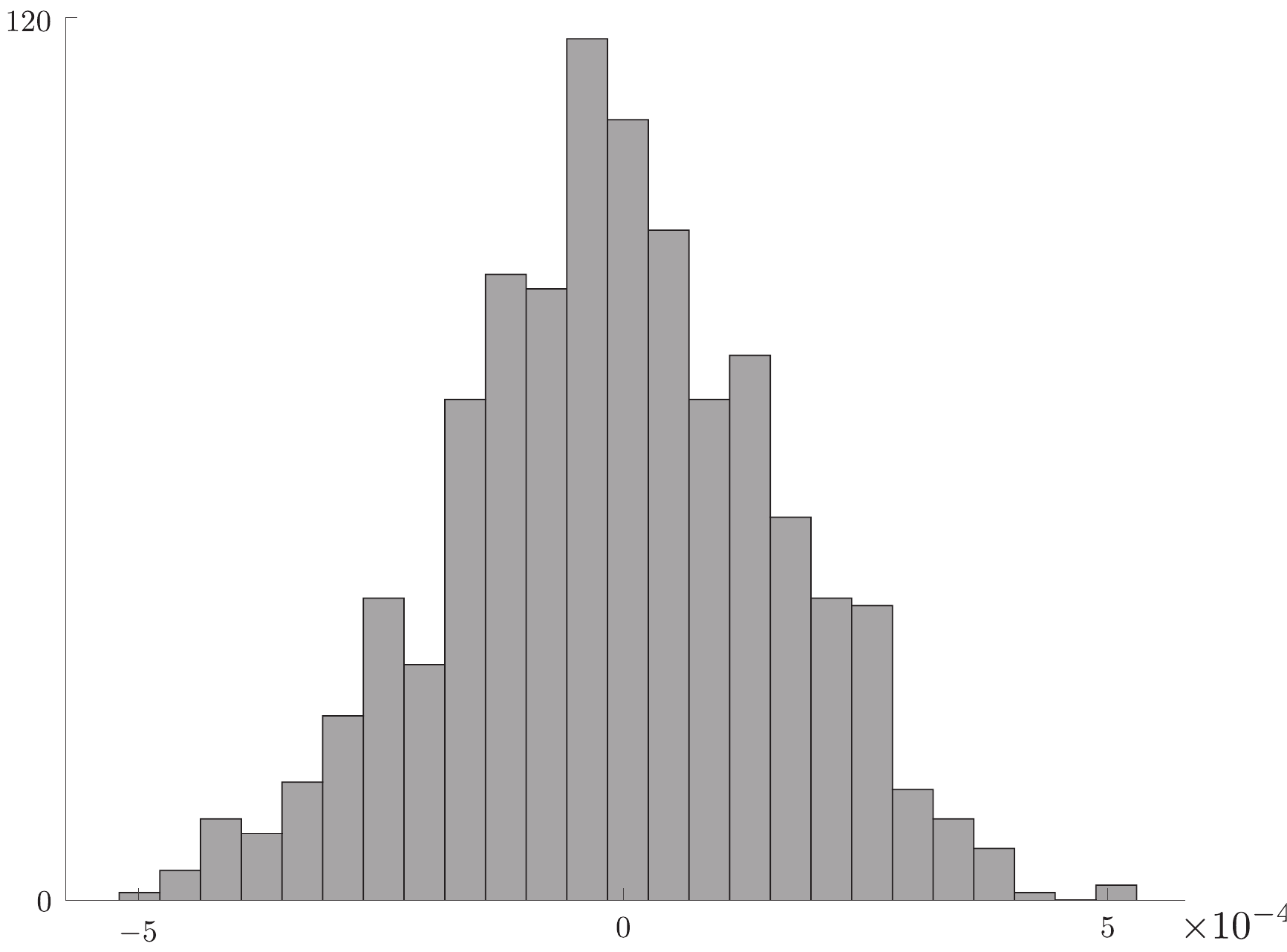}
  \caption{\footnotesize{$\varphi$\, for $RSD_{\lambda}<2\,\%$}}
  \label{fig01a_fel}
\end{subfigure}%
\begin{subfigure}{.33\textwidth}
  \centering
   \includegraphics[scale=0.3]{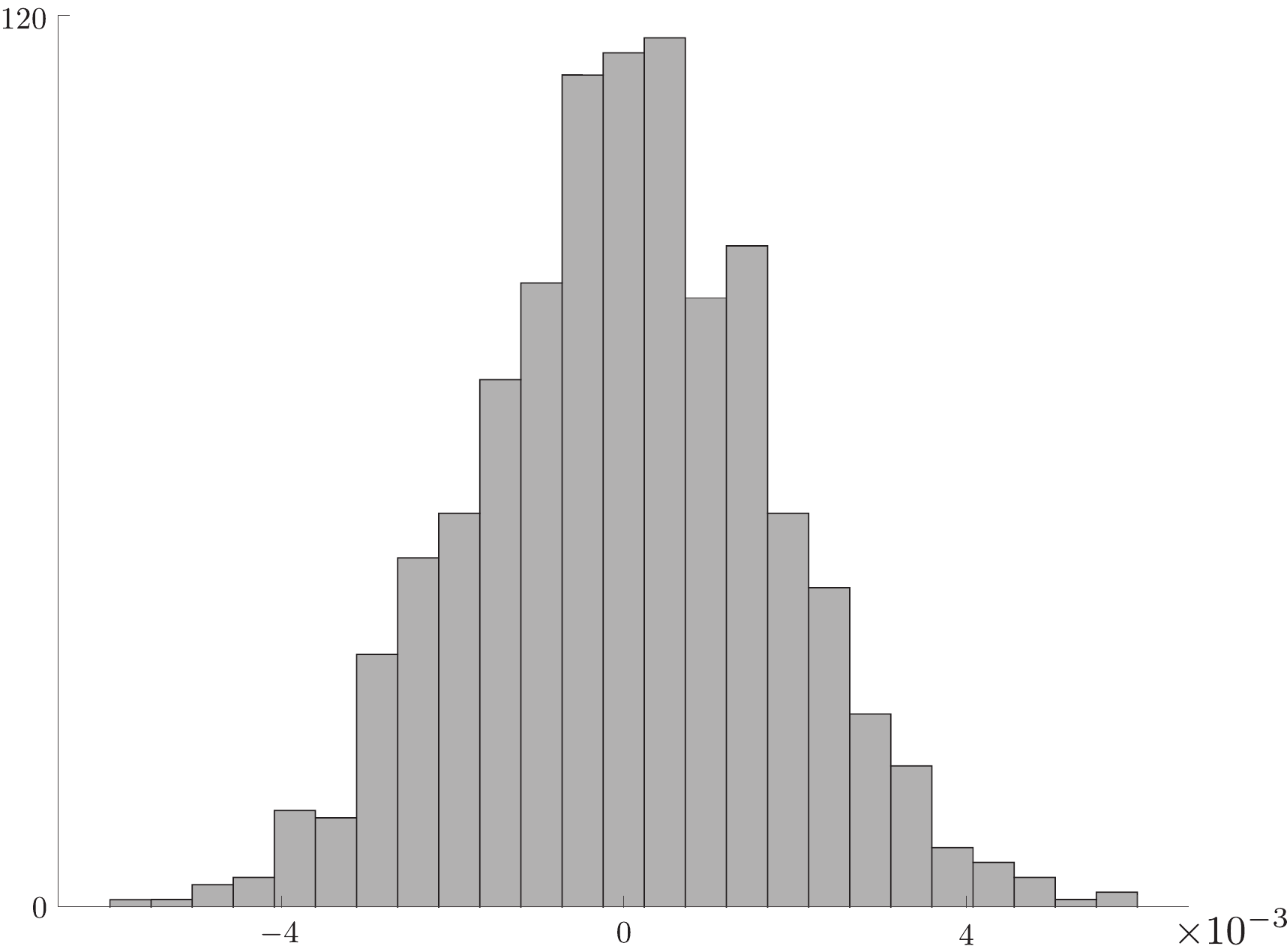}
  \caption{\footnotesize{$\varphi$\, for $RSD_{\lambda}\in(2\,\%\,,\,20\,\%)$}}
  \label{fig01b_fel}
\end{subfigure}%
\begin{subfigure}{.33\textwidth}
  \centering
   \includegraphics[scale=0.3]{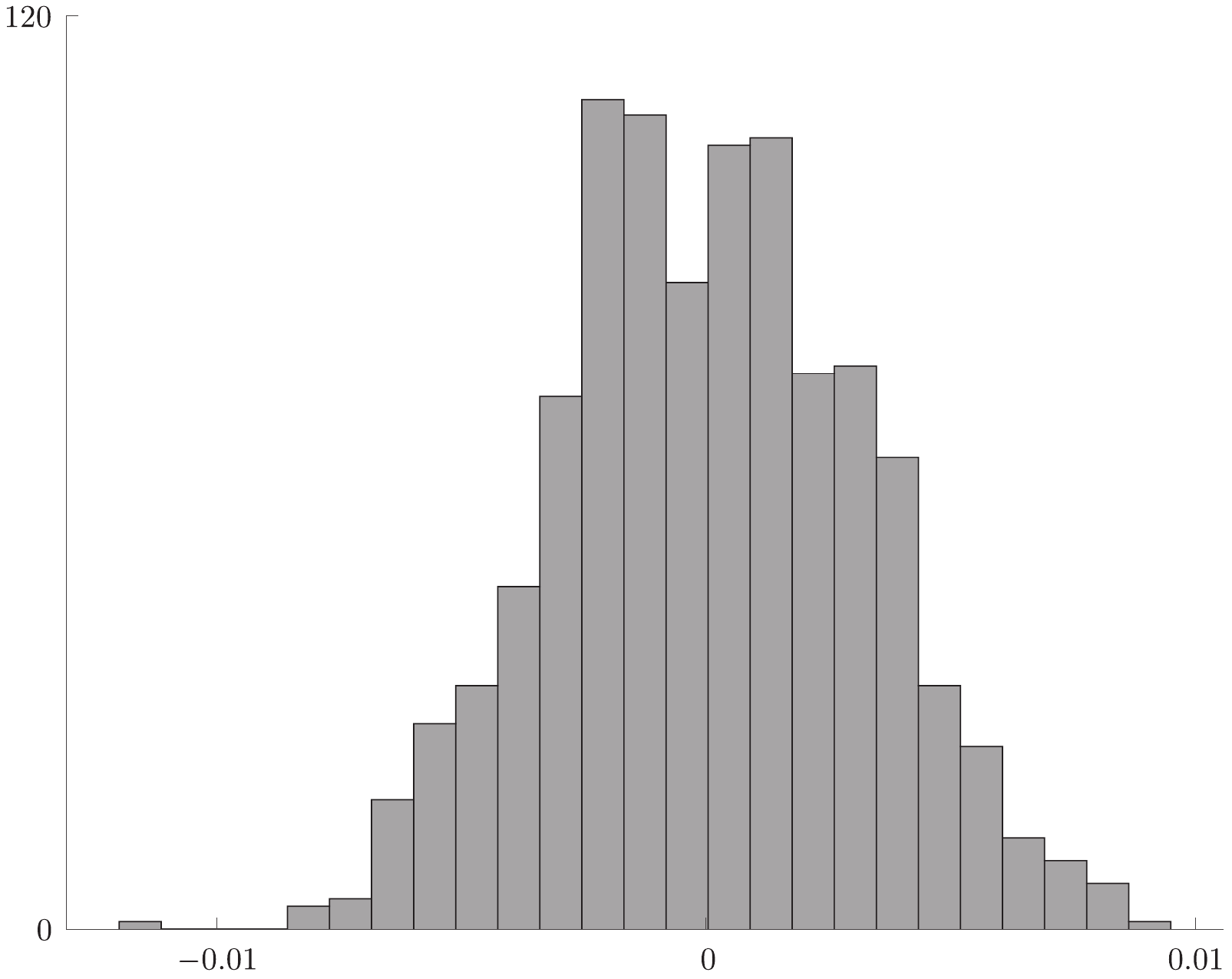}
  \caption{\footnotesize{$\varphi$\, for $RSD_{\lambda}>20\,\%$}}
  \label{fig01c_fel}
\end{subfigure}

\qquad

\centering
\begin{subfigure}{.33\textwidth}
  \centering
   \includegraphics[scale=0.3]{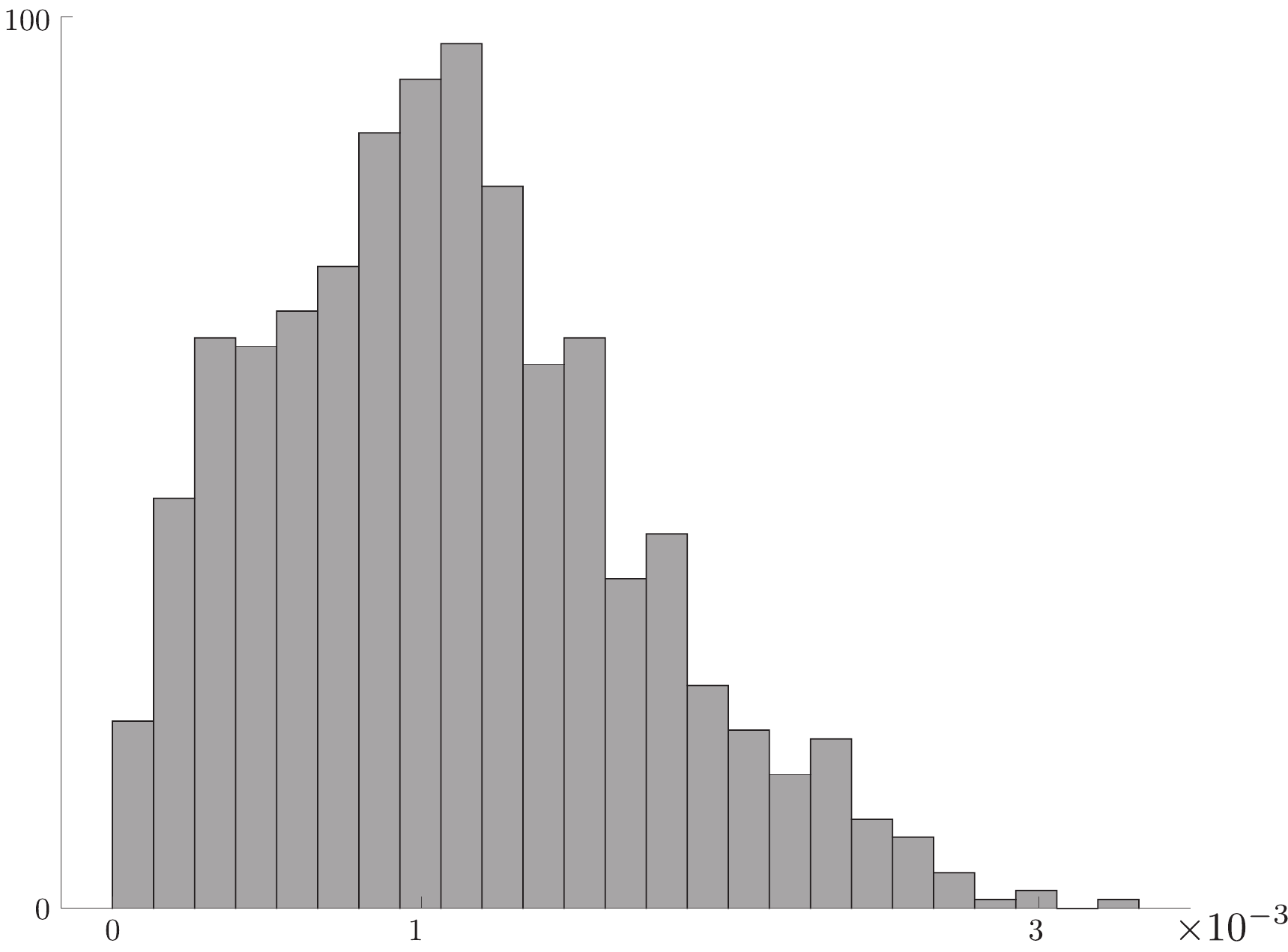}
  \caption{\footnotesize{$\epsilon$\, for $RSD_{\lambda}<2\,\%$}}
  \label{fig01d_fel}
\end{subfigure}%
\begin{subfigure}{.33\textwidth}
  \centering
   \includegraphics[scale=0.3]{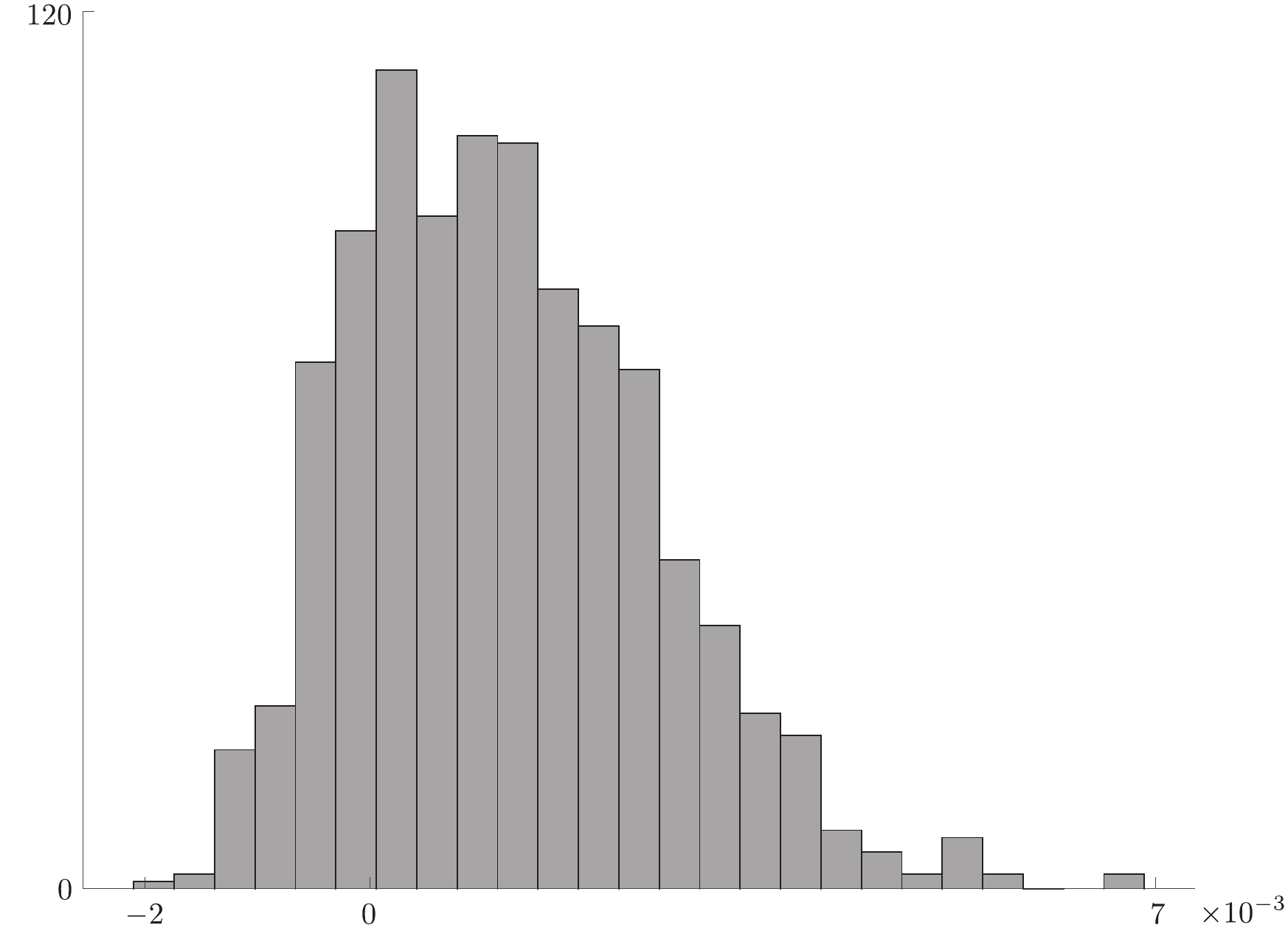}
  \caption{\footnotesize{$\epsilon$\, for $RSD_{\lambda}\in(2\,\%\,,\,20\,\%)$}}
  \label{fig01e_fel}
\end{subfigure}%
\begin{subfigure}{.33\textwidth}
  \centering
   \includegraphics[scale=0.3]{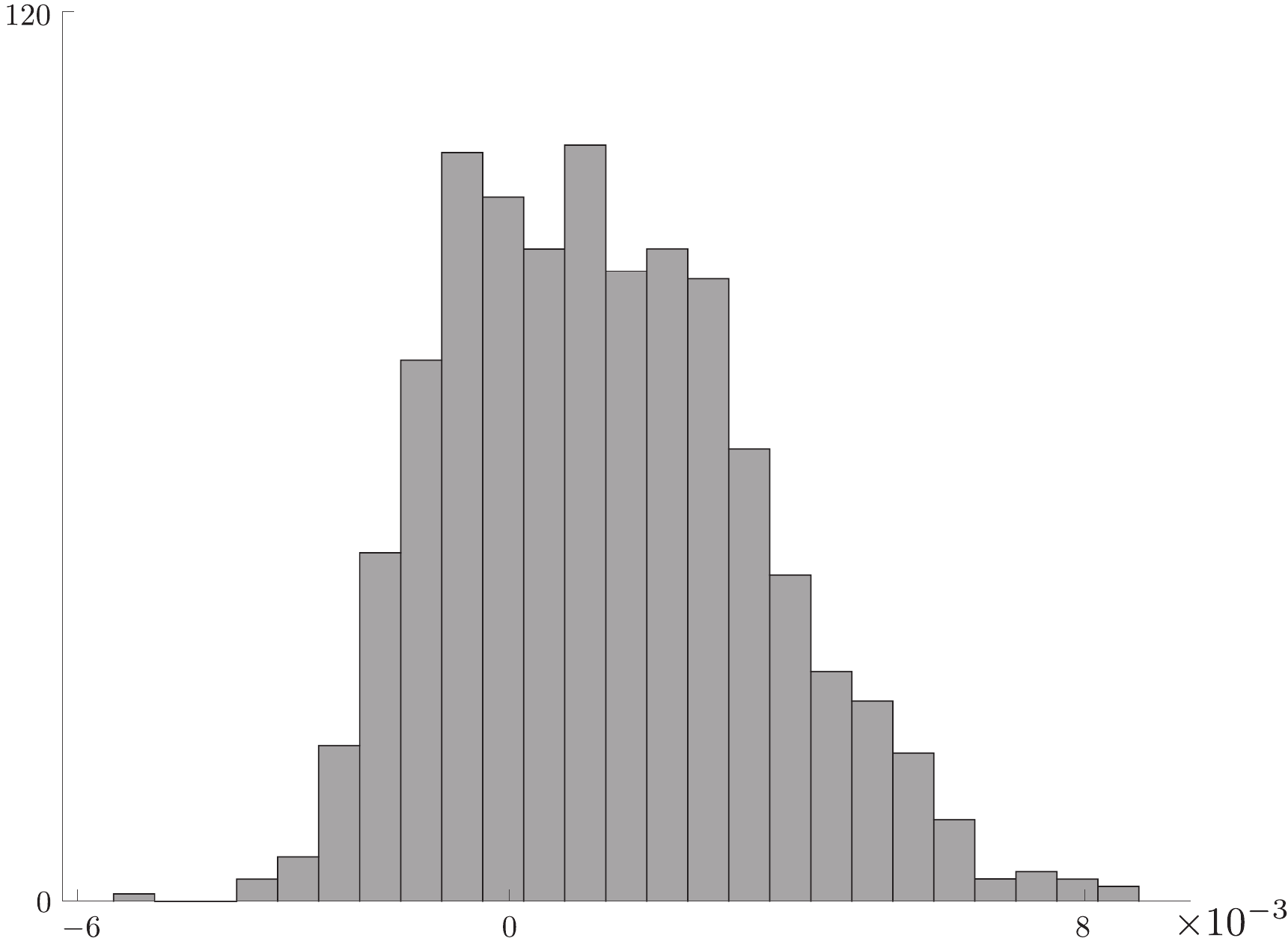}
  \caption{\footnotesize{$\epsilon$\, for $RSD_{\lambda}>20\,\%$}}
  \label{fig01f_fel}
\end{subfigure}

\qquad

\centering
\begin{subfigure}{.33\textwidth}
  \centering
   \includegraphics[scale=0.3]{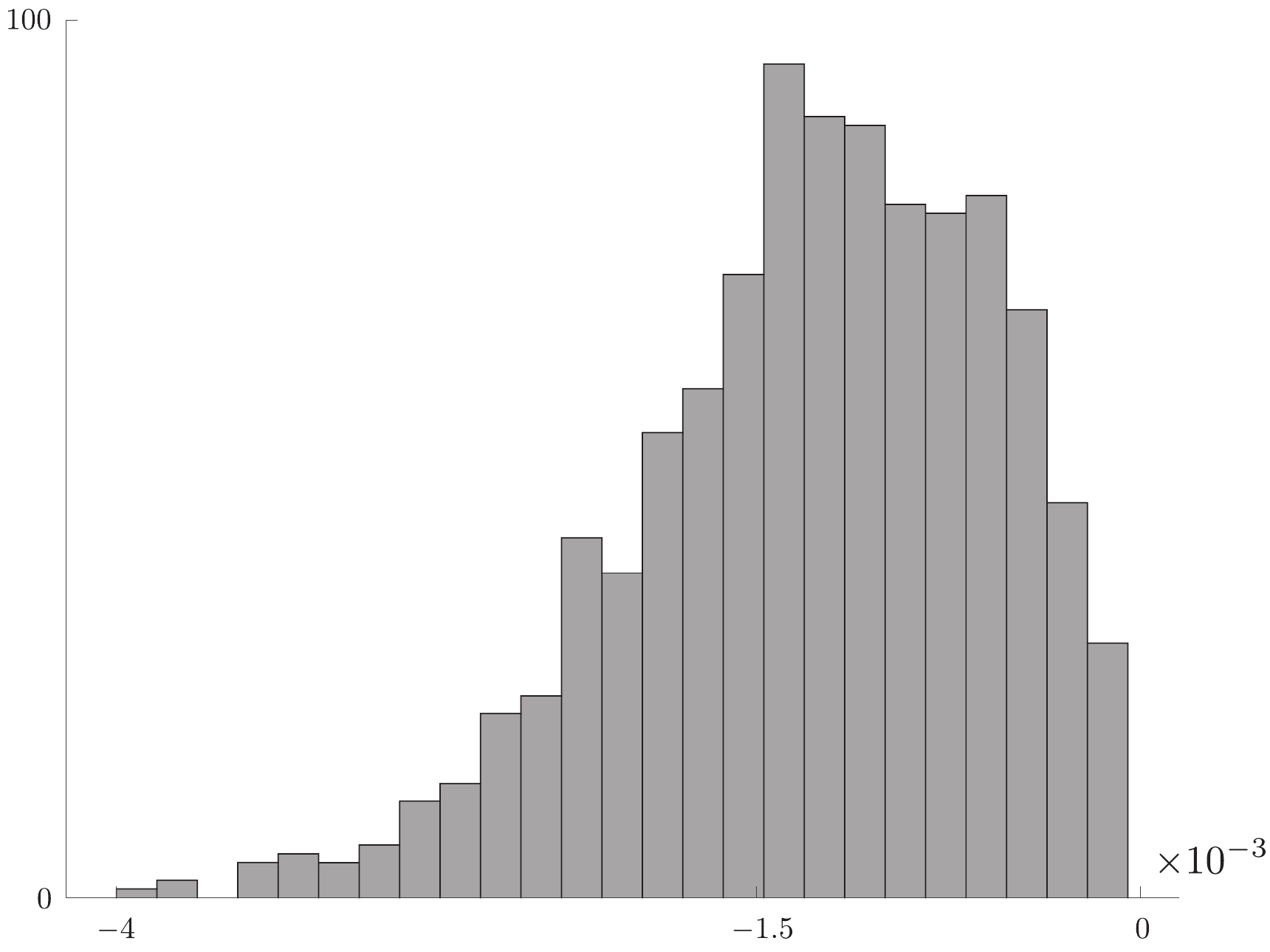}
  \caption{\footnotesize{$\delta$\, for $RSD_{\lambda}<2\,\%$}}
  \label{fig01g_fel}
\end{subfigure}%
\begin{subfigure}{.33\textwidth}
  \centering
   \includegraphics[scale=0.3]{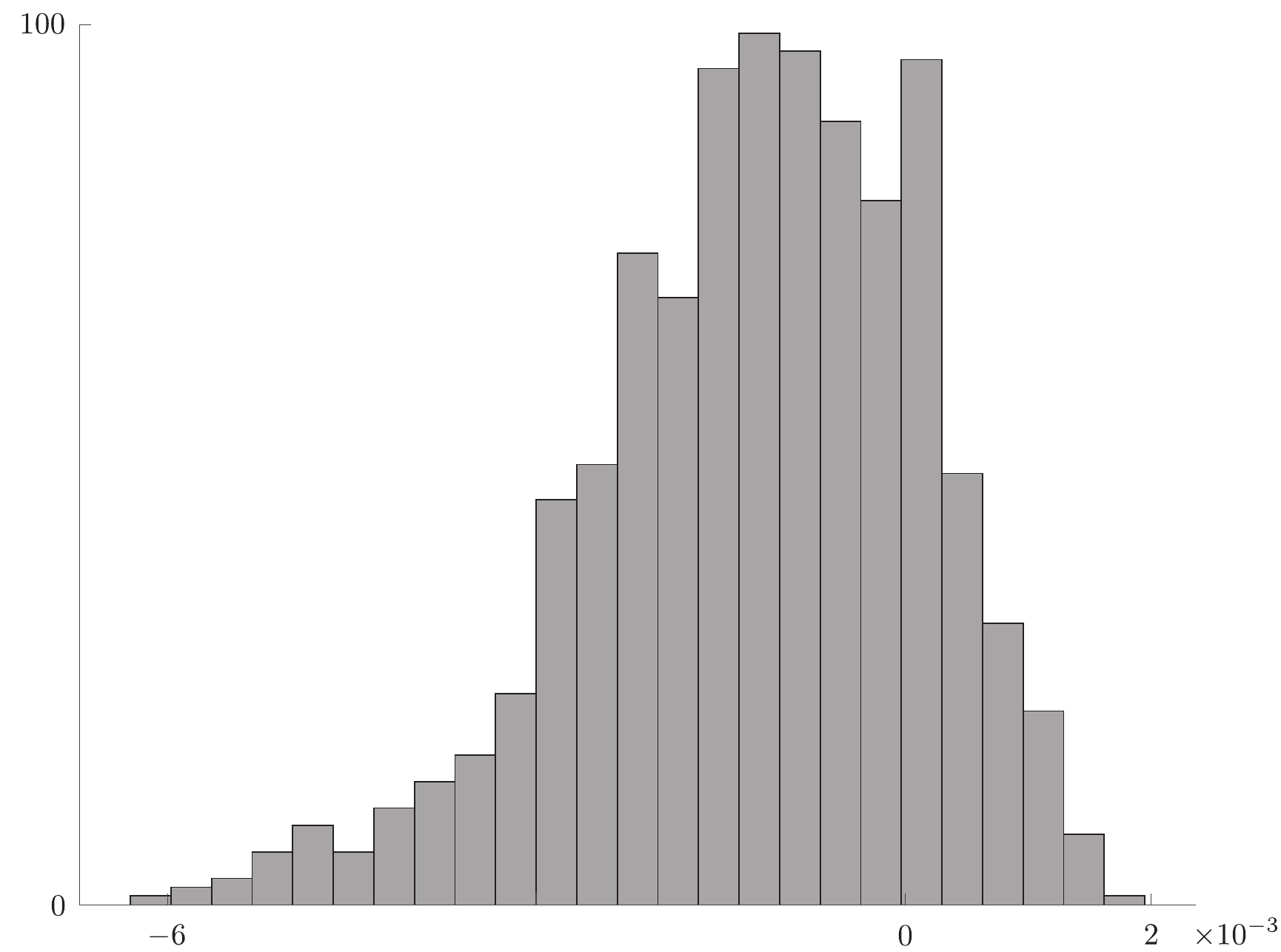}
  \caption{\footnotesize{$\delta$\, for $RSD_{\lambda}\in(2\,\%\,,\,20\,\%)$}}
  \label{fig01h_fel}
\end{subfigure}%
\begin{subfigure}{.33\textwidth}
  \centering
   \includegraphics[scale=0.3]{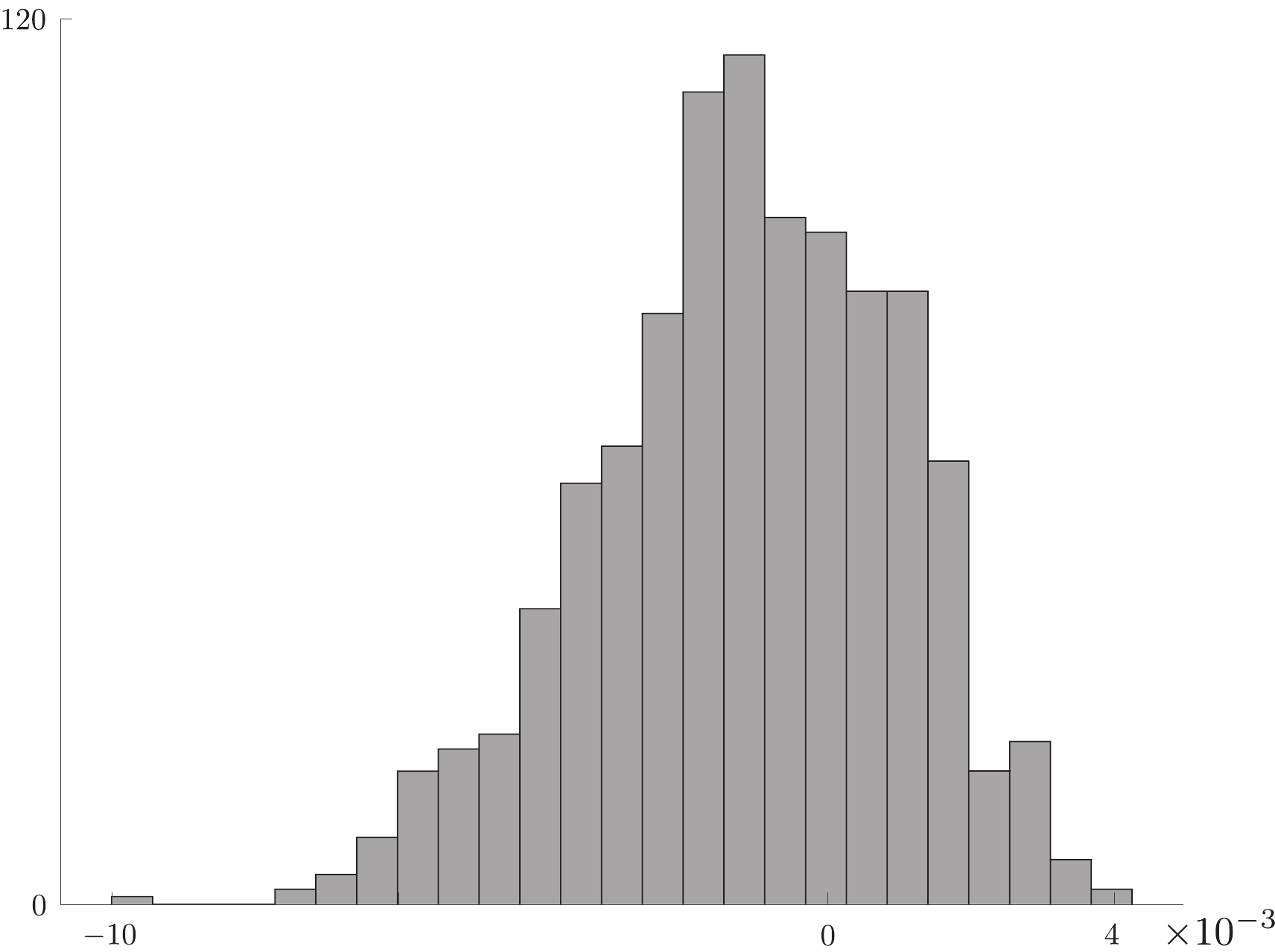}
  \caption{\footnotesize{$\delta$\, for $RSD_{\lambda}>20\,\%$}}
  \label{fig01i_fel}
\end{subfigure}
\caption{\small{Distribution of $\varphi$, $\epsilon$ and $\delta$ for $1000$ examples of equivalent TI felsic media with $RSD_{\mu}>2\,\%\,$. }}
\label{fig:dis_mb_fel}
\end{figure}

\begin{figure}[!htbp]
\centering
\begin{subfigure}{.33\textwidth}
  \centering
   \includegraphics[scale=0.3]{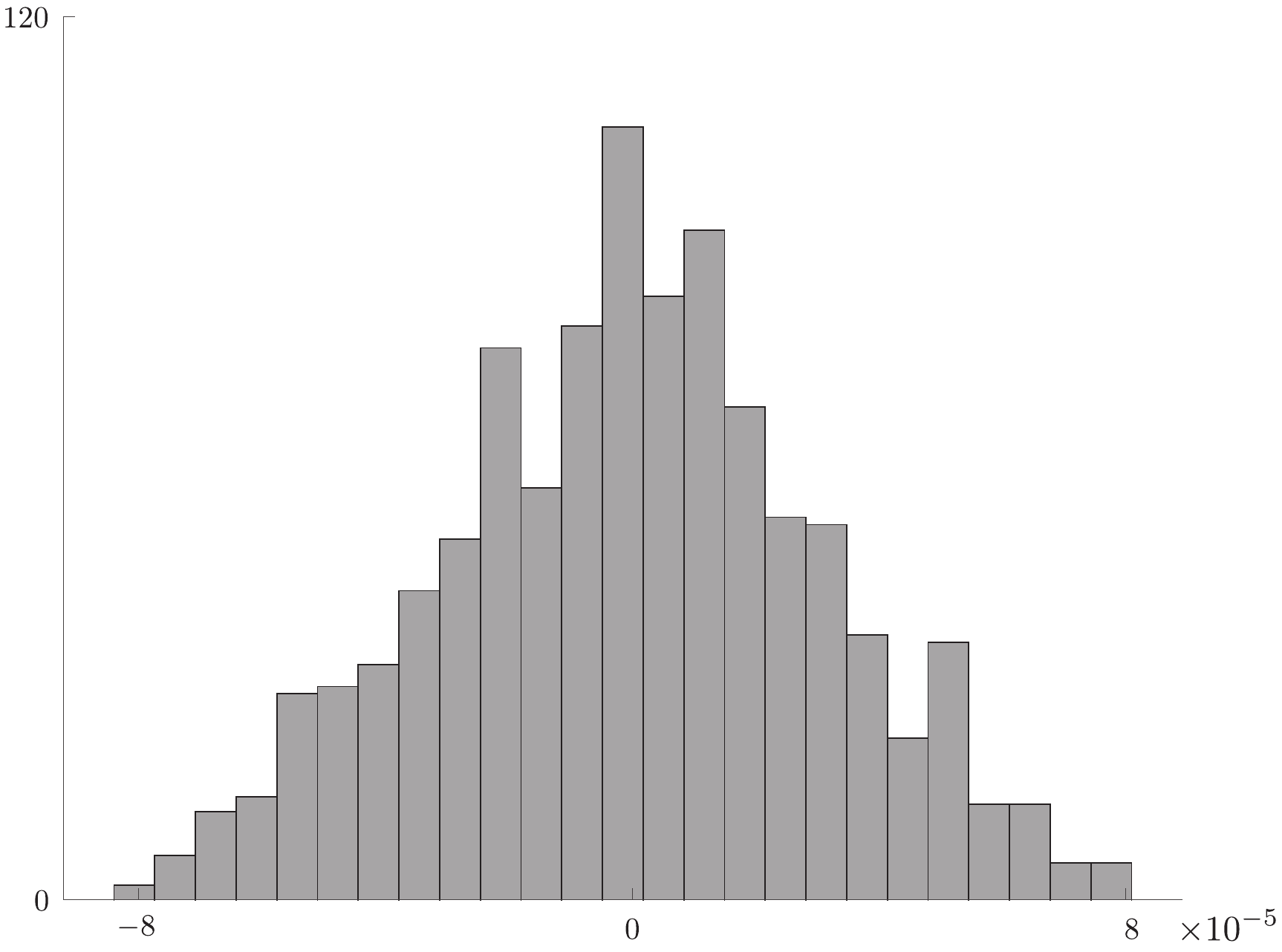}
  \caption{\footnotesize{$\varphi$\, for $RSD_{\lambda}<2\,\%$}}
  \label{fig02a_fel}
\end{subfigure}%
\begin{subfigure}{.33\textwidth}
  \centering
   \includegraphics[scale=0.3]{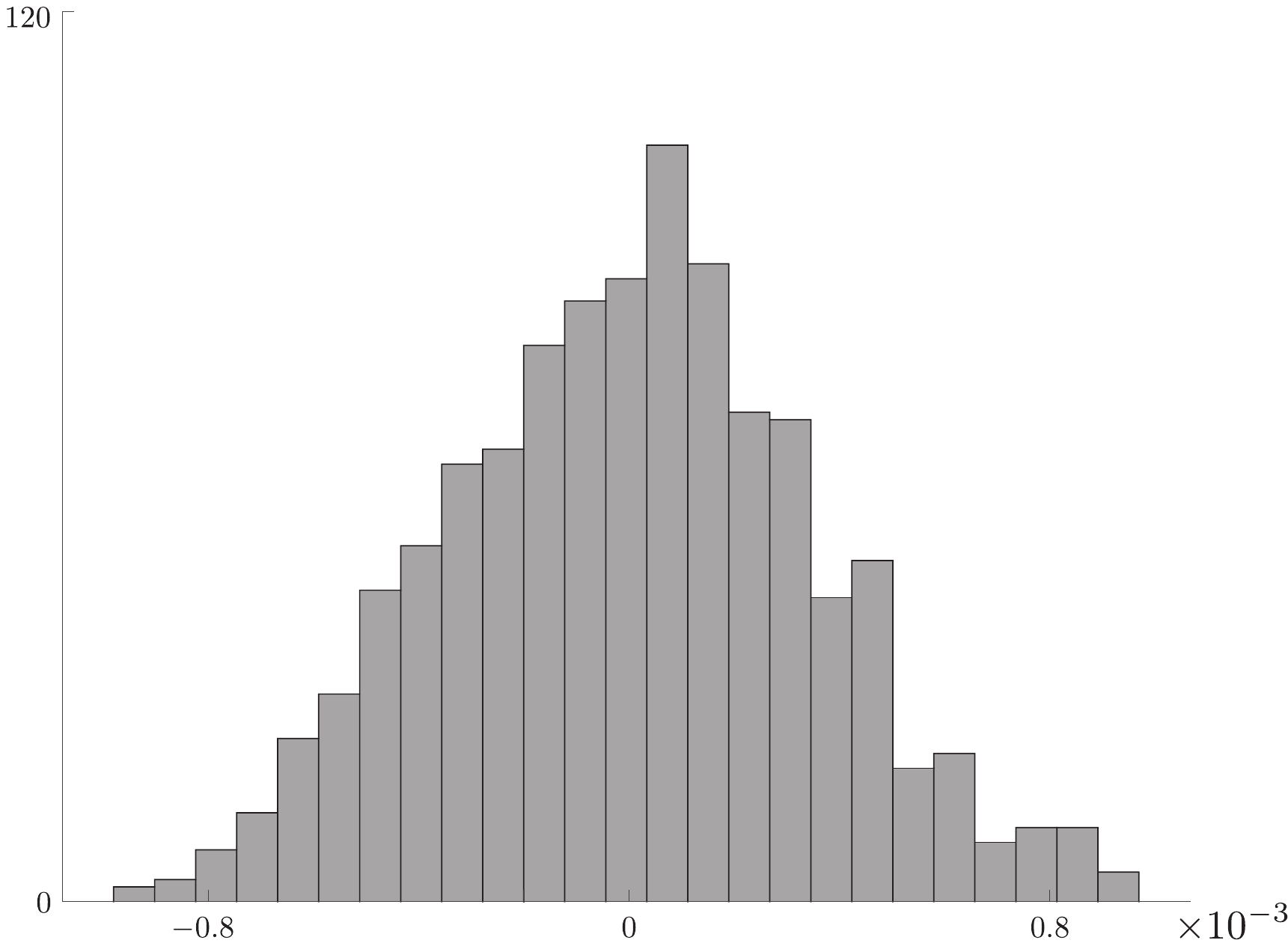}
  \caption{\footnotesize{$\varphi$\, for $RSD_{\lambda}\in(2\,\%\,,\,20\,\%)$}}
  \label{fig02b_fel}
\end{subfigure}%
\begin{subfigure}{.33\textwidth}
  \centering
   \includegraphics[scale=0.3]{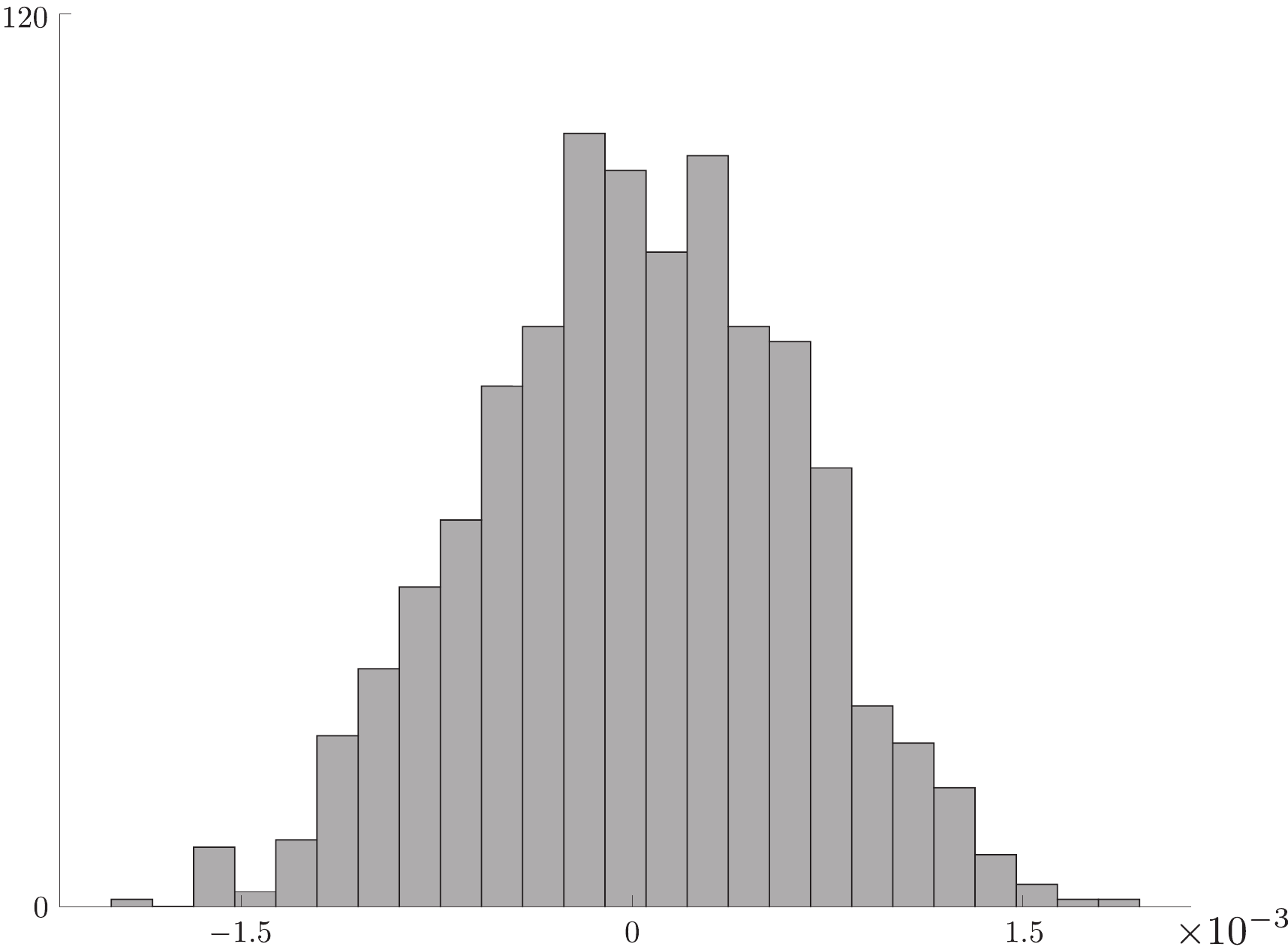}
  \caption{\footnotesize{$\varphi$\, for $RSD_{\lambda}>20\,\%$}}
  \label{fig02c_fel}
\end{subfigure}

\qquad

\centering
\begin{subfigure}{.33\textwidth}
  \centering
   \includegraphics[scale=0.3]{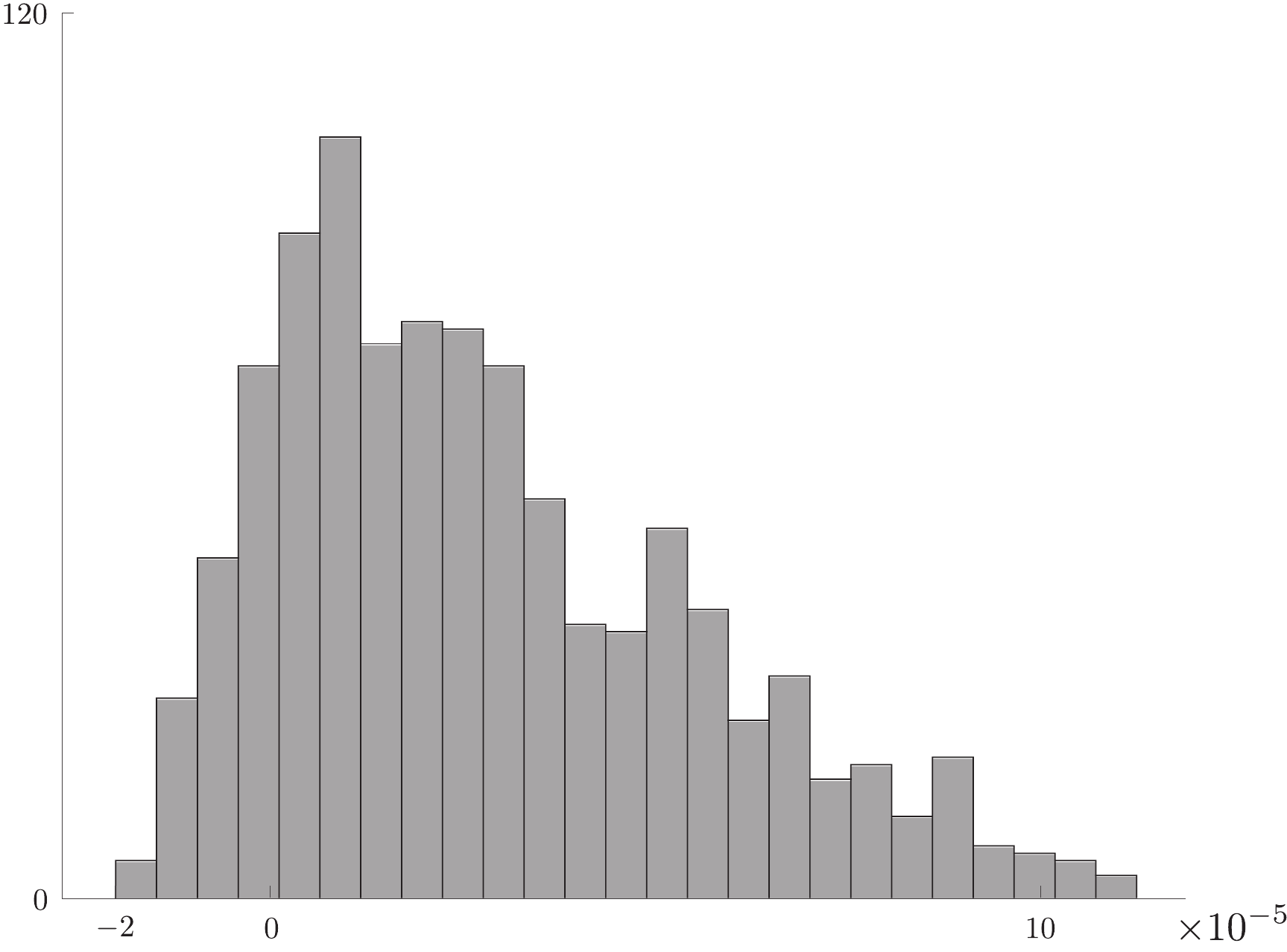}
  \caption{\footnotesize{$\epsilon$\, for $RSD_{\lambda}<2\,\%$}}
  \label{fig02d_fel}
\end{subfigure}%
\begin{subfigure}{.33\textwidth}
  \centering
   \includegraphics[scale=0.3]{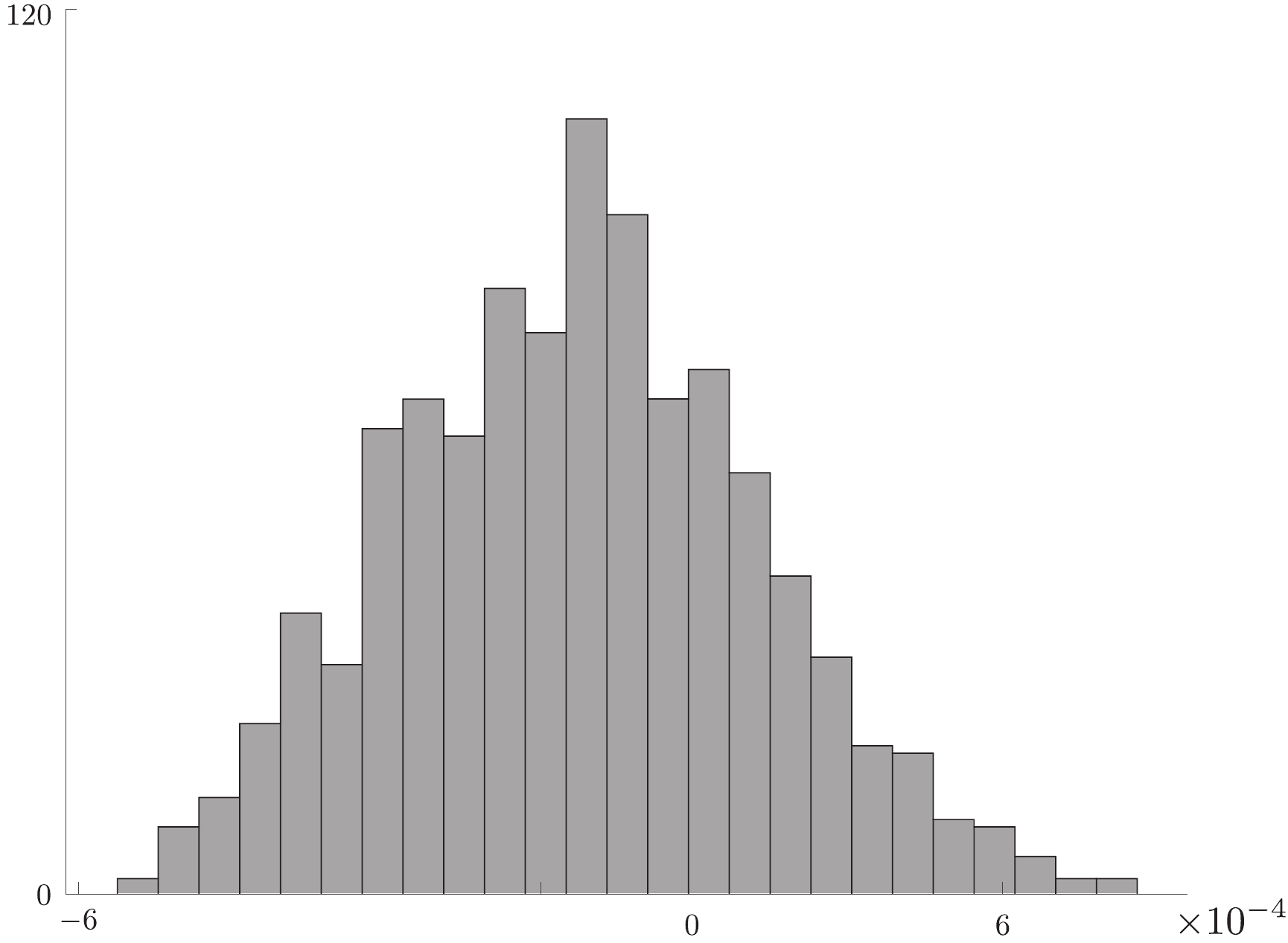}
  \caption{\footnotesize{$\epsilon$\, for $RSD_{\lambda}\in(2\,\%\,,\,20\,\%)$}}
  \label{fig02e_fel}
\end{subfigure}%
\begin{subfigure}{.33\textwidth}
  \centering
   \includegraphics[scale=0.3]{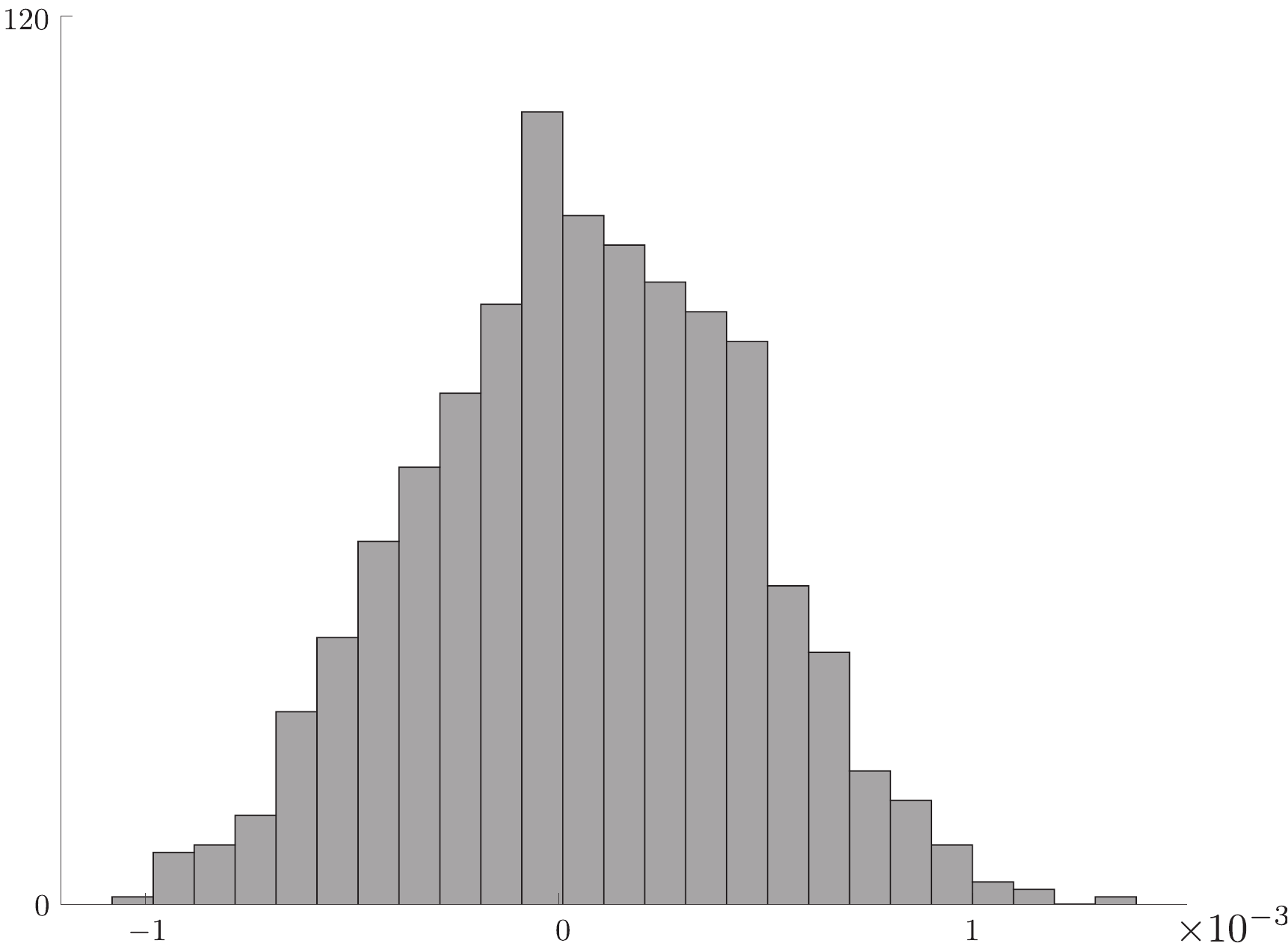}
  \caption{\footnotesize{$\epsilon$\, for $RSD_{\lambda}>20\,\%$}}
  \label{fig02f_fel}
\end{subfigure}

\qquad

\centering
\begin{subfigure}{.33\textwidth}
  \centering
   \includegraphics[scale=0.3]{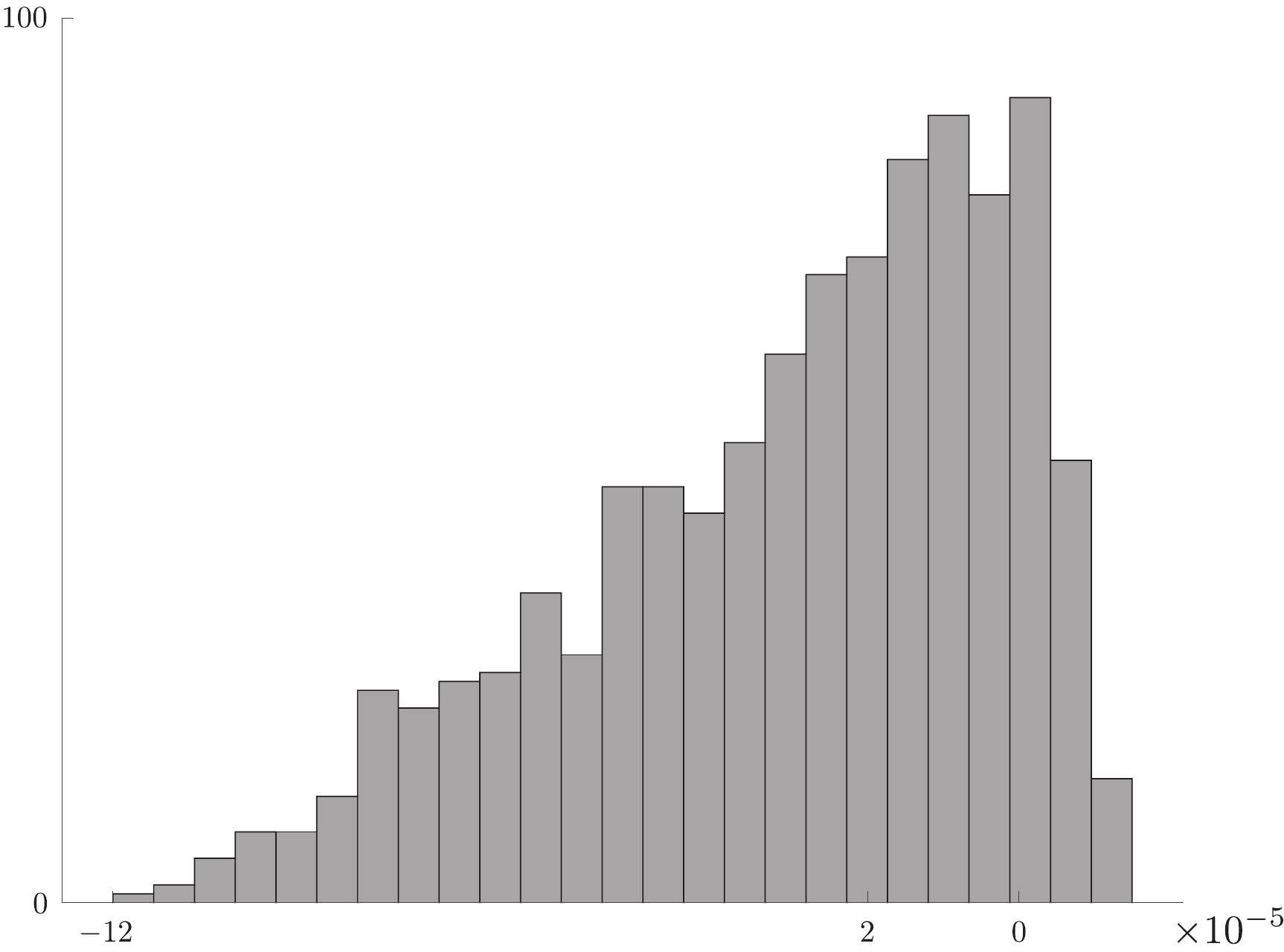}
  \caption{\footnotesize{$\delta$\, for $RSD_{\lambda}<2\,\%$}}
  \label{fig02g_fel}
\end{subfigure}%
\begin{subfigure}{.33\textwidth}
  \centering
   \includegraphics[scale=0.3]{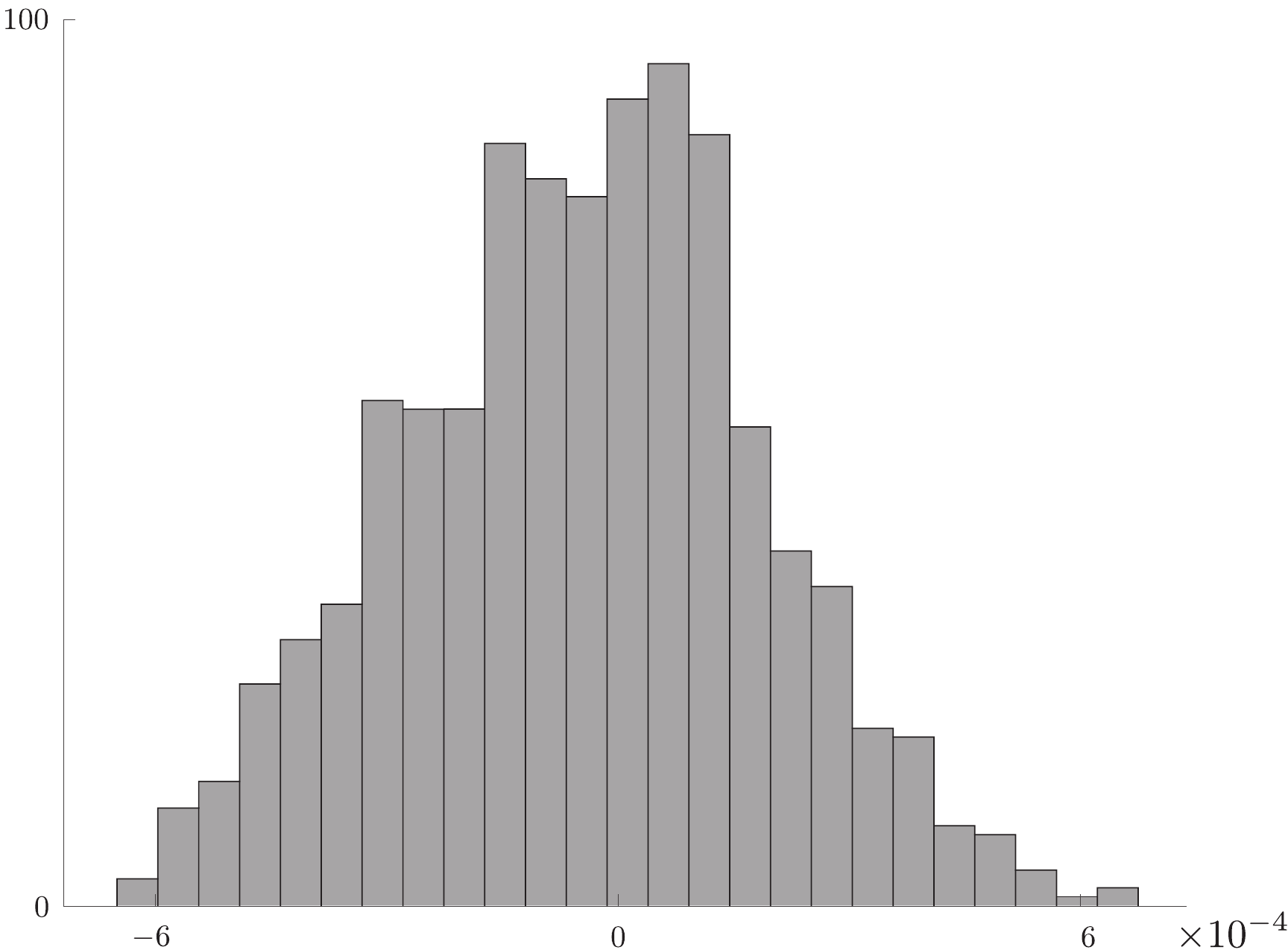}
  \caption{\footnotesize{$\delta$\, for $RSD_{\lambda}\in(2\,\%\,,\,20\,\%)$}}
  \label{fig02h_fel}
\end{subfigure}%
\begin{subfigure}{.33\textwidth}
  \centering
   \includegraphics[scale=0.3]{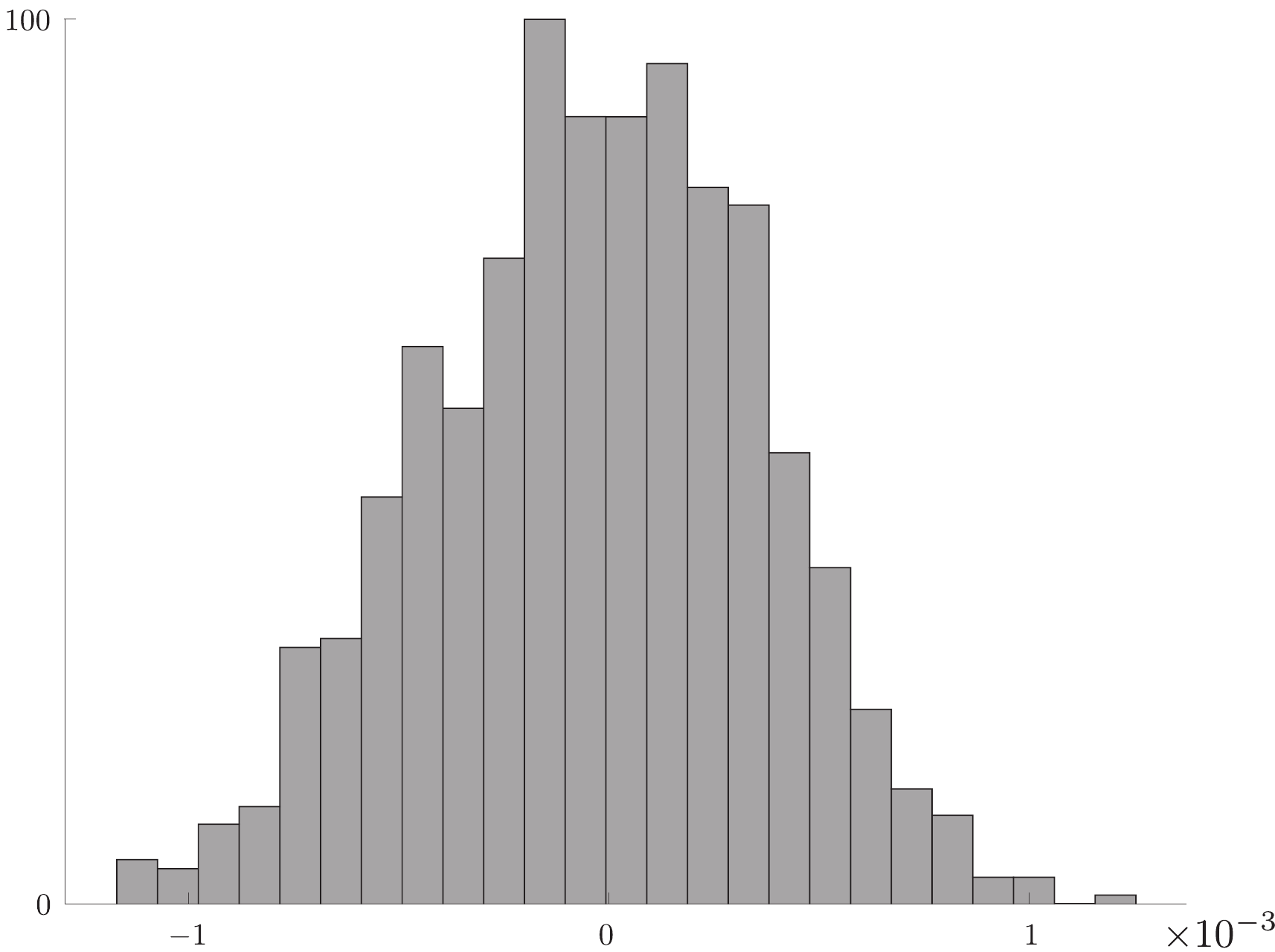}
  \caption{\footnotesize{$\delta$\, for $RSD_{\lambda}>20\,\%$}}
  \label{fig02i_fel}
\end{subfigure}
\caption{\small{Distribution of $\varphi$, $\epsilon$ and $\delta$ for $1000$ examples of equivalent TI felsic media with $RSD_{\mu}<2\,\%\,$.}}
\label{fig:dis_ms_fel}
\end{figure}

\begin{figure}[!htbp]
\centering
\begin{subfigure}{.4\textwidth}
  \centering
   \includegraphics[scale=0.35]{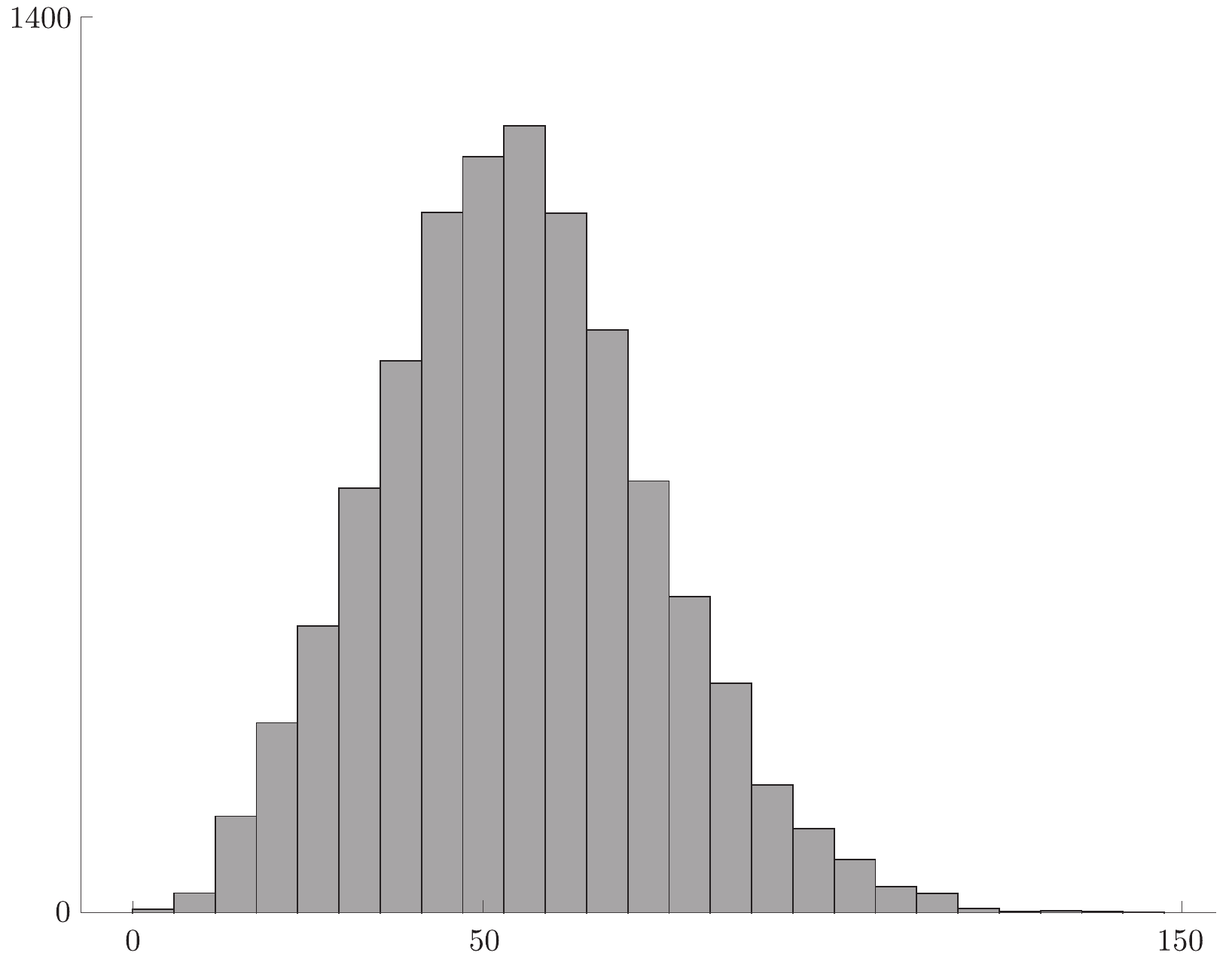}
  \caption{\footnotesize{Variations of $\mu$ $(RSD_{\mu}\,[\%])$}}
  \label{fig:rsdmsan}
\end{subfigure}%
\begin{subfigure}{.4\textwidth}
  \centering
   \includegraphics[scale=0.35]{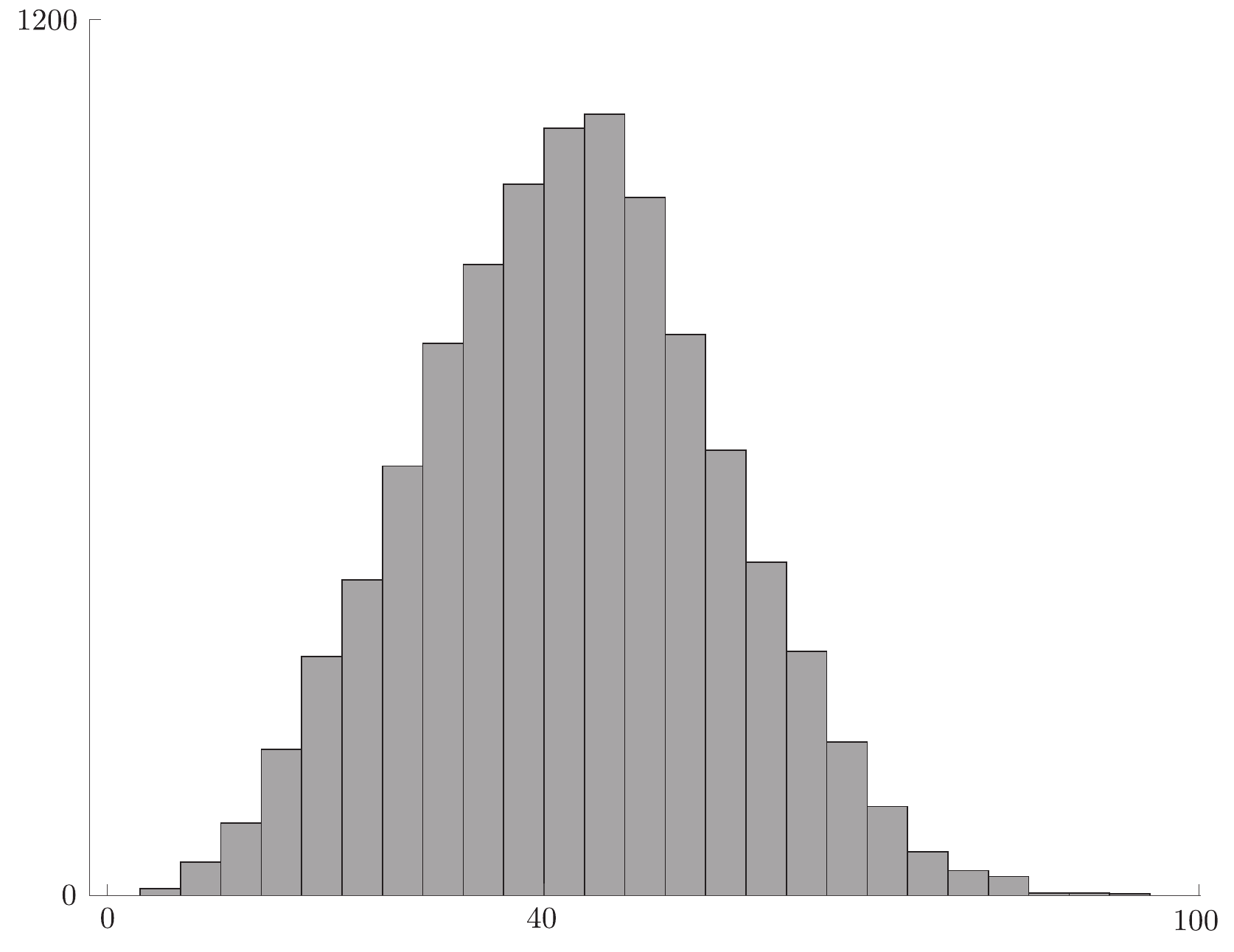}
  \caption{\footnotesize{Variations of $\lambda$ $(RSD_{\lambda}\,[\%])$}}
  \label{fig:rsdlsan}
\end{subfigure}%
\caption{\small{Distributions of random variations of $\mu$ and $\lambda$ for $10000$ TI media relevant to layered sandstones.}}
\label{fig:rsdsan}
\end{figure}

\begin{figure}[!htbp]
\centering
\begin{subfigure}{.4\textwidth}
  \centering
   \includegraphics[scale=0.35]{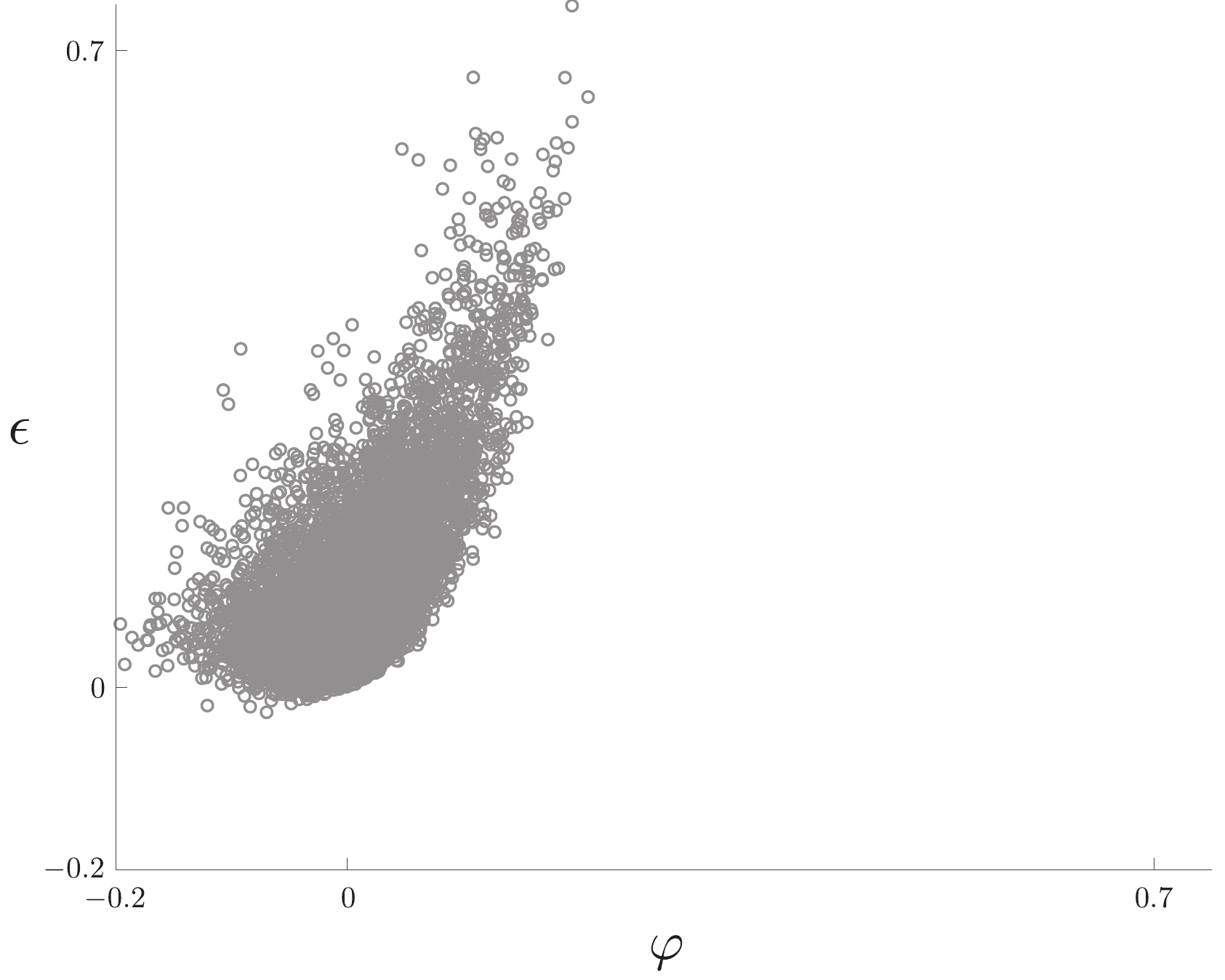}
  \caption{\footnotesize{$\varphi$ versus $\epsilon$}}
  \label{fig:rel_ml_phieps_san}
\end{subfigure}%
\begin{subfigure}{.4\textwidth}
  \centering
   \includegraphics[scale=0.35]{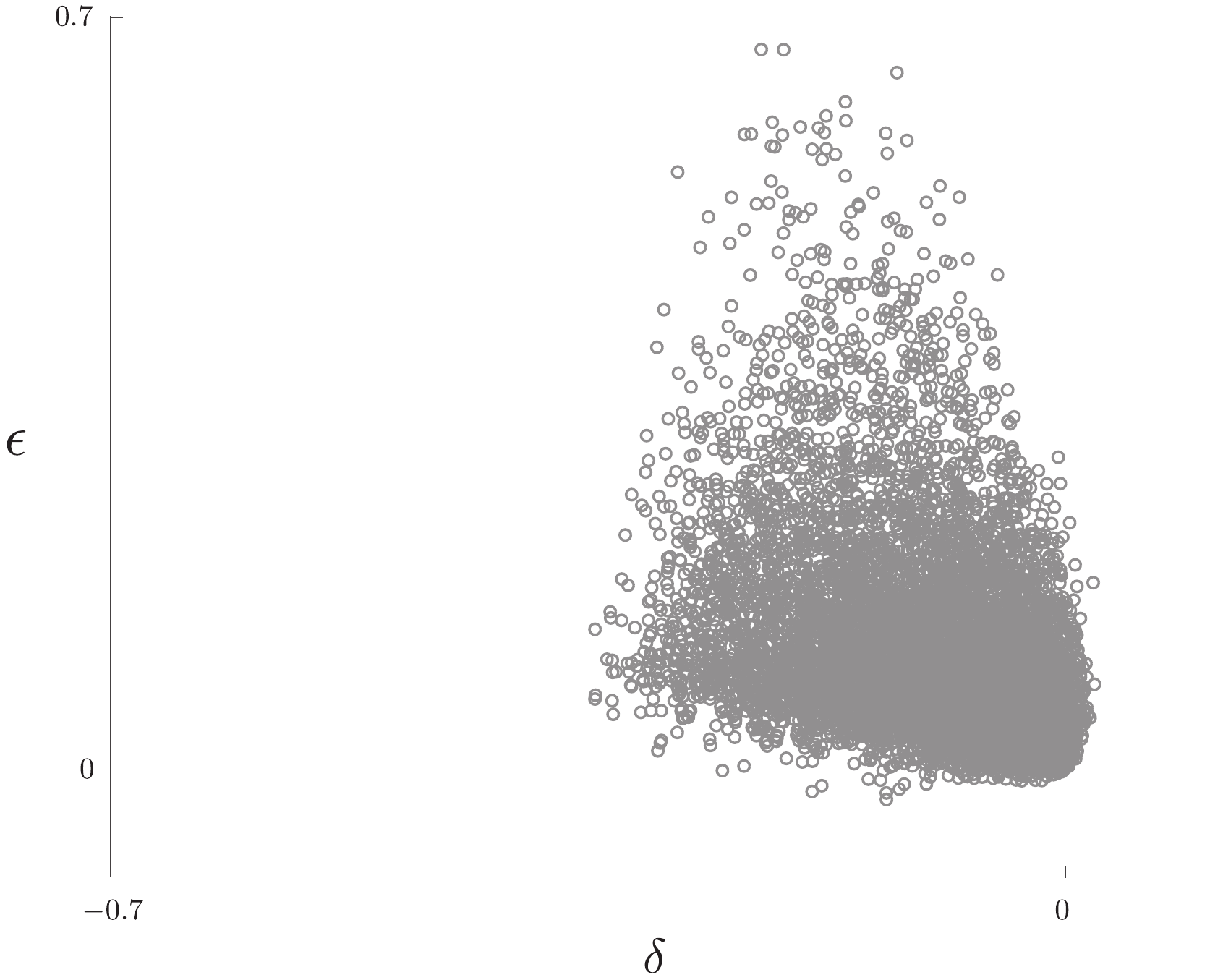}
  \caption{\footnotesize{$\delta$ versus $\epsilon$}}
  \label{fig:rel_ml_deleps_san}
\end{subfigure}%
\caption{\small{Cross-plots of anisotropy parameters for $10000$ examples of equivalent TI sandstones.}}
\label{fig:rel_ml_san}
\end{figure}

\begin{figure}[!htbp]
\centering
\begin{subfigure}{.4\textwidth}
\centering
   \includegraphics[scale=0.35]{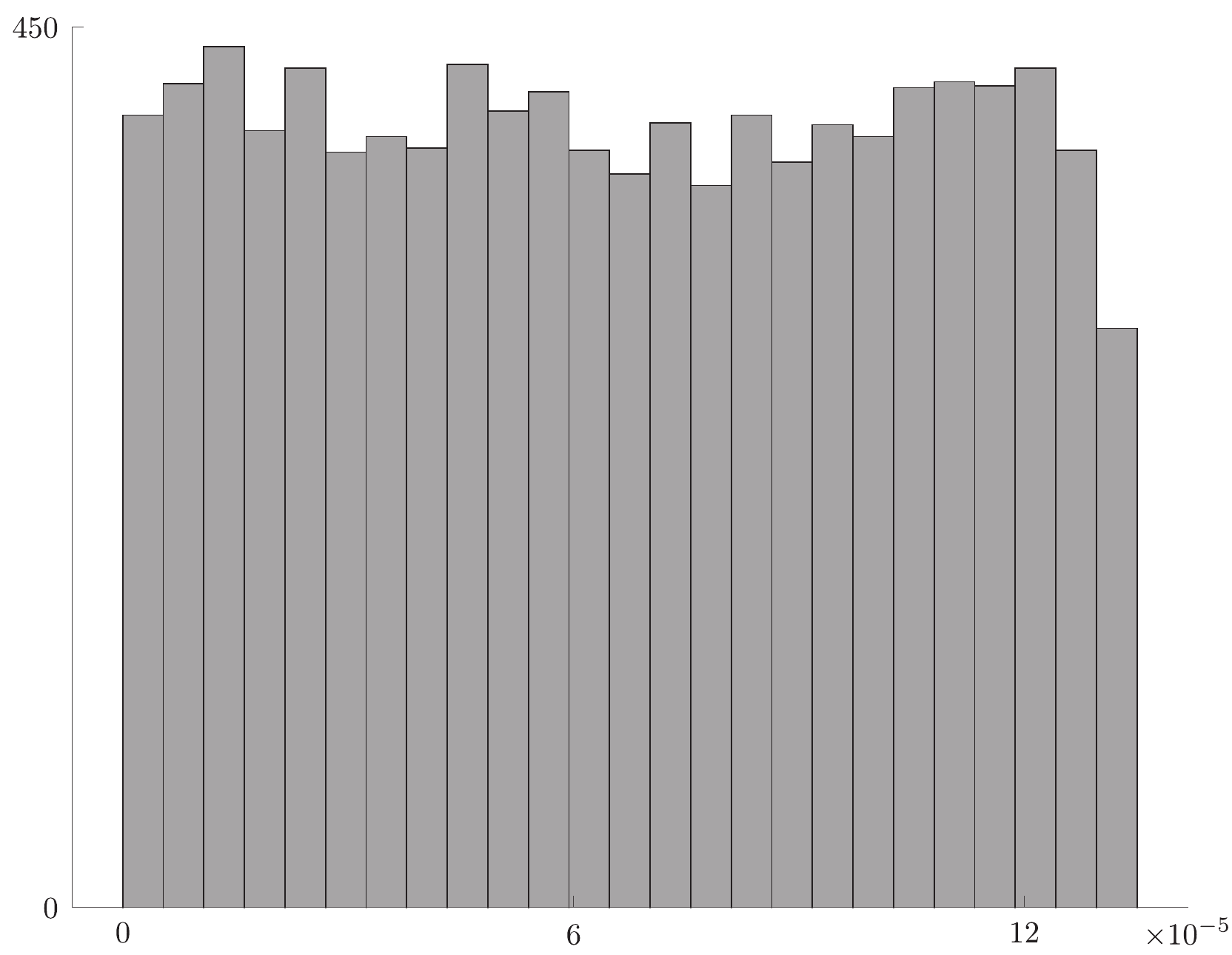}
 \caption{\small{$RSD_{\mu}<2\,\%\,$}}
\label{fig:gamma_ms_san}
\end{subfigure}
\begin{subfigure}{.4\textwidth}
\centering
   \includegraphics[scale=0.35]{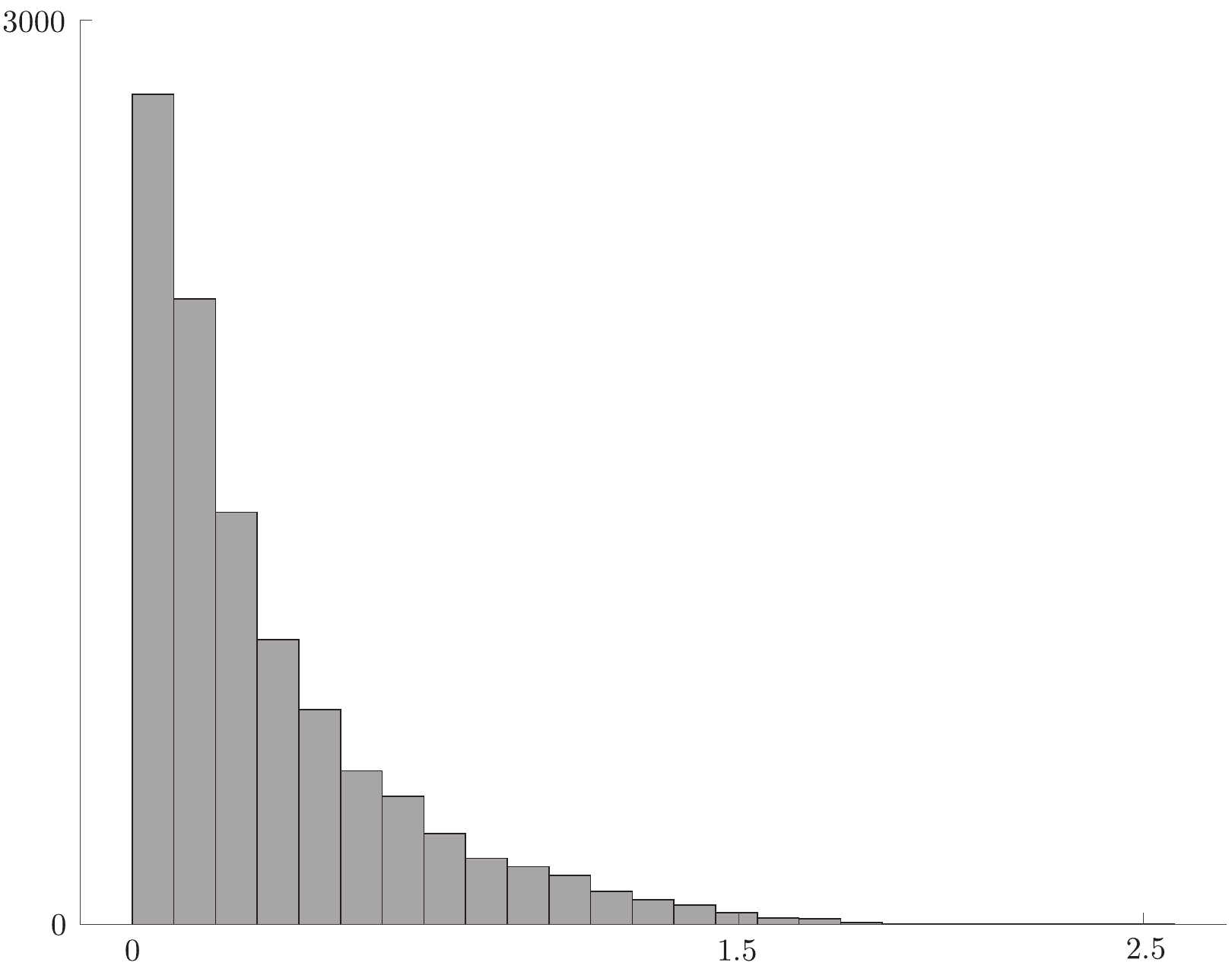}
 \caption{\small{$RSD_{\mu}>2\,\%\,$}}
\label{fig:gamma_mb_san}
\end{subfigure}
\caption{\small{Distribution of $\gamma$ for $10000$ examples of equivalent TI sandstones.}}
\label{fig:gamma_san}
\end{figure}

\begin{figure}[!htbp]
\centering
\begin{subfigure}{.33\textwidth}
  \centering
   \includegraphics[scale=0.3]{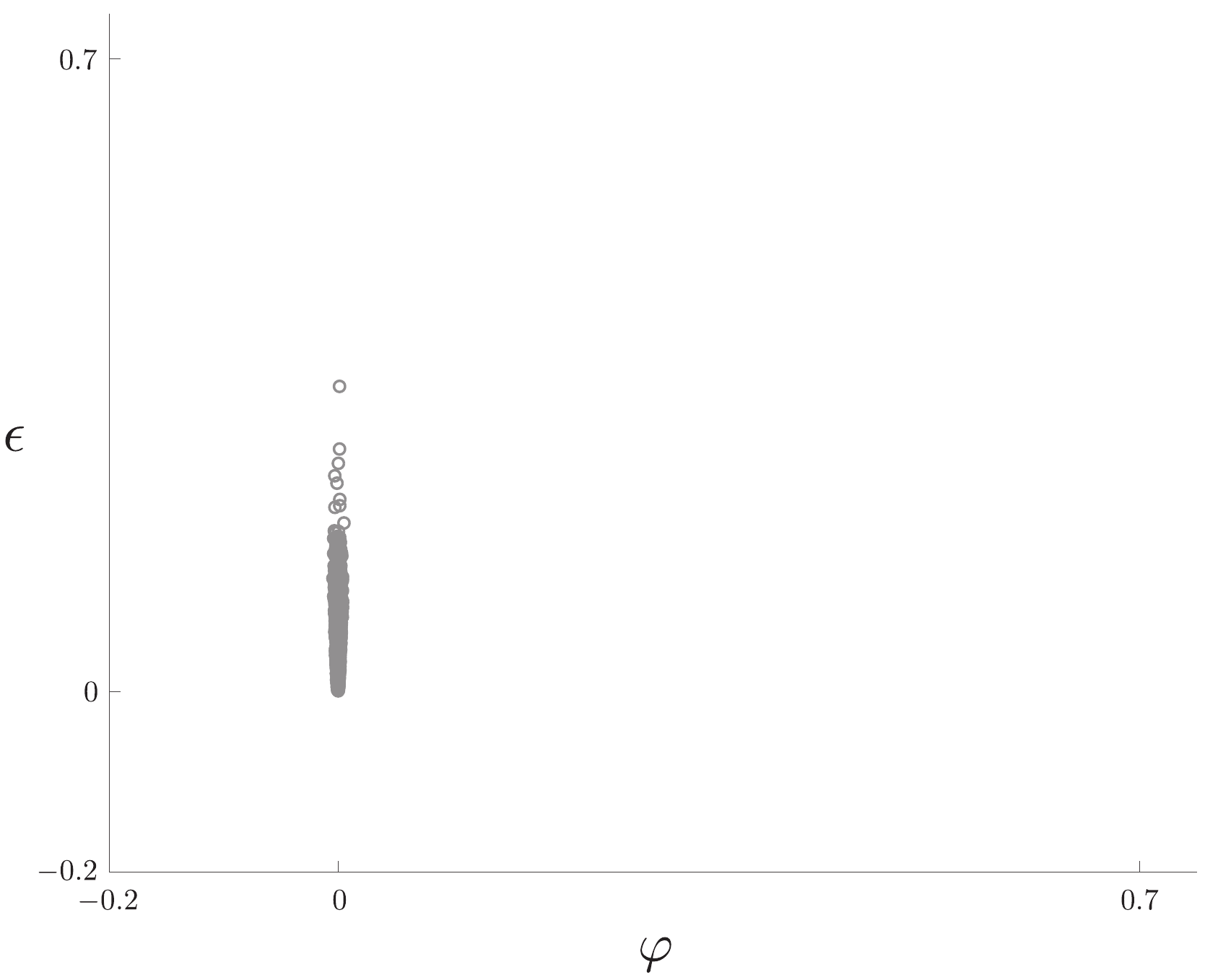}
  \caption{\footnotesize{$RSD_{\lambda}<2\,\%$}}
  \label{fig:rel_phieps_mb_ls_san}
\end{subfigure}%
\begin{subfigure}{.33\textwidth}
  \centering
   \includegraphics[scale=0.3]{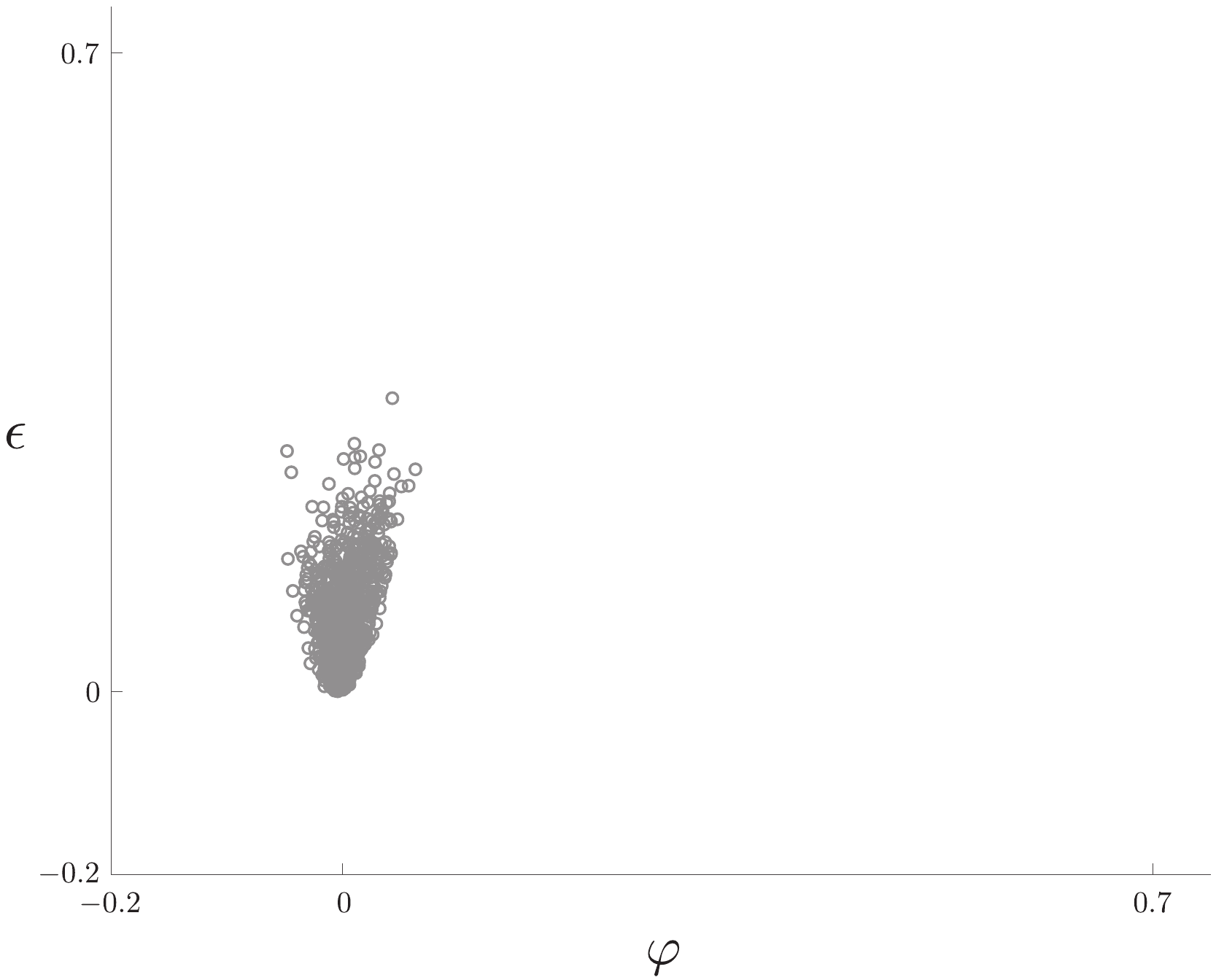}
  \caption{\footnotesize{$RSD_{\lambda}\in\left(2\,\%\,,\,20\,\%\right)$}}
  \label{fig:rel_phieps_mb_2l20_san}
\end{subfigure}%
\begin{subfigure}{.33\textwidth}
  \centering
   \includegraphics[scale=0.3]{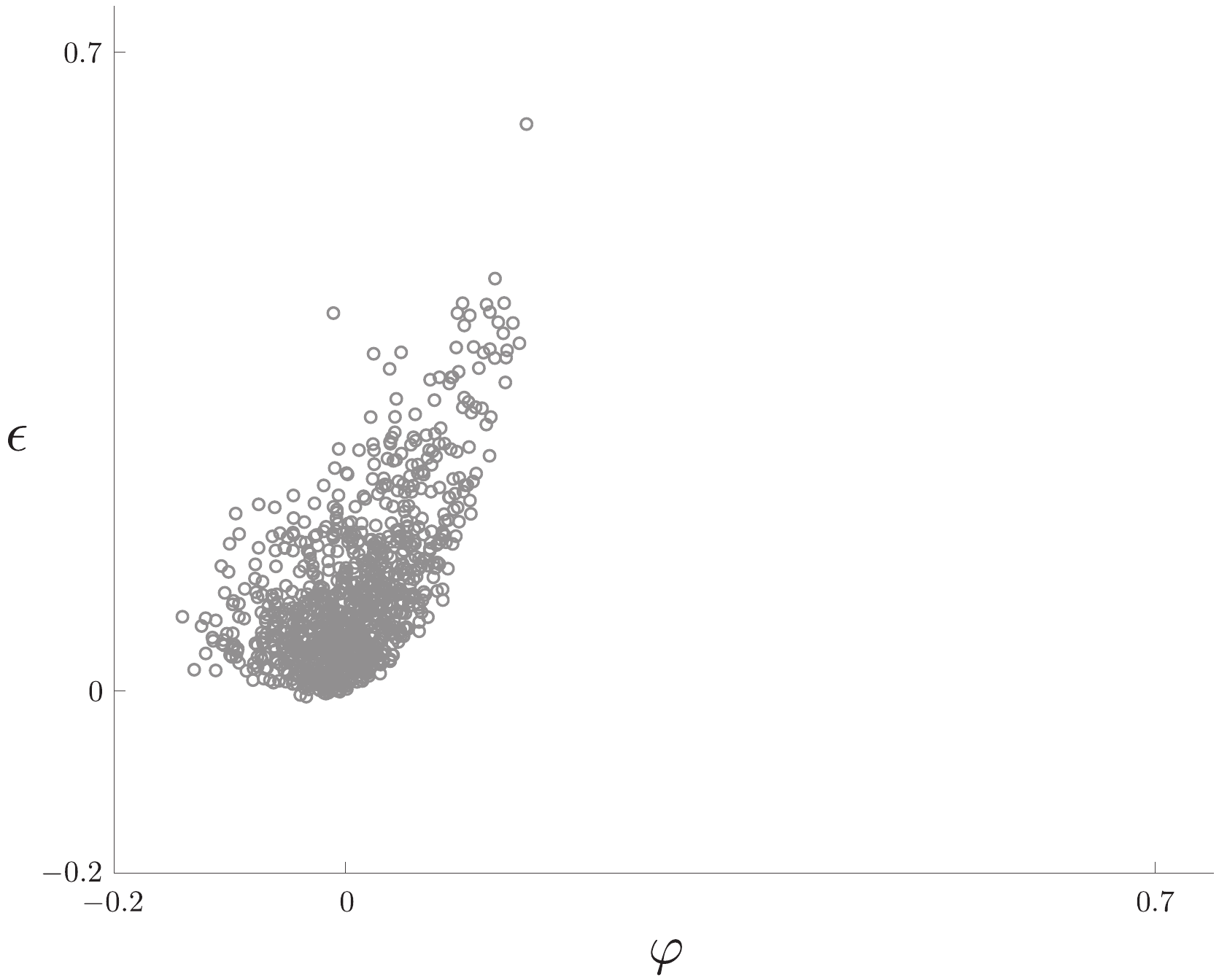}
  \caption{\footnotesize{$RSD_{\lambda}>20\,\%$}}
  \label{fig:rel_phieps_mb_lvb_san}
\end{subfigure}
\caption{\small{Cross-plots of $\varphi$ versus $\epsilon$ for $1000$ examples of equivalent TI sandstones with $RSD_{\mu}>2\,\%\,$.}}
\label{fig:rel_phieps_mb_san}
\end{figure}

\begin{figure}[!htbp]
\centering
\begin{subfigure}{.33\textwidth}
  \centering
   \includegraphics[scale=0.3]{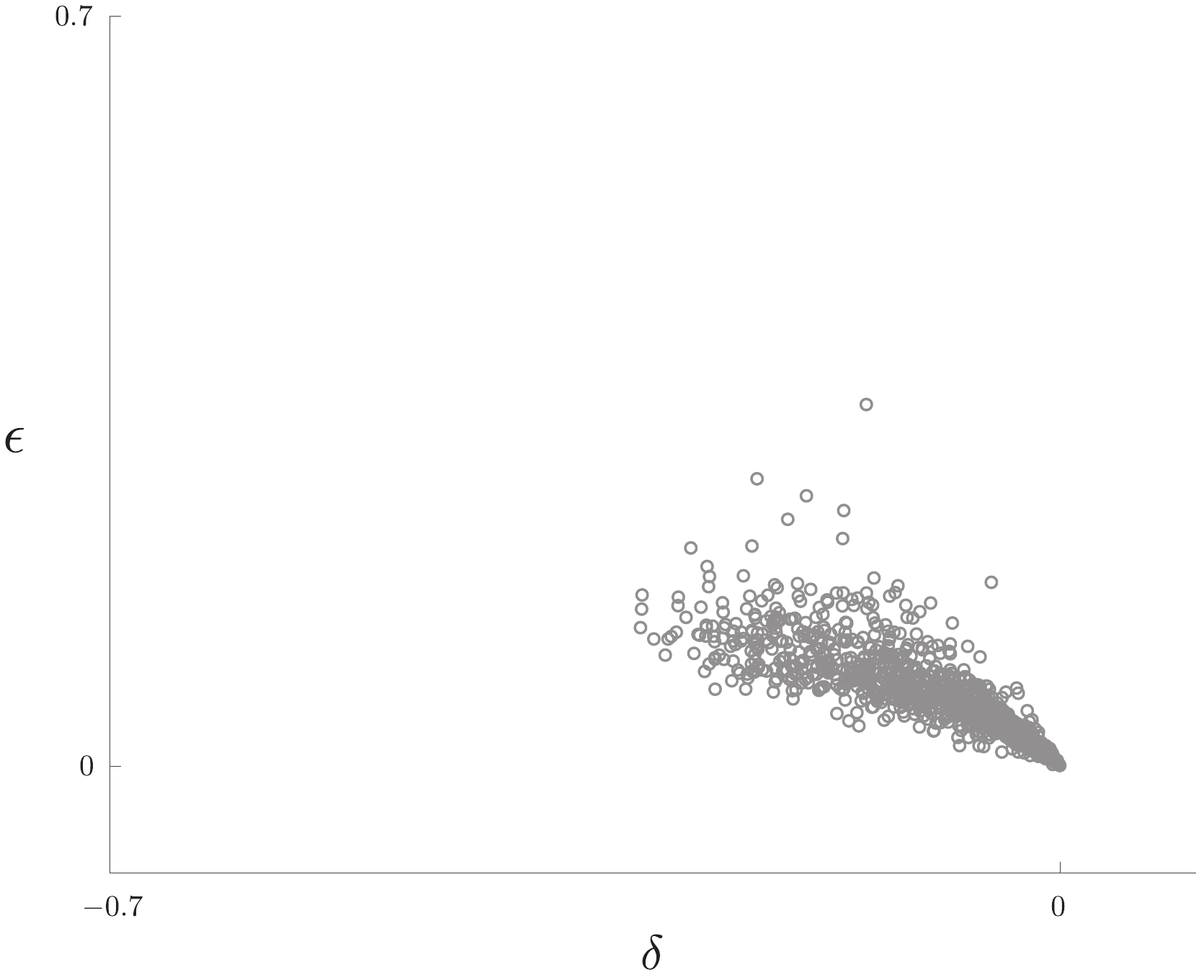}
  \caption{\footnotesize{$RSD_{\lambda}<2\,\%$}}
  \label{fig:rel_deleps_mb_ls_san}
\end{subfigure}%
\begin{subfigure}{.33\textwidth}
  \centering
   \includegraphics[scale=0.3]{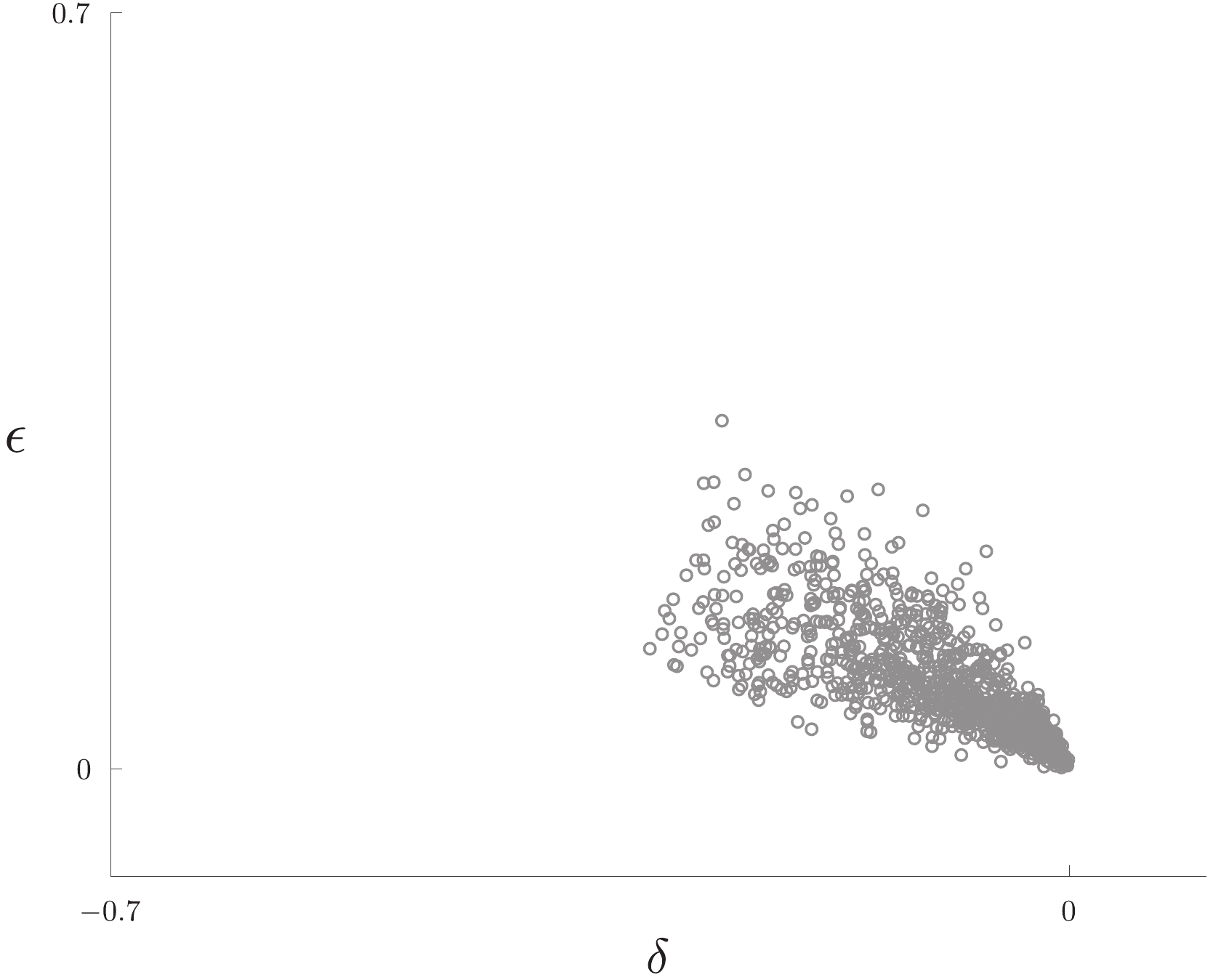}
  \caption{\footnotesize{$RSD_{\lambda}\in\left(2\,\%\,,\,20\,\%\right)$}}
  \label{fig:rel_deleps_mb_2l20_san}
\end{subfigure}%
\begin{subfigure}{.33\textwidth}
  \centering
   \includegraphics[scale=0.3]{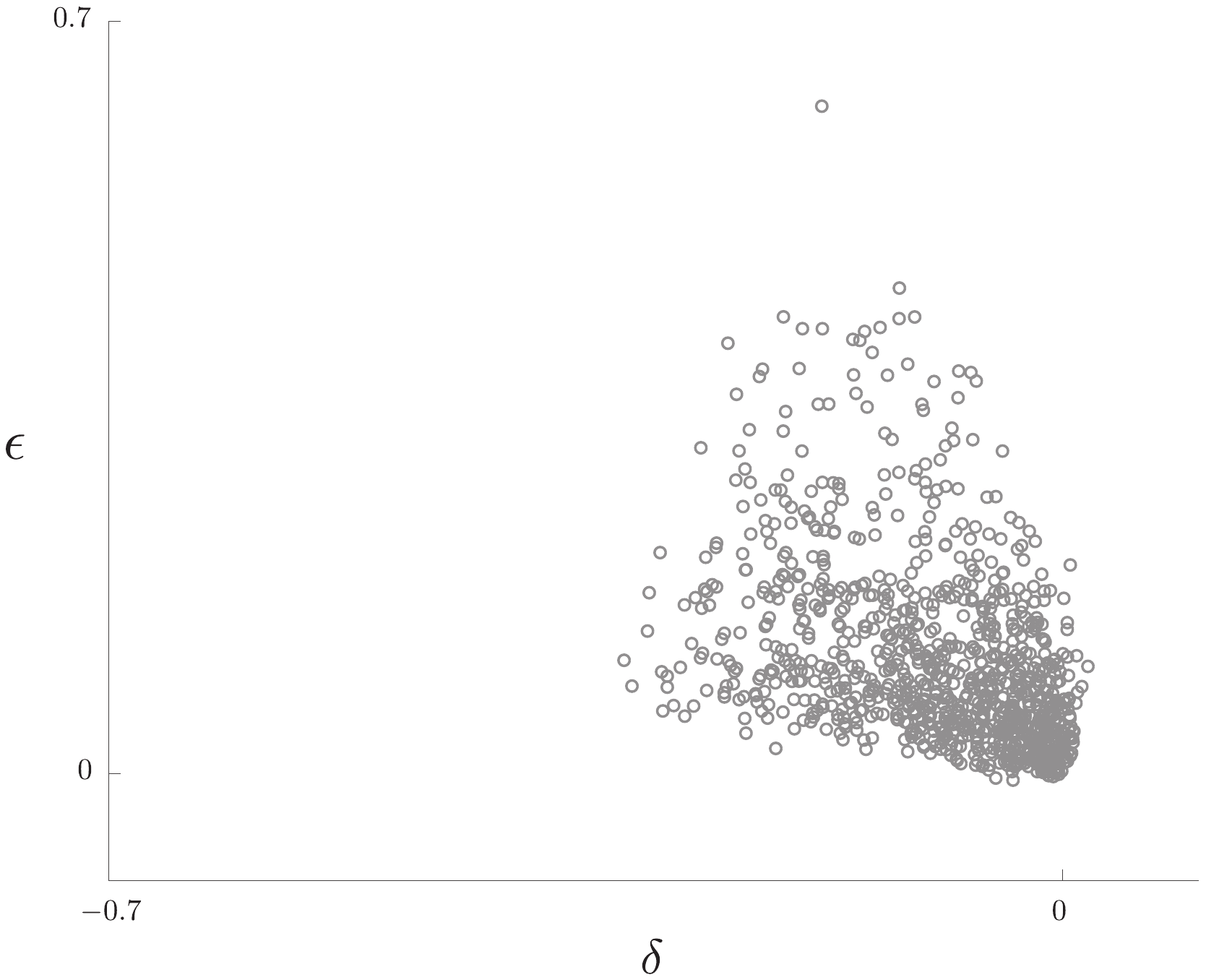}
  \caption{\footnotesize{$RSD_{\lambda}>20\,\%$}}
  \label{fig:rel_deleps_mb_lvb_san}
\end{subfigure}
\caption{\small{Cross-plots of $\delta$ versus $\epsilon$ for $1000$ examples of equivalent TI sandstones with $RSD_{\mu}>2\,\%\,$.}}
\label{fig:rel_deleps_mb_san}
\end{figure}

\begin{figure}
\centering
\begin{subfigure}{.33\textwidth}
  \centering
   \includegraphics[scale=0.3]{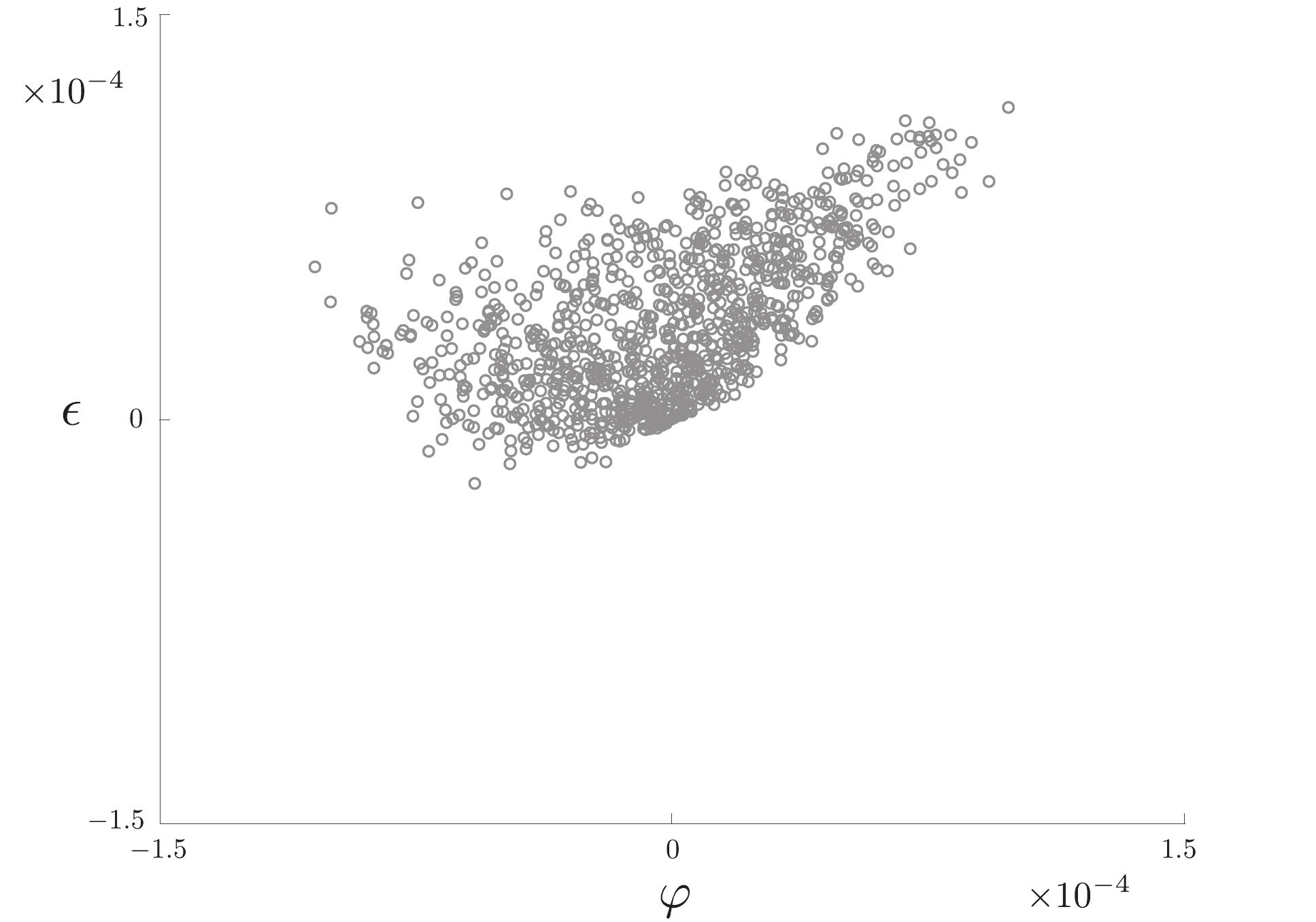}
  \caption{\footnotesize{$RSD_{\lambda}<2\,\%$}}
  \label{fig:rel_phieps_ms_ls_san}
\end{subfigure}%
\begin{subfigure}{.33\textwidth}
  \centering
   \includegraphics[scale=0.3]{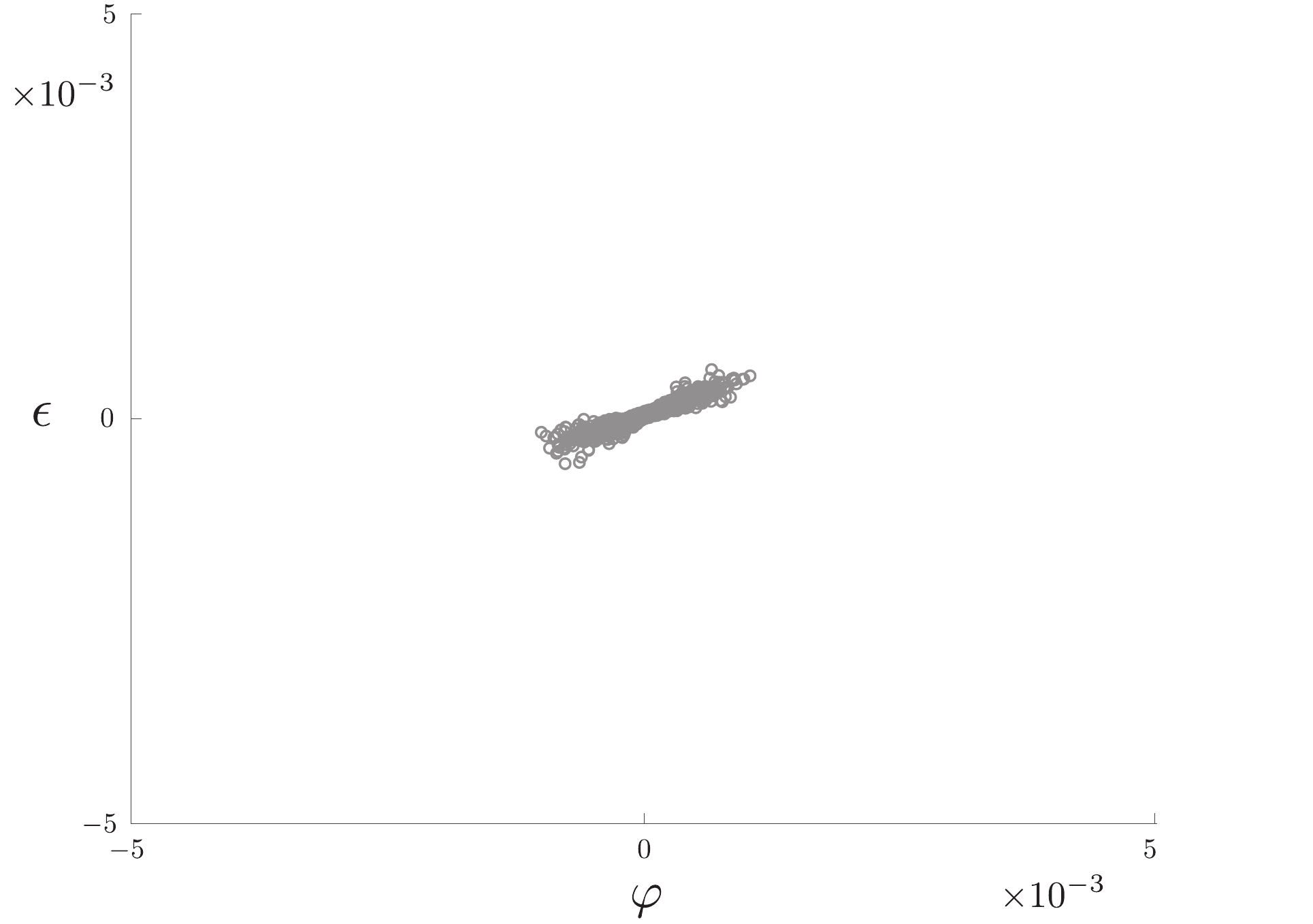}
  \caption{\footnotesize{$RSD_{\lambda}\in\left(2\,\%\,,\,20\,\%\right)$}}
  \label{fig:rel_phieps_ms_2l20_san}
\end{subfigure}%
\begin{subfigure}{.33\textwidth}
  \centering
   \includegraphics[scale=0.3]{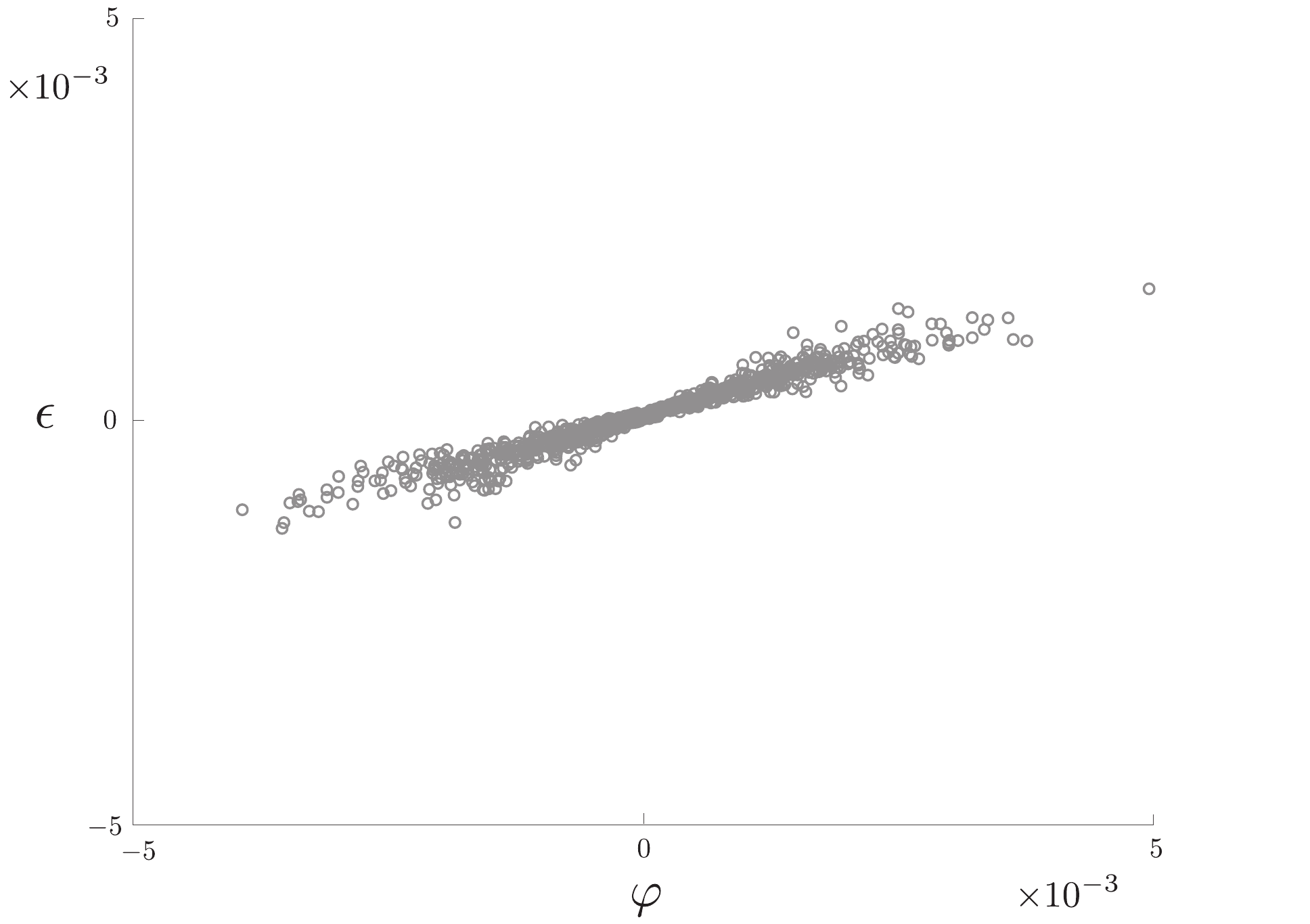}
  \caption{\footnotesize{$RSD_{\lambda}>20\,\%$}}
  \label{fig:rel_phieps_ms_lvb_san}
\end{subfigure}
\caption{\small{Cross-plots of $\varphi$ versus $\epsilon$ for $1000$ examples of equivalent TI sandstones with $RSD_{\mu}<2\,\%\,$.}}
\label{fig:rel_phieps_ms_san}
\end{figure}

\begin{figure}
\centering
\begin{subfigure}{.33\textwidth}
  \centering
   \includegraphics[scale=0.3]{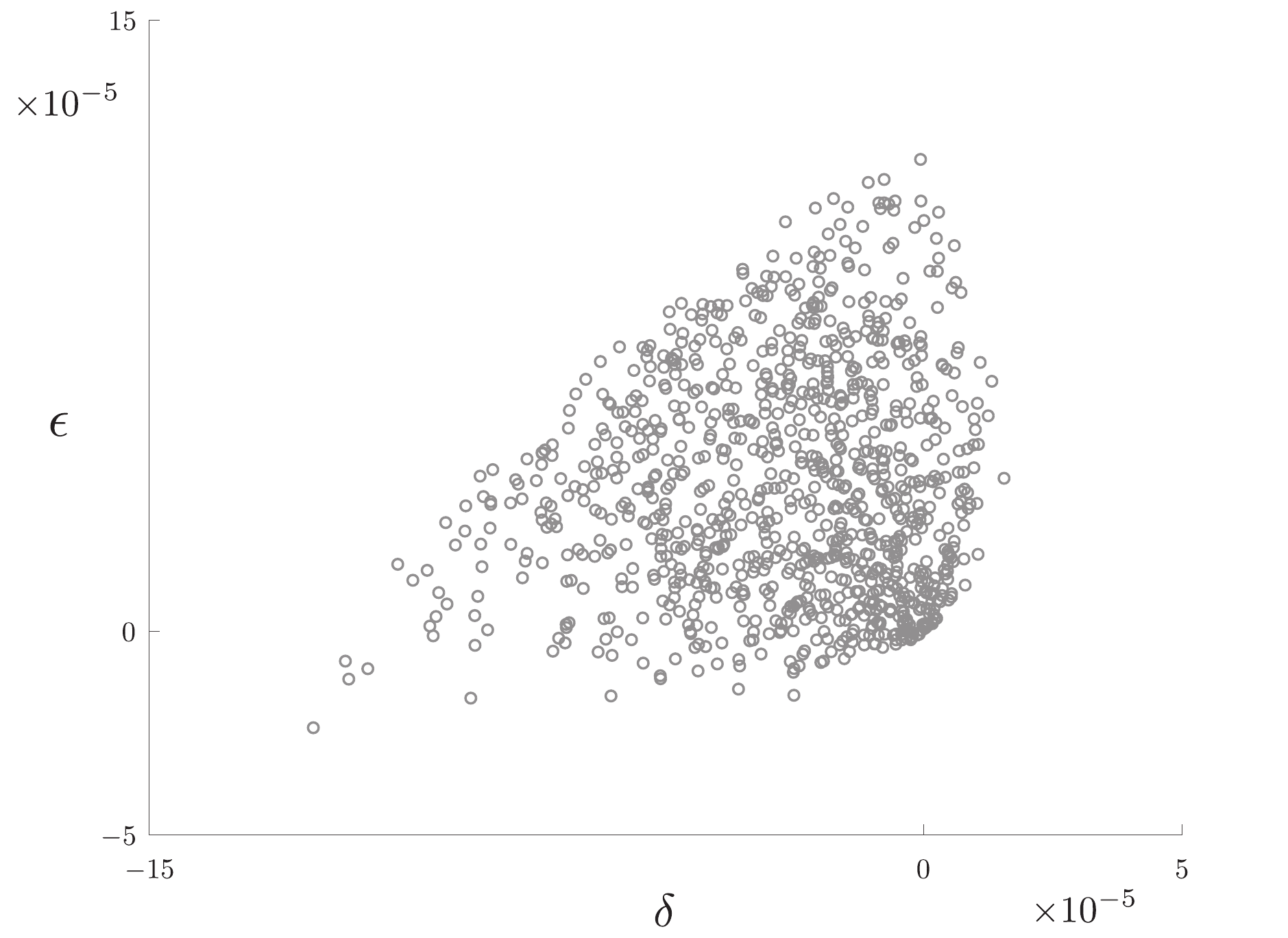}
  \caption{\footnotesize{$RSD_{\lambda}<2\,\%$}}
  \label{fig:rel_deleps_ms_ls_san}
\end{subfigure}%
\begin{subfigure}{.33\textwidth}
  \centering
   \includegraphics[scale=0.3]{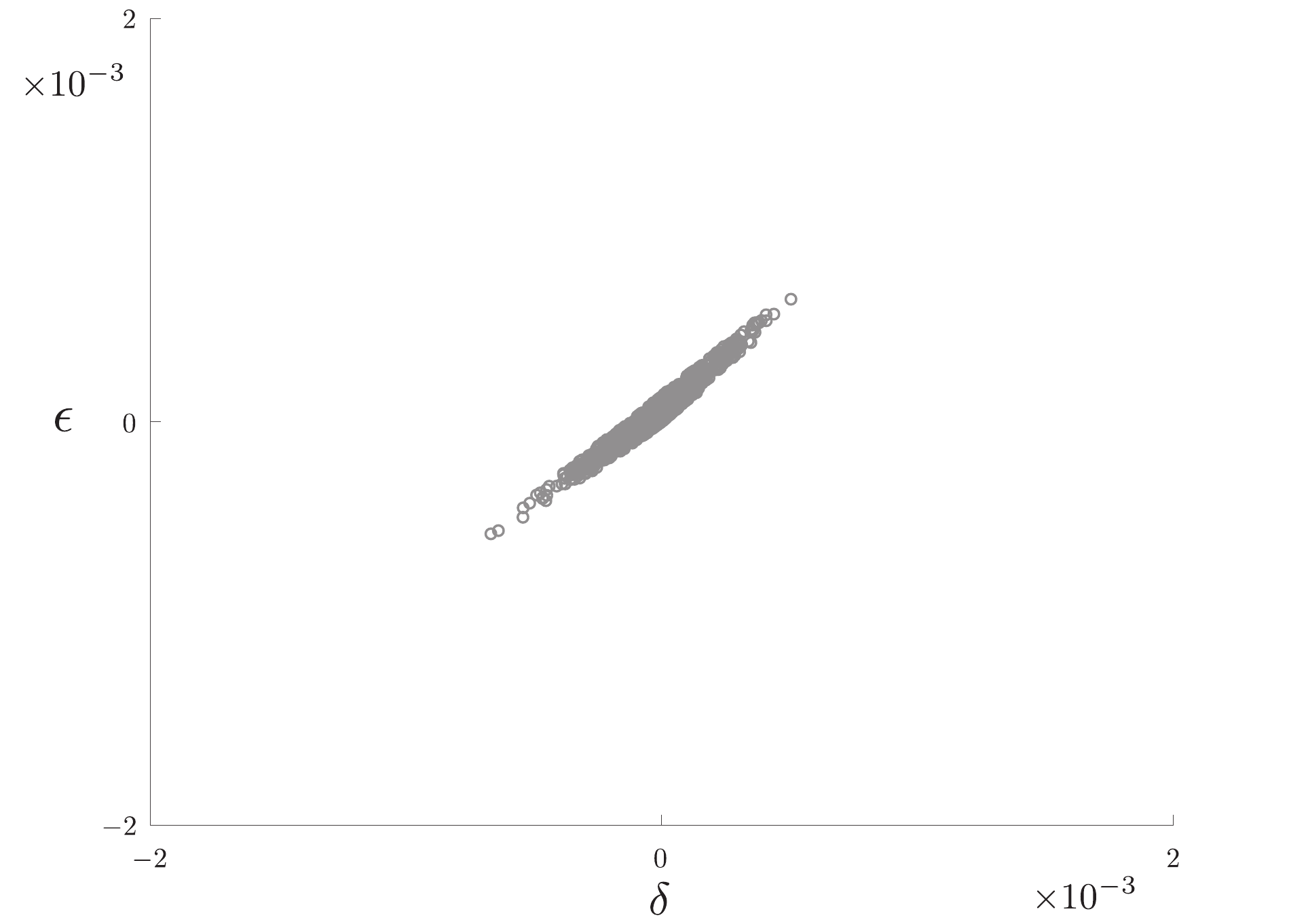}
  \caption{\footnotesize{$RSD_{\lambda}\in\left(2\,\%\,,\,20\,\%\right)$}}
  \label{fig:rel_deleps_ms_2l20_san}
\end{subfigure}%
\begin{subfigure}{.33\textwidth}
  \centering
   \includegraphics[scale=0.3]{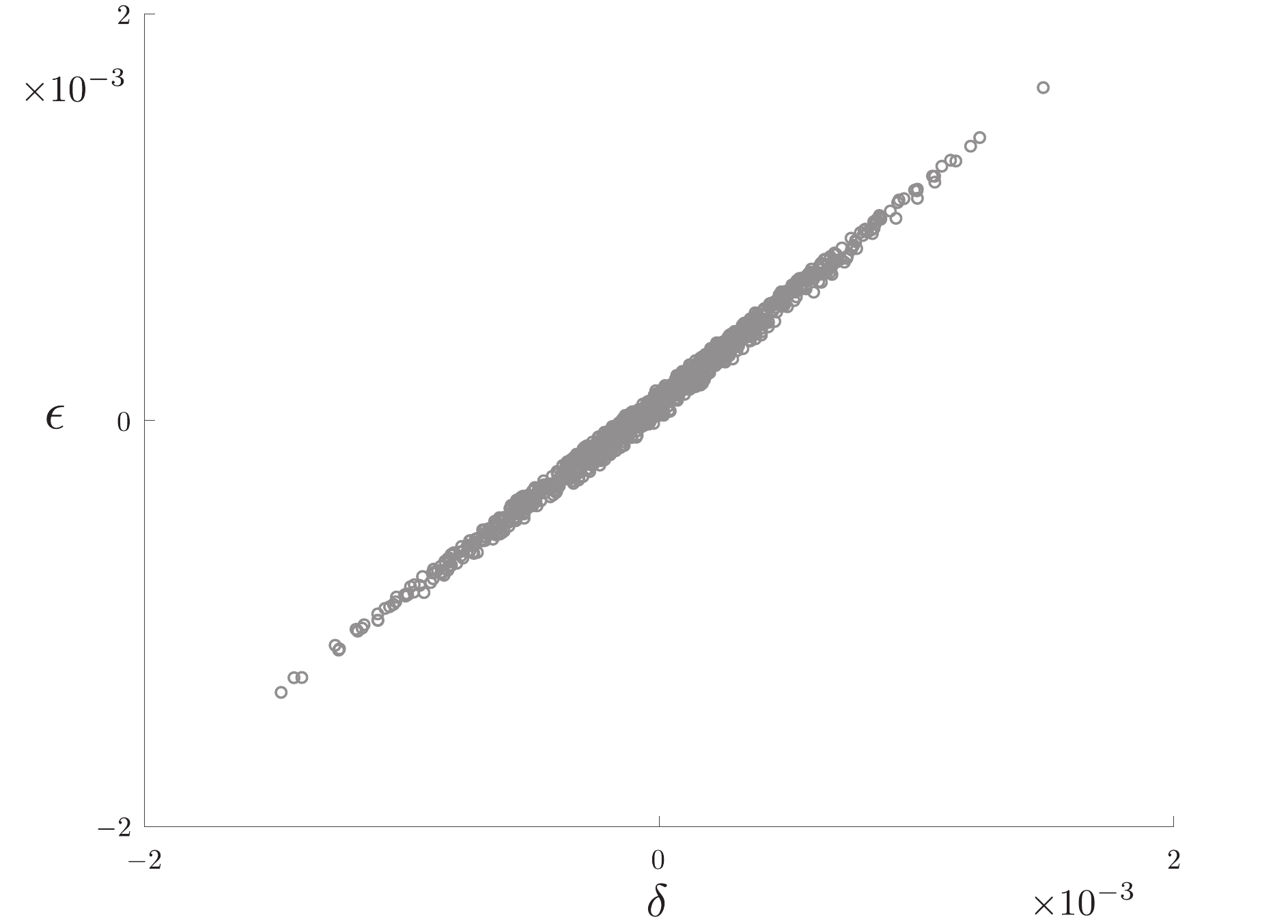}
  \caption{\footnotesize{$RSD_{\lambda}>20\,\%$}}
  \label{fig:rel_deleps_ms_lvb_san}
\end{subfigure}
\caption{\small{Cross-plots of $\delta$ versus $\epsilon$ for $1000$ examples of equivalent TI sandstones with $RSD_{\mu}<2\,\%\,$.}}
\label{fig:rel_deleps_ms_san}
\end{figure}

\begin{figure}[!htbp]
\centering
\begin{subfigure}{.33\textwidth}
  \centering
   \includegraphics[scale=0.3]{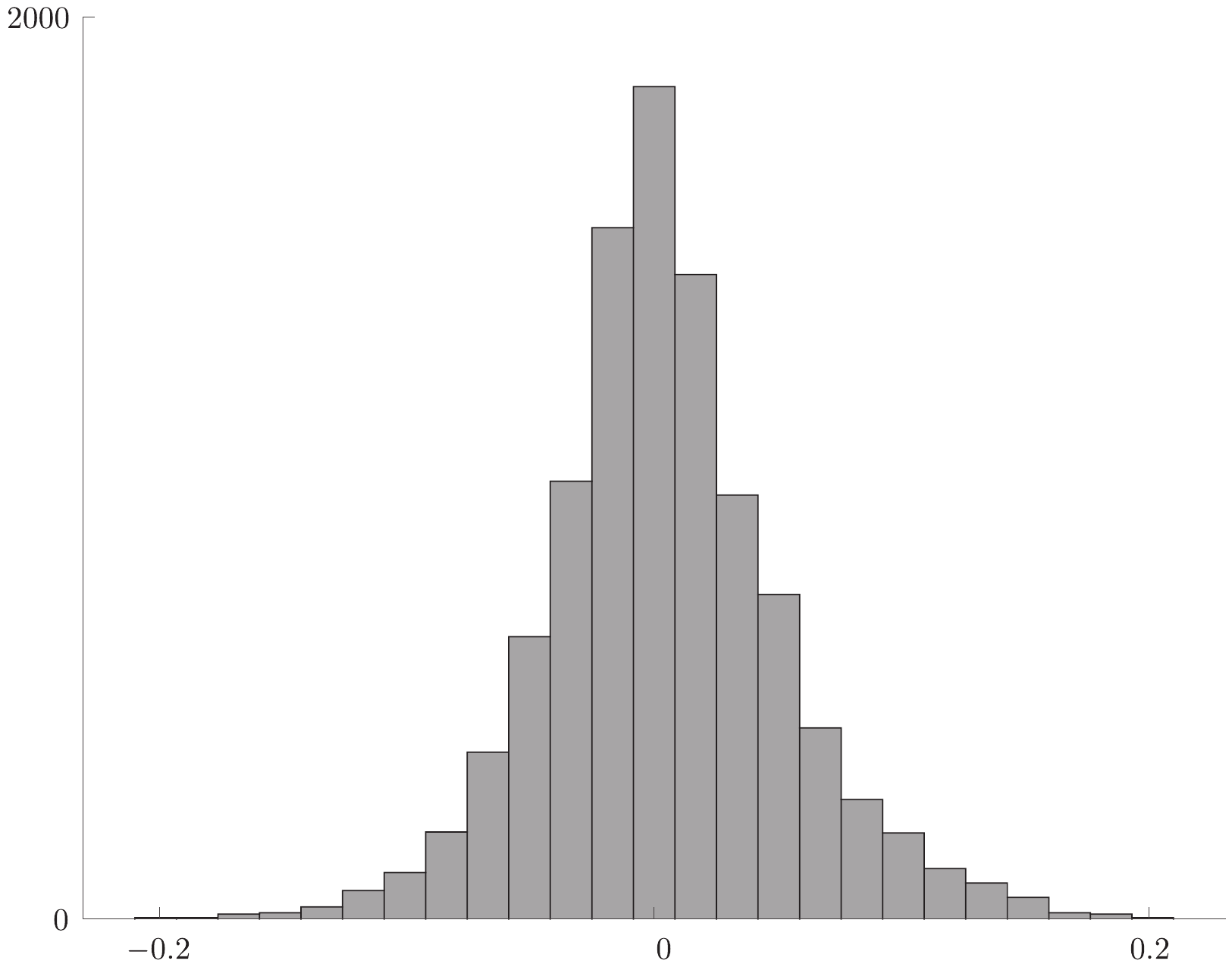}
  \caption{\footnotesize{Distribution of $\varphi$ }}
  \label{fig:disphi_san}
\end{subfigure}%
\begin{subfigure}{.33\textwidth}
  \centering
   \includegraphics[scale=0.3]{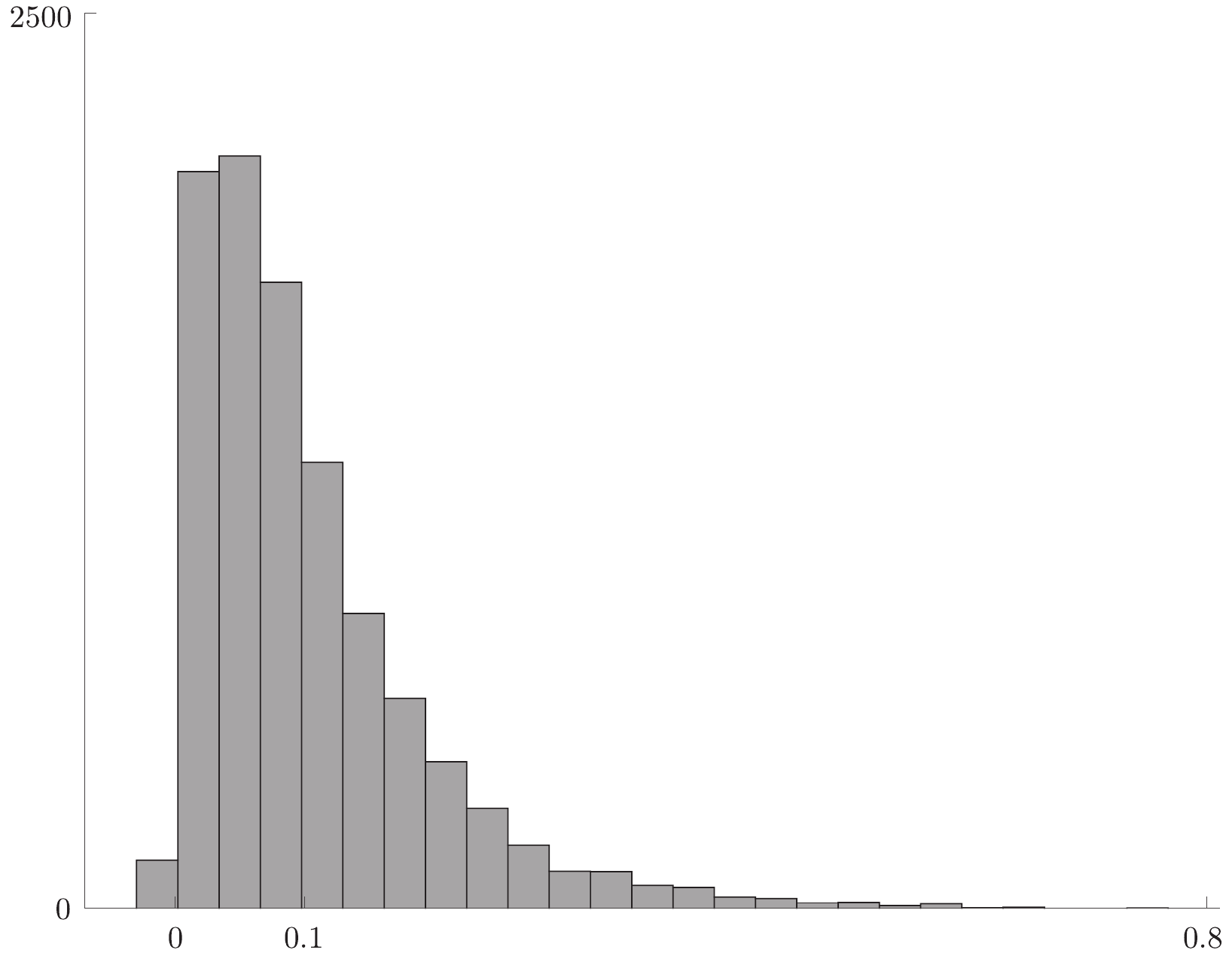}
  \caption{\footnotesize{Distribution of $\epsilon$ }}
  \label{fig:diseps_san}
\end{subfigure}%
\begin{subfigure}{.33\textwidth}
  \centering
   \includegraphics[scale=0.3]{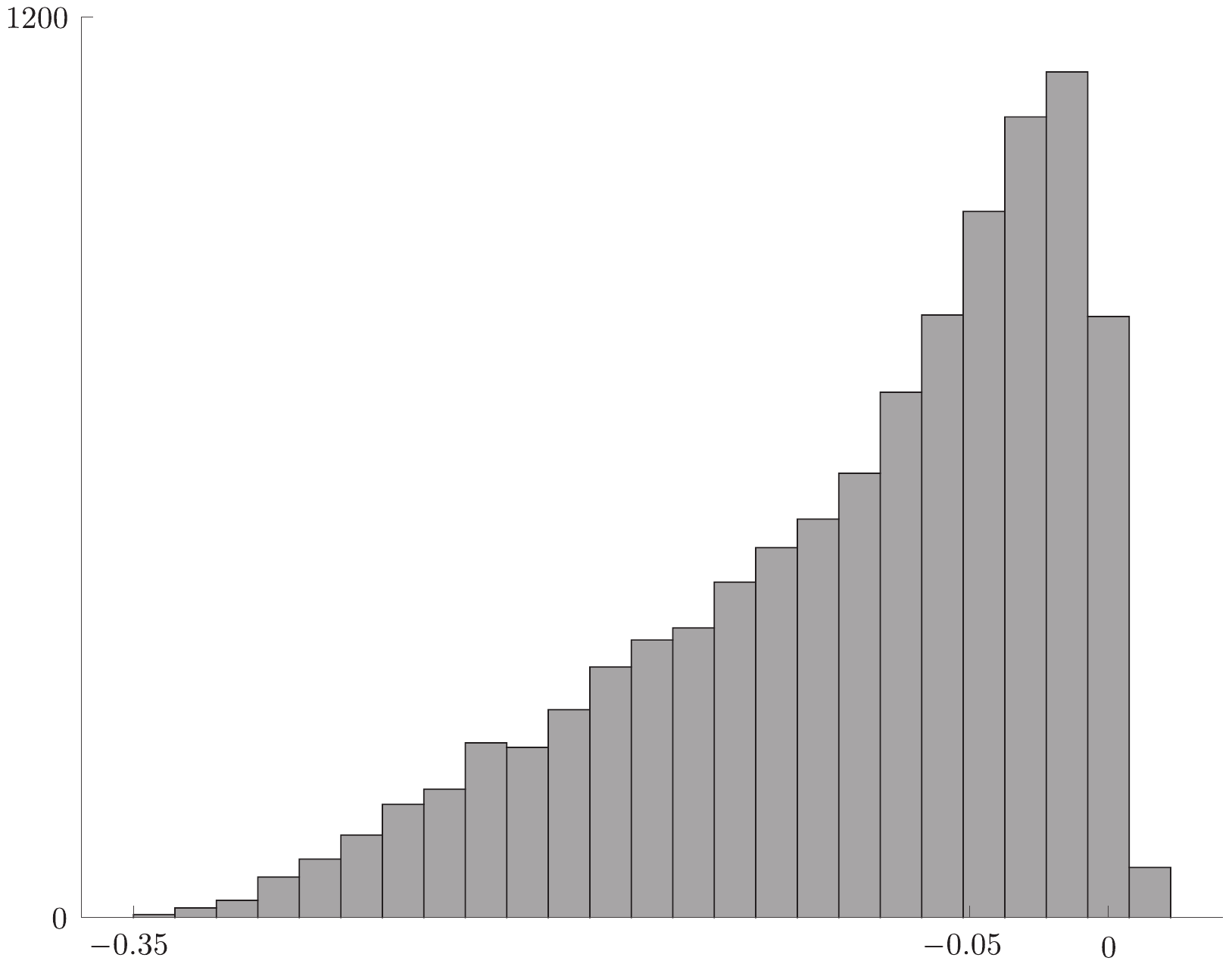}
  \caption{\footnotesize{Distribution of $\delta$ }}
  \label{fig:disdel_san}
\end{subfigure}
\caption{\small{Distribution of anisotropy parameters for $10000$ TI media, relevant to layered sandstones.}}
\label{fig:dis_fel}
\end{figure}

\begin{figure}[!htbp]
\centering
\begin{subfigure}{.33\textwidth}
  \centering
   \includegraphics[scale=0.3]{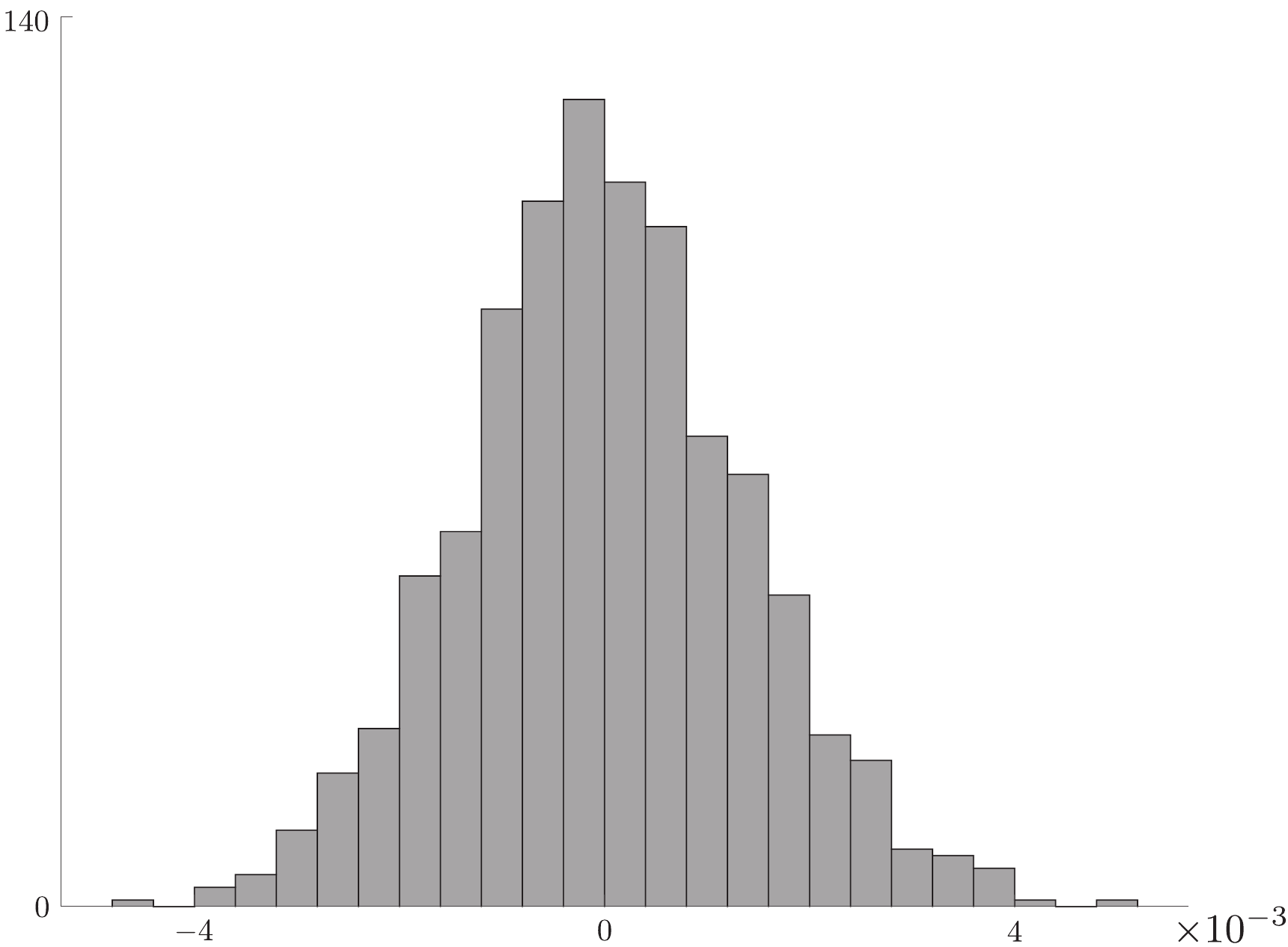}
  \caption{\footnotesize{$\varphi$\, for $RSD_{\lambda}<2\,\%$}}
  \label{fig01a_san}
\end{subfigure}%
\begin{subfigure}{.33\textwidth}
  \centering
   \includegraphics[scale=0.3]{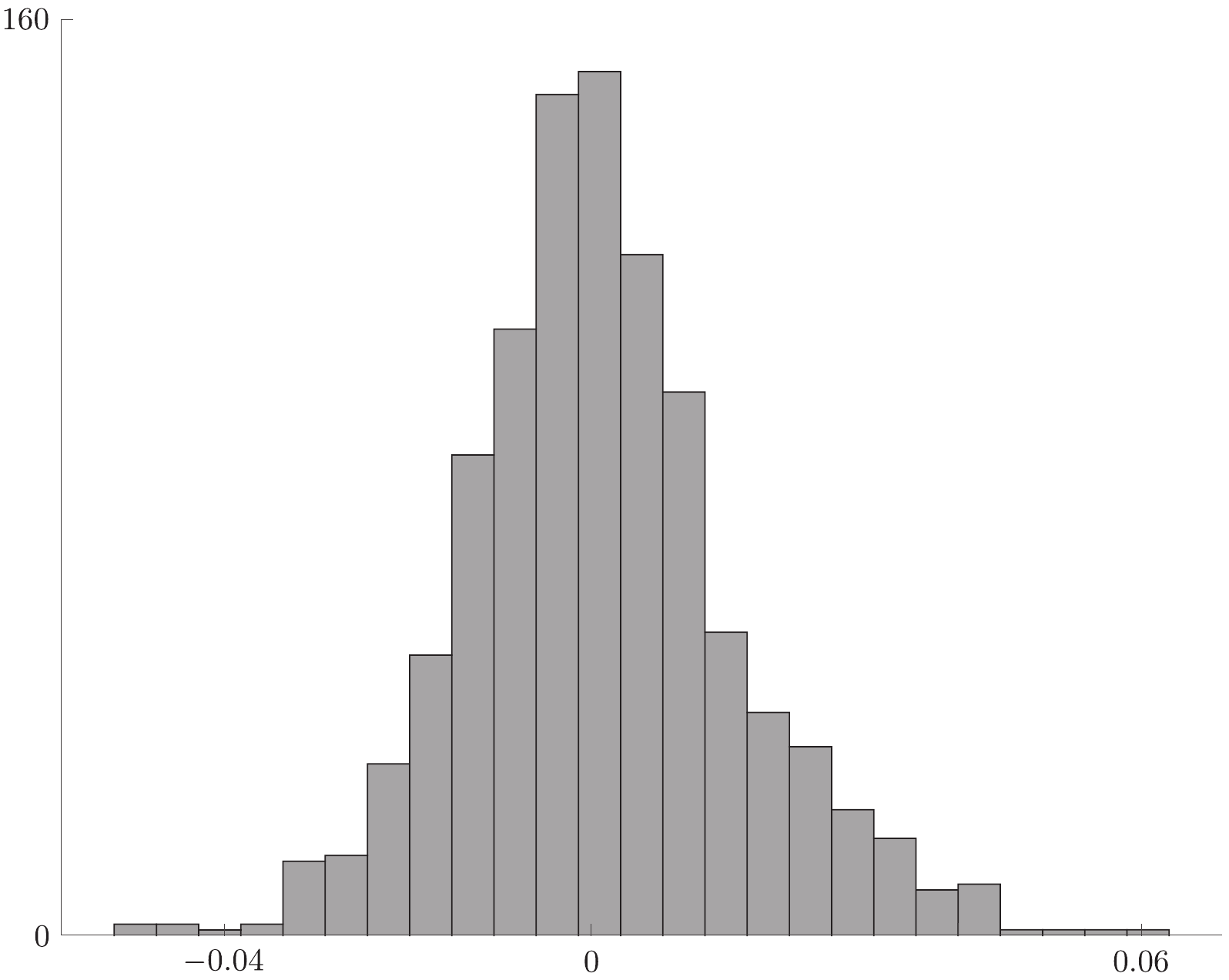}
  \caption{\footnotesize{$\varphi$\, for $RSD_{\lambda}\in(2\,\%\,,\,20\,\%)$}}
  \label{fig01b_san}
\end{subfigure}%
\begin{subfigure}{.33\textwidth}
  \centering
   \includegraphics[scale=0.3]{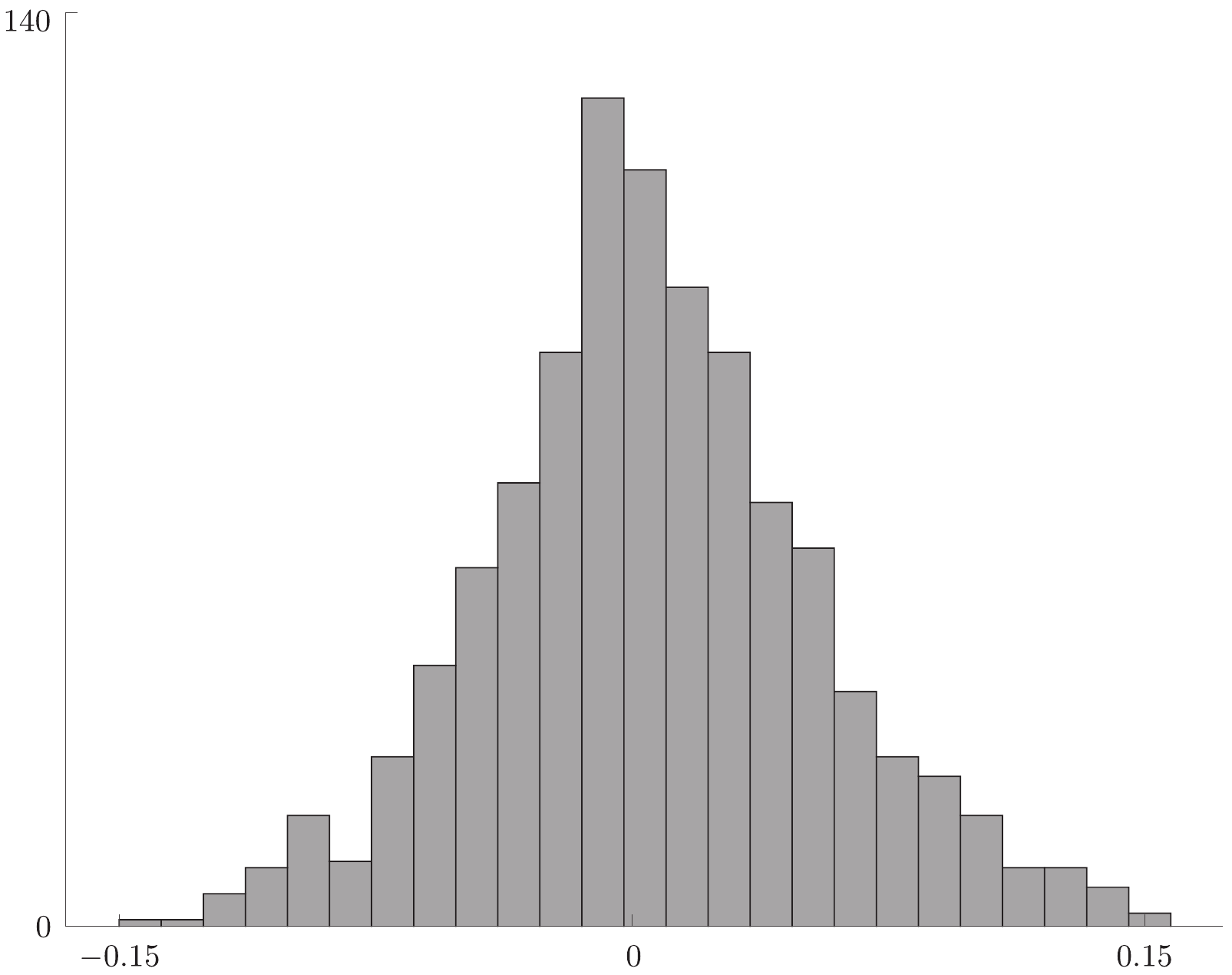}
  \caption{\footnotesize{$\varphi$\, for $RSD_{\lambda}>20\,\%$}}
  \label{fig01c_san}
\end{subfigure}

\qquad

\centering
\begin{subfigure}{.33\textwidth}
  \centering
   \includegraphics[scale=0.3]{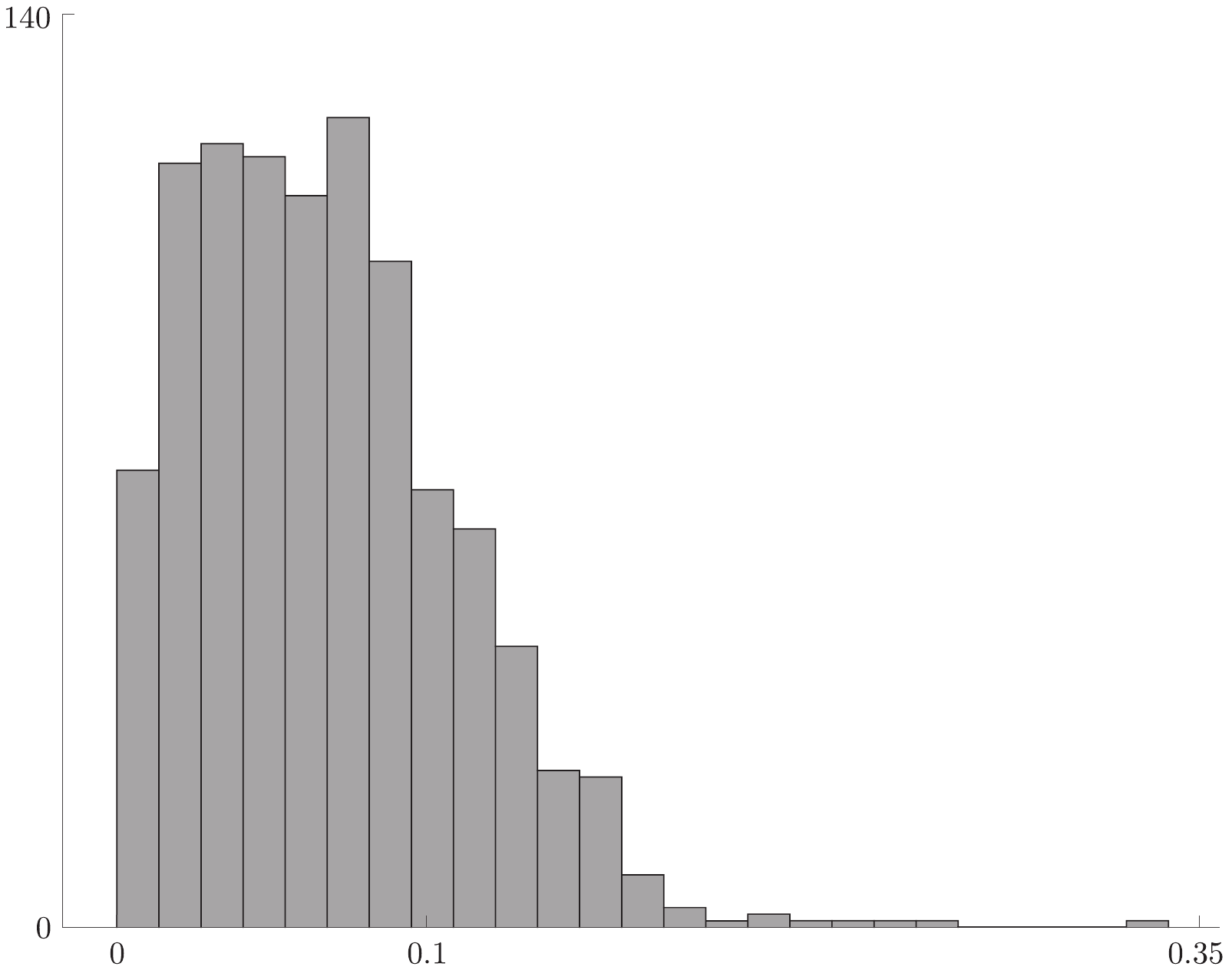}
  \caption{\footnotesize{$\epsilon$\, for $RSD_{\lambda}<2\,\%$}}
  \label{fig01d_san}
\end{subfigure}%
\begin{subfigure}{.33\textwidth}
  \centering
   \includegraphics[scale=0.3]{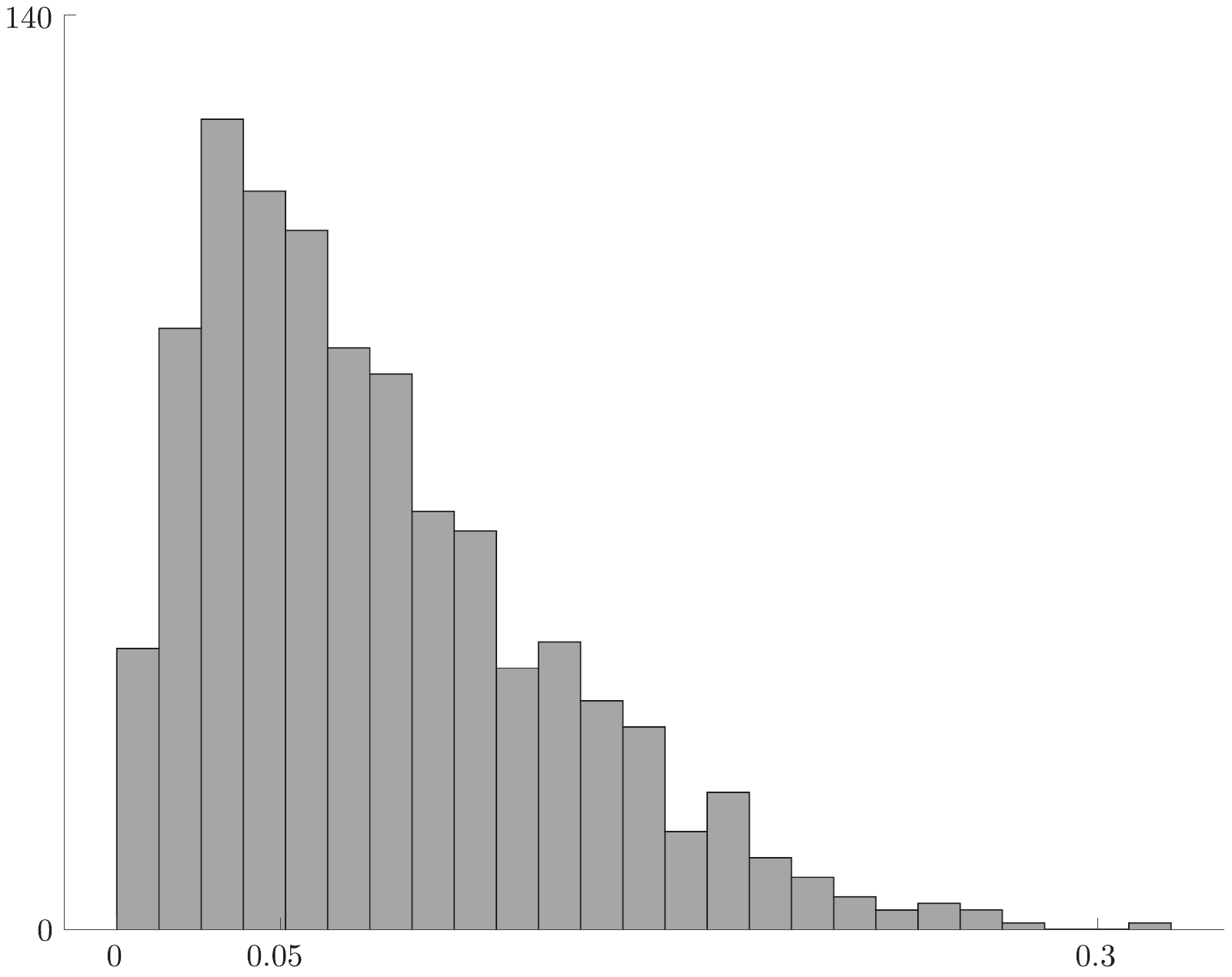}
  \caption{\footnotesize{$\epsilon$\, for $RSD_{\lambda}\in(2\,\%\,,\,20\,\%)$}}
  \label{fig01e_san}
\end{subfigure}%
\begin{subfigure}{.33\textwidth}
  \centering
   \includegraphics[scale=0.3]{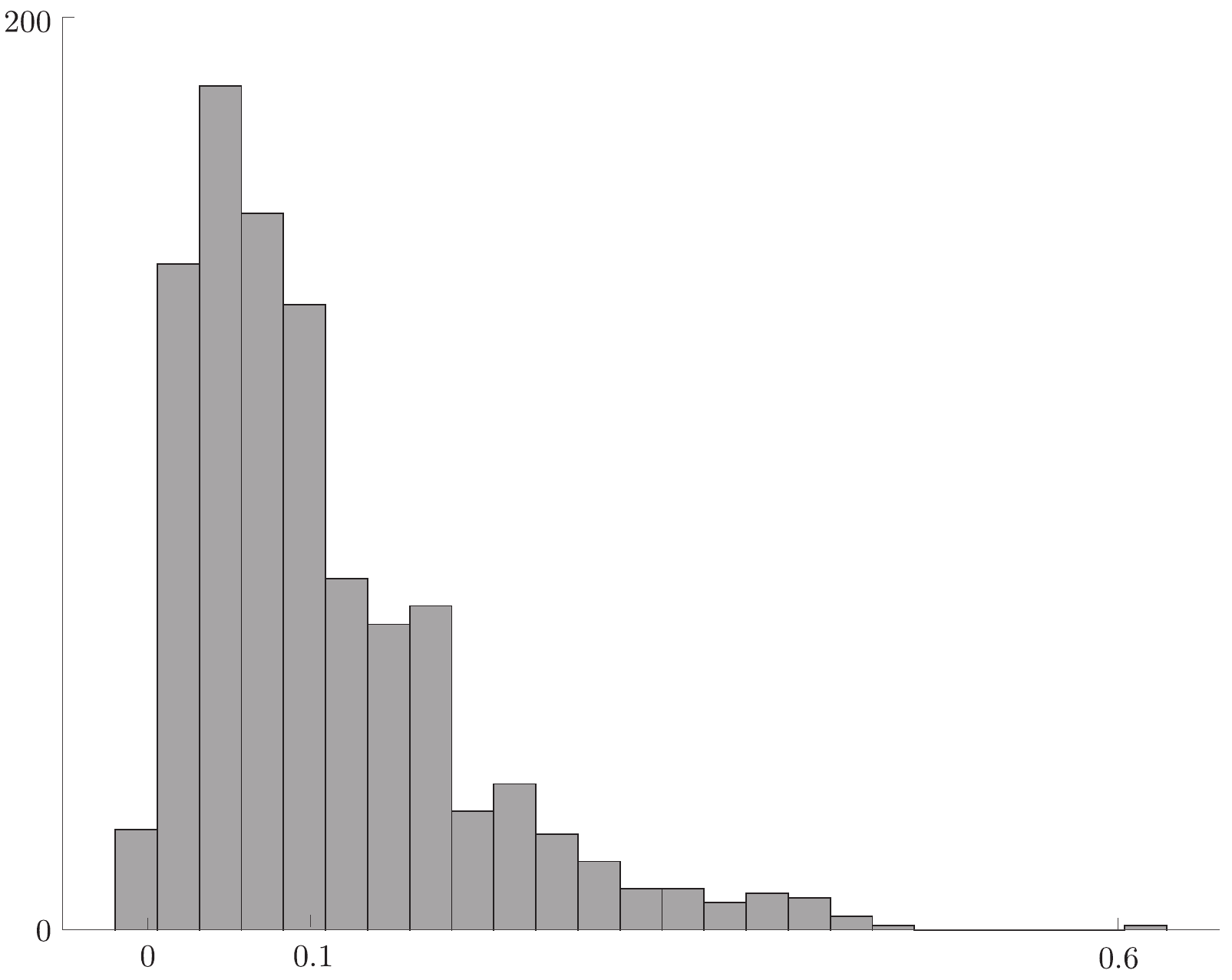}
  \caption{\footnotesize{$\epsilon$\, for $RSD_{\lambda}>20\,\%$}}
  \label{fig01f_san}
\end{subfigure}

\qquad

\centering
\begin{subfigure}{.33\textwidth}
  \centering
   \includegraphics[scale=0.3]{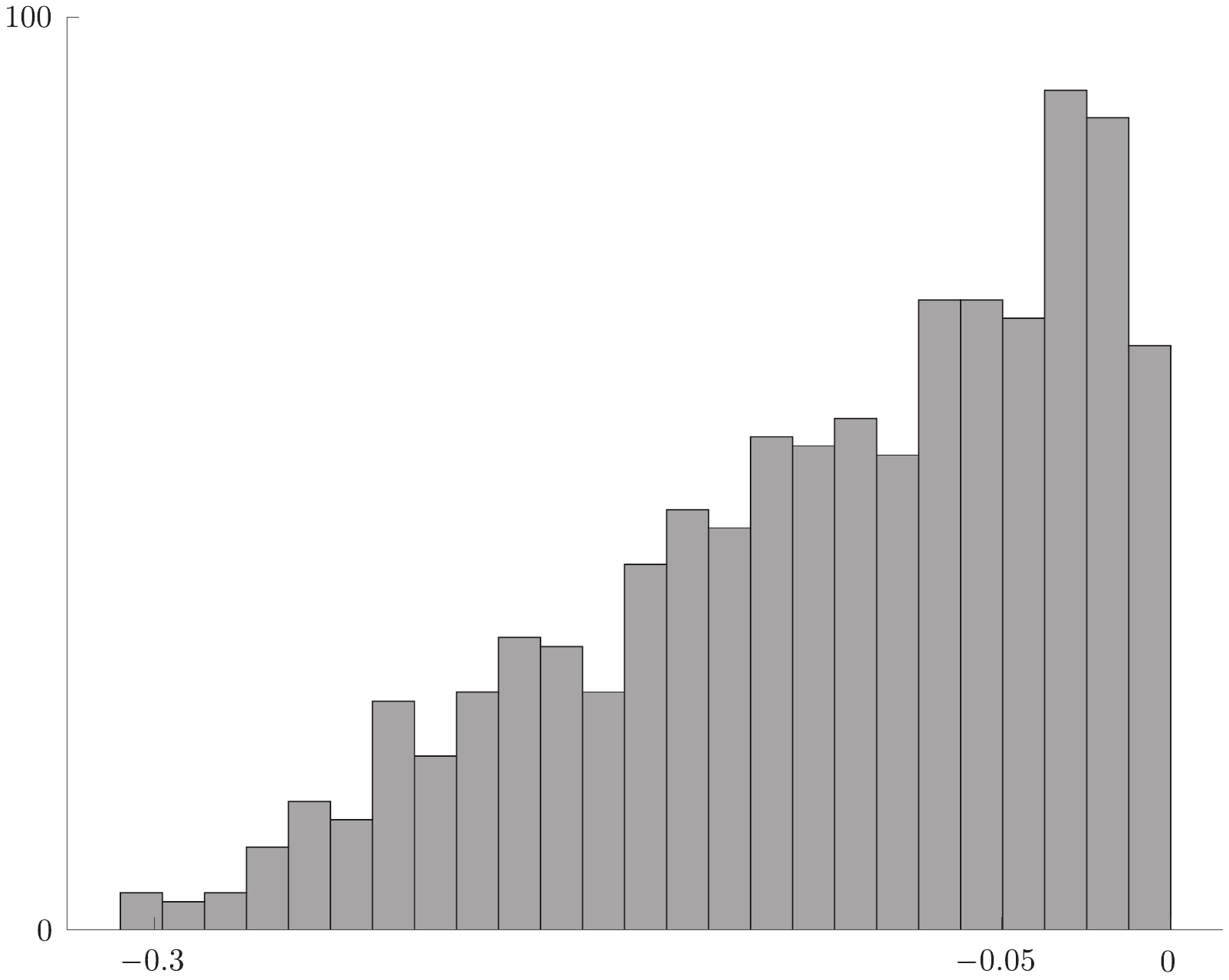}
  \caption{\footnotesize{$\delta$\, for $RSD_{\lambda}<2\,\%$}}
  \label{fig01g_san}
\end{subfigure}%
\begin{subfigure}{.33\textwidth}
  \centering
   \includegraphics[scale=0.3]{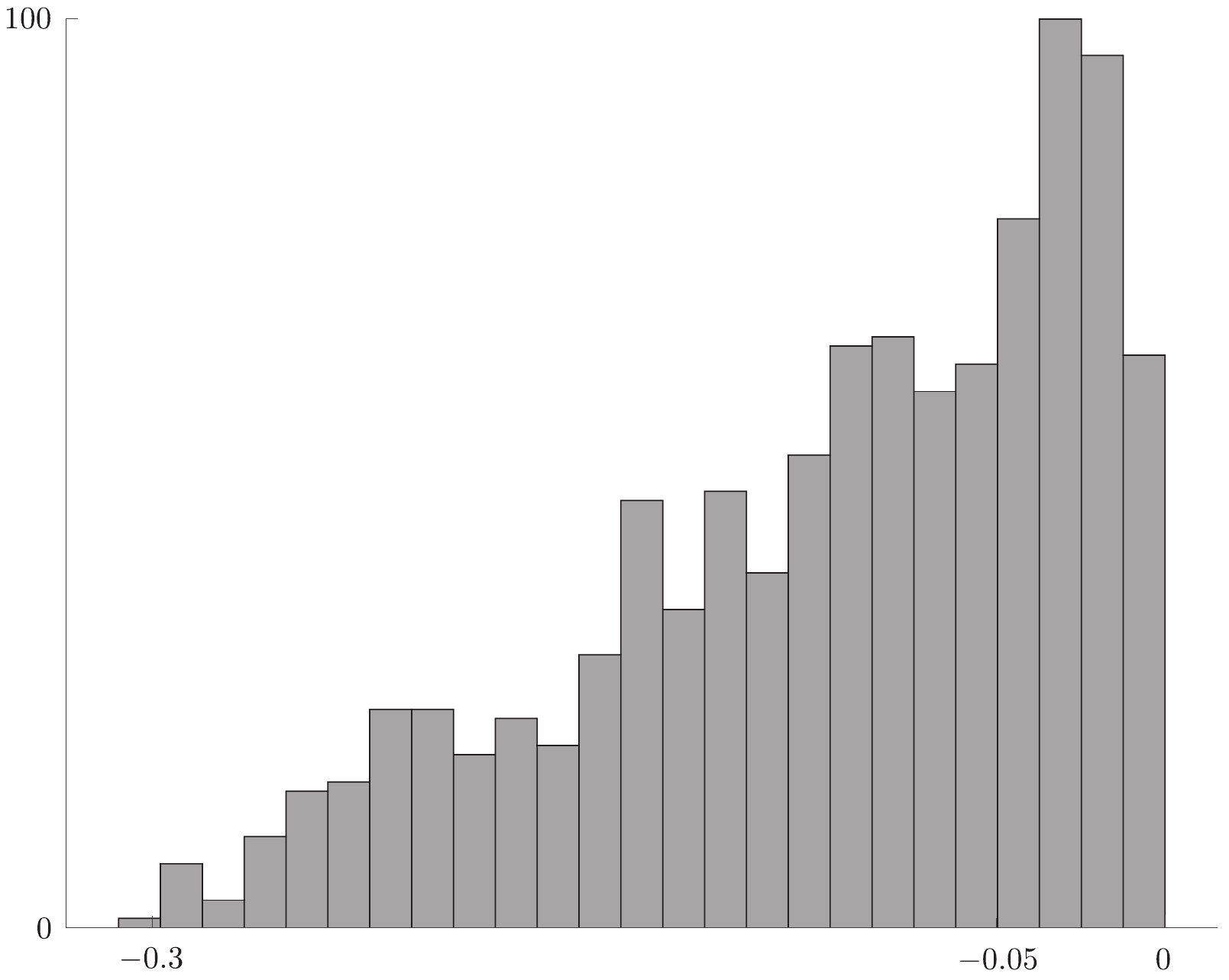}
  \caption{\footnotesize{$\delta$\, for $RSD_{\lambda}\in(2\,\%\,,\,20\,\%)$}}
  \label{fig1h_san}
\end{subfigure}%
\begin{subfigure}{.33\textwidth}
  \centering
   \includegraphics[scale=0.3]{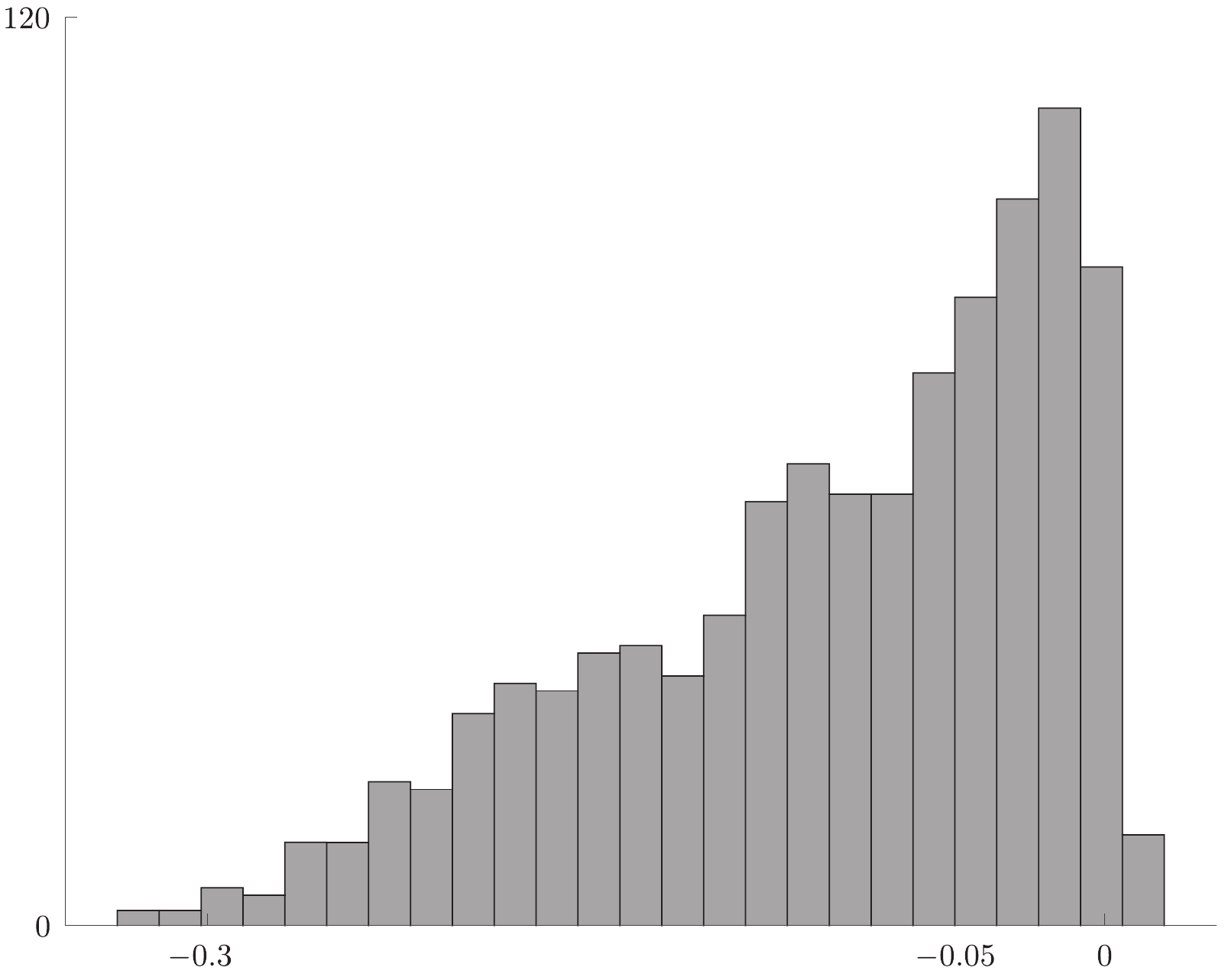}
  \caption{\footnotesize{$\delta$\, for $RSD_{\lambda}>20\,\%$}}
  \label{fig01i_san}
\end{subfigure}
\caption{\small{Distribution of $\varphi$, $\epsilon$ and $\delta$ for $1000$ examples of equivalent TI sandstones with $RSD_{\mu}>2\,\%\,$. }}
\label{fig:dis_mb_san}
\end{figure}

\begin{figure}[!htbp]
\centering
\begin{subfigure}{.33\textwidth}
  \centering
   \includegraphics[scale=0.3]{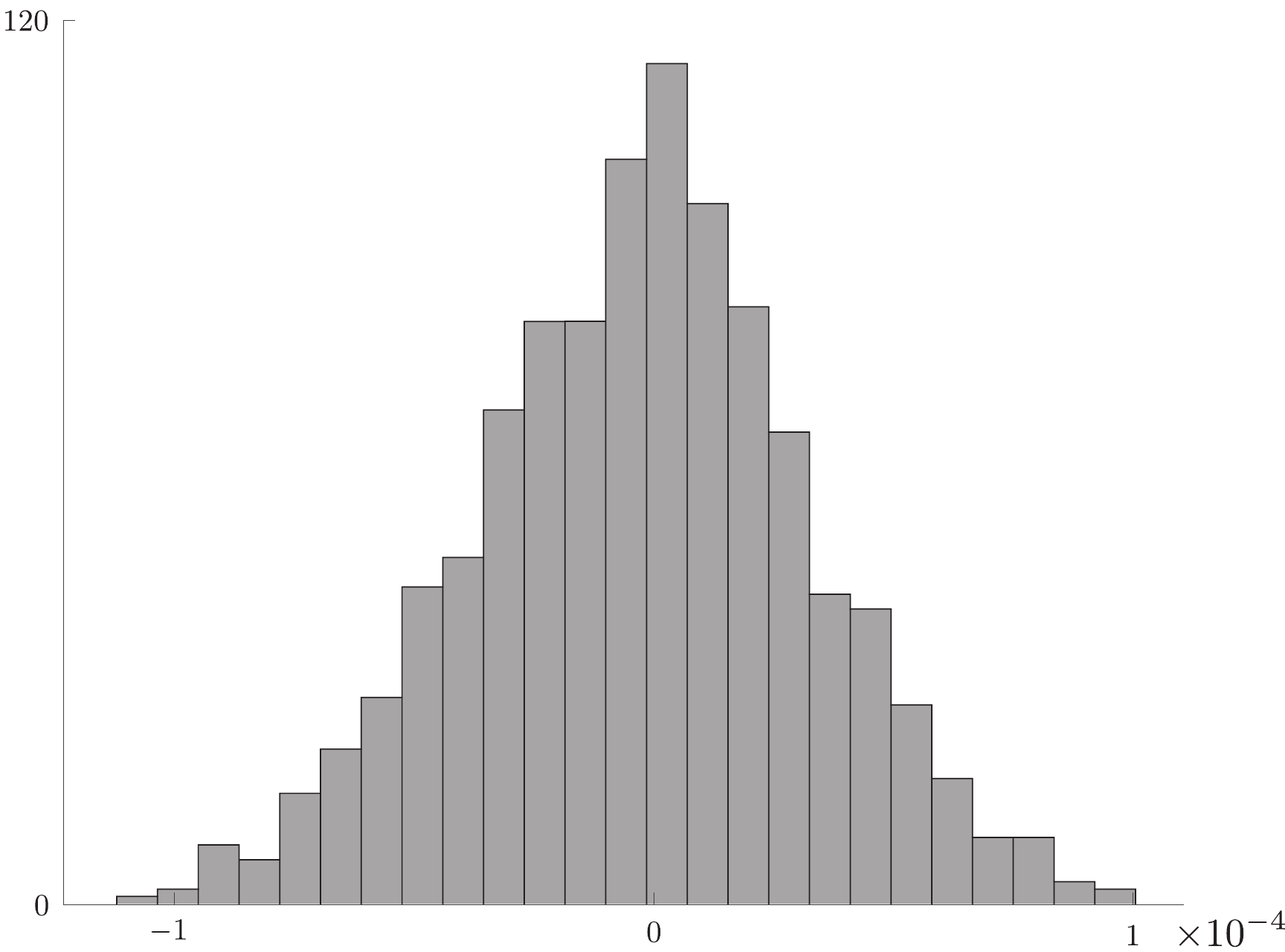}
  \caption{\footnotesize{$\varphi$\, for $RSD_{\lambda}<2\,\%$}}
  \label{fig02a_san}
\end{subfigure}%
\begin{subfigure}{.33\textwidth}
  \centering
   \includegraphics[scale=0.3]{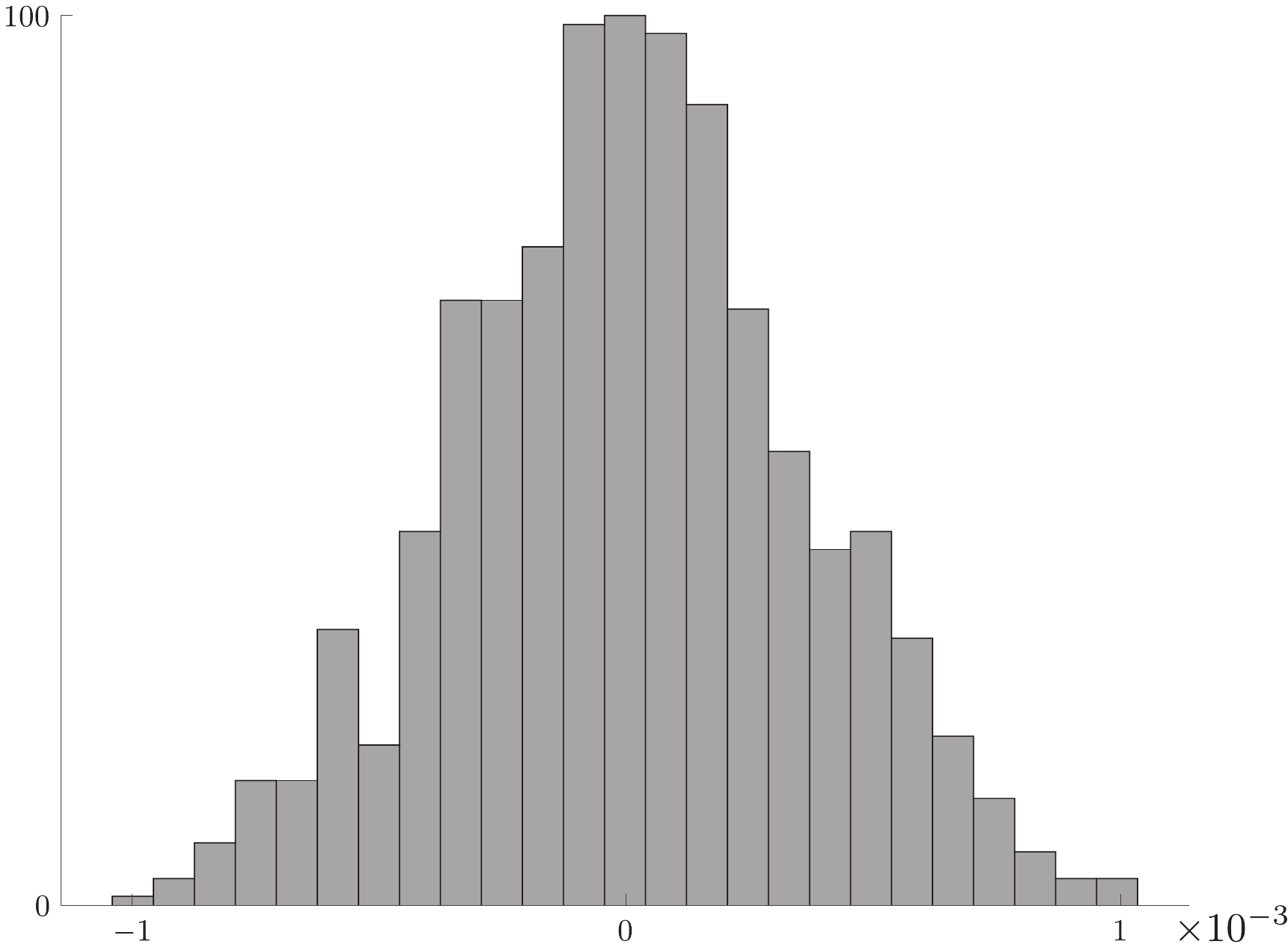}
  \caption{\footnotesize{$\varphi$\, for $RSD_{\lambda}\in(2\,\%\,,\,20\,\%)$}}
  \label{fig02b_san}
\end{subfigure}%
\begin{subfigure}{.33\textwidth}
  \centering
   \includegraphics[scale=0.3]{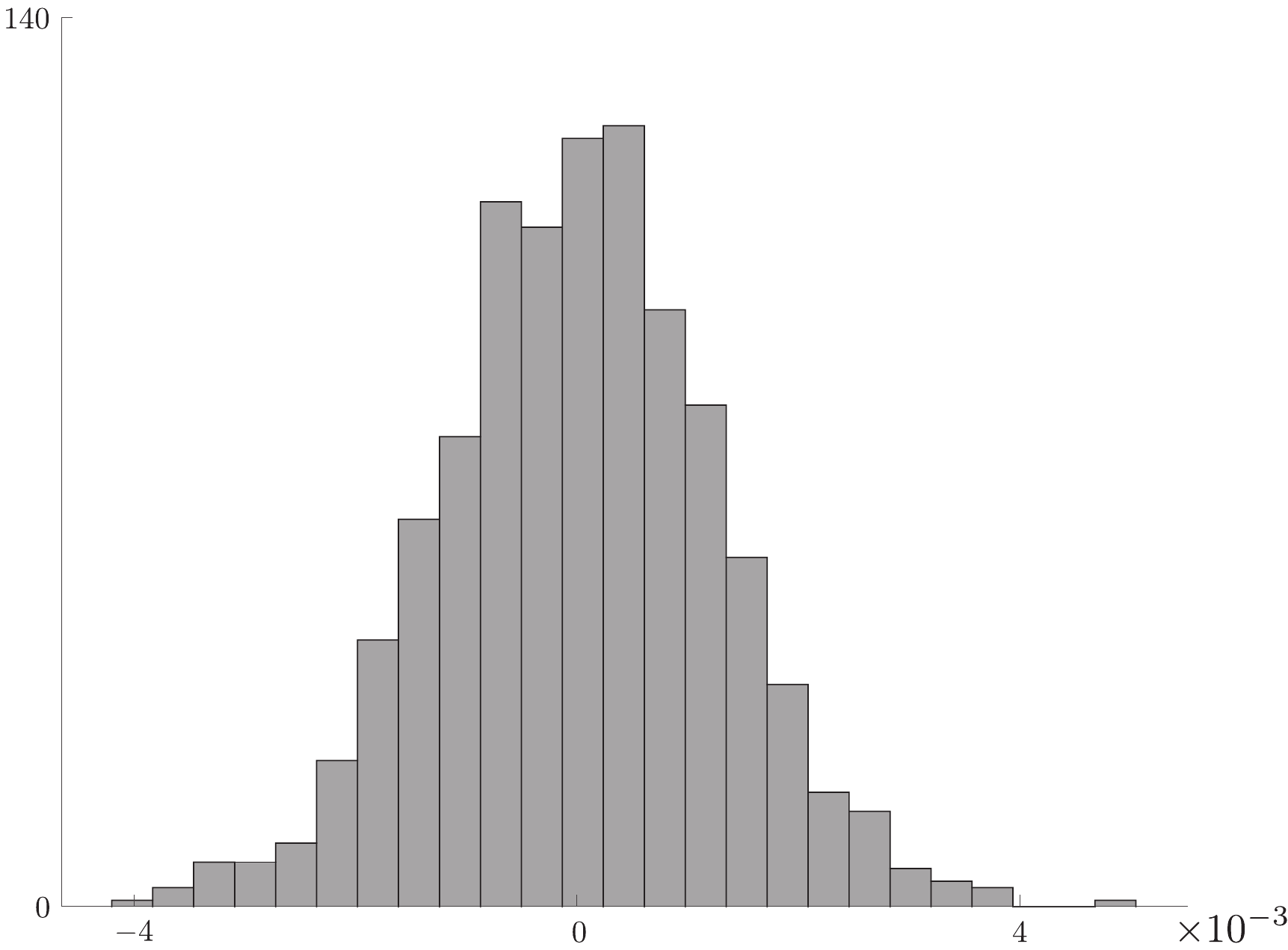}
  \caption{\footnotesize{$\varphi$\, for $RSD_{\lambda}>20\,\%$}}
  \label{fig02c_san}
\end{subfigure}

\qquad

\centering
\begin{subfigure}{.33\textwidth}
  \centering
   \includegraphics[scale=0.3]{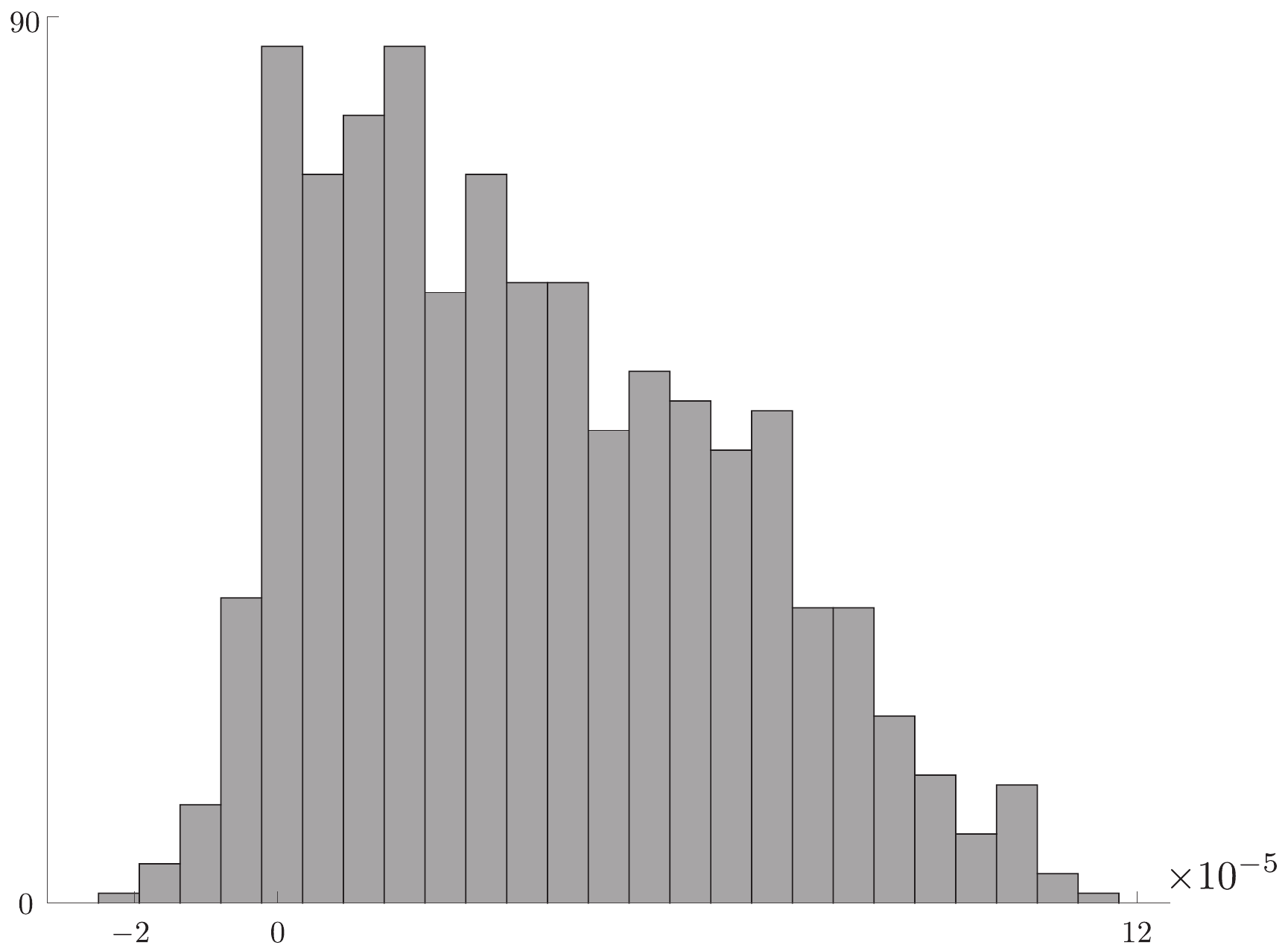}
  \caption{\footnotesize{$\epsilon$\, for $RSD_{\lambda}<2\,\%$}}
  \label{fig02d_san}
\end{subfigure}%
\begin{subfigure}{.33\textwidth}
  \centering
   \includegraphics[scale=0.3]{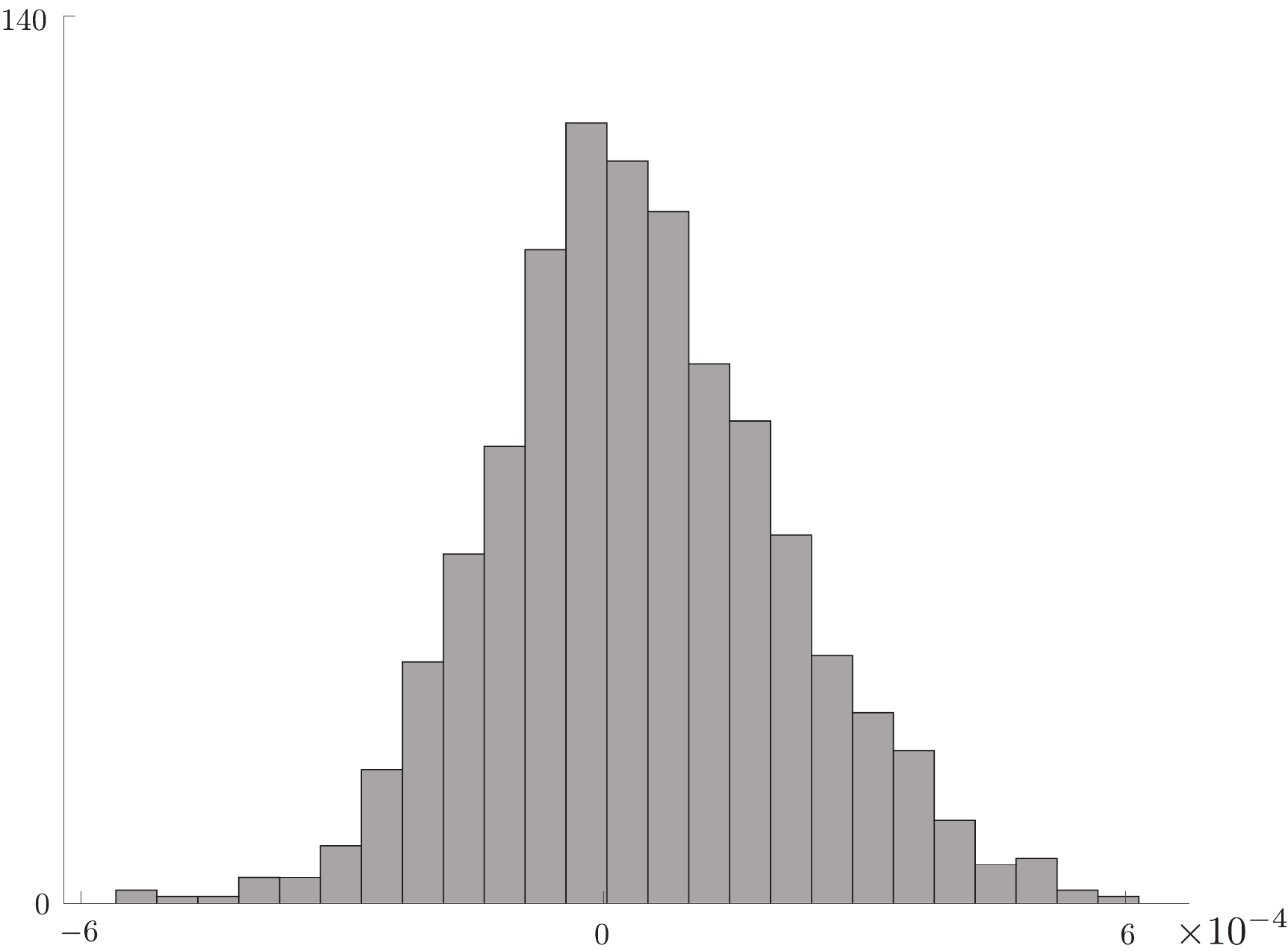}
  \caption{\footnotesize{$\epsilon$\, for $RSD_{\lambda}\in(2\,\%\,,\,20\,\%)$}}
  \label{fig02e_san}
\end{subfigure}%
\begin{subfigure}{.33\textwidth}
  \centering
   \includegraphics[scale=0.3]{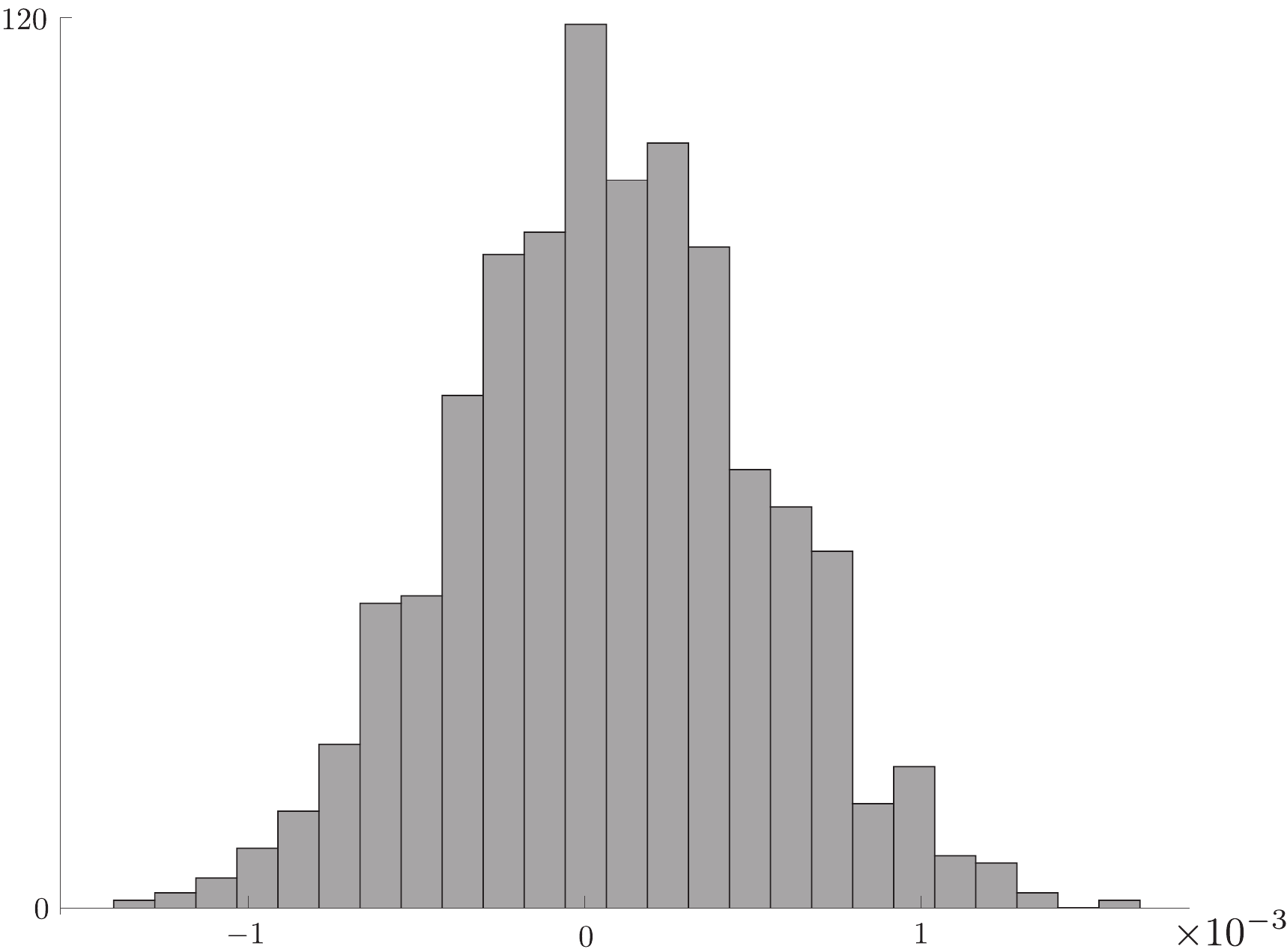}
  \caption{\footnotesize{$\epsilon$\, for $RSD_{\lambda}>20\,\%$}}
  \label{fig02f_san}
\end{subfigure}

\qquad

\centering
\begin{subfigure}{.33\textwidth}
  \centering
   \includegraphics[scale=0.3]{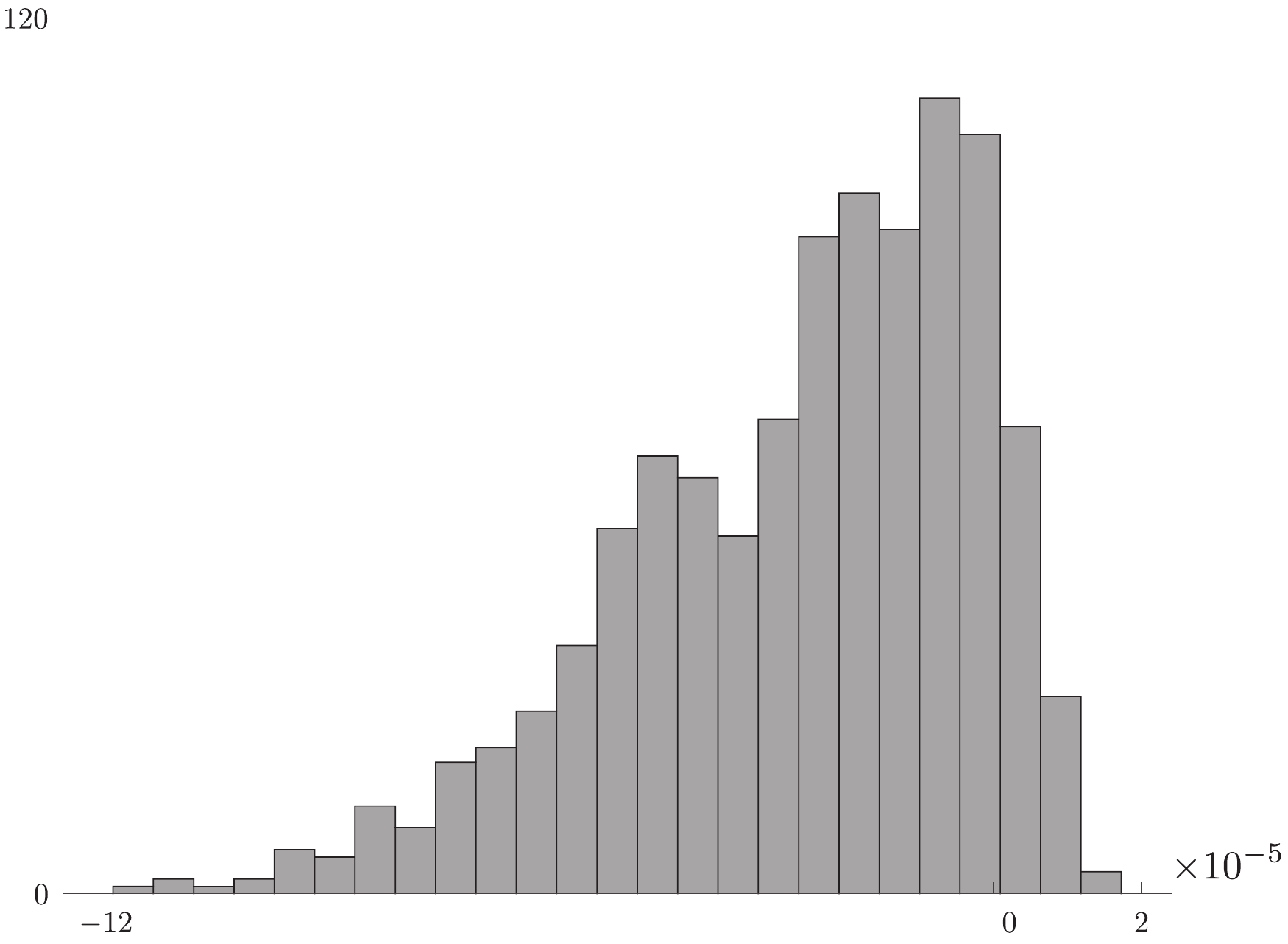}
  \caption{\footnotesize{$\delta$\, for $RSD_{\lambda}<2\,\%$}}
  \label{fig02g_san}
\end{subfigure}%
\begin{subfigure}{.33\textwidth}
  \centering
   \includegraphics[scale=0.3]{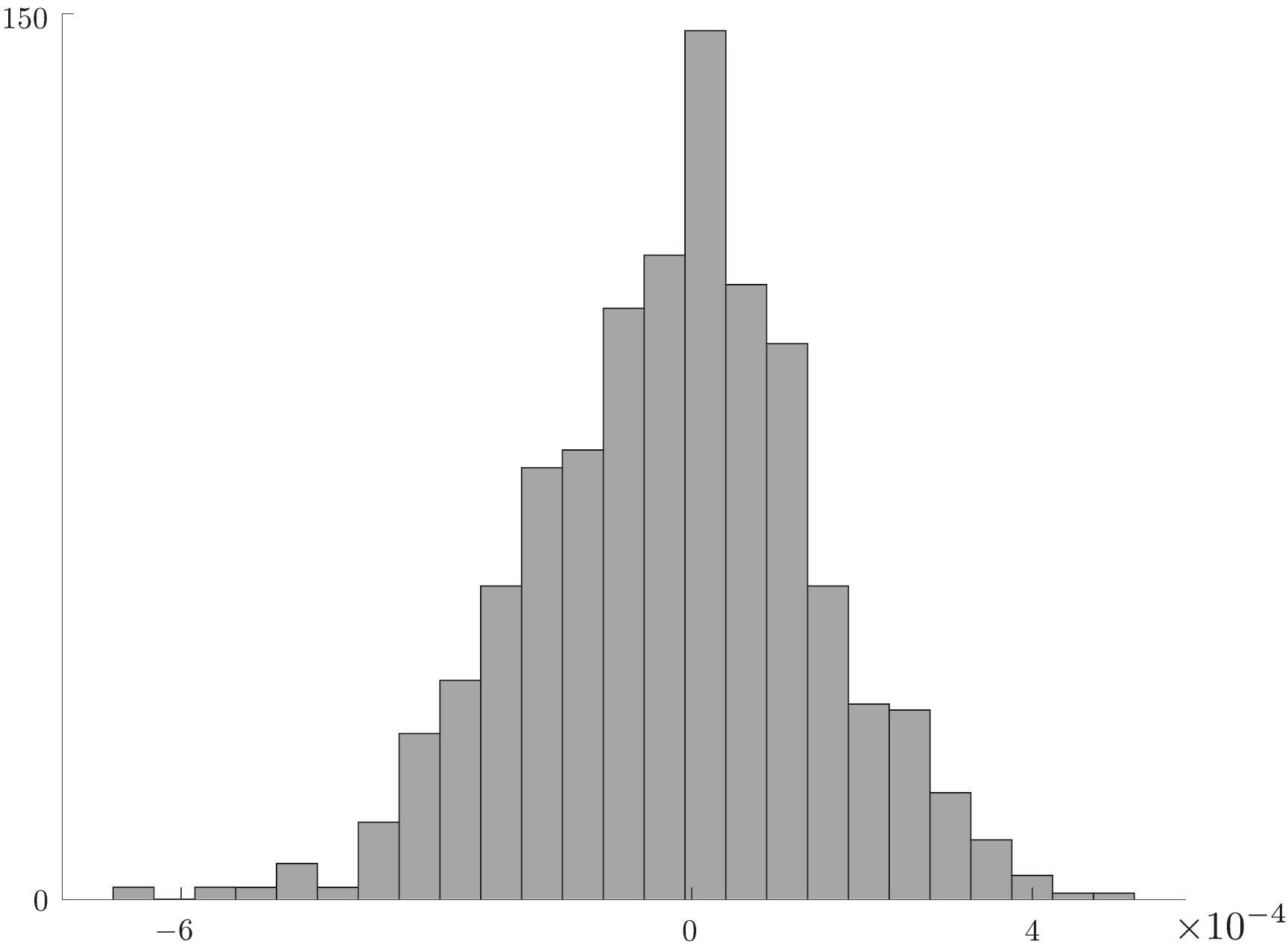}
  \caption{\footnotesize{$\delta$\, for $RSD_{\lambda}\in(2\,\%\,,\,20\,\%)$}}
  \label{fig02h_san}
\end{subfigure}%
\begin{subfigure}{.33\textwidth}
  \centering
   \includegraphics[scale=0.3]{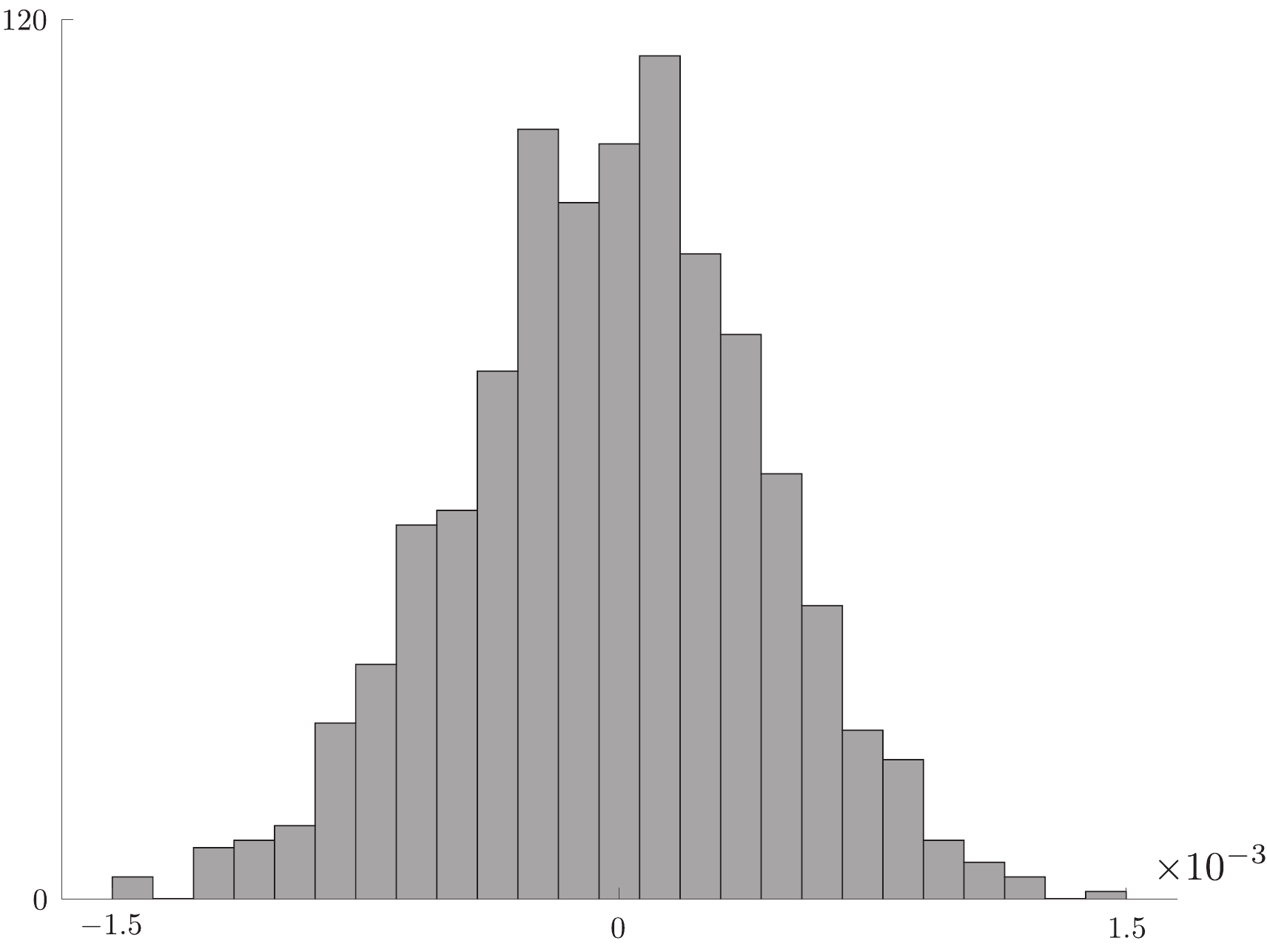}
  \caption{\footnotesize{$\delta$\, for $RSD_{\lambda}>20\,\%$}}
  \label{fig02i_san}
\end{subfigure}
\caption{\small{Distribution of $\varphi$, $\epsilon$ and $\delta$ for $1000$ examples of equivalent TI sandstones with $RSD_{\mu}<2\,\%\,$.}}
\label{fig:dis_ms_san}
\end{figure}

To conclude, different ranges of Lam\'e parameters cause different distributions of their relative variations and different ranges of anisotropy parameters. 
Successively, it entails the growth or loss of certain fluid indicator's efficiency.
Nevertheless, these changes are relatively small, thus, our discussion concerned fluid indicators or distributions for mafic rocks, still remains valid and actual for felsic rocks or sandstones.
The only one significant change is that in the case of $RSD_{\mu}>2\,\%$, instead of $|\varphi|>10^{-3}$, for felsic rocks we should use $|\varphi|>0.5\cdot10^{-3}$, while for sandstones, $|\varphi|>5\cdot10^{-3}$.
Figures~\ref{fig:rsdfel}--\ref{fig:dis_ms_fel} regard felsic rocks; they present relations among $\varphi$, $\epsilon$ and $\delta$, distributions of these parameters, as well as distributions of $\gamma$ or variations of $\mu$ and $\lambda$. 
Sandstones are considered in analogical way in Figures~\ref{fig:rsdsan}--\ref{fig:dis_ms_san}.
\newpage
%%%%%%%%%%%%%%%%%%%%%%%%%%%%%%%%%%%%%%%%%%%%%%%%%%%%%%%%%%%%%%%%%%%%%%%%%%%%%%%%%%%%%%%%
\clearpage
\section{Properties of $\varphi$, $\epsilon$ and $\delta$, relevant to various layered rocks} \label{ap:two}
%%%%%%%%%%%%%%%%%%%%%%%%%%%%%%%%%%%%%%%%%%%%%%%%%%%%%%%%%%%%%%%%%%%%%%%%%%%%%%%%%%%%%%%%
%\begin{sidewaystable}[!htb]
%\begin{sidewaystable}[ph!]
\begin{table}[!htbp]
\centering
\rotatebox{90}{%
%\begin{varwidth}{\textheight}
\begin{minipage}{0.92\textheight}
%\captionsetup{width=10cm}
\resizebox{1.05\textwidth}{!}{%
%\caption{\small{Dominants and ranges of $\varphi$, $\epsilon$ and $\delta$, relevant to various layered rocks. Results for $RSD_{\mu}<2\,\%$ and $RSD_{\lambda}<2\,\%$, relevant to each type of rocks, and results for $RSD_{\mu}<2\,\%$ and $RSD_{\lambda}\in(2\,\%\,,\,20\,\%)$, relevant to sandstones; are based on $1000$ random examples of TI media equivalent to three isotropic layers.
%Results for the first column on the right, are based on $10000$ random examples of TI media equivalent to five isotropic layers.
%For the rest of rocks and ranges of variations of $\mu$ and $\lambda$, results are based on $1000$ random examples of TI media equivalent to five isotropic layers.
%}}
%\scalebox{0.5}{%
\begin{tabular}{|c|c|c|c|c|c|c|c|c|c|c|c|c|c|c|c|}
  \cline{3-16}
 \multicolumn{2}{c|}{} &  \multicolumn{6}{|c|}{$RSD_{\mu}<2\,\%$}  &  \multicolumn{6}{|c|}{$RSD_{\mu}>2\,\%$} &  \multicolumn{2}{|c|}{$RSD_{\mu}\in\forall$}  \\ \cline{3-16}
 
 \multicolumn{2}{c|}{} & \multicolumn{2}{|c|}{$RSD_{\lambda}<2\,\%$}& \multicolumn{2}{|c|}{$RSD_{\lambda}\in\left(2\,\%\,,\,20\,\%\right)$} & \multicolumn{2}{|c|}{$RSD_{\lambda}>20\,\%$} & \multicolumn{2}{|c|}{$RSD_{\lambda}<2\,\%$} & \multicolumn{2}{|c|}{$RSD_{\lambda}\in\left(2\,\%\,,\,20\,\%\right)$} & \multicolumn{2}{|c|}{$RSD_{\lambda}>20\,\%$} & \multicolumn{2}{|c|}{$RSD_{\lambda}\in\forall$} \\ \cline{3-16}
 
  \multicolumn{2}{c|}{} & {\small{dominant}} & {\small{range}}& {\small{dominant}} &  {\small{range}} & {\small{dominant}} &  {\small{range}} & {\small{dominant}} &  {\small{range}} & {\small{dominant}} &  {\small{range}} & {\small{dominant}} &  {\small{range}} & {\small{dominant}} &  {\small{range}} \\
  
 \multicolumn{2}{c|}{} & {\scriptsize{$\left[\cdot10^{-5}\right]$}} &  {\scriptsize{$\left[\cdot10^{-3}\right]$}} & {\scriptsize{$\left[\cdot10^{-5}\right]$}} & {\scriptsize{$\left[\cdot10^{-3}\right]$}} & {\scriptsize{$\left[\cdot10^{-5}\right]$}} & {\scriptsize{$\left[\cdot10^{-3}\right]$}} & {\scriptsize{$\left[\cdot10^{-3}\right]$}} & {\scriptsize{$\left[\cdot10^{-3}\right]$}} & {\scriptsize{$\left[\cdot10^{-3}\right]$}} & {\scriptsize{$\left[\cdot10^{-3}\right]$}} & {\scriptsize{$\left[\cdot10^{-3}\right]$}} & {\scriptsize{$\left[\cdot10^{-3}\right]$}} & {\scriptsize{$\left[\cdot10^{-3}\right]$}} & {\scriptsize{$\left[\cdot10^{-3}\right]$}} \\ \hline
 
\parbox[t]{2mm}{\multirow{3}{*}{\rotatebox[origin=c]{90}{mafic}}}&$\varphi$&$0.227$& $\left(-0.081\,,\,0.075\right)$&$-13.17$&$\left(-0.842\,,\,0.926\right)$&$-8.566$&$\left(-1.091\,,\,1.169\right)$&$-0.105$&$\left(-0.867\,,\,1.149\right)$&$-0.002$&$\left(-8.096\,,\,9.605\right)$&$1.070$&$\left(-12.67\,,\,11.61\right)$&$0.296$&$\left(-11.68\,,\,12.07\right)$ \\ 
&$\epsilon$&$0.916$&$\left(-0.024\,,\,0.105\right)$&$-4.327$&$\left(-0.569\,,\,0.721\right)$&$-3.503$&$\left(-0.738\,,\,0.841\right)$&$4.104$&$\left(0.118\,,\,13.62\right)$&$1.978$&$\left(-1.559\,,\,16.26\right)$&$1.639$&$\left(-3.328\,,\, 17.64\right)$&$1.904$&$\left(-2.767\,,\,17.87\right)$ \\ 
&$\delta$&$-0.892$&$\left(-0.125\,,\,0.016\right)$&$-10.17$&$\left(-0.691\,,\,0.588\right)$&$-21.79$&$\left(-0.874\,,\,0.702\right)$&$-2.966$&$\left(-13.33\,,\,-0.070\right)$&$-3.514$&$\left(-15.45\,,\,1.603\right)$&$-1.520$&$<\left(-19.37\,,\,2.377\right)$&$-1.580$&$\left(-18.83\,,\,2.275\right)$ \\ \hline

\parbox[t]{2mm}{\multirow{3}{*}{\rotatebox[origin=c]{90}{felsic}}}&$\varphi$&$-0.160$&$\left(-0.082\,,\,0.080\right)$&$7.456$&$\left(-0.945\,,\,0.963\right)$&$-17.95$&$\left(-1.949\,,\,1.937\right)$&$-0.036$&$\left(-0.511\,,\,0.519\right)$&$0.495$&$\left(-5.903\,,\,5.811\right)$& $-2.089$&$\left(-11.79\,,\,9.432\right)$&$-0.423$&$\left(-11.42\,,\,12.61\right)$ \\ 
&$\epsilon$&$0.885$&$\left(-0.018\,,\,0.111\right)$&$6.013$&$\left(-0.506\,,\,0.761\right)$&$-3.963$&$\left(-0.997\,,\,1.385\right)$&$1.132$&$\left(0.009\,,\,3.313\right)$&$0.231$&$\left(-1.866\,,\,6.737\right)$&$1.067$&$\left(-5.285\,,\,8.514\right)$&$0.956$&$\left(-5.034\,,\,11.29\right)$ \\ 
&$\delta$&$0.130$&$\left(-0.116\,,\,0.015\right)$&$6.237$&$\left(-0.643\,,\,0.667\right)$&$-15.15$&$\left(-1.128\,,\, 1.246\right)$&$-1.394$&$\left(-3.924\,,\,-0.060\right)$&$-1.179$&$\left(-6.087\,,\,1.854\right)$&$-1.188$&$\left(-9.556\,,\,4.168\right)$&$-0.720$&$\left(-10.29\,,\,5.222\right)$ \\ \hline

\parbox[t]{2mm}{\multirow{3}{*}{\rotatebox[origin=c]{90}{sands}}}&$\varphi$&$0.259$&$\left(-0.105\,,\,0.099\right)$&$0.313$&$\left(-1.001\,,\,1.034\right)$&$43.49$&$\left(-3.937\,,\,4.956\right)$&$-0.191$&$\left(-4.511\,,\, 5.121\right)$&$0.725$&$\left(-48.25\,,\,62.68\right)$&$-8.924$&$\left(-140.9\,,\,156.5\right)$&$-0.032$&$\left(-204.2\,,\, 209.2\right)$ \\ 
&$\epsilon$&$1.784$&$\left(-0.024\,,\,0.116\right)$&$-2.004$&$\left(-0.555\,,\,0.608\right)$&$0.629$&$\left(-1.340\,,\,1.638\right)$&$74.41$&$\left(0.528\,,\,337.5\right)$&$31.70$&$\left(1.026\,,\,322.1\right)$&$44.71$&$\left(-6.271\,,\,621.0\right)$&$49.21$&$\left(-27.26\,,\,749.2\right)$ \\ 
&$\delta$&$-0.737$&$\left(-0.118\,,\,0.016\right)$&$1.587$&$\left(-0.667\,,\,0.506\right)$&$12.25$&$\left(-1.470\,,\,1.493\right)$&$-31.08$&$\left(-309.2\,,\,-0.287\right)$&$-30.70$&$\left(-306.4\,,\,-0.734\right)$&$-14.68$&$\left(-322.0\,,\,18.80\right)$&$-14.54$&$\left(-345.3\,,\,21.31\right)$ \\ \hline
\end{tabular}\unmarkedfntext{
\hspace*{-0.7cm}{\bf Table 18:}
{Dominants and ranges of $\varphi$, $\epsilon$ and $\delta$, relevant to various layered rocks. Results for $RSD_{\mu}<2\,\%$ and $RSD_{\lambda}<2\,\%$, relevant to each type of rocks, and results for $RSD_{\mu}<2\,\%$ and $RSD_{\lambda}\in(2\,\%\,,\,20\,\%)$, relevant to sandstones; are based on $1000$ random examples of TI media equivalent to three isotropic layers.
Results for the first column on the right, are based on $10000$ random examples of TI media equivalent to five isotropic layers.
For the rest of rocks and ranges of variations of $\mu$ and $\lambda$, results are based on $1000$ random examples of TI media equivalent to five isotropic layers.
}
}
}
\label{tab:bigtable}
%\end{varwidth}
\end{minipage}
}
\end{table}
\addtocounter{table}{1}
%\end{sidewaystable}
%%%%%%%%%%%%%%%%%%%%%%%%%%%%%%%%%%%%%%%%%%%%%%%%%%%%%%%%%%%%%%%%%%%%%%%%%%%%%%%%%%%%%%%%
\clearpage
\section{Relations among $\varphi$, $\epsilon$ and $\delta$, relevant to various layered rocks} \label{ap:three}
%%%%%%%%%%%%%%%%%%%%%%%%%%%%%%%%%%%%%%%%%%%%%%%%%%%%%%%%%%%%%%%%%%%%%%%%%%%%%%%%%%%%%%%%
\begin{table}[!htb]
%\resizebox{\textwidth}{!}{%
\scalebox{0.7}{%
\begin{tabular}{|c|c|c|c|c|c|c|c|c|}
  \cline{3-9}
 \multicolumn{2}{c|}{} &  \multicolumn{3}{|c|}{$RSD_{\mu}<2\,\%$}  &  \multicolumn{3}{|c|}{$RSD_{\mu}>2\,\%$} &  $RSD_{\mu}\in\forall$  \\ \cline{3-9}
 
 \multicolumn{2}{c|}{} & {\small{$RSD_{\lambda}<2\,\%$}}& {\small{$RSD_{\lambda}\in\left(2\,\%\,,\,20\,\%\right)$ }}& {\small{$RSD_{\lambda}>20\,\%$}} &{\small{$RSD_{\lambda}<2\,\%$}}& {\small{$RSD_{\lambda}\in\left(2\,\%\,,\,20\,\%\right)$}} & {\small{$RSD_{\lambda}>20\,\%$}} &{\small{$RSD_{\lambda}\in\forall$}} \\ \cline{1-9}
  
\parbox[t]{2mm}{\multirow{12}{*}{\rotatebox[origin=c]{90}{mafic rocks}}}
&$\varphi>\epsilon$                                                                        &$2.00$&$32.4$&$37.3$                                                  &$0.00$&$0.30$&$0.90$        &$0.32$\\ 
&$\varphi<\delta$                                                                            &$2.30$&$32.1$&$40.6$                                                  &$0.00$&$0.10$&$0.70$        &$0.38$\\ 
&$|\varphi|>|\epsilon|$                                                                     &$34.3$&$80.5$&$87.2$                                                  &$0.00$&$26.7$&$32.8$        &$25.7$ \\ 
&$|\varphi|>|\delta|$                                                                         &$33.4$&$77.2$&$86.1$                                                 &$0.00$&$24.1$&$32.4$         &$24.2$\\
&$|\epsilon|\wedge|\delta|>10^{-4}$                                                &$0.00$&$51.8$&$68.7$                                                  &$99.9$&$97.1$&$95.4$         &$97.1$\\
&$\epsilon<0$                                                                                 &$15.5$&$44.8$&$47.1$                                                  &$0.00$&$7.30$&$13.8$         &$6.87$\\
&$\delta>0$                                                                                     &$11.9$&$41.3$&$43.1$                                                  &$0.00$&$4.30$&$11.4$         &$5.21$\\
&$|\varphi|>10^{-4}$                                                                        &$\mathbf{0.00}$&$\mathbf{76.3}$&$\mathbf{85.0}$       &$76.4$&$97.4$&$99.1$         &$97.4$\\
&$|\varphi|>0.5\cdot10^{-3}$                                                            &$0.00$&$10.7$&$32.5$                                                 &$12.0$&$85.4$&$92.5$        &$86.0$\\
&$|\varphi|>10^{-3}$                                                                       &$0.00$&$0.00$&$1.30$            &$\mathbf{0.20}$&$\mathbf{74.2}$&$\mathbf{82.1}$      &$72.5$\\
&$|\varphi|>5\cdot10^{-3}$                                                              &$0.00$&$0.00$&$0.00$                                                   &$0.00$&$9.40$&$26.1$        &$10.5$\\ 
&$|\delta|>|\epsilon|$                                                                       &$53.8$&$51.3$&$52.2$                                                   &$86.9$&$54.2$&$48.5$         &$53.9$\\ \hline

\parbox[t]{2mm}{\multirow{12}{*}{\rotatebox[origin=c]{90}{felsic rocks}}}
&$\varphi>\epsilon$                                                                        &$3.60$&$35.8$&$43.4$                                                    &$0.00$&$3.80$&$15.6$          &$12.9$\\ 
&$\varphi<\delta$                                                                            &$2.70$&$32.6$&$40.9$                                                    &$0.00$&$3.20$&$18.0$          &$13.9$\\ 
&$|\varphi|>|\epsilon|$                                                                    &$37.7$&$81.0$&$90.4$                                                     &$0.80$&$39.3$&$58.3$          &$53.4$ \\ 
&$|\varphi|>|\delta|$                                                                        &$36.1$&$81.0$&$91.0$                                                     &$0.30$&$38.1$&$60.3$          &$54.4$\\
&$|\epsilon|\wedge|\delta|>10^{-4}$                                               &$0.00$&$53.6$&$73.0$                                                     &$98.6$&$87.9$&$93.5$          &$91.9$\\
&$\epsilon<0$                                                                                 &$13.0$&$42.0$&$44.6$                                                    &$0.00$&$20.2$&$31.5$          &$27.4$\\ 
&$\delta>0$                                                                                     &$13.1$&$43.8$&$47.9$                                                   &$0.00$&$19.8$&$31.9$          &$28.2$\\
&$|\varphi|>10^{-4}$                                                                        &$\mathbf{0.00}$&$\mathbf{77.2}$&$\mathbf{87.5}$        &$55.4$&$77.2$&$98.0$          &$97.0$\\
&$|\varphi|>0.5\cdot10^{-3}$                                                           &$0.00$&$15.2$&$43.7$              &$\mathbf{0.20}$&$\mathbf{76.2}$&$\mathbf{89.3}$    &$84.8$\\
&$|\varphi|>10^{-3}$                                                                        &$0.00$&$0.00$&$10.3$                                                  &$0.00$&$55.5$&$77.3$           &$70.0$\\
&$|\varphi|>5\cdot10^{-3}$                                                              &$0.00$&$0.00$&$0.00$                                                  &$0.00$&$0.50$&$11.1$           &$7.88$\\ 
&$|\delta|>|\epsilon|$                                                                      &$51.4$&$48.9$&$47.8$                                                   &$70.7$&$49.1$&$49.1$           &$50.1$\\ \hline

\parbox[t]{2mm}{\multirow{12}{*}{\rotatebox[origin=c]{90}{sandstones}}}
&$\varphi>\epsilon$                                                                        &$5.50$&$43.7$&$46.6$                                                      &$0.00$&$0.00$&$1.10$              &$0.99$\\ 
&$\varphi<\delta$                                                                            &$14.1$&$42.6$&$47.7$                                                       &$0.00$&$0.10$&$3.80$             &$3.13$\\ 
&$|\varphi|>|\epsilon|$                                                                     &$38.0$&$88.9$&$94.3$                                                      &$0.00$&$1.20$&$14.9$             &$15.2$ \\ 
&$|\varphi|>|\delta|$                                                                         &$46.2$&$93.6$&$97.6$                                                      &$0.00$&$1.40$&$25.5$             &$21.3$\\
&$|\epsilon|\wedge|\delta|>10^{-4}$                                                &$0.00$&$41.5$&$76.1$                                                      &$100$&$100$&$99.9$               &$99.9$\\
&$\epsilon<0$                                                                                  &$8.00$&$42.4$&$43.3$                                                     &$0.00$&$0.00$&$0.70$             &$0.78$\\
&$\delta>0$                                                                                      &$11.0$&$47.2$&$47.7$                                                     &$0.00$&$0.00$&$4.30$             &$3.48$\\
&$|\varphi|>10^{-4}$                                                                         &$\mathbf{0.10}$&$\mathbf{75.8}$&$\mathbf{92.3}$          &$93.4$&$99.3$&$99.9$            &$99.8$\\
&$|\varphi|>0.5\cdot10^{-3}$                                                            &$0.00$&$17.2$&$68.8$                                                    &$69.6$&$96.4$&$99.1$             &$98.9$\\
&$|\varphi|>10^{-3}$                                                                         &$0.00$&$0.20$&$41.9$                                                    &$44.9$&$92.8$&$97.9$             &$97.9$\\
&$|\varphi|>5\cdot10^{-3}$                                                               &$0.00$&$0.00$&$0.00$               &$\mathbf{0.10}$&$\mathbf{68.2}$&$\mathbf{91.7}$        &$88.9$\\ 
&$|\delta|>|\epsilon|$                                                                       &$41.5$&$47.0$&$47.7$                                                      &$74.6$&$58.9$&$44.1$             &$45.6$\\ \hline

\end{tabular}
}
\caption{\small{Percentage of occurrence of relations among $\varphi$, $\epsilon$ and $\delta$, relevant to various layered rocks. The most efficient fluid indicators that show the change of variations of $\lambda$, are highlighted in bold. Results for $RSD_{\mu}<2\,\%$ and $RSD_{\lambda}<2\,\%$, relevant to each type of rocks, and results for $RSD_{\mu}<2\,\%$ and $RSD_{\lambda}\in(2\,\%\,,\,20\,\%)$, relevant to sandstones; are based on $1000$ random examples of TI media equivalent to three isotropic layers.
Results for the first column on the right, are based on $10000$ random examples of TI media equivalent to five isotropic layers.
For the rest of rocks and ranges of variations of $\mu$ and $\lambda$, results are based on $1000$ random examples of TI media equivalent to five isotropic layers.}}
\label{tab:bigtable2}
\end{table}
%%%%%%%%%%%%%%%%%%%%%%%%%%%%%%%%%%%%%%%%%%%%%%%%%%%%%%%%%%%%%%%%%%%%%%%%%%%%%%%%%%%%%%%%
\end{document}